\documentclass[runningheads]{llncs}

\usepackage{eccv}

\usepackage{eccvabbrv}
\usepackage{multirow}
\usepackage{graphicx}
\usepackage{booktabs}
\usepackage{pgfplots}
\usepackage{adjustbox}
\usepackage[utf8]{inputenc}
\usepackage{tikz}
\usepackage{bm}
\usepackage{xr}
\usepackage{soul}
\usetikzlibrary{shapes.geometric, arrows.meta, calc,backgrounds}
\usetikzlibrary{positioning,shapes,arrows,arrows.meta,fit,backgrounds,calc}

\usepackage[accsupp]{axessibility}  % Improves PDF readability for those with disabilities.

\usepackage{hyperref}

\usepackage{orcidlink}

\begin{document}

\title{Distributed Semantic Segmentation with\\ Efficient Joint Source and Task Decoding} 

\titlerunning{Distributed Semantic Segmentation with Efficient Joint Decoding}

\author{Danish Nazir\inst{1,2}\orcidlink{0000-0001-6364-8427} \and
Timo Bartels\inst{1}\orcidlink{0009-0002-4917-8849} \and
Jan Piewek\inst{2}\orcidlink{0009-0000-4047-2115} \and
Thorsten Bagdonat\inst{2}\orcidlink{0000-0002-4008-6721} \and
Tim Fingscheidt\inst{1}\orcidlink{0000-0002-8895-5041} 
}

\authorrunning{D.~Nazir et al.}

\institute{Technische Universität Braunschweig, Braunschweig, Germany \\
\email{\{danish.nazir,timo.bartels,t.fingscheidt\}@tu-bs.de}
\and
Group Innovation, Volkswagen AG, Wolfsburg, Germany
\email{\{danish.nazir,jan.piewek,thorsten.bagdonat\}@volkswagen.de} }

\maketitle

%should be between 70 and 150 words
\begin{abstract}
Distributed computing in the context of deep neural networks (DNNs) implies the execution of one part of the network on edge devices and the other part typically on a large-scale cloud platform. Conventional methods propose to employ a serial concatenation of a learned image and source encoder, the latter projecting the image encoder output (bottleneck features) into a quantized representation for bitrate-efficient transmission. In the cloud, a respective source decoder reprojects the quantized representation to the original feature representation, serving as an input for the downstream task decoder performing, e.g., semantic segmentation. In this work, we propose \textit{joint} source and task decoding, as it allows for a \textit{smaller network size in the cloud}.~This further enables the scalability of such services in large numbers without requiring extensive computational load on the cloud per channel.~We demonstrate the effectiveness of our method by achieving a distributed semantic segmentation SOTA over a wide range of bitrates on the mean intersection over union metric, while using only $9.8 \%$ ... $11.59 \%$ of cloud DNN parameters used in previous SOTA on the COCO and Cityscapes datasets.
\end{abstract}

\section{Introduction}

Deep neural networks (DNNs) have been very successful in performing machine perception tasks, including semantic segmentation \cite{long2015fully,FCN2,contextualinfo1,contextualinfo2}.~Due to a surge in camera-enabled edge devices with limited computational capabilities across various application fields, such as smart transportation, agriculture, and manu\-facturing, improving the task efficiency of powerful segmentation DNNs is very important\cite{tang2021dffnet,anand2021agrisegnet,chakravarthy2022dronesegnet}.~However, since there is a trend towards an increasing size and complexity of advanced DNN architectures, it is difficult to execute them on edge devices due to both power and computational complexity limitations \cite{eshratifar2019jointdnn}.~Ac\-cord\-ing\-ly, current approaches \cite{ahuja2023neural,matsubara2020quantize,matsubara2022supervised,chakravarthy2022dronesegnet} propose to distribute the execution of segmentation DNNs into two parts, where one part of the DNN is executed on the edge device and the other on a large-scale cloud platform.

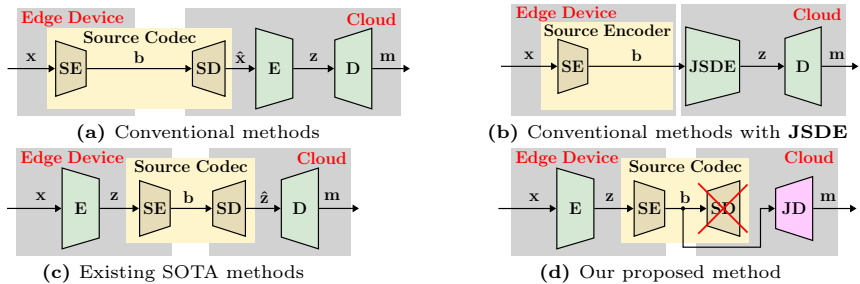
\begin{figure}[t!]
  \hspace{0.055em}
  \begin{subfigure}{0.54\textwidth}
     \centering
  \resizebox{\linewidth}{!}{ % Define some colors
 \definecolor{encoder_decoder}{RGB}{213, 232, 212}
 \definecolor{rectangle}{RGB}{255, 245, 204}
 \definecolor{bottleneck_color}{RGB}{230, 221, 184}
\definecolor{bg}{RGB}{210, 210, 210}
 % Define styles
%  \tikzstyle{data} = [draw, cylinder, line width=1pt, shape border rotate=90, outer
% sep=0, inner sep=5, minimum size=30, align=center, fill=white]

 \tikzstyle{arrow} = [-Triangle, line width=1pt]
 \tikzstyle{label} = [text width= 0.4cm, align=center]
 \tikzstyle{encoder} = [trapezium,
    trapezium angle=55,
    trapezium stretches=true,
    minimum width=2.4cm,
    minimum height=1.1cm,
    trapezium right angle=80,
    trapezium left angle=80,
    line width=1pt,
    shape border rotate=270,
    text centered,
    draw=black]
 \tikzstyle{bottleneck_encoder} = [trapezium,
    trapezium angle=55,
    trapezium stretches=true,
    %minimum width=2.4cm,
    %minimum height=1.1cm,
    %trapezium right angle=83.5,
    %trapezium left angle=83.5,
     minimum width=1.65cm,
    minimum height=0.75cm,
    trapezium right angle=80,
    trapezium left angle=80,
    line width=1pt,
    shape border rotate=270,
    text centered,
    draw=black]
 \tikzstyle{bottleneck_decoder} = [trapezium,
    trapezium angle=55,
    trapezium stretches=true,
    minimum width=1.65cm,
    minimum height=0.75cm,
    trapezium right angle=80,
    trapezium left angle=80,
    line width=1pt,
    shape border rotate=90,
    text centered,
    draw=black]

  \tikzstyle{decoder} = [trapezium,
    trapezium angle=55,
    trapezium stretches=true,
    minimum width=2.4cm,
    minimum height=1.1cm,
    trapezium right angle=80,
    trapezium left angle=80,
    line width=1pt,
    shape border rotate=90,
    text centered,
    draw=black]

\begin{tikzpicture}[font=\Large,node distance = 1.5cm]
\node (point) {};
\node (fake) [label, left of=point ,xshift=-0.7cm] {};

\node[draw, rectangle, minimum width=3.5cm, minimum height=3.2cm,  line width=1pt,draw=none, fill=bg,yshift=0.2cm] (background) at (0.15,0) {};
\node[font=\Large,  label, text width=4cm, above of=background, align=center ,yshift=-0.22cm, xshift=-0.1cm,text=red ] {$\mathbf{Edge \ Device}$};

\node[draw, rectangle, minimum width=6.4cm, minimum height=3.2cm,  line width=1pt,draw=none, fill=bg,yshift=0.2cm] (background2) at (6.6,0) {};
\node[font=\Large,  label, text width=4cm, above of=background2, align=center ,yshift=-0.22cm, xshift=2.34cm,text=red ] {$\mathbf{Cloud}$};

\node[draw, rectangle, minimum width=5.5cm, minimum height=2.4cm, yshift=0cm, line width=1pt,draw=none, fill=rectangle] (myRectangle) at (2.05,0) {};

\node (encoder_1) [bottleneck_encoder , right of=point, xshift=-1.5cm, fill=bottleneck_color] { $\mathbf{SE}$} ;
\draw [arrow] (fake) to node[midway, xshift=0cm,yshift=0.35cm] {$\mathbf{x}$} (encoder_1);
%\node[font=\Large,  label, text width=4cm, above of=myRectangle, align=center ,xshift=0.45cm,yshift=-0.7cm ] {$\mathbf{Source}$ \\ $\mathbf{Codec}$};
%\node[font=\Large,  label, text width=4cm, above of=myRectangle, align=center ,xshift=0cm,yshift=-0.451cm ] {$\mathbf{Source \ Codec}$};
\node[font=\Large,  label, text width=4cm, above of=myRectangle, align=center ,yshift=-0.55cm ] {$\mathbf{Source \ Codec}$};

\node (be_decoder) [bottleneck_decoder , right of=encoder_1, xshift=2.6cm, fill=bottleneck_color] { $\mathbf{SD}$} ;
\draw [arrow] (encoder_1) to node[midway, xshift=0cm,yshift=0.35cm] {$\mathbf{b}$} (be_decoder);
\node (encoder_2) [encoder , right of=be_decoder, xshift=0.45cm, fill=encoder_decoder] { $\mathbf{E}$} ;
\draw [arrow] (be_decoder) to node[midway, xshift=0cm,yshift=0.35cm] {$\mathbf{\hat{x}}$} (encoder_2);
% \draw [arrow] (encoder_1) to node[midway, xshift=0cm,yshift=0.35cm] {$\mathbf{\hat{r}}$} (encoder_2);
\node (decoder_1) [decoder , right of=encoder_2, xshift=0.85cm, fill=encoder_decoder] { $\mathbf{D}$} ;
\draw [arrow] (encoder_2) to node[midway, xshift=0cm,yshift=0.35cm] {$\mathbf{z}$} (decoder_1);
 
\node (m) [label, right of=decoder_1,xshift=0.5cm ] {};
 \draw [arrow] (decoder_1) to node[midway, xshift=-0.1cm,yshift=0.35cm] {$\mathbf{m}$} (m);

\end{tikzpicture}}
  % \scalebox{0.45}{
  %   {\input{Figures/conventional_methods}}
  %   }
    \caption{Conventional methods    }
    \label{fig:conventional}
  \end{subfigure}
 \hspace{-0.489em}
\begin{subfigure}{0.485\textwidth}
   \centering
  \resizebox{\linewidth}{!}{ % Define some colors
 \definecolor{encoder_decoder}{RGB}{213, 232, 212}
 \definecolor{rectangle}{RGB}{255, 245, 204}
 \definecolor{bottleneck_color}{RGB}{230, 221, 184}
\definecolor{bg}{RGB}{210, 210, 210}
 % Define styles
%  \tikzstyle{data} = [draw, cylinder, line width=1pt, shape border rotate=90, outer
% sep=0, inner sep=5, minimum size=30, align=center, fill=white]

 \tikzstyle{arrow} = [-Triangle, line width=1pt]
 \tikzstyle{label} = [text width= 0.4cm, align=center]
 \tikzstyle{encoder} = [trapezium,
    trapezium angle=55,
    trapezium stretches=true,
    minimum width=2.4cm,
    minimum height=1.1cm,
    trapezium right angle=80,
    trapezium left angle=80,
    line width=1pt,
    shape border rotate=270,
    text centered,
    draw=black]
 \tikzstyle{bottleneck_encoder} = [trapezium,
    trapezium angle=55,
    trapezium stretches=true,
    %minimum width=2.4cm,
    %minimum height=1.1cm,
    %trapezium right angle=83.5,
    %trapezium left angle=83.5,
     minimum width=1.65cm,
    minimum height=0.75cm,
    trapezium right angle=80,
    trapezium left angle=80,
    line width=1pt,
    shape border rotate=270,
    text centered,
    draw=black]
 \tikzstyle{bottleneck_decoder} = [trapezium,
    trapezium angle=55,
    trapezium stretches=true,
    minimum width=1.65cm,
    minimum height=0.75cm,
    trapezium right angle=80,
    trapezium left angle=80,
    line width=1pt,
    shape border rotate=90,
    text centered,
    draw=black]

  \tikzstyle{decoder} = [trapezium,
    trapezium angle=55,
    trapezium stretches=true,
    minimum width=2.4cm,
    minimum height=1.1cm,
    trapezium right angle=80,
    trapezium left angle=80,
    line width=1pt,
    shape border rotate=90,
    text centered,
    draw=black]

\begin{tikzpicture}[font=\Large,node distance = 1.5cm]
\node (point) {};
\node (fake) [label, left of=point ,xshift=-1cm] {};

\node[draw, rectangle, minimum width=5cm, minimum height=3.4cm,  line width=1pt,draw=none, fill=bg,yshift=0.2cm] (background) at (0.62,0) {};
\node[font=\Large,  label, text width=4cm, above of=background, align=center ,yshift=-0.1cm, xshift=-0.82cm,text=red ] {$\mathbf{Edge \ Device}$};

\node[draw, rectangle, minimum width=5cm, minimum height=3.4cm,  line width=1pt,draw=none, fill=bg,yshift=0.2cm] (background2) at (5.8,0) {};
\node[font=\Large,  label, text width=4cm, above of=background2, align=center ,yshift=-0.1cm, xshift=1.63cm,text=red ] {$\mathbf{Cloud}$};

\node[draw, rectangle, minimum width=4cm, minimum height=2.67cm,  line width=1pt,draw=none, fill=rectangle] (myRectangle) at (1.05,0) {};

\node (encoder_1) [bottleneck_encoder , right of=point, xshift=-1.5cm, fill=bottleneck_color] { $\mathbf{SE}$} ;
\draw [arrow] (fake) to node[midway, xshift=0cm,yshift=0.35cm] {$\mathbf{x}$} (encoder_1);
%\node[font=\Large,  label, text width=4cm, above of=myRectangle, align=center ,xshift=0.45cm,yshift=-0.7cm ] {$\mathbf{Source}$ \\ $\mathbf{Codec}$};
%\node[font=\Large,  label, text width=4cm, above of=myRectangle, align=center ,xshift=0cm,yshift=-0.451cm ] {$\mathbf{Source \ Codec}$};
\node[font=\Large,  label, text width=4cm, above of=myRectangle, align=center ,yshift=-0.4cm ] {$\mathbf{Source \ Encoder}$};

\node (encoder_2) [encoder , right of=myRectangle, xshift=1.7cm, fill=encoder_decoder,  trapezium right angle=84.8,  trapezium left angle=84.8] { $\mathbf{JSDE}$} ;
\draw [arrow] (encoder_1) to node[midway, xshift=0cm,yshift=0.35cm] {$\mathbf{b}$} (encoder_2);
\node (decoder_1) [decoder , right of=encoder_2, xshift=1.25cm, fill=encoder_decoder] { $\mathbf{D}$} ;
\draw [arrow] (encoder_2) to node[midway, xshift=0cm,yshift=0.35cm] {$\mathbf{z}$} (decoder_1);

\node (m) [label, right of=decoder_1,xshift=0.5cm ] {};
\draw [arrow] (decoder_1) to node[midway, xshift=-0.1cm,yshift=0.35cm] {$\mathbf{m}$} (m);

\end{tikzpicture}}
  % \scalebox{0.45}{
  %   {\input{Figures/conventional_methods}}
  %   }
    \caption{Conventional methods with $\bf{JSDE}$    }
    \label{fig:conventional2}
  \end{subfigure} 
  \medskip
  \medskip
 \hspace{0.025em}
  \begin{subfigure}{0.487\textwidth}
    \centering
    % \scalebox{0.5}{
    % {\input{Figures/codec_baseline}}
    % }
      \resizebox{\linewidth}{!}{%\documentclass{standalone}

% \usepackage{graphicx} % Required for inserting images

% \usepackage[utf8]{inputenc}
% \usepackage{tikz}
% \usetikzlibrary{shapes.geometric, arrows}
%  \usetikzlibrary{shapes.geometric, arrows.meta, calc,backgrounds}

%\begin{document}

 % Define some colors
 \definecolor{encoder_decoder}{RGB}{213, 232, 212}
 \definecolor{rectangle}{RGB}{255, 245, 204}
 \definecolor{bottleneck_color}{RGB}{230, 221, 184}
\definecolor{bg}{RGB}{210, 210, 210}
 % Define styles
%  \tikzstyle{data} = [draw, cylinder, line width=1pt, shape border rotate=90, outer
% sep=0, inner sep=5, minimum size=30, align=center, fill=white]

 \tikzstyle{arrow} = [-Triangle, line width=1pt]
 \tikzstyle{label} = [text width= 0.4cm, align=center]
 \tikzstyle{encoder} = [trapezium,
    trapezium angle=55,
    trapezium stretches=true,
    minimum width=2.4cm,
    minimum height=1.1cm,
    trapezium right angle=80,
    trapezium left angle=80,
    line width=1pt,
    shape border rotate=270,
    text centered,
    draw=black]

 \tikzstyle{bottleneck_encoder} = [trapezium,
    trapezium angle=55,
    trapezium stretches=true,
    minimum width=1.65cm,
    minimum height=0.75cm,
    trapezium right angle=80,
    trapezium left angle=80,
    line width=1pt,
    shape border rotate=270,
    text centered,
    draw=black]
 \tikzstyle{bottleneck_decoder} = [trapezium,
    trapezium angle=55,
    trapezium stretches=true,
    minimum width=1.65cm,
    minimum height=0.75cm,
    trapezium right angle=80,
    trapezium left angle=80,
    line width=1pt,
    shape border rotate=90,
    text centered,
    draw=black]

  \tikzstyle{decoder} = [trapezium,
    trapezium angle=55,
    trapezium stretches=true,
    minimum width=2.4cm,
    minimum height=1.1cm,
    trapezium right angle=80,
    trapezium left angle=80,
    line width=1pt,
    shape border rotate=90,
    text centered,
    draw=black]    

\begin{tikzpicture}[font=\Large,node distance = 1.5cm]
\node (point) {};
\node (fake) [label, left of=point ,xshift=-1cm,] {};

\node[draw, rectangle, minimum width=4.6cm, minimum height=3.2cm,  line width=1pt,draw=none, fill=bg,yshift=0.2cm] (background) at (0.4,0) {};
\node[font=\Large,  label, text width=4cm, above of=background, align=center ,yshift=-0.22cm, xshift=-0.65cm,text=red ] {$\mathbf{Edge \ Device}$};

\node[draw, rectangle, minimum width=4.2cm, minimum height=3.2cm,  line width=1pt,draw=none, fill=bg,yshift=0.2cm] (background2) at (5.9,0) {};
\node[font=\Large,  label, text width=4cm, above of=background2, align=center ,yshift=-0.22cm, xshift=1.25cm,text=red ] {$\mathbf{Cloud}$};

\node (encoder_1) [encoder , right of=point, xshift=-1.5cm, fill=encoder_decoder] { $\mathbf{E}$} ;
\draw [arrow] (fake) to node[midway, xshift=0.2cm,yshift=0.35cm] {$\mathbf{x}$} (encoder_1);

\node (z) [label,yshift=0.35cm,xshift=-0.5cm,right of=encoder_1 ] {$\mathbf{z}$};
%\draw [arrow] (encoder_1) to (z);
% 
\node[draw, rectangle, minimum width=3.8cm, minimum height=2.5cm,  line width=1pt,draw=none, fill=rectangle] (myRectangle) at (3.27,0.25) {};

\node[font=\Large,  label, text width=4cm, above of=myRectangle, align=center ,yshift=-0.5cm ] {$\mathbf{Source \ Codec}$};

\node (be_encoder) [bottleneck_encoder , right of=encoder_1,xshift=+0.7cm, fill=bottleneck_color] { $\mathbf{SE}$} ;
\draw [arrow] (encoder_1) to (be_encoder);

%\draw [arrow] (z) to (be_encoder);

\node (y) [label, right of=be_encoder,yshift=0.35cm,xshift=-0.5cm ] {$\mathbf{b}$};

\node (be_decoder) [bottleneck_decoder , right of=be_encoder,xshift=0.7cm, fill=bottleneck_color] { $\mathbf{SD}$} ;
 %\node [waypoint] at ($(be_encoder)!0.454!(be_decoder)$) {};
\draw [arrow] (be_encoder) to (be_decoder);

%  \draw [red,line width=1,shorten <= -3.5mm, shorten >= -3mm](be_decoder.north west) -- (be_decoder.south east);
%   \draw [red,line width=1,shorten <= -3mm, shorten >= -3.5mm](be_decoder.north east) -- (be_decoder.south west);

\node (z_hat) [label, right of=be_decoder,  yshift=0.35cm,xshift=-0.45cm ] {$\mathbf{\hat{z}}$};

\node (decoder_1) [decoder , right of=be_decoder, xshift=0.6cm, fill=encoder_decoder] { $\mathbf{D}$} ;
\draw [arrow] (be_decoder) to (decoder_1);

  %\draw[arrow] (y) -- ++(0,-1.5) -- ++(2.3,0) -- ++(0,1.15) --(decoder_1) ;

\node (m) [label, right of=decoder_1,xshift=0.57cm ] {};
\draw [arrow] (decoder_1) to node[midway, xshift=-0.2cm,yshift=0.35cm] {$\mathbf{m}$} (m);

% Define a command for connecting to nodes with rectangular paths
 \newcommand\connect[2]{\path[draw,arrow] (#1) |- ($(#1)!4/10!(#2)$) -| (#2)}
\end{tikzpicture}

%\end{document}
}
    \caption{Existing SOTA methods }
    \label{fig:baseline}
  \end{subfigure} 
\hspace{1.2em}
  \begin{subfigure}{0.487\textwidth}
   \centering
    %\scalebox{0.5}{\input{Figures/compressed_domain}}
    \resizebox{\linewidth}{!}{%\documentclass{standalone}

% \usepackage{graphicx} % Required for inserting images

% \usepackage[utf8]{inputenc}
% \usepackage{tikz}
% \usetikzlibrary{shapes.geometric, arrows}
%  \usetikzlibrary{shapes.geometric, arrows.meta, calc,backgrounds}

%\begin{document}

 % Import some libraries
 \usetikzlibrary{shapes.geometric, arrows.meta, calc,backgrounds}

 % Define some colors
 \definecolor{encoder_decoder}{RGB}{213, 232, 212}
 \definecolor{rectangle}{RGB}{255, 245, 204}
 \definecolor{bottleneck_color}{RGB}{230, 221, 184}
 \definecolor{_decoder}{RGB}{255, 204, 255} 
 \definecolor{bg}{RGB}{210, 210, 210}
 % Define styles
%  \tikzstyle{data} = [draw, cylinder, line width=1pt, shape border rotate=90, outer
% sep=0, inner sep=5, minimum size=30, align=center, fill=white]

 \tikzstyle{arrow} = [-Triangle, line width=1pt]
 \tikzstyle{label} = [text width= 0.2cm, align=center]
 \tikzstyle{waypoint}=[fill,circle,minimum size=3.5pt,inner sep=0pt]
 \tikzstyle{encoder} = [trapezium,
    trapezium angle=55,
    trapezium stretches=true,
    minimum width=2.4cm,
    minimum height=1.1cm,
    trapezium right angle=80,
    trapezium left angle=80,
    line width=1pt,
    shape border rotate=270,
    text centered,
    draw=black]

 \tikzstyle{bottleneck_encoder} = [trapezium,
    trapezium angle=55,
    trapezium stretches=true,
    minimum width=1.65cm,
    minimum height=0.75cm,
    trapezium right angle=80,
    trapezium left angle=80,
    line width=1pt,
    shape border rotate=270,
    text centered,
    draw=black]
 \tikzstyle{bottleneck_decoder} = [trapezium,
    trapezium angle=55,
    trapezium stretches=true,
    minimum width=1.65cm,
    minimum height=0.75cm,
    trapezium right angle=80,
    trapezium left angle=80,
    line width=1pt,
    shape border rotate=90,
    text centered,
    draw=black]

  \tikzstyle{decoder} = [trapezium,
    trapezium angle=55,
    trapezium stretches=true,
    minimum width=1.9cm,
    minimum height=1.1cm,
    trapezium right angle=80,
    trapezium left angle=80,
    line width=1pt,
    shape border rotate=90,
    text centered,
    draw=black]    

\begin{tikzpicture}[node distance = 1.5cm,font=\Large]
\node (point) {};

\node (fake) [label, left of=point ,xshift=-1cm,] {};

\node[draw, rectangle, minimum width=4.6cm, minimum height=3.2cm,  line width=1pt,draw=none, fill=bg,yshift=0.2cm] (background) at (0.4,0) {};
\node[font=\Large,  label, text width=4cm, above of=background, align=center ,yshift=-0.22cm, xshift=-0.65cm,text=red ] {$\mathbf{Edge \ Device}$};

\node[draw, rectangle, minimum width=4.2cm, minimum height=3.2cm,  line width=1pt,draw=none, fill=bg,xshift=-1.2cm,yshift=0.2cm] (background2) at (6.9,0) {};
\node[font=\Large,  label, text width=4cm, above of=background2, align=center ,yshift=-0.18cm, xshift=1.25cm,text=red ] {$\mathbf{Cloud}$};

\node (encoder_1) [encoder , right of=point, xshift=-1.5cm, fill=encoder_decoder] { $\mathbf{E}$} ;
\draw [arrow] (fake) to node[midway, xshift=0.2cm,yshift=0.35cm] {$\mathbf{x}$} (encoder_1);

\node (z) [label,yshift=0.35cm,xshift=-0.5cm,right of=encoder_1 ] {$\mathbf{z}$};
%\draw [arrow] (encoder_1) to (z);
% 
\node[draw, rectangle, minimum width=3.8cm, minimum height=2.5cm,  line width=1pt,draw=none, fill=rectangle] (myRectangle) at (3.27,0.25) {};

\node[font=\Large,  label, text width=4.5cm, above of=myRectangle, align=center ,yshift=-0.5cm ] {$\mathbf{Source \ Codec}$};

\node (be_encoder) [bottleneck_encoder , right of=encoder_1,xshift=+0.7cm, fill=bottleneck_color] { $\mathbf{SE}$} ;
\draw [arrow] (encoder_1) to (be_encoder);

%\draw [arrow] (z) to (be_encoder);

\node (y) [label, right of=be_encoder,yshift=0.35cm,xshift=-0.5cm ] {$\mathbf{b}$};

\node (be_decoder) [bottleneck_decoder , right of=be_encoder,xshift=0.7cm, fill=bottleneck_color] { $\mathbf{SD}$} ;
 \node [waypoint] at ($(be_encoder)!0.454!(be_decoder)$) {};
\draw [arrow] (be_encoder) to (be_decoder);

  \draw [red,line width=1.5,shorten <= -3.5mm, shorten >= -3mm](be_decoder.north west) -- (be_decoder.south east);
   \draw [red,line width=1.5,shorten <= -3mm, shorten >= -3.5mm](be_decoder.north east) -- (be_decoder.south west);

\node (z_hat) [label, right of=be_decoder,  xshift=-0.1cm ] {};

\node (decoder_1) [decoder , right of=be_decoder, xshift=0.6cm, fill=_decoder] { $\mathbf{JD}$} ;

  \draw[arrow] (y) -- ++(0,-1.5) -- ++(2.3,0) -- ++(0,1.15) --(decoder_1) ;

\node (m) [label, right of=decoder_1,xshift=0.57cm ] {};
\draw [arrow] (decoder_1) to node[midway, xshift=-0.2cm,yshift=0.35cm] {$\mathbf{m}$} (m);

% Define a command for connecting to nodes with rectangular paths
 \newcommand\connect[2]{\path[draw,arrow] (#1) |- ($(#1)!4/10!(#2)$) -| (#2)}
 % [... CODE ..]
\end{tikzpicture}

%\end{document}
}
    \caption{Our proposed method  }
    \label{fig:compressed}
  \end{subfigure}
  \caption{ \textbf{High-level comparison} of our proposed approach with existing SOTA approaches in distributed semantic segmentation.~Here, $\bf{SE}$ and $\bf{SD}$ represent the source encoder and decoder, respectively.~Blocks $\bf{E}$ and $\bf{D}$ are the image encoder and task decoder, respectively.~Further, $\bf{JSDE}$ is the joint source decoder and image encoder, while $\bf{JD}$ represents the proposed joint source and task decoder, in short: joint decoder.} 
  \label{fig:high_level}
\end{figure}

As shown in Figure \ref{fig:conventional}, the conventional solution is to deploy a source encoder $\bf{SE}$ on the edge device and a source decoder $\bf{SD}$ along with image encoder $\bf{E}$ and task decoder $\bf{D}$ in the cloud.~Both, traditional codecs, such as JPEG \cite{jpeg} and HEVC \cite{HEVC}, and learned image codecs \cite{balle2018variational,minnen2018joint,qmapc} can be employed as source codec.~The $\bf{SE}$ outputs a quantized compressed bitstream $\bf{b}$, which is transmitted to the cloud.~Although the conventional solution involves deployment of low-complex $\bf{SE}$ on an edge device, executing the three functions $\bf{SD}$, $\bf{E}$, and $\bf{D}$ in the cloud prohibits scalability of such service.~Furthermore, image compression requires a higher bitrate \cite{ahuja2023neural,matsubara2022supervised} and it also compromises potential data protection requirements as images can be reproduced anywhere from the bitstream.

As shown in Figure \ref{fig:conventional2}, one can merge $\bf{SD}$ and $\bf{E}$, while the edge device computational complexity stays the same as in Figure \ref{fig:conventional}\cite{torfason2018towards,liu2022improving,wang2022image}.~However, this approach may reduce the computational complexity in the cloud by utilizing a joint source decoder and image encoder $\bf{JSDE}$.~Similar to the conventional solution described in Figure \ref{fig:conventional}, it also faces the data protection problem and still requires a relatively high bitrate.

As shown in Figure \ref{fig:baseline}, the current state-of-the-art (SOTA) approach \cite{ahuja2023neural} deploys $\bf{E}$ and a low-complex $\bf{SE}$ on the edge device.~Ahuja et al. \cite{ahuja2023neural} aim at compressing the bottleneck features $\bf{z}$ based on the same principles as learned image codecs\cite{singh2020end}.~Compared to conventional approaches as described in Figures \ref{fig:conventional} and \ref{fig:conventional2}, this incurs a slightly higher computational complexity on the edge device but with the advantage that this method is more bitrate efficient.~Additionally, it requires a lower computational complexity in the cloud and the data protection problem is also largely solved, as, in the case of a semantic segmentation task, only the segmentation masks can be retrieved on the receiver side, but the original image $\bf{x}$ cannot be reconstructed with sufficient fidelity.

As shown in Figure \ref{fig:compressed}, inspired by the idea of joint functions in Figure \ref{fig:conventional2}, we propose to perform \textit{joint source and task decoding} using $\bf{JD}$, joint decoding.~It allows even lower computational complexity in the cloud, while achieving superior rate-distortion (RD) performance. Further, it enables to scale such service to millions of edge devices, keeping the edge device's computational complexity of our proposed approach the same as the so-far SOTA (Figure \ref{fig:baseline}).

\label{sect:figures}

Our contribution with this work on distributed semantic segmentation is threefold.~First, we propose to perform joint source and task decoding, which results in a computationally highly efficient cloud DNN, enabling us to perform distributed semantic segmentation at a very large scale without requiring extensive computational load per channel. Secondly, instead of applying training-time over-parameterization in the image encoder as suggested by \cite{mobileone,repvgg,fastvit}, we employ it in our proposed joint decoder $\bf{JD}$, which takes quantized bottle\-neck features as input, to further enhance the performance in the high-bitrate regime. Finally, we set a new distributed semantic segmentation SOTA benchmark over a wide range of bitrates on the mean intersection over union (mIoU) metric, while using only $9.8 \%$ of the cloud DNN parameters on COCO, and $11.59 \%$ on the Cityscapes dataset, compared to previous SOTA \cite{ahuja2023neural}.

\section{Related Works}

In this section, we start by discussing general DNN-based semantic segmentation methods, followed by previous research in the distributed setting.
\subsection{Semantic Segmentation}
\label{sect:general_semantic_segmentation}
As a fundamental computer vision task, semantic segmentation aims to classify each pixel of an image with a set of categories. Since the advent of DNNs \cite{krizhevsky2012imagenet,resnet50}, fully convolutional networks (FCNs) \cite{long2015fully,FCN2} have achieved SOTA on various segmentation benchmarks \cite{pascalvoc,cocodataset,cityscapes,ade20k}. After that, researchers shifted their focus on refining FCNs through various aspects, including exploitation of contextual information \cite{contextualinfo1,contextualinfo2}, introduction of image pyramids \cite{chen2017deeplab,chen2016attention} and pooling \cite{grauman2005pyramid,lazebnik2009spatial}, development of enhanced receptive fields \cite{wu2016bridging,dai2017deformable}. Nowadays, transformer-based \cite{xie2021segformer,Guo2022SegNeXtRC,Kang_2024_WACV} and multi-modal foundation models \cite{wang2022image,wang2023internimage} dominate the semantic segmentation benchmarks.

%changed a significant number of -> several

\subsection{Distributed Semantic Segmentation}
\subsubsection{General\normalfont:}\label{subsec:distributed_semantic_segmentation} Over the past few years, several approaches proposed to distribute the execution of the semantic segmentation DNN over a client/server architecture\cite{torfason2018towards,liu2022improving,wang2022learning}. All of them share a common idea, which is to utilize a pre-trained learned image compression method as source codec \cite{balle2018variational} and use the source encoder output, i.e., the quantized compressed latent space, to perform joint source decoding and image encoding through $\bf{JSDE}$.~As shown in Figure \ref{fig:conventional2}, the output of $\bf{JSDE}$ is sent to a task decoder $\bf{D}$ to produce a semantic segmentation map.~Both $\bf{JSDE}$ and $\bf{D}$ together form a semantic segmentation DNN, operating on a compressed bitstream $\bf{b}$.~Typically, in the first step, the source codec is trained, and in the second step, the semantic segmentation DNN is trained on the output of the pre-trained source encoder.~The choice for a semantic segmentation DNN in conventional approaches \cite{torfason2018towards,liu2022improving,wang2022learning} is $\texttt{DeepLabv3}$\cite{deeplabv3}, but a problem arises, as the size of the quantized compressed latent space is not suitable for its \texttt{ResNet-50} encoder.~As a solution, they either use transposed convolutional layers \cite{torfason2018towards,liu2022improving} or pixel shuffling \cite{wang2022learning} for upsampling it to meet the \texttt{ResNet-50} input dimensions, while removing the initial two layers. Furthermore, some approaches \cite{liu2022semantic,liu2022improving} employ feature selection mechanisms and supervision based on knowledge distillation (KD) to further enhance the RD performance.~Their semantic segmentation DNN is fully executed in the cloud, while the source encoder is deployed on the edge device with limited computational demands.
\subsubsection{Low-complexity encoder-decoder setup\normalfont:}  In contrast to the conventional distributed semantic segmentation approaches outlined above, current SOTA methods \cite{ahuja2023neural,matsubara2022supervised,feng2022image,datta2022low} propose to apply the source codec \textit{within} the \texttt{DeepLabv3} \texttt{ResNet-50} encoder or between the \texttt{ResNet-50} and \texttt{DeepLabv3} task decoder as shown in Figure \ref{fig:baseline}, to compress features instead of focusing on image compression.~This results in two benefits: (1) There is no need for a pre-trained learned image codec, as compression will be applied directly on feature level and trained in an end-to-end manner.~(2) The source decoder ($\bf{SD}$) becomes smaller, thus enhancing the scalability of the service.~We build upon the setup of current SOTA approaches and focus on making the cloud DNNs even more efficient by performing \textit{joint source and task decoding} with a single joint decoder DNN $\bf{JD}$, cf. Figure \ref{fig:compressed}.~This further increases the scalability of performing semantic segmentation in a distributed computing paradigm.~Furthermore, we improve performance in the high-bitrate regime by introducing re-parameterizable branches in the joint decoder $\bf{JD}$ via the re-parameterization trick on the quantized features \cite{repvgg,mobileone}.

%\subsubsection{Low-complexity decoders:}  

\section{Method}

In this section, we first describe the feature compression method.~Afterwards, we discuss our proposed approach for joint source and task decoding followed by training-time over-parameterization. 

\subsection{Bottleneck Feature Compression with Variational Models}

\begin{figure}[t!]
  \hspace{-1.7em}
  \resizebox{1.03\linewidth}{!}{\input{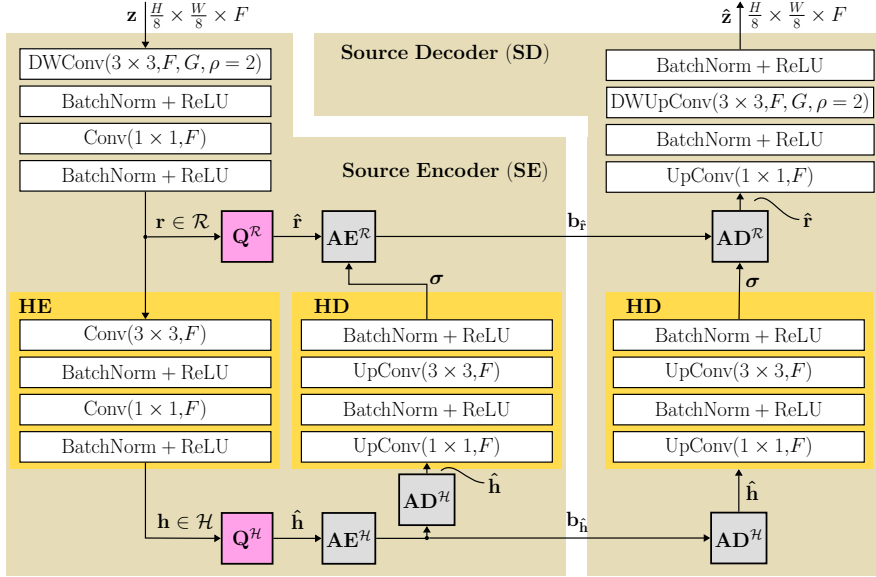}} 
  \caption{Hyperprior architecture of source encoder $\bf{SE}$ and source decoder $\bf{SD}$, see \cite{ahuja2023neural,balle2018variational}.}
  \label{fig:source_codec}
\end{figure}

Image compression works \cite{factorizedprior,balle2018variational,zhang2018advances} have shown that the rate-distortion (RD) objective of lossy compression can be formulated as a variational autoencoder (VAE) \cite{kingma2013auto}.~Similarly, recent works \cite{feng2022image,matsubara2022supervised,ahuja2023neural} have adapted the same framework and proposed to use a source codec for compressing the bottleneck features $\bf{z=E(x;\bm{\theta}^{\mathrm{E}})}$ from image encoder $\bf{E}$ with parameters $\bm{\theta}^{\mathrm{E}}$. Here, $\bf{x} = (\bf{x}_{\textit{i}}) \in \mathbb{I}$$^{H \!\times\! W \!\times\! C}$ is a normalized image of height $H$, width $W$, and $C=3$ color channels, with pixel $\bf{x}_{\textit{i}} \in \mathbb{I}$$^C$, pixel index $i \in \mathcal{I}$, pixel index set $\mathcal{I} = \{ 1,...,H \cdot W \}$ and $\mathbb{I}=[0,1]$.~As shown in Figure \ref{fig:source_codec}, the source codec consists of the bottleneck source encoder $\bf{SE}$ with parameters $\bm{\theta}^{\mathrm{SE}}$ and the bottleneck source decoder $\bf{SD}$ with parameters $\bm{\theta}^{\mathrm{SD}}$, which follow a hyperprior \cite{balle2018variational} architecture.~The $\bf{SE}$ utilizes a quantizer $\bf{Q}^{\mathcal{R}}$ to produce a latent bitstream $\bf{b_{\hat{r}}=SE^{(r)}(z;\bm{\theta}^{\mathrm{SE}})}$, which is entropy-coded using arithmetic coding \cite{AE}.~However, quantization is a non-differentiable operation, making it impossible to backpropagate during training.~Therefore, we need to relax the problem to cast it to a VAE by either replacing the quantization by the smooth approximation of the gradients \cite{theis2017lossy} or by additive uniform noise with a probability density function (PDF) of width one, over the interval ($-\frac{1}{2},\frac{1}{2}$) \cite{factorizedprior}, given that the elements of $\bf{r} \in \mathbb{R}$$^{d}$ and the uniform noise are in 32 bit floating point number representation.~In our work, we follow the latter method, but switch back to true quantization during inference.~The source decoder proceeds by reconstructing the bottleneck features $\bf{\hat{z}}=\bf{SD}(\bf{SE}(\bf{z};\bm{\theta}^{\mathrm{SE}});\bm{\theta}^{\mathrm{SD}})$, which are passed to task decoder $\bf{D}$ resulting in the network prediction $\bm{y}=(y_{i,s})$$\ \in \mathbb{I}$$^{H \times W \times S}$, with classes $\mathcal{S}=\{1,2,..,S\}$, class index $s \in \mathcal{S}$, and number of classes $S$.~The final semantic segmentation map is given as $\bm{m}=(m_{i})$ with $m_{i}=\mathrm{arg \ max}_{s \in \mathcal{S}} \ y_{i,s}$.~Both, $\bf{SE}$ and $\bf{SD}$, contain grouped convolutions and grouped transposed convolutions \cite{ahuja2023neural}, which are denoted by $\mathrm{DWConv(\textit{h} \times \textit{h},\textit{F},\textit{G},\rho=2)}$ and $\mathrm{DWUpConv(\textit{h} \times \textit{h},\textit{F},\textit{G},\rho=2)}$, respectively, where $\mathrm{\textit{h} \times \textit{h}}$ is the kernel size, $F$ is the number of kernels, $G$ is the number of groups present in the layer, and stride $\rho=2$.~They also contain reg\-ular convolutional and transposed convolutional layers, which are represented by $\mathrm{Conv(\textit{h} \times \textit{h},\textit{F})}$ and $\mathrm{UpConv(\textit{h} \times \textit{h},\textit{F})}$, respectively.

The elements of bitstream $\bf{b_{\hat{r}}}$ reveal a substantial amount of statistical dependencies and correlations.~Accordingly, an enhancement bitstream $\bf{b_{\hat{h}}}$ is being introduced in the source codec to model a time-variant standard deviation vector $\bf{\boldsymbol{\sigma}}$ for arithmetic coding $\bf{AE}^{\mathcal{H}}$ \cite{balle2018variational}.~As shown in Figure \ref{fig:source_codec}, it is produced by a hyperprior encoder and a quantizer $\bf{Q}^{\mathcal{H}}$ as $\bf{b_{\hat{h}}= SE^{(h)}(r;\bm{\theta}^{\mathrm{HE}})}$.~The quantization and entropy coding process to obtain $\bf{b_{\hat{h}}}$ is similar as with $\bf{b_{\hat{r}}}$.~Due to the introduction of $\bf{b_{\hat{h}}}$, each element in $\bf{\hat{r}}$ can now be modeled as zero-mean Gaussian with its individual standard deviation $\sigma_{i}$ using a hyperprior decoder $\bf{HD}$ as $\bf{\boldsymbol{\sigma}=(\sigma_{\textit{i}})=HD(\hat{h};\bm{\theta}^{\mathrm{HD}})}$ with parameters $\bm{\theta}^{\mathrm{HD}}$.~Note that the entire source encoder can be written as $(\mathbf{b_{\hat{r}},b_{\hat{h}}})=\mathbf{{SE(z;\bm{\theta}^{\mathrm{SE}}})}$ and the respective source decoder as $\mathbf{\hat{z}}=\mathbf{{SD(b_{\hat{r}},b_{\hat{h}};\bm{\theta}^{\mathrm{SD}}})}$.~The expected bitrate during training can be defined as

\begin{equation}
J^{\mathrm{rate}} = \mathbb{E}_{\bf{x} \sim \mathrm{p_{train}}}  \biggl[ \frac{  \mathrm{-log_{2}} (\mathrm{P_{\bf{\hat{r}}}}(\bf{\hat{r}} | \bf{\hat{h}}))  \mathrm{-log_{2}} (\mathrm{P_{\bf{\hat{h}}}}(\bf{\hat{h}}))  }{H \cdot W} \biggr],
\label{Eq:rate}
\end{equation}
with $\mathbb{E}_{\bf{x} \sim \mathrm{p_{train}} }$ representing the expectation over a minibatch in the dataset, $\mathrm{P_{\bf{\hat{r}}}}$ and $\mathrm{P_{\bf{\hat{h}}}}$ denoting the discrete probability distributions over the quantized latent spaces $\bf{\hat{r}}$ and $\bf{\hat{h}}$, respectively.

Since the task decoder $\bf{D}$ performs semantic segmentation, the distortion objective is formulated as a cross-entropy loss. 
\begin{equation}
J^{\mathrm{dist}} = \mathbb{E}_{\bf{x} \sim \mathrm{p_{train}}}  \left[ \frac{1}{|\mathcal{I}|}  \sum_{\substack{i \in \mathcal{I} }} \sum_{\substack{s \in \mathcal{S} }} \Bar{y}_{i,s}\cdot\mathrm{log}(y_{i,s}) \right].
\label{Eq:CE}
\end{equation}
Here, $ \Bar{\bm{y}}=(\Bar{y}_{i,s}) \in \{0,1 \}^{H \times W \times S}$ is the one-hot-encoded ground truth and we have $\forall i \in \mathcal{I}:\sum_{s \in \mathcal{S}}y_{i,s}=1$,\quad$\sum_{s \in \mathcal{S}}\Bar{y}_{i,s}=1$.~By combining (\ref{Eq:rate}) and (\ref{Eq:CE}), we obtain the RD trade-off as follows: 
\begin{equation}
J = \alpha \cdot J^{\mathrm{dist}} + (1 - \alpha) \cdot  J^{\mathrm{rate}}.
\label{Eq:RD}
\end{equation}
%change here such as object detection and image classification -> such as object detection etc. 
The RD trade-off is controlled by the hyperparameter $\alpha \in (0,1)$. Further, in the case of other downstream tasks such as object detection etc., $\bf{JD}$ and $J^{\mathrm{dist}}$ in (\ref{Eq:RD}) can be replaced by a task-specific decoder and loss, respectively.

\subsection{Proposed Joint Source and Task Decoder ($\bf{JD}$)}
\begin{figure}[t!]
    \hspace{-4.5em}
    \resizebox{1.12\linewidth}{!}{\input{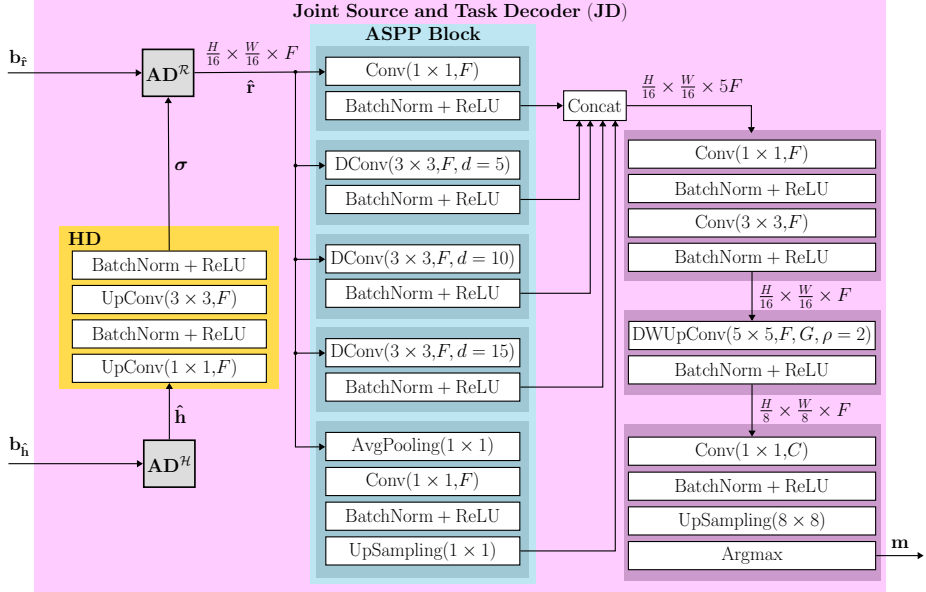}}
    %\resizebox{\linewidth}{!}{\input{Figures/compress_domain_decoder}}
    
  \caption{\textbf{Proposed architecture of the joint source and task decoder} ($\mathbf{JD}$, see Fig. \ref{fig:compressed}).~\textit{Training} details of the blue convolutional blocks within the ASPP block are shown in Figure \ref{fig:overparam}.}
  \label{fig:CD}
\end{figure}

As shown in Figure \ref{fig:compressed}, and unlike existing approaches \cite{ahuja2023neural,feng2022image} in distributed semantic segmentation, we propose to omit the reconstruction of $\bf{\hat{z}}$ and redirect the quantized latent space variable $\bf{\hat{r}}$ to the proposed joint source and task decoder $\bf{JD(b_{\hat{r}},b_{\hat{h}};\bm{\theta}^{\mathrm{JD}})}$ with parameters $\bm{\theta}^{\mathrm{JD}}$. Omitting the reconstruction target $\bf{\hat{z}}$ does not impede the learning process as shown in (\ref{Eq:RD}).~The aim of $\bf{JD}$ is to perform both source and task decoding at once.~Therefore, the bottleneck source decoder $\bf{SD}$ is not needed anymore, only the former decoder $\bf{D}$ is to be expanded towards a joint source and task decoder to produce the semantic segmentation mask as $\bf{m}\bf{ =JD(b_{\hat{r}},b_{\hat{h}};\bm{\theta}^{\mathrm{JD}}) \in \mathbb{I}}$$^{H \times W \times S}$ with classes $\mathcal{S}=\{1,2,..,S\}$, class index $s \in \mathcal{S}$, and number of classes $S$.
The network structure of $\bf{JD}$ is shown in Figure \ref{fig:CD}.~The source encoder generates $\bf{b_{\hat{h}}}$, which serves as an input to $\mathbf{JD}$, alongside the hyperprior bitstream $\bf{b_{\hat{r}}}$.~Similar to $\bf{D}$, also $\mathbf{JD}$ contains an atrous spatial pyramid pooling (ASPP) block \cite{deeplabv3}.~It consists of dilated convolutions, which are denoted by $\mathrm{DConv(\textit{h} \times \textit{h},\textit{F},\textit{d})}$, where $\mathrm{\textit{h} \times \textit{h}}$ is the kernel size, $F$ is the number of kernels present in the layer and $d$ is the dilation rate.~Separated from the dilated convolutions, it also contains upsampling layers, which are represented by $\mathrm{UpSampling(\textit{u} \times \textit{v})}$, where $\textit{u} \times \textit{v}$ is the desired spatial size of the input features.~Further, $\mathbf{JD}$ also contains a single grouped transposed convolution, which is denoted by $\mathrm{DWUpConv(\textit{h} \times \textit{h},\textit{F},\textit{G},\rho=2)}$, where $\mathrm{\textit{h} \times \textit{h}}$ is the kernel size, $F$ is the number of kernels, $G$ is the number of groups present in the layer, and stride $\rho=2$.~However, the key difference between $\bf{D}$ and $\bf{JD}$ remains that $\bf{JD}$ can perform joint source and task decoding.~For an easy comparison, we also provide the full diagram of $\bf{SD}$ and $\bf{D}$ of the so-far SOTA (Figure \ref{fig:baseline}, \cite{ahuja2023neural}) in the Supplement Figure 9.~Furthermore, our RD optimization problem remains the same as described in (\ref{Eq:RD}).~Interestingly, we do not enforce $\bf{JD}$ to learn source decoding by utilizing any extra loss terms.

% \ref{fig:decoder}

{\subsection{Training-Time Over-Parameterization}}
\label{section:over-parameterization}
As a consequence of performing joint source and task decoding in $\mathbf{JD}$, the number of network parameters is significantly smaller compared to the task decoder $\bf{D}$.~For low bitrates, the proposed $\bf{JD}$ outperforms existing SOTA methods, but without further measures we would observe a slight quality decrease at higher bitrates.~However, simply increasing the size of $\bf{JD}$ is not desirable, since it must be highly efficient to enable large-scale services in the cloud.~Accordingly, to reach superior performance also at high bitrates, we propose to perform training-time over-parameterization \cite{mobileone,fastvit,repvgg} by introducing re-parameterizable branches \cite{reparam_ding2021diverse,repvgg} without skip connections in the ASPP block of $\mathbf{JD}$.~Typically, this is only applied to the encoder $\bf{E}$.~However, since the ASPP block also acts as a feature extractor by extracting features from $\bf{\hat{r}}$ at multiple scales through various dilation rates $d$, it is also possible to perform over-parameterization to the ASPP subblocks.~Note that the over-parameterization is applied to $\bf{JD}$ only during training.~At inference, the network topology of ASPP block remains as in Figure \ref{fig:CD}.

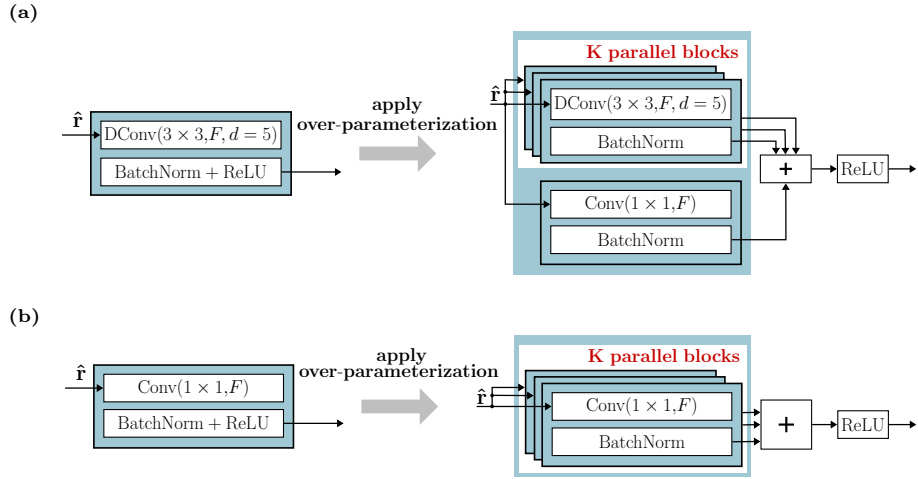
\begin{figure}[t!]
    %\hspace{-2.4em}
    % First subfigure
    \begin{subfigure}[t]{\textwidth}
        \captionsetup{position=top,singlelinecheck=off,justification=raggedright}
        \caption{}
        %\resizebox{1.04\linewidth}{!}{\input{Figures/overparam}}
        \resizebox{1\linewidth}{!}{ \definecolor{encoder_decoder}{RGB}{213, 232, 212}
  \definecolor{_rectangle_block}{RGB}{205, 154, 205}
 \definecolor{_rectangle}{RGB}{159, 200, 212}
  \definecolor{_rectangle2}{RGB}{159, 200, 212}
  \definecolor{_rectangle3}{RGB}{159, 200, 212}
\definecolor{_rectangle_overparam}{RGB}{159, 200, 212}
 \definecolor{bottleneck_color}{RGB}{230, 221, 184}
 \definecolor{conv}{RGB}{255, 255, 255}
 \definecolor{kblocks}{RGB}{202,0, 0}
 \definecolor{kblocks_outline}{RGB}{255,100,100}

 % Define styles
%  \tikzstyle{data} = [draw, cylinder, line width=1pt, shape border rotate=90, outer
% sep=0, inner sep=5, minimum size=30, align=center, fill=white]

 \tikzstyle{arrow} = [-Triangle, line width=1.2pt]
 \tikzstyle{waypoint}=[fill,circle,minimum size=3.5pt,inner sep=0pt]
 
\tikzstyle{process} = [rectangle,minimum width=5cm, minimum height=1cm, text
centered, draw=black, line width = 1pt, fill=conv]

\newcommand{\Plus}{\mathord{\begin{tikzpicture}[baseline=0ex, line width=2, scale=0.2]
\draw (1,0) -- (1,2);
\draw (0,1) -- (2,1);
\end{tikzpicture}}}
\begin{tikzpicture}[font=\LARGE,node distance = 1.5cm]

\node (point) {};

\node[draw, rectangle, minimum width=6.5cm, minimum height=2.7cm,  line width=1.5pt,draw=black, fill=_rectangle,yshift=-0.2cm] (myRectangle)  {};

\node (conv1) [process, above of=point,yshift=-1.1cm,minimum width=5.2cm, minimum height=0.9cm] {$\mathrm{DConv(3 \times 3, }F,d=5)$};

\node (_point) [label, left of=conv1,xshift=-2.8cm,yshift=0cm ] {};
\draw[arrow] (_point) to node[midway, xshift=-0.1cm,yshift=0.5cm, font=\fontsize{22.5}{20}\selectfont] {$\mathbf{\hat{r}}$}   (conv1) ;

\node (batchnorm) [process,minimum width=5.85cm,yshift=-0.8cm, minimum height=0.9cm] at (point)  {$\mathrm{BatchNorm + ReLU}$};
\node (_point2) [label, right of=batchnorm,xshift=3.5cm,yshift=0cm ] {};
\draw[arrow] (batchnorm) to node[midway, xshift=0cm,yshift=0.5cm] {}   (_point2) ;

\node[ rectangle, right of=myRectangle, minimum width=7.7cm, minimum height=7.9cm,  line width=0pt,draw=none, fill=_rectangle_overparam,xshift=12.8cm] (overparam)  {};

\node(arrow2)[single arrow, draw=none, fill=lightgray, right of=myRectangle, xshift=5.1cm,
      minimum width = 1.2cm, single arrow head extend=5.5pt,
      minimum height=25mm] {}; % length of arrow

\node(text)[above of=arrow2,yshift=0.05cm,xshift=0.1cm,font=\LARGE]{{$\mathbf{apply}$}};
\node(text2)[below of=text,yshift=0.86cm,align=center,font=\LARGE]{$\mathbf{over\text{-}parameterization}$};

\node[draw, rectangle, minimum width=7.4cm, minimum height=4.15cm, yshift=0.1cm,xshift=0.02cm,line width=1.5pt,draw=none, fill=white,above of=overparam] (conv_block)  {};

\node[draw, rectangle, above of=overparam, minimum width=6.2cm, minimum height=2.7cm,  line width=1.5pt,draw=black, fill=_rectangle3,yshift=-0.02cm,xshift=-0.365cm] (conv_rect3)  {};
\node[draw, rectangle, above of=overparam,  minimum width=6.2cm, minimum height=2.7cm,,  line width=1.5pt,draw=black, fill=_rectangle2,yshift=-0.235cm,xshift=-0.11cm] (conv_rect2)  {};
\node[draw, rectangle, above of=overparam, minimum width=6.5cm, minimum height=2.7cm,  line width=1.5pt,draw=black, fill=_rectangle,yshift=-0.452cm,xshift=0.3cm] (conv_rect)  {};

\node (conv2) [process, above of=overparam,minimum width=5.2cm,yshift=0.1cm,xshift=0.3cm] {$\mathrm{DConv(3 \times 3,}F,d=5)$};
\node (batchnorm2_) [process,minimum height=0.9cm, minimum width=5.85cm ,yshift=0.4cm,xshift=0.3cm] at (overparam)  {$\mathrm{BatchNorm}$};

\node[font=\LARGE, draw=black, label, text width=5.5cm, text=kblocks, fill=none, draw=none, above of=overparam, align=center , xshift=1cm,   yshift=1.75cm, line width=1.5pt ] {$\mathbf{K}\mathrm{\bf{ \ parallel  \ blocks}}$};

\node[draw, rectangle, below of=overparam, minimum width=6.5cm, minimum height=2.7cm,  line width=1.5pt,draw=black, fill=_rectangle,yshift=-0.75cm,xshift=0.3cm] (conv_one)  {};

\node (conv1by1_) [process,minimum width=5.85cm, minimum height=0.9cm, above of=conv_one,yshift=-0.9cm,xshift=0cm] {$\mathrm{Conv(1 \times 1,}F)$};

\node (batchnorm2) [process, minimum width=5.85cm , minimum height=0.9cm  ,yshift=-0.55cm,xshift=0cm]  at (conv_one)    {$\mathrm{BatchNorm}$};

\node(plus)[right of=overparam,draw, fill=white,line width = 1pt, rectangle,xshift=3.5cm,minimum width=1.648cm,minimum height=0.9cm,yshift=-0.5cm] (plus) {$\Plus$};

\draw[arrow] ([yshift=1mm]conv_rect.east) -- ++(1.75,0) -- ([yshift=4.7mm,xshift=11.7mm]plus.west) node[midway, above right] {};
\draw[arrow] (batchnorm2_) -- ++(4.38,0) -- ([yshift=4.7mm,xshift=5.25mm]plus.west) node[midway, above right] {};
\draw[arrow] ([yshift=-2.8mm]conv_rect.east) -- ++(0.9,0)  -- ++(0.5,0)   --(plus) ;
\draw[arrow] (batchnorm2) -- ++(4.205,0)  -- ++(0.5,0)  --(plus) ;

\node (relu2) [process,right of=plus,minimum width=1.2cm,yshift=0.1cm,xshift=1cm, yshift=-0.1cm,minimum height=0.9cm] {$\mathrm{ReLU}$};
\node (_point) [label, right of=relu2,xshift=0.35cm,yshift=0cm ] {};
 \draw[arrow] (relu2) to node[midway, xshift=0cm,yshift=0.5cm] {}   (_point) ;

 \draw[arrow] (plus) -- (relu2) ;

\node (input) [label, left of=conv2,xshift=-3.5cm,yshift=0cm, font=\LARGE ] {};

\node (waypoint_input) [waypoint,yshift=0.115cm,xshift=-0.4cm] at ($(input)!0.2!(conv_rect)$) {};

\node (waypoint_input2) [waypoint,yshift=0.47cm,xshift=-0.32cm] at ($(input)!0.2!(conv_rect2)$) {};
%\node (waypoint_input3) [waypoint,yshift=-0.72cm,xshift=-0.32cm] at ($(input)!0.2!(conv_rect3)$) {};

%\draw [arrow] (waypoint_input) -- ++(0,0) -- ++(0,0.35)--  (conv_rect2);
\draw [arrow] (waypoint_input) -- ++(0,0) -- ++(0,0.39)--   ([yshift=7.2mm,xshift=0mm]conv_rect2.west);

\draw [arrow] (input) to node[midway, xshift=-0.80cm,yshift=0.35cm,font=\fontsize{22.5}{20}\selectfont] {$\mathbf{\hat{r}}$}  (conv2);

%\draw[arrow] (input2) to node[midway, xshift=-1.05cm,yshift=0.35cm] {$\mathbf{\hat{r}}$}   (conv1by1_2) ;

%\draw [arrow] (waypoint_input) -- ++(0,0) -- ++(0,-0.65)--  (conv_rect3);
\draw [arrow] (waypoint_input2) -- ++(0,0) -- ++(0,0.39)--   ([yshift=9mm,xshift=0mm]conv_rect3.west);

\draw [arrow] (waypoint_input) -- ++(0,0) -- ++(0,-3.25)--  (conv1by1_.west);

\end{tikzpicture}}
        \label{fig:overparam_dilated}
    \end{subfigure}
    
    %\hspace{-2.4em}
    \begin{subfigure}[t]{\textwidth}
        \captionsetup{position=top, singlelinecheck=off,justification=raggedright}
        %\caption*{\hspace{2.6em}\textbf{b)}}
        \caption{}
        \resizebox{1\linewidth}{!}{ \definecolor{encoder_decoder}{RGB}{213, 232, 212}
  \definecolor{_rectangle_block}{RGB}{205, 154, 205}
 \definecolor{_rectangle}{RGB}{159, 200, 212}
  \definecolor{_rectangle2}{RGB}{159, 200, 212}
  \definecolor{_rectangle3}{RGB}{159, 200, 212}
\definecolor{_rectangle_overparam}{RGB}{159, 200, 212}
 \definecolor{bottleneck_color}{RGB}{230, 221, 184}
 \definecolor{conv}{RGB}{255, 255, 255}
 \definecolor{kblocks}{RGB}{202,0, 0}
 \definecolor{kblocks_outline}{RGB}{255,100,100}

 % Define styles
%  \tikzstyle{data} = [draw, cylinder, line width=1pt, shape border rotate=90, outer
% sep=0, inner sep=5, minimum size=30, align=center, fill=white]

 \tikzstyle{arrow} = [-Triangle, line width=1.2pt]
 \tikzstyle{waypoint}=[fill,circle,minimum size=3.5pt,inner sep=0pt]
 
\tikzstyle{process} = [rectangle,minimum width=5cm, minimum height=1cm, text
centered, draw=black, line width = 1pt, fill=conv]

\newcommand{\Plus}{\mathord{\begin{tikzpicture}[baseline=0ex, line width=2, scale=0.2]
\draw (1,0) -- (1,2);
\draw (0,1) -- (2,1);
\end{tikzpicture}}}
\newcommand{\newplus}{\mathord{\begin{tikzpicture}[baseline=0ex, line width=2, scale=0.25]
\draw (1,0) -- (1,2);
\draw (0,1) -- (2,1);
\end{tikzpicture}}}

\begin{tikzpicture}[font=\LARGE,node distance = 1.5cm]

\node (point2)[yshift=-7.5cm] {};

\node[draw, rectangle, minimum width=6.5cm, minimum height=2.7cm,  line width=1.5pt,draw=black, fill=_rectangle,yshift=0.65cm] at (point2) (myRectangle2)  {};
\node (conv1_2) [process, above of=point2,yshift=-0.25cm,minimum width=5.85cm, minimum height=0.9cm] {$\mathrm{Conv(1 \times 1, }F)$};
\node (batchnorm_2) [process, minimum width=5.85cm,yshift=0.1cm, minimum height=0.9cm] at (point2)  {$\mathrm{BatchNorm + ReLU}$};

\node[ rectangle, right of=myRectangle2, minimum width=7.7cm, minimum height=4.6cm,  line width=0pt,draw=none, fill=_rectangle_overparam,xshift=12.8cm] (overparam2)  {};

\node[draw, rectangle, minimum width=7.4cm, minimum height=4.15cm, yshift=-0.07cm,xshift=0.02cm,line width=1.5pt,draw=none, fill=white] at (overparam2)  (conv_block2)  {};
\node[font=\LARGE, draw=black, label, text width=5.5cm, text=kblocks, fill=none, draw=none, above of=conv_block2, align=center , xshift=1cm,   yshift=0.16cm, line width=1.5pt ] {$\mathbf{K}\mathrm{\bf{ \ parallel  \ blocks}}$};

\node[draw, rectangle,  minimum width=6.2cm, minimum height=2.7cm,  line width=1.5pt,draw=black, fill=_rectangle3,yshift=-0.18cm,xshift=-0.39cm]  at (overparam2) (conv_rect3)  {};
\node[draw, rectangle,   minimum width=6.2cm, minimum height=2.7cm,,  line width=1.5pt,draw=black, fill=_rectangle2,yshift=-0.385cm,xshift=-0.1cm] at (overparam2) (conv_rect2)  {};
\node[draw, rectangle,  minimum width=6.5cm, minimum height=2.7cm,  line width=1.5pt,draw=black, fill=_rectangle,yshift=-0.6cm,xshift=0.3cm] at (overparam2) (conv_rect)  {};

\node (conv1by1_2) [process,minimum width=5.85cm, minimum height=0.9cm, above of=conv_rect,yshift=-0.9cm,xshift=0cm] {$\mathrm{Conv(1 \times 1,}F)$};

\node (batchnorm2) [process, minimum width=5.85cm , minimum height=0.9cm  ,yshift=-0.55cm,xshift=0cm]  at (conv_rect)    {$\mathrm{BatchNorm}$};

\node(arrow3)[single arrow, draw=none, fill=lightgray, right of=myRectangle2, xshift=5.1cm,
      minimum width = 1.2cm, single arrow head extend=5.5pt,
      minimum height=25mm] {}; % length of arrow

\node(text_)[above of=arrow3,yshift=0.15cm,xshift=0.18cm,font=\LARGE]{{$\mathbf{apply}$}};
\node(text2_)[below of=text_,yshift=0.96cm,align=center,font=\LARGE]{$\mathbf{over\text{-}parameterization}$
};

\node(plus)[right of=conv_rect3,draw, fill=white,line width = 1pt, rectangle,xshift=3.89cm,minimum width=1.648cm,minimum height=1.6cm,yshift=-0.41cm] (plus2) {$\newplus$};

%\draw [arrow] (batchnorm2) -- (plus2);

\draw[arrow] ([yshift=4mm]conv_rect.east) -- ++(0,0) -- ([yshift=4mm,xshift=0mm]plus2.west) node[midway, above right] {};
\draw[arrow] ([yshift=0mm]conv_rect.east) -- ++(0.0,0) -- ([yshift=0mm,xshift=0mm]plus2.west) node[midway, above right] {};
%\draw[arrow] ([yshift=7.1mm]conv_rect.east) -- ++(0.4,0)  -- ([yshift=2.85mm,xshift=0mm]plus2.west) node[midway, above right] {};

\draw [arrow] (batchnorm2.east) -- ([yshift=-5.5mm,xshift=0mm]plus2.west);

\node (relu2) [process,right of=plus2,minimum width=1.2cm,yshift=0.1cm,xshift=1cm, yshift=-0.1cm,minimum height=0.9cm] {$\mathrm{ReLU}$};
\draw [arrow] (plus2) -- (relu2);
\node (_point2) [label, right of=relu2,xshift=0.35cm,yshift=0cm ] {};

\draw[arrow] (relu2) to node[midway, xshift=0cm,yshift=0.5cm] {}   (_point2) ;
\node (input2) [label, left of=conv1by1_2,xshift=-4cm,yshift=0cm, font=\LARGE ] {};

\draw[arrow] (input2) to node[midway, xshift=-1.05cm,yshift=0.35cm,font=\fontsize{22.5}{20}\selectfont] {$\mathbf{\hat{r}}$}   (conv1by1_2) ;
\node (waypoint_input) [waypoint,yshift=0.08cm,xshift=-0.4cm] at ($(input2)!0.2!(conv_rect2)$) {};

\draw [arrow] (waypoint_input) -- ++(0,0) -- ++(0,0.36)--  ([yshift=7.4mm,xshift=0mm]conv_rect2.west);
\node (waypoint_input2) [waypoint,yshift=0.36cm,xshift=-0.48cm] at ($(input2)!0.2!(conv1by1_2)$) {};
\draw [arrow] (waypoint_input2) -- ++(0,0) -- ++(0,0.26)--   ([yshift=8mm,xshift=0mm]conv_rect3.west);

\node (_point5) [label, left of=conv1_2,xshift=-2.8cm,yshift=0cm ] {};
\draw[arrow] (_point5) to node[midway, xshift=-0.1cm,yshift=0.5cm, font=\fontsize{22.5}{20}\selectfont] {$\mathbf{\hat{r}}$}   (conv1_2) ;

\node (_point6) [label, right of=batchnorm_2,xshift=3.5cm,yshift=0cm ] {};
\draw[arrow] (batchnorm_2) to node[midway, xshift=0cm,yshift=0.5cm] {}   (_point6) ;

\end{tikzpicture}}
        \label{fig:overparam_pointwise}
    \end{subfigure}
    
    \caption{Proposed over-parameterization of the $\bf{JD}$ ASPP subblocks in Figure \ref{fig:CD}. }
    \label{fig:overparam}
\end{figure}

Figure \ref{fig:overparam} presents how to apply training-time over-parameterization to the subblocks of the ASPP block.~As depicted in Figure \ref{fig:overparam_dilated}, for all three dilated convolutional subblocks with dilation rates $d \in (5,10,15)$ of the ASPP block, the dilated convolutional layer along with batchnorm is duplicated into $K$ parallel blocks with indices $k \in \mathcal{K} = \{1,..,K\}$.~Furthermore, a single pointwise convolutional block \cite{fastvit,mobileone} is also applied in parallel to the $K$ blocks of the dilated convolutional subblocks, resulting in $K+1$ parallel branches.~The outputs of the in total $K+1$ parallel branches is summed up and followed by the $\textrm{ReLU}$ activation function.~As shown in Figure \ref{fig:overparam_pointwise}, for the remaining two convolutional subblocks of the ASPP block (top and bottom), since they contain only pointwise convolutions, the output of $K$ parallel subblocks is summed up \cite{fastvit,mobileone} and the $\textrm{ReLU}$ activation function is applied on the summed output.

\section{Experimental Overview}

Following current SOTA methods \cite{ahuja2023neural,matsubara2022supervised,feng2022image}, we also chose \texttt{DeepLabv3} \cite{deeplabv3} with a \texttt{ResNet-50}\cite{resnet50} encoder for conducting all our experiments. \texttt{DeepLabv3} does not use encoder-decoder skip connections, which enables bitrate-efficient transmission between the edge device and the cloud platform, making it optimal for distributed deployment. Since our proposed $\mathbf{JD}$ takes the final output of the bottleneck source encoder, we chose the split point right after the \texttt{ResNet-50} backbone as shown in Supplement Figure 10, which presents the entire edge device processing (transmitter side) in our investigations.~For a fair comparison to our approach, we also use the same split point to train the SOTA baseline \cite{ahuja2023neural}.~All of our experiments and evaluation metrics are implemented using \texttt{PyTorch} \cite{paszke2019pytorch}, the \texttt{MMSegmentation} toolbox \cite{mmseg2020}, and the \texttt{Compress-AI} \cite{begaint2020compressai} library.~Following previous works \cite{ahuja2023neural,matsubara2022supervised,feng2022image}, we perform our ablation studies and results benchmarking on the validation set.~In the following, we introduce the datasets, training and evaluation settings, and the metrics.
%\ref{fig:resnet_with_se}
\subsection{Datasets}
%change here chose to provide -> provide
%change here large dataset size -> large size 
%change so-far in Figure 6 description
We report the results on well-established indoor and outdoor datasets for semantic segmentation, including COCO-2017 \cite{cocodataset} and Cityscapes \cite{cityscapes}.~Typically, in distributed semantic segmentation, COCO-2017 is considered to be most challenging due to its diverse object categories, crowded scenes, scale variation and large size \cite{cocodataset,matsubara2022supervised}. It also includes pixel-level annotations for \texttt{things} and \texttt{stuff} categories, enabling the generation of high-quality segmentation masks. Many prior works (e.g., \hspace{1sp}\cite{ahuja2023neural,matsubara2022supervised,singh2020end}) report results on it.~Note that another relevant approach \cite{feng2022image} also reported results on Cityscapes.~Therefore, in order to be comparable with existing works, we provide results on both datasets. The number of images and split information in respective datasets is shown in Table \ref{table:datasets}.

\begin{table}[t!]
  \caption{\textbf{Datasets \& splits} used in our experiments}
  \label{tab:headings}
  \centering
  \begin{tabular}{@{}cccc@{}}
    \toprule
     \centering Dataset &  Official subsets & \#Images & Symbol \\
    \midrule
   \centering \multirow{2}{*}{COCO-2017 \cite{cocodataset}}   & train & 118,287 & $\mathcal{D}_{\mathrm{COCO}}^{\mathrm{train2017}}$ \\  [0.2ex]
   \centering    & val & 5,000 & $\mathcal{D}_{\mathrm{COCO}}^{\mathrm{val2017}}$ \\ 
     \midrule
\centering \multirow{2}{*}{Cityscapes \cite{cityscapes} }   & train & 2,975 & $\mathcal{D}_{\mathrm{CS}}^{\mathrm{train}}$ \\ 
  \centering    & val & 500 & $\mathcal{D}_{\mathrm{CS}}^{\mathrm{val}}$ \\
  \bottomrule
  \end{tabular}
\label{table:datasets}
\end{table}
\subsection{Experimental Design, Training and Metrics}
We trained and evaluated all of our models on an \texttt{NVIDIA V100} GPU.~The training resolution for the COCO-2017 dataset is $513{\times}513$ \cite{ahuja2023neural}, whereas for the Cityscapes dataset it is $769{\times}769$ \cite{feng2022image}.~Furthermore, we select $21$ classes in COCO-2017 based on the classes in PASCAL-VOC-2012 \cite{pascalvoc} dataset, as suggested by important prior works \cite{ahuja2023neural,matsubara2022supervised,singh2020end,paszke2019pytorch}. For the COCO dataset, the number of feature maps $F$ in the ASPP head is $256$\cite{ahuja2023neural,matsubara2022supervised}, while for Cityscapes it is set to $512$\cite{feng2022image}. We follow the same training setup as in \cite{ahuja2023neural}.~However, since $\bf{JD}$ has to perform joint source and task decoding, we update its weights along with $\bf{SE}$ in an end-to-end manner during training.~As a starting point in the training, both $\bf{JD}$ and $\bf{SE}$ weights are randomly initialized and we do not use any pre-trained weights.~The training setup and all hyperparameters are described in more detail in the Supplement Section 1.

%\ref{section:training_details}

To ensure fair comparison, a recent SOTA learned image codec from Song et al. \cite{qmapc} is fine-tuned to the respective dataset.~After fine-tuning, the model weights for Song et al.\cite{qmapc} are fixed.~For both datasets, we use the pre-trained \texttt{No\-Compression} $\texttt{DeepLabv3}$ model with a \texttt{ResNet-50} \cite{resnet50} encoder and fine-tune it based on the compressed images from traditional codecs $\bf{\hat{x}}=\bf{SD(\bf{b})}$ and learned codec $\bf{\hat{x}}=\bf{SD(\bf{b};\bm{\theta}^{\mathrm{SD}}})$.~The conventional method baselines (Figure \ref{fig:conventional}) include the traditional codecs JPEG \cite{jpeg}, HEVC \cite{HEVC}, and a learned codec by Song et al.\ \cite{qmapc}.~Note that for COCO, the results for JPEG and HEVC are from Matsuba et al. \cite{matsubara2022supervised}, whereas for Cityscapes, we obtain the results for JPEG ourselves, while HEVC results are from Feng et al.\ \cite{feng2022image}.

%Section \ref{subsec:distributed_semantic_segmentation}

For the conventional method with the $\bf{JSDE}$ baseline (Figure \ref{fig:conventional2}), we follow the training recipe from Liu et al.\ \cite{liu2022improving} to train the method proposed by Torfason et al.\ \cite{torfason2018towards}.~The $\bf{JSDE}$ follows a $\texttt{ResNet-50}$ topology, but to utilize the compressed latent variable $\bf{\hat{r}}$ as input, the initial convolutions are replaced by transposed convolutions \cite{torfason2018towards}, cf. Section \ref{subsec:distributed_semantic_segmentation}.~We use the learned image codec by Ball{\'e} et al.\ \cite{balle2018variational} from the $\texttt{Compress-AI}$ model zoo, which is pre-trained on the Vimeo-90K \cite{vimeo} dataset.~After that, we perform joint fine-tuning of $\bf{SE}$, $\bf{JSDE}$ and $\bf{D}$ on the respective datasets.

%change here standard mIoU metric -> mIoU metric
%change here adopted widely in -> prevalent
To evaluate the RD performance of all our models, the rate is defined following (\ref{Eq:rate}) as bits per pixel (bpp). To measure distortion, we report the mIoU metric, which is prevalent in distributed semantic segmentation methods\cite{ahuja2023neural,matsubara2022supervised,liu2022semantic,torfason2018towards}. Furthermore, we report the number of floating-point operations per image (FLOPs) and the number of parameters.  

\section{Results and Discussion}
\label{results}

In this section, we will first discuss the rate-distortion (RD) performance of our proposed approach  and compare it against recent SOTA approaches.~Afterwards, we evaluate different model settings through ablation studies to design our final joint decoder ($\bf{JD}$) for both COCO and Cityscapes datasets.

\subsection{Comparison With SOTA Methods}

We reproduced the existing SOTA baseline (Figure \ref{fig:baseline}) from Ahuja et al. \cite{ahuja2023neural} on COCO and also obtain it for the Cityscapes dataset. Furthermore, for both datasets, our method's results and those of Ahuja et al. \cite{ahuja2023neural} are averaged over three different random seeds. However, for different random seeds, resulting RD results are not necessarily exactly equal and may have a slight variation even for the same $\alpha$ in (\ref{Eq:RD}). Therefore, to obtain more reliable results, we averaged the RD values over three different seeds for exactly the same $\alpha$.
\begin{table}[t!]
  \caption{\textbf{Comparison of FLOPs and $\#$params (cloud DNNs)}}
  \centering
  \begin{tabular}{@{}lr@{\hskip 0.1in}r@{\hskip 0.1in}r@{\hskip 0.1in}r@{}}
    \toprule
  \multirow{2}{*}{Methods} &\multicolumn{2}{c}{\textbf{COCO}}&\multicolumn{2}{c}{\textbf{Cityscapes}} \\
    \cmidrule(lr){2-3}\cmidrule(lr){4-5}
      &FLOPs (G)& $\#$params (M) & FLOPs (G)& $\#$params (M) \\  
    \midrule
    Song et al.\ \cite{qmapc} (\ref{fig:conventional})  & 1386 & 67.21 & 2834 & 93.30 \\      
    Torfason et al.\ \cite{torfason2018towards} (\ref{fig:conventional2})  & 734 & 43.23 & 1153 & 69.32 \\          
    Ahuja et al.\ \cite{ahuja2023neural} (\ref{fig:baseline})  & 521 & 16.78 & 1366 & 42.44 \\      
    Ours (\ref{fig:compressed}) & \textbf{10} & \textbf{1.66} & \textbf{39} & \textbf{4.92} \\      
    \bottomrule
  \end{tabular}
\label{table:flops_params_comparison}
\end{table}

\begin{figure}[t!]
  \begin{subfigure}{0.5\textwidth}

    \resizebox{\linewidth}{!}{\begin{tikzpicture}
 \tikzstyle{arrow} = [-Triangle, line width=1pt]
\definecolor{color0}{rgb}{0.63921568627451,0.603921568627451,0.552941176470588}
\definecolor{color1}{rgb}{0.913725490196078,0.588235294117647,0.47843137254902}
\definecolor{color2}{rgb}{0.294117647058824,0,0.509803921568627}
\definecolor{darkgoldenrod1811370}{RGB}{181,137,0}
\definecolor{darkgreen}{RGB}{0,100,0}
\definecolor{lightgray}{RGB}{211,211,211}
\definecolor{lightgray204}{RGB}{204,204,204}
\definecolor{rosybrown163154141}{RGB}{163,154,141}
\definecolor{voilet}{RGB}{139,69,19}

\begin{axis}[
width=9.5cm,   % Set the width of the plot
legend cell align={left},
legend cell align={left},
legend cell align={left},
legend style={
  fill opacity=1.0,
  draw opacity=1,
  text opacity=1,
  at={(0.35,0.02)},
  anchor=south west,
  draw=white!80!black
},
legend style={
  fill opacity=1.0,
  draw opacity=1,
  text opacity=1,
  at={(0.35,0.02)},
  anchor=south west,
  draw=white!80!black
},
legend style={
  fill opacity=1.0,
  draw opacity=1,
  text opacity=1,
  at={(0.58,0.02)},
  anchor=south west,
  draw=white!80!black
},
tick align=outside,
tick pos=left,
tick label style={font=\large},
label style={font=\large},
x grid style={lightgray},
xlabel={bits per pixel (bpp) },
xmajorgrids,
xmin=0, xmax=1.08,
xtick style={color=black,font=\small},
xtick={0,0.2,0.4,0.6,0.8,1},
xticklabels={0,0.2,0.4,0.6,0.8,1.0},
y grid style={white!82.7450980392157!black},
ylabel={\(\displaystyle \textrm{mIoU on $\mathcal{D}_{\mathrm{COCO}}^{\mathrm{val2017}}$}\) (\%)},
 ylabel style={yshift=-6pt},
ymajorgrids,
ymin=48, ymax=67,
ytick style={color=black},
ytick={48,50,52,54,56,58,60,62,64,66},
yticklabels={
  \(\displaystyle {48}\),
  \(\displaystyle {50}\),
  \(\displaystyle {52}\),
  \(\displaystyle {54}\),
  \(\displaystyle {56}\),
  \(\displaystyle {58}\),
  \(\displaystyle {60}\),
  \(\displaystyle {62}\),
  \(\displaystyle {64}\),
  \(\displaystyle {66}\)
}
]

\addplot [line width=1.5pt,  mark size=3.5, mark options={ draw=black, line width=1, solid, fill=none},  mark=diamond, dotted]
table{%
x  y
0.0800000000000001 61.8
0.49 64.8
0.9 66.1
};
\label{coco_HEVC}

\addplot [line width=1.5pt,  mark size=3.5, mark options={ draw=black, line width=1, solid, fill=none},  mark=triangle, dotted]
table{%
x  y
0.8 61.8
0.9 64.1
0.99 64.5
};

\label{coco_JPEG}

\addplot [line width=1.5pt, blue, dashed, forget plot]
table {%
0 66.4000015258789
1.05
66.4000015258789
};
\label{No_Compression}

\addplot [line width=1.5pt, color0, mark=*, mark size=2, mark options={solid}, forget plot]
table {%
0.0900000333786011 49.5200004577637
0.317999958992004 53.6300010681152
0.440000057220459 54.0400009155273
};
\label{Song}

\addplot [line width=1.5pt, green!39.2156862745098!black, mark=*, mark size=2, mark options={solid}, forget plot]
table {%
0.0729999542236328 56.9599990844727
0.0900000333786011 62.060001373291
0.120000004768372 63.8899993896484
0.190000057220459 65.0199966430664
0.440000057220459 66.3499984741211
};
\label{Ahuja}

\addplot [line width=1.5pt, red, mark=*, mark size=2, mark options={solid}, forget plot]
table {%
0.0470000505447388 62.6500015258789
0.059999942779541 63.8199996948242
0.10099995136261 64.1900024414062
0.129999995231628 64.6500015258789
0.225000023841858 65.870002746582
0.319000005722046 66.2600021362305
0.634999990463257 66.4300015258789
};
\label{ours}

\addplot [line width=1.5pt, color1, mark=*, mark size=2, mark options={solid}, forget plot]
table {%
0.0700000524520874 54.9000015258789
0.129999995231628 59.4000015258789
0.309999942779541 64.9000015258789
0.559999942779541 65.3000030517578
0.850000023841858 65.5
};
\label{entropic_student}

\addplot [line width=1.5pt, color2, mark=*, mark size=2, mark options={solid}, forget plot]
table {%
0.399999976158142 58.2000007629395
0.629999995231628 60.2999992370605
0.990000009536743 60.7999992370605
};
\label{singh}

\addplot [line width=1.5pt, voilet, mark=*, mark size=2, mark options={solid}, forget plot]
table {%
0.199999976158142 58.2000007629395
0.35 60.2000007629395
0.70000009536743 62.7999992370605
};
\label{torfason}

\end{axis}

% \node [draw,fill=white,font=\small] at (rel axis cs: 0.34,0.085){\shortstack[l]{
%     \ref*{coco_HEVC} HEVC \\ \ref*{coco_JPEG} JPEG  }};

\node (HEVC) [draw=none,fill=white,font=\small] at (rel axis cs: 0.4,0.75){\shortstack[l]{
     HEVC$^{\textcolor{red}{*}}$   }};
     
%\draw [arrow]  at (rel axis cs: 0.4,0.75)  to  (rel axis cs: 0.6,0.75); 

\draw[-,line width=1.1] (rel axis cs:0.32,0.82) -- (HEVC);

\node (JPEG) [draw=none,fill=white,font=\small] at (rel axis cs: 0.7,0.861){\shortstack[l]{
     JPEG$^{\textcolor{red}{*}}$   }};
\draw[-,line width=1.1] (rel axis cs:0.8,0.8) -- (JPEG);

\node [draw,fill=white,font=\small] at (rel axis cs: 0.707,0.25){\shortstack[l]{
\ref*{No_Compression} No Compression$^{\textcolor{red}{*}}$ \cite{deeplabv3} \\ 
 \ref*{ours} Ours (\ref{fig:compressed}) \\
\ref*{Ahuja} Ahuja et al. \cite{ahuja2023neural} (\ref{fig:baseline})  \\ 
\ref*{entropic_student} Matsuba et al.$^{\textcolor{red}{*}}$ \cite{matsubara2022supervised} (\ref{fig:baseline}) \\ 
\ref*{singh} Singh et al.$^{\textcolor{red}{*}}$ \cite{singh2020end} (\ref{fig:baseline})   \\
\ref*{Song} Song et al. \cite{qmapc} (\ref{fig:conventional}) \\
\ref*{torfason} Torfason et al. \cite{torfason2018towards} (\ref{fig:conventional2})  }};
%Torfason

\end{tikzpicture}}
    \caption{COCO    }
    \label{fig:coco_results}
  \end{subfigure}%
  %\hfill
  %\hspace{0.02em}
  \hspace{-0.2em}
  \begin{subfigure}{0.5\textwidth}
    \resizebox{\linewidth}{!}{\begin{tikzpicture}

\definecolor{darkgoldenrod1811370}{RGB}{181,137,0}
\definecolor{darkgreen}{RGB}{0,100,0}
\definecolor{lightgray}{RGB}{211,211,211}
\definecolor{lightgray204}{RGB}{204,204,204}
\definecolor{rosybrown163154141}{RGB}{163,154,141}
\definecolor{voilet}{RGB}{139,69,19}
\begin{axis}[
width=9.55cm,   % Set the width of the plot
legend cell align={left},
legend cell align={left},
legend cell align={left},
scaled ticks=false,
tick align=outside,
tick pos=left,
tick label style={font=\large},
label style={font=\large},
x grid style={lightgray},
xlabel={bits per pixel (bpp) },
xmajorgrids,
xmin=0, xmax=0.45,
xtick style={color=black},
x tick label style={/pgf/number format/precision=10},
y grid style={lightgray},
ylabel={\(\displaystyle \textrm{mIoU on $\mathcal{D}_{\mathrm{CS}}^{\mathrm{val}}$}\) (\%)},
ylabel style={yshift=-6pt},
ymajorgrids,
ymin=33, ymax=80,
ytick style={color=black},
ytick={33,38,43,48,53,58,63,68,73,78},
yticklabels={
  \(\displaystyle {33}\),
  \(\displaystyle {38}\),
  \(\displaystyle {43}\),
  \(\displaystyle {48}\),
  \(\displaystyle {53}\),
  \(\displaystyle {58}\),
  \(\displaystyle {63}\),
  \(\displaystyle {68}\),
  \(\displaystyle {73}\),
  \(\displaystyle {78}\)
}
]
\addplot [line width=1.5pt,  mark size=3.5, mark options={ draw=black, line width=1, solid, fill=none},  mark=diamond, dotted]
table{%
x  y
0.0600000000000001 36
0.12 52
0.2 62
};
\label{hevccs}

\addplot [line width=1.5pt,  mark size=3.5, mark options={ draw=black, line width=1, solid, fill=none},  mark=triangle, dotted]
table{%
x  y
0.14 46.7
0.3 61
};
%\addlegendentry{ JPEG};
\label{_JPEG}
\addplot [line width=1.5pt, blue, dashed]
table {%
0 78.8499984741211
0.45 78.8499984741211
};
\label{No_Compression}
\addplot [line width=1.5pt, rosybrown163154141, mark=*, mark size=1.6, mark options={solid}]
table {%
%0.0125999450683594 34.3199996948242
0.0134999752044678 37.2900009155273
0.038599967956543 65.3000030517578
0.0450999736785889 67.5299987792969
0.0536999702453613 69.1999969482422
0.0673999786376953 71.0999984741211
0.0789999961853027 72.6399993896484
0.183200001716614 75.0899963378906
0.280900001525879 78.0100021362305
};
\label{Song}
\addplot [line width=1.5pt,darkgreen, mark=*, mark size=1.6, mark options={solid}]
table {%
%0.00499999523162842 46.9500007629395
%0.0119999647140503 67.7900009155273
0.0190000534057617 70.0999984741211
0.0219999551773071 72.9199981689453
0.031000018119812 75.0699996948242
%0.0429999828338623 76.2799987792969
0.0520000457763672 76.8000030517578
0.0740000009536743 77.3000030517578
0.113999962806702 77.6699981689453
0.28600001335144 78.2300033569336
0.4 78.2700033569336
};
\label{Ahuja_cityscape}
\addplot [line width=1.5pt,  red, mark=*, mark size=1.6, mark options={solid}]
table {%
0.0147 72.71
0.03 76.60
0.0877 77.66
0.1 78.06
0.24 78.22
0.3874 78.24
0.41 78.34
};
\label{ours}
\addplot [line width=1.5pt, darkgoldenrod1811370, mark=*, mark size=1.6, mark options={solid}]
table {%
0.0800000429153442 67
0.120000004768372 72
0.139999985694885 73.8000030517578
0.159999966621399 74.9000015258789
};
\label{Feng}

\addplot [line width=1.5pt, voilet, mark=*, mark size=1.6, mark options={solid}]
table {%
0.0800000429153442 52
0.140000004768372 55
0.270000004768372 55
0.3 60
};
\label{Torfason}

\end{axis}

\node (HEVC) [draw=none,fill=white,font=\small] at (rel axis cs: 0.35,0.67){\shortstack[l]{
     HEVC$^{\textcolor{red}{*}}$   }};
 \draw[-,line width=1.1] (rel axis cs:0.4,0.58) -- (HEVC);
\node (JPEG) [draw=none,fill=white,font=\small] at (rel axis cs: 0.6,0.67){\shortstack[l]{
      JPEG   }};
\draw[-,line width=1.1] (rel axis cs:0.65,0.6) -- (JPEG);
% 

% \node [draw,fill=white,font=\small] at (rel axis cs: 0.36,0.07){\shortstack[l]{
%    \ref*{hevccs} HEVC  \\ \ref*{_JPEG} JPEG  }};

\node [draw,fill=white,font=\small] at (rel axis cs: 0.72,0.21){\shortstack[l]{
\ref*{No_Compression} No Compression \cite{deeplabv3} \\
\ref*{ours} Ours (\ref{fig:compressed})  \\ 
\ref*{Ahuja_cityscape} Ahuja et al. \cite{ahuja2023neural} (\ref{fig:baseline})   \\ 
\ref*{Song} Song et al. \cite{qmapc} (\ref{fig:conventional})  \\ 
\ref*{Feng} Feng et al.$^{\textcolor{red}{*}}$ \cite{feng2022image} (\ref{fig:baseline}) \\ 
\ref*{Torfason} Torfason et al. \cite{torfason2018towards} (\ref{fig:conventional2})}};

\end{tikzpicture}}
    \caption{Cityscapes    }
    \label{fig:cityscapes_results}
  \end{subfigure}
  \caption{ \textbf{Proposed $\bf{JD}$ approach ("Ours") against SOTA approaches} on the mIoU metric for 
(a) $\mathcal{D}_{\mathrm{COCO}}^{\mathrm{val2017}}$
and (b) $\mathcal{D}_{\mathrm{CS}}^{\mathrm{val}}$ datasets. The values denoted by $^{\textcolor{red}{*}}$ are taken from respective papers and the identifiers in parentheses (1x) refer to the type of approach in Figure \ref{fig:conventional} ... \ref{fig:compressed}. On both COCO and Cityscapes datasets, our proposed approach "Ours" achieves better RD trade-off than SOTA baselines at a wide range of bitrates. Note that our proposed approach "Ours" uses $K=1$ (COCO) and $K=3$ (Cityscapes) in the ASPP block.  }
  \label{fig:result_comparison}
\end{figure}

%changed The values denoted by $^{\textcolor{red}{*}}$ are taken from respective papers and the identifiers in parentheses (1x) refer to the type of approach in Figure \ref{fig:conventional} ... \ref{fig:compressed}

Figure \ref{fig:result_comparison} shows the RD performance comparison of our proposed $\bf{JD}$ against recent SOTA and traditional codec baselines including JPEG \cite{jpeg} and HEVC \cite{HEVC}. Following the prior SOTA methods \cite{ahuja2023neural,matsubara2022supervised,feng2022image,torfason2018towards}, we choose \texttt{DeepLabv3} \cite{deeplabv3} with a \texttt{ResNet-50} \cite{resnet50} encoder as the \texttt{No Compression} baseline for both datasets. Further, on COCO, unlike JPEG, HEVC achieves better RD trade-off compared to learned image codec methods \cite{torfason2018towards,qmapc} and compared to the feature compression approach by Singh et al.\cite{singh2020end}.~However, on Cityscapes, both perform worse than learning-based approaches.~The existing SOTA baselines on feature compression by Matsuba et al. \cite{matsubara2022supervised} and particularly Ahuja et al. \cite{ahuja2023neural} outperform conventional approaches (Figures \ref{fig:conventional} and \ref{fig:conventional2}) on both datasets. In the figure, our proposed approach employs over-parameterization for both datasets, see Figure \ref{fig:overparam}. Note that we also include results with a \texttt{ResNet-101} encoder in the Supplement Figure 12 to better prove generalizability of our proposed $\bf{JD}$.

%\ref{fig:result_comparison_101}

\begin{figure}[t!]
  \begin{subfigure}{0.248\textwidth}

    \includegraphics[width=\textwidth]{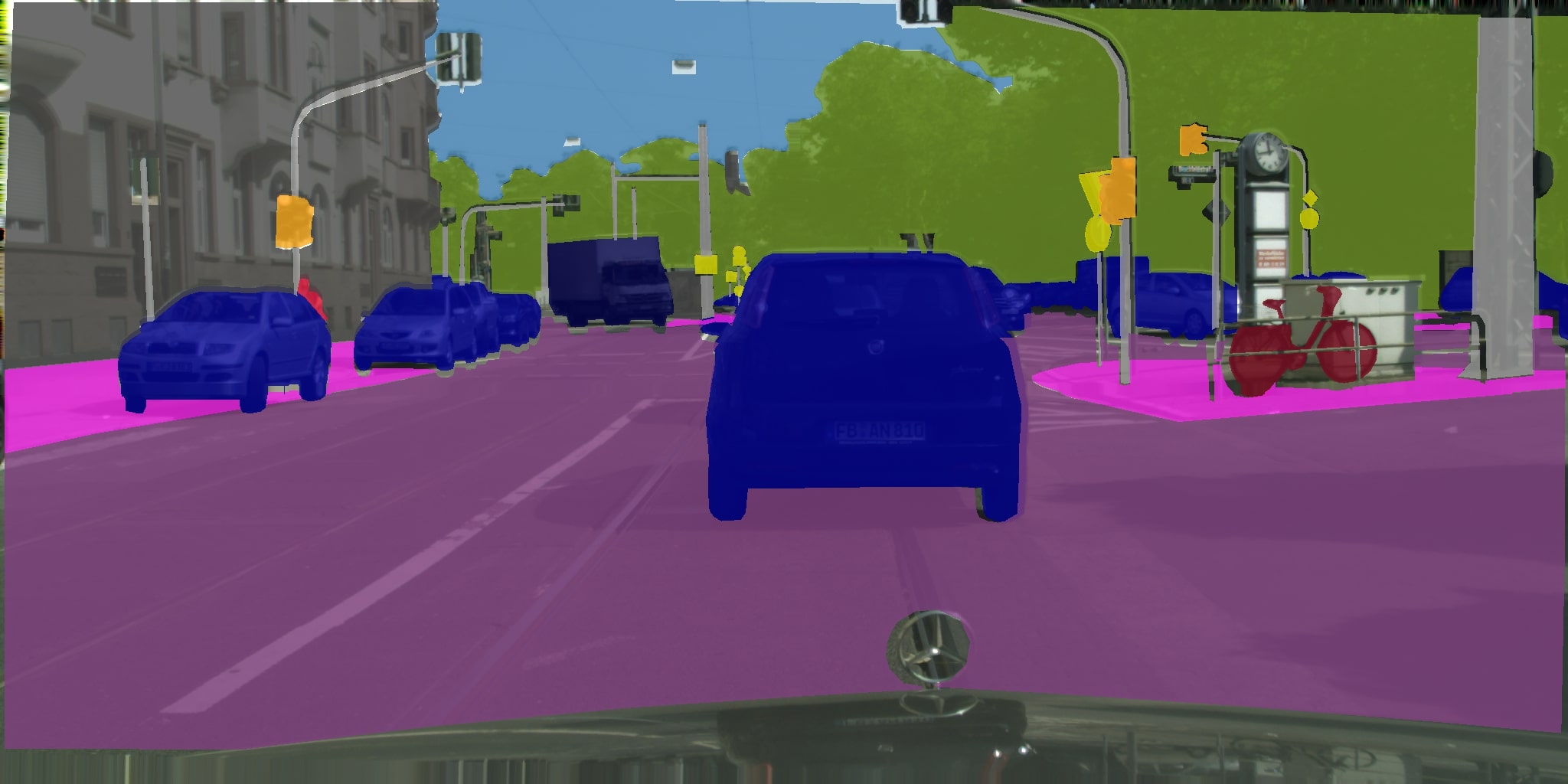} 
    \includegraphics[width=\textwidth]{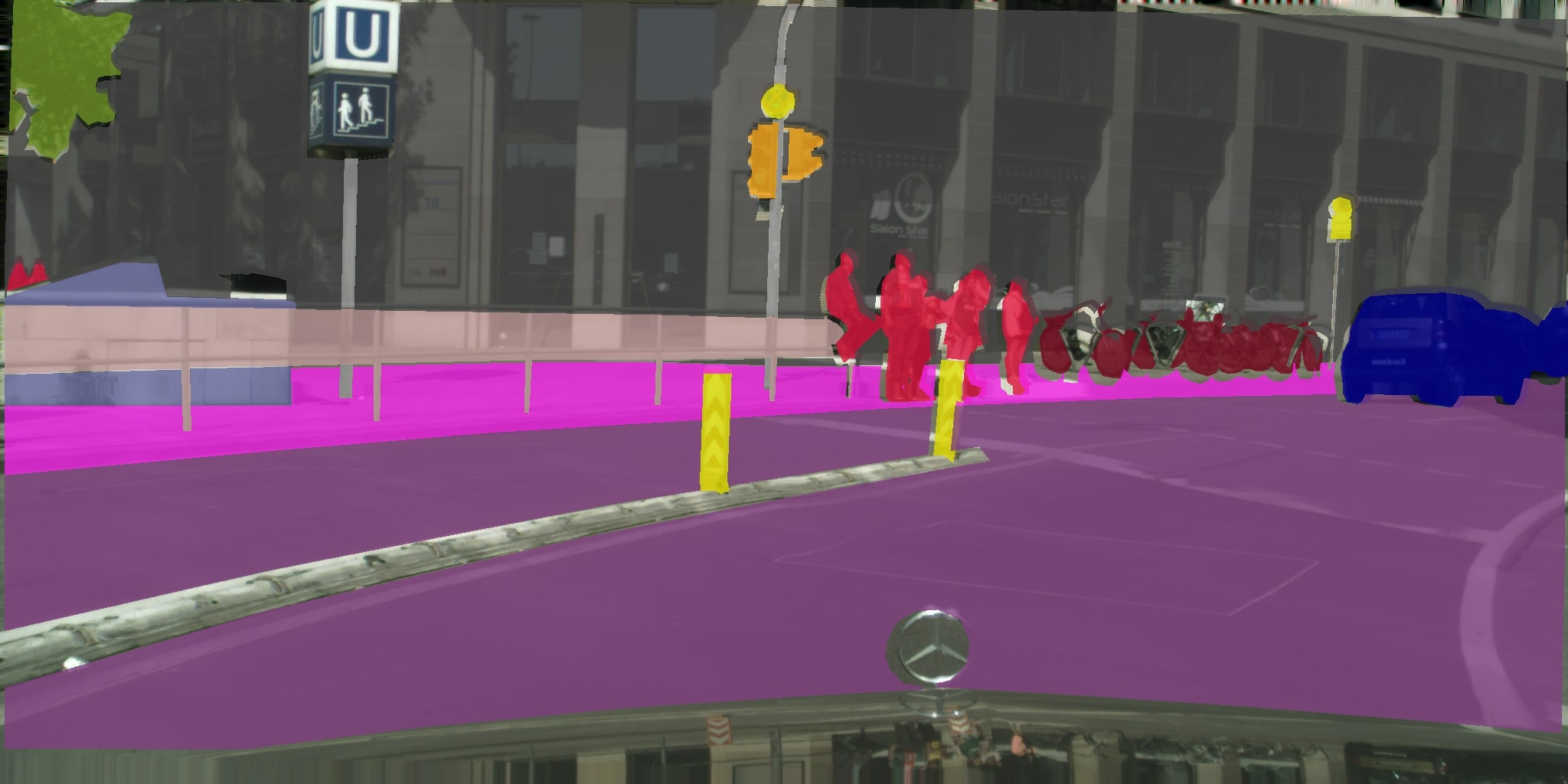}  
    \caption{Ground Truth}
  \end{subfigure}
    \hspace{-0.35em}
  \begin{subfigure}{0.248\textwidth}
    \includegraphics[width=\textwidth]{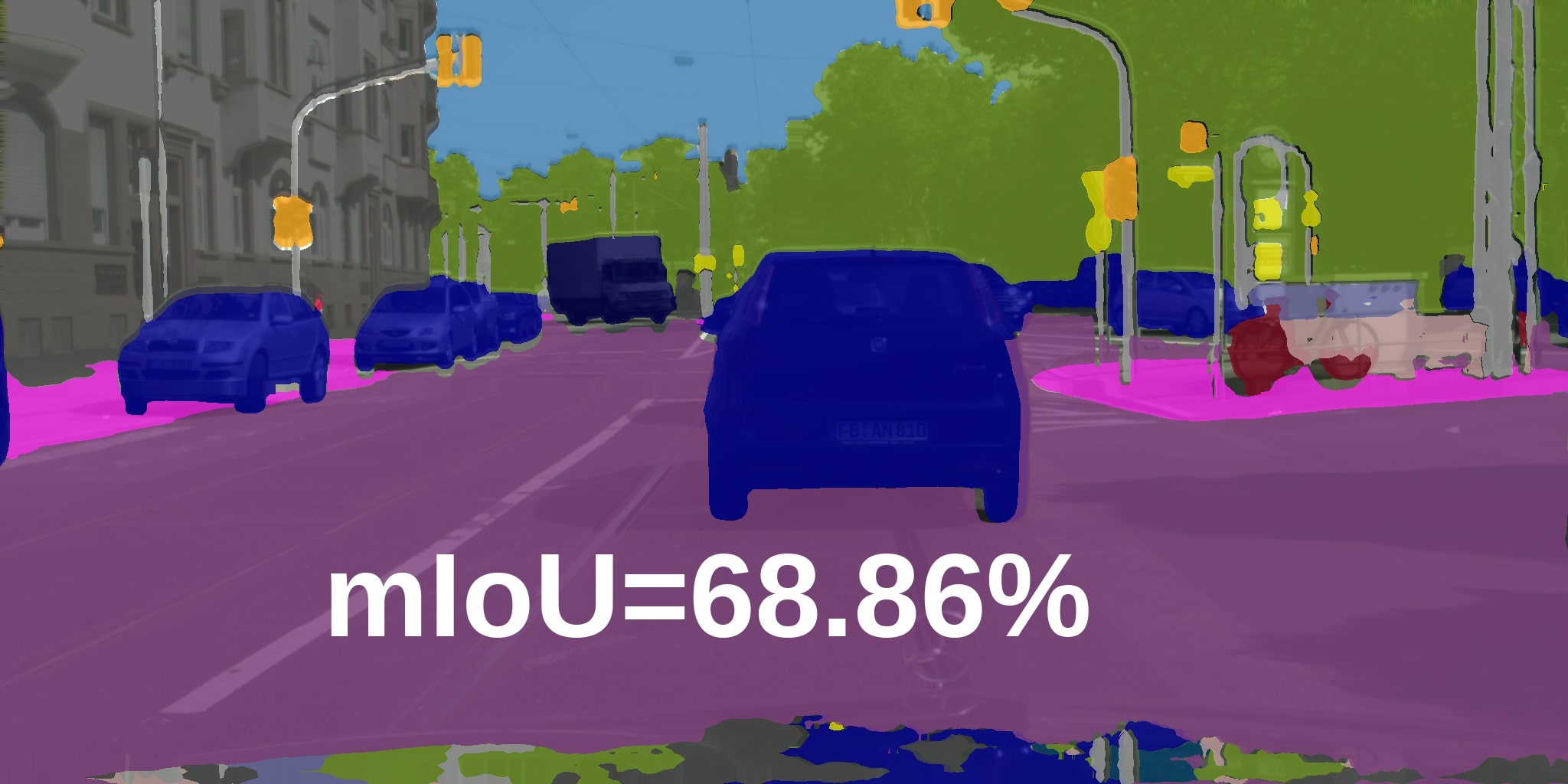} 
        \includegraphics[width=\textwidth]{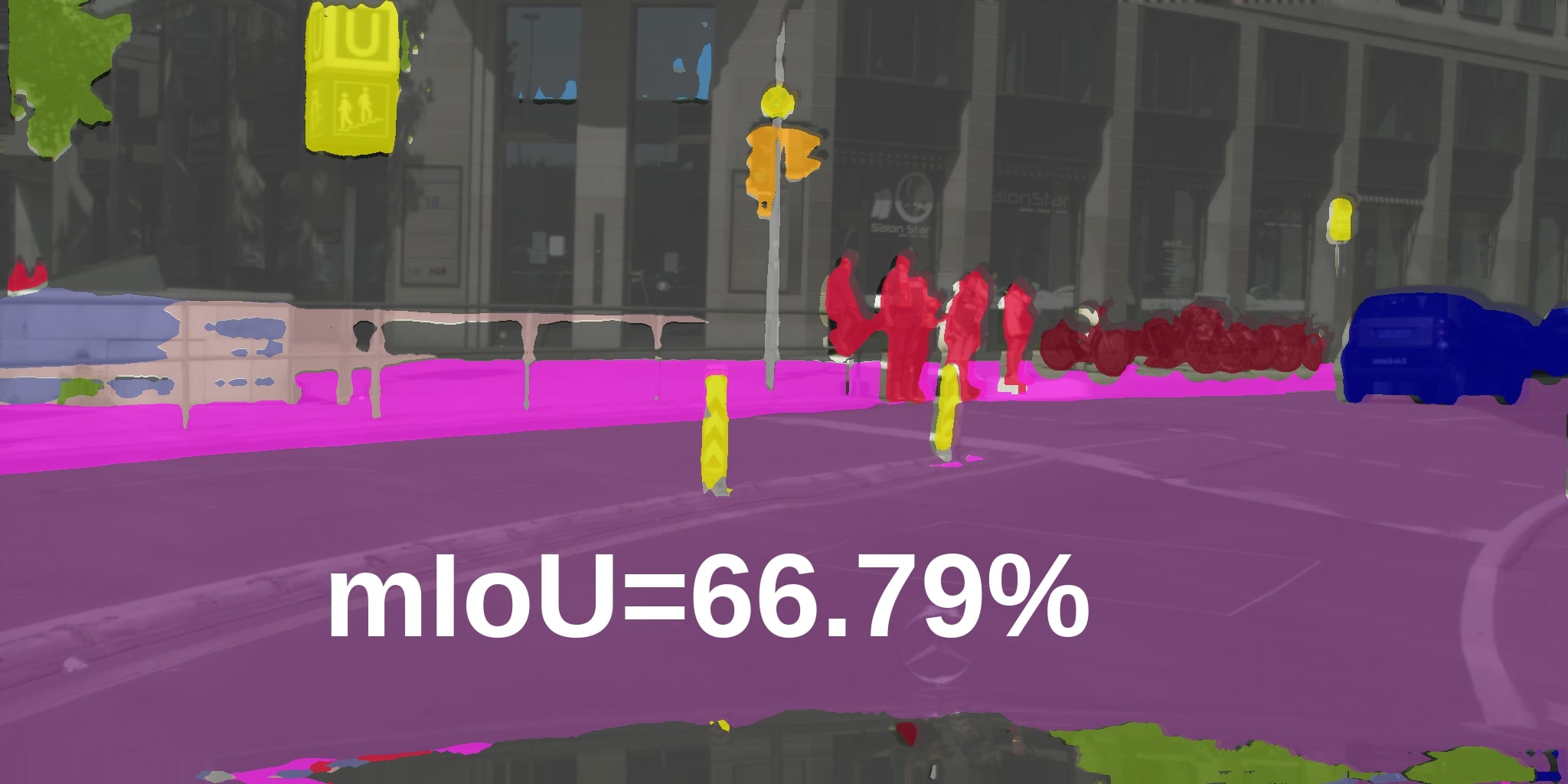} 
 
    \caption{No Compression \cite{deeplabv3}}
  \end{subfigure}
   \hspace{-0.35em}
  \begin{subfigure}{0.248\textwidth}
    \includegraphics[width=\textwidth]{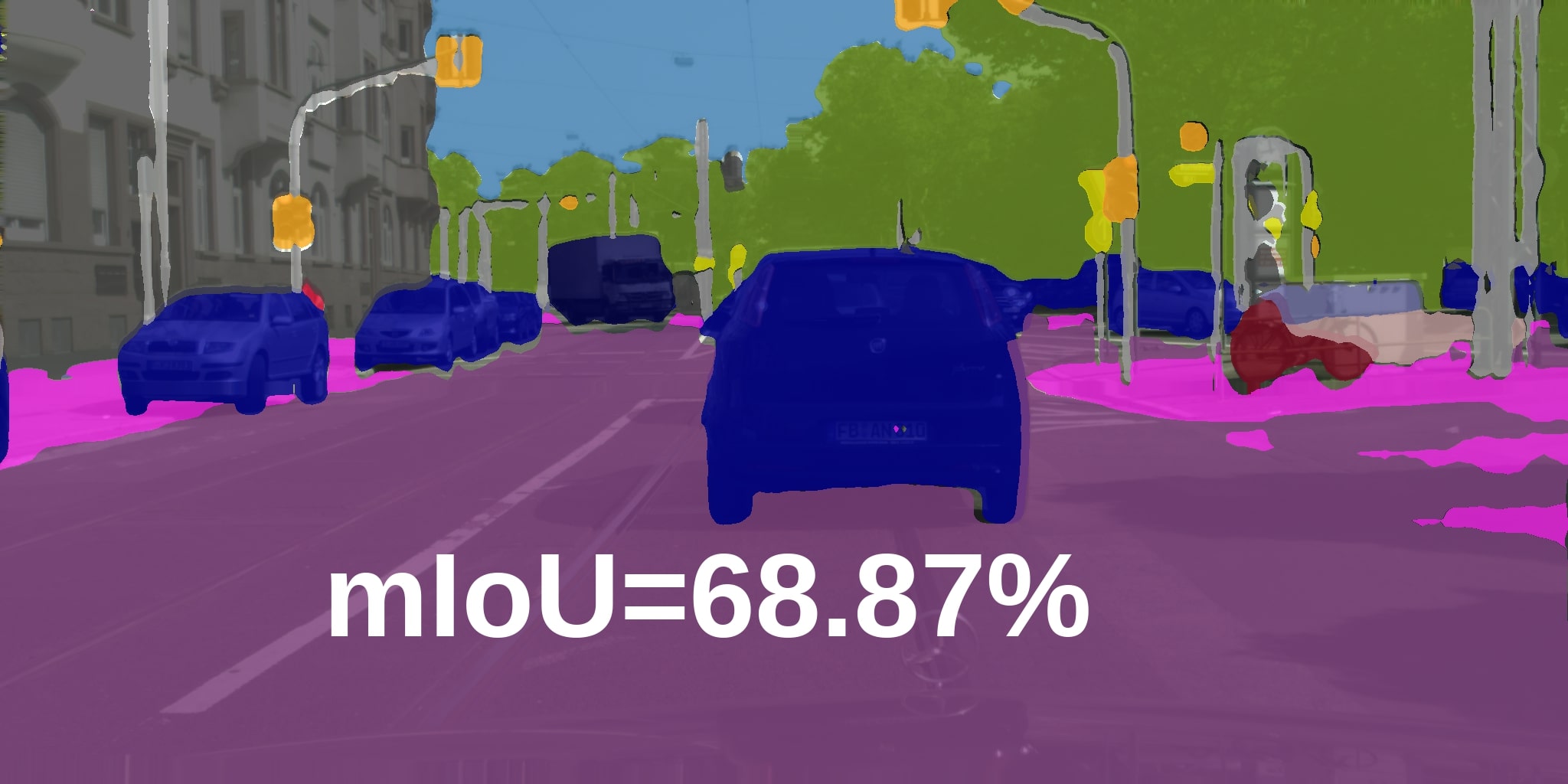} 
        \includegraphics[width=\textwidth]{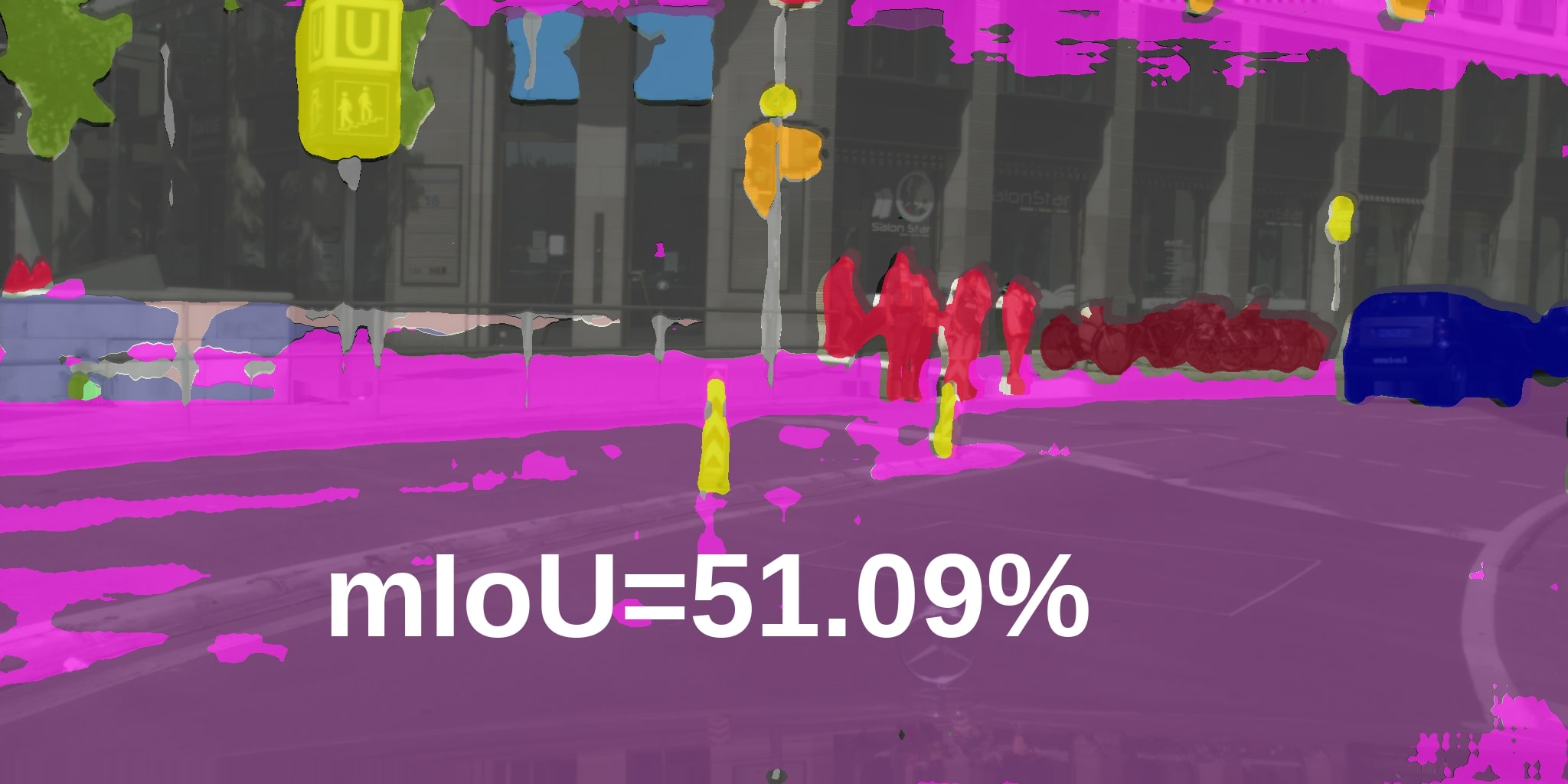}

    \caption{Ahuja \cite{ahuja2023neural} ($0.019$ bpp)}
    \label{fig:ahuja_qualitative}
  \end{subfigure}%
  \hspace{-0.12em}
  \begin{subfigure}{0.248\textwidth}
    \includegraphics[width=\textwidth]{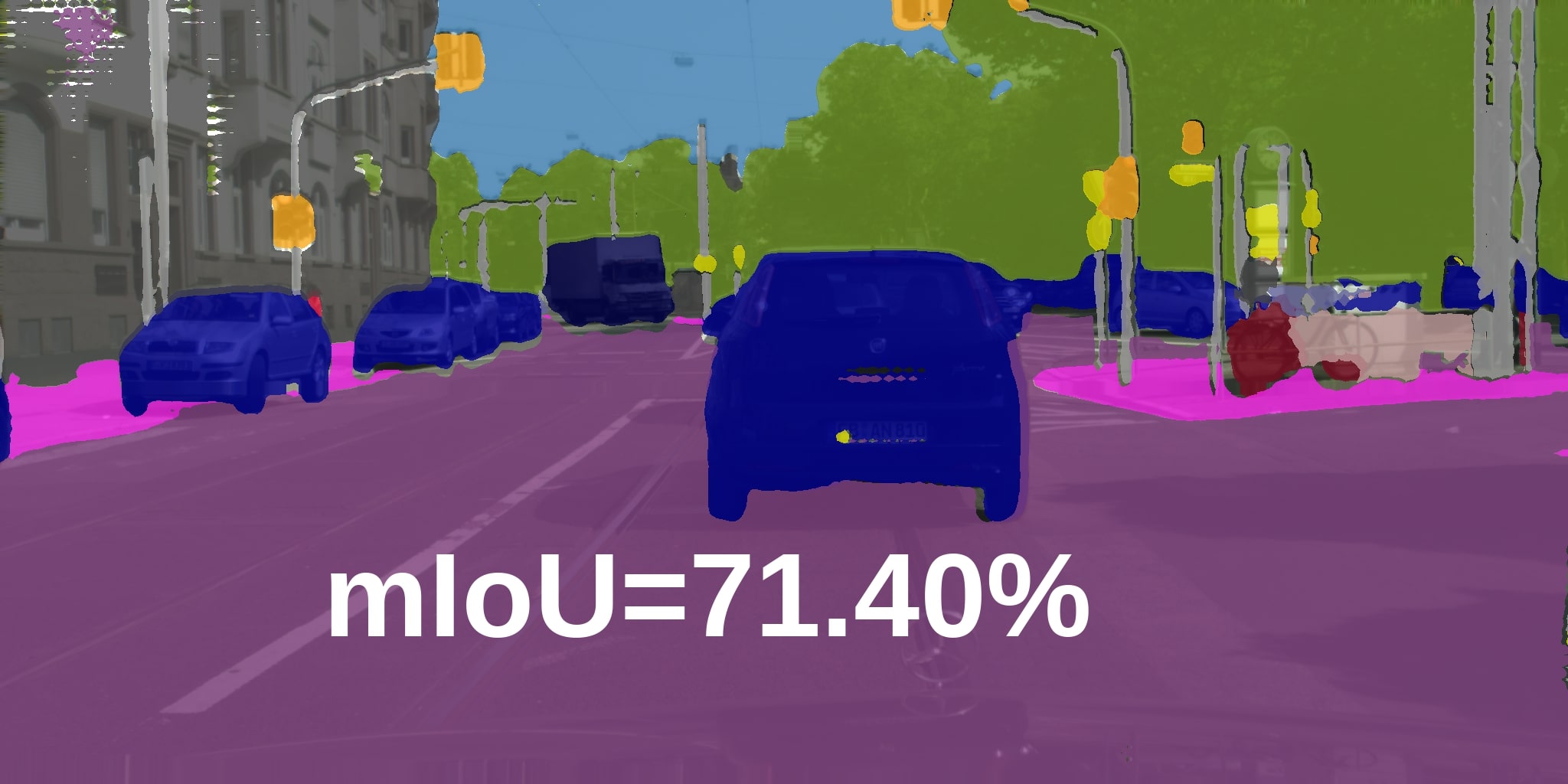} 
        \includegraphics[width=\textwidth]{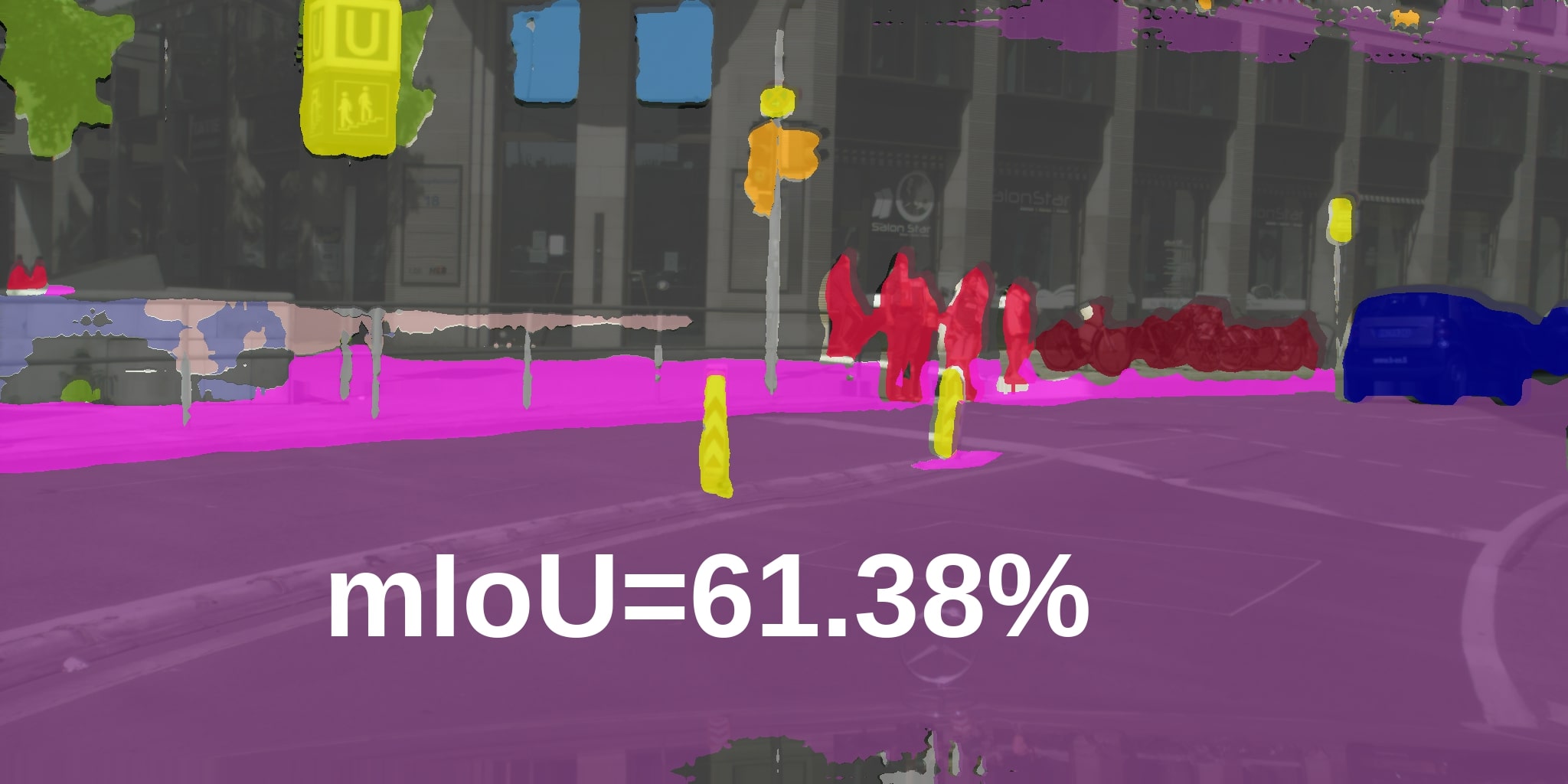}

    \caption{Ours ($0.014 \ \mathrm{bpp}$)}
    \label{fig:ours_qualitative}
  \end{subfigure}

  \caption{\textbf{Qualitative comparison of the proposed $\bf{JD}$ approach ("Ours") against Ahuja et al. \cite{ahuja2023neural} on two Cityscapes samples.} }
  \label{fig:qualitative_result_comparison}
\end{figure}
%\vspace{-0.15cm} % Reduce the amount of space as needed
\begin{figure}[t!]
  \begin{subfigure}{0.515\textwidth}
   \centering
    \resizebox{\linewidth}{!}{\begin{tikzpicture}
\definecolor{ahuja}{RGB}{0,100,0}
%238, 75,43
\definecolor{brightred}{RGB}{238, 75,43}
\definecolor{redbrown}{RGB}{165, 42, 42}
\definecolor{rubyred}{RGB}{224, 17, 95}

% \definecolor{mediumseagreen}{RGB}{238, 75,81}
% \definecolor{mediumgreen}{RGB}{76, 187, 23}
% \definecolor{mediumgreen2}{RGB}{201, 204, 63}

\begin{axis}[
legend cell align={left},
legend cell align={left},
legend cell align={left},
tick align=outside,
tick pos=left,
tick label style={font=\large},
label style={font=\large},
x grid style={white!82.7450980392157!black},
xlabel={bits per pixel (bpp) },
xmajorgrids,
xmin=0, xmax=0.8,
xtick style={color=black},
xtick={0,0.2,0.4,0.6,0.8},
xticklabels={0,0.2,0.4,0.6,0.8},
y grid style={white!82.7450980392157!black},
ylabel={\(\displaystyle \textrm{mIoU on $\mathcal{D}_{\mathrm{COCO}}^{\mathrm{val2017}}$}\) (\%)},
ymajorgrids,
ymin=59.5, ymax=66.6,
 ylabel style={yshift=4pt},
ytick style={color=black},
ytick={59.5,60.5,61.5,62.5,63.5,64.5,65.5,66.5},
yticklabels={
  \(\displaystyle {59.5}\),
  \(\displaystyle {60.5}\),
\(\displaystyle {61.5}\),
  \(\displaystyle {62.5}\),
  \(\displaystyle {63.5}\),
  \(\displaystyle {64.5}\),
  \(\displaystyle {65.5}\),
  \(\displaystyle {66.5}\)
}
]
\addplot [line width=1.5pt, blue, dashed, forget plot]
table {%
0 66.4
1 66.4
};
\label{No_Compression}
\addplot [line width=1.5pt, color=black,  dotted, draw=none,mark=none]
table {%
0.0095 67.5
};
\label{fake}
% \addplot [line width=1.5pt, red, dotted, mark=+, mark size=3.5, mark options={solid,fill opacity=0,line width=1.1}, forget plot]
% table {%
% 0.06 62.33
% 0.09 62.05
% 0.16 63.22
% 0.37 64.83
% 0.74 64.75
% };
% \label{coco_64}
\addplot [line width=1.5pt, red, dotted, mark=*, mark size=3.20, mark options={solid,fill opacity=0,line width=1.1}, forget plot]
table {%
0.05 63.11
0.08 63.1
0.18 63.59
0.33 65.08
0.65 64.97
};
\label{coco_ablation_128}
\addplot [line width=1.5pt, red, dotted, mark=pentagon, mark size=3.50, mark options={solid,line width=1.1,fill opacity=0}, forget plot]
table {%
0.05 63.71
0.06 64.14
0.14 64.59
0.36 65.19
 0.43 65.74
 };
\label{coco_ablation_256}
\addplot [line width=1.5pt, red, dotted, mark=square, mark size=3.050, mark options={solid,fill opacity=0,line width=1.1}, forget plot]
table {%
0.04 61.65
0.06 63.35
0.12 63.68
0.33 66.13
0.71 65.79
};
\label{coco_ablation_512}
\addplot [line width=1.5pt, ahuja, mark=*, mark size=1.6, mark options={solid}, forget plot]
table {%
0.09 62.06
0.12 63.89
0.19 65.02
0.44 66.35
};
\label{coco_ablation_Ahuja}

\addplot [line width=1.5pt, red, mark=*, mark size=2.0, mark options={solid}, forget plot]
table {%
0.0470000505447388 62.6500015258789
0.059999942779541 63.8199996948242
0.10099995136261 64.1900024414062
0.129999995231628 64.6500015258789
0.225000023841858 65.870002746582
0.319000005722046 66.2600021362305
0.634999990463257 66.4300015258789
};
\label{coco_ablation_ours}

\end{axis}

\node [draw,fill=white,font=\small] at (rel axis cs: 0.34,0.26){\shortstack[l]{
\textbf{FLOPs (G)} \\
\ref*{coco_ablation_Ahuja} 517 \\
\ref*{coco_ablation_512} \hspace{0.5em}39 \\
\ref*{coco_ablation_256} \hspace{0.5em}10 \\
\ref*{coco_ablation_ours} \hspace{0.5em}10 \\  
\ref*{coco_ablation_128}\hspace{1em} 5 
%\ref*{coco_64} \hspace{1em}2  
}};

\node [draw,fill=white,font=\small, align=right] at (rel axis cs: 0.753,0.23){\shortstack[l]{
\textbf{$\#$ Params (M)} \\
\ref*{coco_ablation_Ahuja} 16.78, $F=256$  \\
\ref*{coco_ablation_512} \hspace{0.5em}4.91, $F=512$ \\ 
\ref*{coco_ablation_256} \hspace{0.5em}1.66,  $F=256$\\
\ref*{coco_ablation_ours} \hspace{0.5em}1.66,  $F=256$\\
\ref*{coco_ablation_128} \hspace{0.5em}0.43, $F=128$
%\ref*{coco_64} \hspace{0.5em}0.13 
}};

\node [draw, fill=white, font=\small, align=center] at (rel axis cs: 0.73, 0.6) {
    \shortstack[l]{\hspace{2.5em} % Adjust the spacing as needed
      \textbf{Methods} \\
      \ref*{No_Compression} No Compression \\
      \ref*{coco_ablation_ours} Ours \\
      %\raisebox{-0.5ex}{\makebox[1.9em][l]{\rule[0.5ex]{1.9em}{.55pt}}} Ours \\
      \ref*{coco_ablation_Ahuja} Ahuja et al. \cite{ahuja2023neural}
    }
  };
\end{tikzpicture}}
    \caption{COCO    }
    \label{fig:coco_ablation}
  \end{subfigure}
   \hspace{-1em}
  \begin{subfigure}{0.495\textwidth}
     \centering
    \resizebox{\linewidth}{!}{\input{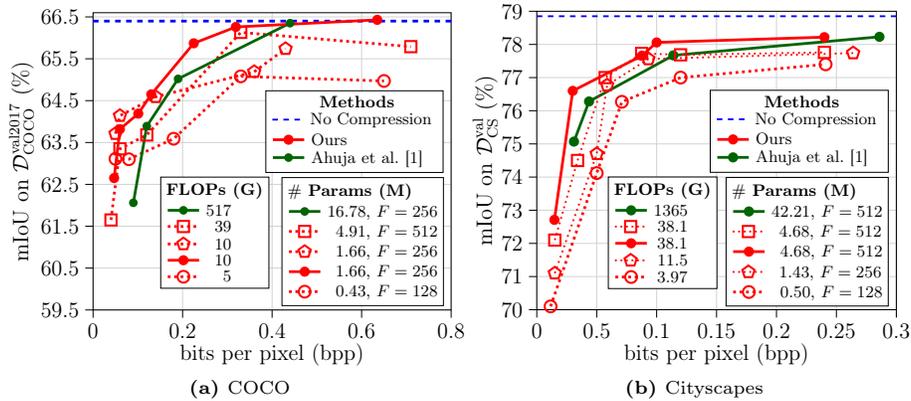}}
    \caption{Cityscapes   }
    \label{fig:cityscapes_ablation}
  \end{subfigure}
  \caption{\textbf{Ablation study on the number $F$ of feature maps of our proposed $\bf{JD}$ approach vs.\ the so-far SOTA approach Ahuja et al.\ \cite{ahuja2023neural}:} \textnormal{Computational complexity per image and model size on (a) $\mathcal{D}_{\mathrm{COCO}}^{\mathrm{val2017}}$ and (b) $\mathcal{D}_{\mathrm{CS}}^{\mathrm{val}}$ datasets.~Each marker type for our approach refers to a different number of feature maps $F$ in the ASPP block of $\mathbf{JD}$. All the curves---except "Ours"---are without over-parameterization in the ASPP block. This is why there is another $F=256$ (COCO) and $F=512$ (Cityscapes) configuration, respectively. Note that all methods with red curves use ($d=5$, $d=10$, $d=15$) in the ASPP block, respectively.}}
  \label{fig:parameter_selection}
\end{figure}

Overall, our proposed $\bf{JD}$ outperforms all existing baselines, including the recent SOTA baseline Ahuja et al.\cite{ahuja2023neural}, on the COCO dataset.~Further, it also achieves the \texttt{No Compression} performance at a bitrate of $0.65 \ \mathrm{bpp}$.~On Cityscapes, our method obtains superior performance in comparison to all baselines, while being comparable to Ahuja et al. \cite{ahuja2023neural} over a wide range of bitrates. As shown in Table \ref{table:flops_params_comparison}, $\bf{JD}$ utilizes only $1.9\%$ ... $2.8\%$ of the FLOPs and $9.8\%$ ... $11.59\%$ of the number of parameters in comparison to the decoder functions $\bf{SD}$ and $\bf{D}$ by Ahuja et al.\ \cite{ahuja2023neural} on COCO and Cityscapes, respectively. As further information, edge device and total computational complexity (edge device plus cloud) are given in Supplement Tables 4 and 5.

%\ref{table:flops_edge} \ref{table:flops_complete}

Figure \ref{fig:qualitative_result_comparison} presents a qualitative comparison of our proposed approach and the SOTA by Ahuja et al. \cite{ahuja2023neural} at very low bitrates, on two Cityscapes samples.~As shown in Figures \ref{fig:ahuja_qualitative} and \ref{fig:ours_qualitative}, the proposed joint source and task decoding with training-time over-parameterization results in a significant reduction of misclassifications, especially in the predictions of object classes such as \textit{sidewalks}, \textit{road} and \textit{poles}. More qualitative results are available in Supplement Section 5. 

%\ref{section:qualitative}

\subsection{Ablation Studies}
%changed We start by varying the number of feature maps $F$ in the ASPP block to investigate the effect of computational complexity on RD performance,followed by a dilation rate case study.~Finally, we ablate over the training-time over-parameterization hyperparameter $K$ to select the number of repetitions. 

In the following, we perform ablation studies to investigate the robustness of the proposed $\bf{JD}$.~We start by varying the number of feature maps $F$ in the ASPP block to investigate the effect of computational complexity on RD performance.~Finally, we ablate over the training-time over-parameterization hyperparameter $K$ to select the number of repetitions. Note that we also ablate over multiple dilation rates to select the dilation rate $d$ in the ASPP dilated convolutional subblocks of $\bf{JD}$ as shown in Supplement Section 6.

%\ref{section:dilation}

\begin{figure}[t!]
  \begin{subfigure}{0.5\textwidth}
    \resizebox{\linewidth}{!}{\begin{tikzpicture}
\definecolor{brightred}{RGB}{238, 75,43}
\definecolor{redbrown}{RGB}{165, 42, 42}
\definecolor{rubyred}{RGB}{224, 17, 95}
\definecolor{terracotta}{RGB}{169, 92, 104}
\begin{axis}[
legend cell align={left},
legend cell align={left},
legend style={
  fill opacity=1.0,
  draw opacity=1,
  text opacity=1,
  at={(0.5,0.02)},
  anchor=south west,
  draw=white!80!black
},
legend style={
  fill opacity=1.0,
  draw opacity=1,
  text opacity=1,
  at={(0.5,0.02)},
  anchor=south west,
  draw=white!80!black
},
tick align=outside,
tick pos=left,
tick label style={font=\large},
label style={font=\large},
x grid style={white!82.7450980392157!black},
xlabel={bits per pixel (bpp) },
xmajorgrids,
xmin=0, xmax=0.7,
xtick style={color=black},
xtick={0,0.2,0.4,0.6},
xticklabels={0,0.2,0.4,0.6},
y grid style={white!82.7450980392157!black},
ylabel={\(\displaystyle \textrm{mIoU on $\mathcal{D}_{\mathrm{COCO}}^{\mathrm{val2017}}$}\) (\%)},
ylabel style={yshift=-7pt},
ymajorgrids,
ymin=61, ymax=66.6,
ytick style={color=black},
ytick={61,62,63,64,65,66},
yticklabels={
  \(\displaystyle {61}\),
  \(\displaystyle {62}\),
  \(\displaystyle {63}\),
  \(\displaystyle {64}\),
  \(\displaystyle {65}\),
  \(\displaystyle {66}\)
}
]
\addplot [line width=1.5pt, blue, dashed, forget plot]
table {%
0 66.4
0.7 66.4
};
\label{k_no_compression}

\addplot [line width=1.5pt, red, mark=*, mark size=2, mark options={solid}, forget plot]
table {%
0.0470000505447388 62.6500015258789
0.059999942779541 63.8199996948242
0.10099995136261 64.1900024414062
0.129999995231628 64.6500015258789
0.225000023841858 65.870002746582
0.319000005722046 66.2600021362305
0.634999990463257 66.4300015258789
};
\label{k_1}
\addplot [line width=1.5pt,  red, dotted, mark=square, mark size=3.05, mark options={solid,fill opacity=0,line width=1.1}, forget plot]
table {%
0.047 62.60
0.08 64.02
0.24 65.47
0.32 66.06
0.65 66.13
};
\label{k_2}
% \addplot [line width=1.5pt, red, dotted, mark=+, mark size=3.5, mark options={solid,fill opacity=0,line width=1.1}, forget plot]
% table {%
% 0.05 62.43
% 0.15 63.65
% 0.2 64.87
% 0.28 66.12
% 0.6 66.13
% };
% \label{k_3}
% \addplot [line width=1.5pt, red, dotted, mark=x, mark size=3.5, mark options={solid,fill opacity=0,line width=1.1}, forget plot]
% table {%
% 0.06 62.04
% 0.11 63.65
% 0.21 64.87
% 0.29 66.08
% 0.62 66.28
% };
% \label{k_4}
% \addplot [line width=1.5pt, red, dotted, mark=star, mark size=3.5, mark options={solid,fill opacity=0,line width=1.1}, forget plot]
% table {%
% 0.08 62.93
% 0.12 63.1
% 0.26 65.1
% 0.3 65.33
% 0.64 65.8
% };
% \label{k_5}
% 
\end{axis}

 \node [draw, fill=white, font=\small, align=center] at (rel axis cs: 0.68, 0.55) {
    \shortstack[l]{\hspace{2.5em} % Adjust the spacing as needed
      \textbf{Methods} \\
      \ref*{k_no_compression} No Compression$^{\textcolor{red}{*}}$ \cite{deeplabv3} \\
       \ref*{k_1} Ours 
    }
  };

\node [draw,fill=white,font=\small] at (rel axis cs: 0.772,0.30){\shortstack[l]{
 \textbf{Repetitions}$\ (K)$ \\ \ref*{k_1} 1  \\ \ref*{k_2} 2 %\\
% \ref*{k_3} 3 \\ \ref*{k_4} 4 \\ \ref*{k_5} 5
}};

\end{tikzpicture}}
    \caption{COCO }
    \label{fig:coco_ablation_repeatition}
  \end{subfigure}
  \begin{subfigure}{0.5\textwidth}%0.465,0.95
    \resizebox{\linewidth}{!}{\input{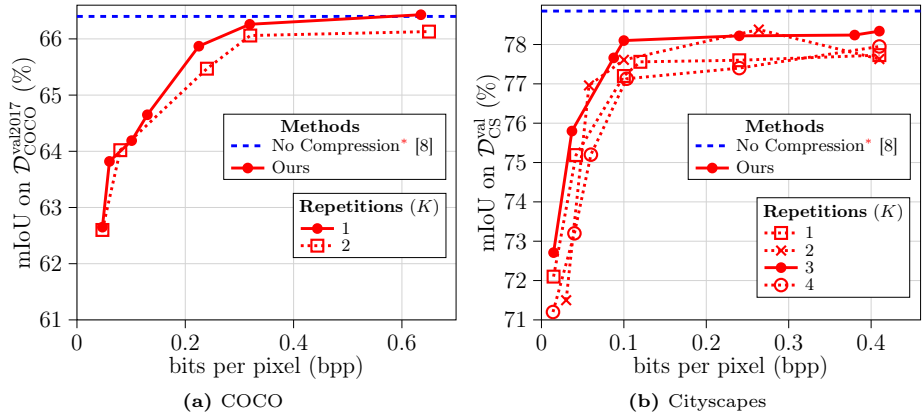}}
    \caption{Cityscapes}
    \label{fig:cityscapes_ablation_repeatition}
  \end{subfigure}
  \caption{\textbf{Ablation study on the number of repetitions ($K$) for over-parameterization explained in Figure \ref{fig:overparam} in the proposed $\mathbf{JD}$ } on (a) $\mathcal{D}_{\mathrm{COCO}}^{\mathrm{val2017}}$ and (b) $\mathcal{D}_{\mathrm{CS}}^{\mathrm{val}}$.~Each marker type shows a different $K$ in $\mathbf{JD}$. Note that all methods use dilations ($d=5$, $d=10$, $d=15$) and $F=256$ (COCO) and $F=512$ (Cityscapes) in the ASPP block, respectively.  
  }
  \label{fig:repeatetion_ablation}
\end{figure}

\subsubsection{Parameter selection\normalfont:}Figure \ref{fig:parameter_selection} refers to the change in the number of output feature maps $F$ in all convolutional layers of the $\bf{JD}$ ASPP block shown in Figure \ref{fig:CD}.~It also results in the change of the size in the following pointwise convolution, which takes the concatenated output of the ASPP block. For a single setting, $F$ remains the same for all convolutional layers in the ASPP block.~By varying $F$, it is possible to achieve different $\bf{JD}$ computational  complexities per image.~We ablate over $F \in \{128,256,512\}$ for both datasets. All curves, except "Ours", are without over-parameterization in the ASPP block. 

As shown in Figure \ref{fig:coco_ablation}, $F=256$ is the best configuration for low bitrates.~However, on higher bitrates, $F=512$ performs better. To make a fair comparison with current baseline approaches \cite{ahuja2023neural,matsubara2022supervised}, we select $F=256$ for further ablation studies on the COCO dataset. On Cityscapes, as shown in Figure \ref{fig:cityscapes_ablation}, we select $F=512$ as it produces the best performance both at low and high bitrates. Note that the baselines on Cityscapes \cite{feng2022image,ahuja2023neural} also utilize $F=512$.

\subsubsection{Number of repetitions in $\bf{JD}$\normalfont:}Figure \ref{fig:repeatetion_ablation} depicts the ablation study to select the number of repetitions $K$ for the over-parameterization (Figure \ref{fig:overparam}) in the ASPP convolutional subblocks of $\bf{JD}$.~Each marker type shows a different number of repetitions in the subblocks of the ASPP block.~We perform up to two repetitions on COCO and up to four repetitions on Cityscapes, respectively.~As shown in Figure \ref{fig:coco_ablation_repeatition}, we observe that $K=1$ produces a similar performance to $K=2$ at low bitrates, but  $K=1$ is better at high bitrates on COCO. As shown in  Figure \ref{fig:cityscapes_ablation_repeatition}, $K=3$ outperforms other curves in RD performance at most bit\-rates. Note that for COCO, our method at $K=1$ already beats the previous SOTA \cite{ahuja2023neural}, therefore we didn't perform a higher number of repetitions.

\section{Conclusions}

In this work, we show how \textit{joint} source and task decoding can result in a highly efficient cloud DNN, while maintaining the same edge device computational complexity as the current SOTA for distributed semantic segmentation. This allows us to scale distributed semantic segmentation up to a large number of edge devices, without putting a high computational burden onto the cloud. Further, instead of applying training-time over-parameterization in the image encoder as suggested by \cite{mobileone,repvgg,fastvit}, we demonstrate that by utilizing it in our proposed joint decoder $\bf{JD}$, we further improve the performance in the high-bitrate regime. We achieve SOTA performance over a wide range of bitrates on the mean intersection over union metric, while using only $9.8 \%$ ... $11.59 \%$ of the cloud DNN parameters used in the previous SOTA on the semantically diverse COCO and Cityscapes datasets.

\section{Limitations and Future Work}

Since not all general semantic segmentation architectures are suitable for distributed deployment, current works only employ the $\texttt{DeepLabv3}$ network topology. Even though it is still considered as a very strong distributed semantic segmentation baseline \cite{matsubara2022supervised}, it is not a SOTA method in general semantic segmentation, as discussed in Section \ref{sect:general_semantic_segmentation}. As part of future work, we and the community should aim to bridge this gap by adapting existing general semantic segmentation approaches for such a distributed setting.

%\clearpage  % TODO REVIEW/FINAL: This \clearpage needs to be removed from both review and camera-ready versions.

\bibliographystyle{splncs04}
\bibliography{main}
% \addtocounter{figure}{-1}
% \refstepcounter{figure}\label{LASTFIGURE}
\end{document}

\title{Supplementary: \\ Distributed Semantic Segmentation with\\ Efficient Joint Source and Task Decoding} 

\titlerunning{Distributed Semantic Segmentation with Efficient Joint Decoding}

\author{Danish Nazir\inst{1,2}\orcidlink{0000-1111-2222-3333} \and
Timo Bartels\inst{1}\orcidlink{1111-2222-3333-4444} \and
Jan Piewek\inst{2}\orcidlink{2222--3333-4444-5555} \and
Thorsten Bagdonat\inst{2}\orcidlink{1111-2222-3333-4444} \and
Tim Fingscheidt\inst{1}\orcidlink{0000-0002-8895-5041} 
}

\authorrunning{D.~Nazir et al.}

\institute{Technische Universität Braunschweig, Braunschweig, Germany \\
\email{\{danish.nazir,timo.bartels,t.fingscheidt\}@tu-bs.de}
\and
Group Innovation, Volkswagen AG, Wolfsburg, Germany
\email{\{danish.nazir,jan.piewek,thorsten.bagdonat\}@volkswagen.de} }

\maketitle

In the supplementary material, we provide a detailed overview of training and evaluation settings along with hyperparameters in Section \ref{section:training_details}.~Further, in Section \ref{section:topology_details}, we provide the topological details of the baseline approach and of our edge device processing (transmitter), followed by computational complexity discussion in Section \ref{section:computational_complexity}. Section \ref{section:resnet_101} provides additional rate-distortion (RD) results with a \texttt{ResNet-101} encoder. Section \ref{section:qualitative} provides qualitative results on both COCO and Cityscapes datasets. Finally, Section \ref{section:dilation} provides a dilation rate ablation study of our proposed joint decoder ($\bf{JD}$) on both COCO and Cityscapes datasets.

\section{Training Details}
\label{section:training_details}
In the following section, we present a detailed description of the employed hyper\-parameters.~Note that, for the ease of use and scalability, we integrated the $\texttt{Compress-}\texttt{AI}$ library into the $\texttt{MMSegmentation}$ toolbox to make a hybrid version and used it for conducting all of our experiments. 
\subsubsection{Optimization strategy\textnormal:} For all of our trainings, on both datasets, we utilize two optimize\-rs, includ\-ing the main optimizer and an auxiliary optimizer as suggested by $\texttt{Compress-}\texttt{AI}$ \cite{begaint2020compressai,balle2018variational}.~The $\texttt{EntropyBottleneck.quantiles}$ parameter is optimized by the auxiliary optimizer, whereas remaining parameters are optimized by the main optimizer.~We select \texttt{Adam} for both main and auxiliary optimizers.~Further, except learning rate, all of the other parameters of both optimizers are exactly the same for both datasets.
\subsubsection{Hyperparameters\textnormal:}We use $\texttt{Python}$ v.3.8.10, $\texttt{PyTorch}$ v.1.9.1 , $\texttt{MMSegmentation}$ v.1.0.0rc3 and $\texttt{Compress-AI}$ v.0.8 for all our experiments.~All of the hyperparameter details for reproducing our results are given in Table \ref{table:hyperparameters}.~Further, to conduct all our trainings, we utilized 4 \texttt{NVIDIA-V100} GPUs and report the effective batch size.~The polynomial learning rate schedule is given as follows:

\setcounter{equation}{3}
\begin{equation}
\eta(\tau) = \eta_{0}  \bigg(1- \frac{\tau}{\tau_{\mathrm{max}}}  \biggr)^{0.9},
\label{Eq:polynomial}
\end{equation}
with $\eta(\tau)$ being the learning rate at optimizer step $\tau$ and $\eta_{0}$ represents the initial learning rate given in Table \ref{table:hyperparameters}.~The maximum number of iterations/epochs are given by $\tau_{\mathrm{max}}$.

\setcounter{table}{2}
\begin{table}[t!]
  \caption{\textbf{Settings and hyperparameters} used for training on COCO and on Cityscapes.}
  \label{tab:headings}
  \centering
  \begin{tabular}{@{}lcc@{}}
    \toprule
     \centering \textbf{Setting/Hyperparameter} &  Cityscapes & COCO  \\
    \midrule
     
     $\#$ of training iterations  & 80,000  & -  \\ 
    
     $\#$ of epochs   &  - & 42  \\ 

     Initial LR ($\eta_{0}$) (main)   & 0.001 & 0.01  \\ 

     Initial LR ($\eta_{0}$) (aux)  & 0.001 & 0.001  \\ 

     Learning rate (LR) schedule $\eta_{0}(\tau)$   & polynomial (\ref{Eq:polynomial})  & polynomial (\ref{Eq:polynomial})   \\ 

     Batch size   & 8 & 16  \\ 

     Random init   & Kaiming initialization  & Kaiming initialization    \\ 

     Clip grad type   & norm  & norm  \\ 

     Clip grad value   & 1.0  & 1.0  \\ 
     
     Optimizer parameters $\beta_{1}$, $\beta_{2}$   & 0.9, 0.99 & 0.9, 0.99  \\ 

     Weight decay   & 0 & 0  \\ 
  \bottomrule
  \end{tabular}
\label{table:hyperparameters}
\end{table}
\setcounter{figure}{8}
\begin{figure}[t!]
    \hspace{-4.8em}
    \resizebox{1.12\linewidth}{!}{\input{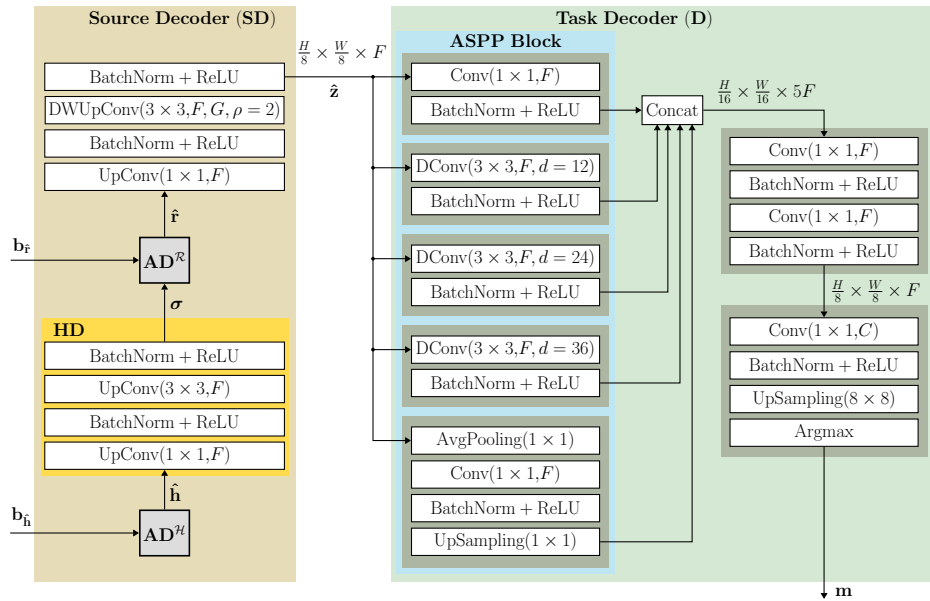}} 
  \caption{\textbf{Reciever-sided architecture of Ahuja et al. \cite{ahuja2023neural}} with source decoder $\bf{SD}$ and task decoder $\bf{D}$ (see Figure 1c, and compare it to Figures 1d and 3)}
  \label{fig:decoder}
\end{figure}

%\ref{fig:baseline} and compare it to Figures \ref{fig:compressed} and \ref{fig:CD}

\begin{figure}[t!]
    \hspace{-2em}
    \resizebox{1.1\linewidth}{!}{\input{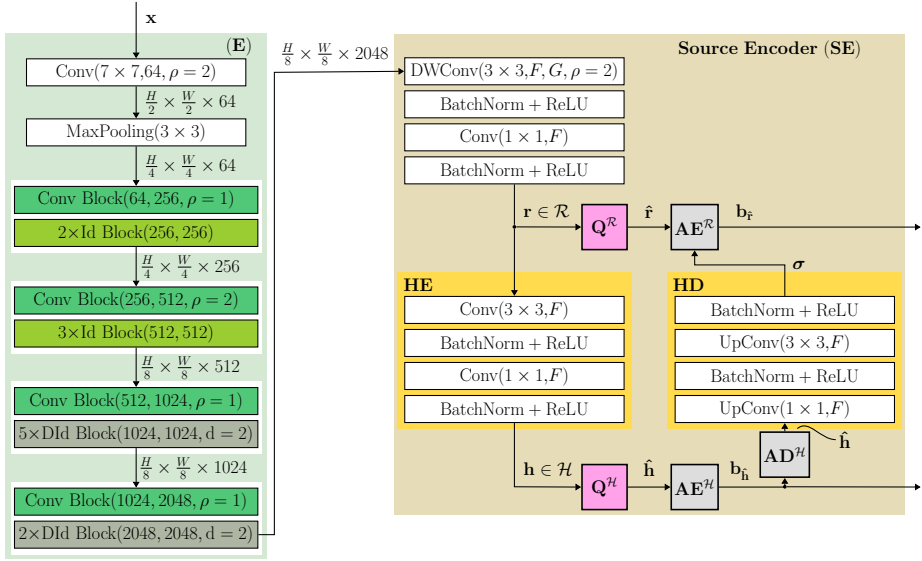}} 
  \caption{Details of the \textbf{edge device processing (transmitter)} diagram with a semantic segmentation image encoder $\bf{E}$ and a source encoder $\bf{SE}$ in our experiments.~This transmitter structure is used to generate the baseline results by Ahuja et al. \cite{ahuja2023neural} and by our proposal.~The subblocks of $\bf{E}$ are drawn in Figure \ref{fig:image_encoder_subblocks}.~For cloud processing (receiver), see Figure \ref{fig:decoder} for the baseline method by Ahuja et al. \cite{ahuja2023neural} and Figure 3 for our proposed approach.}
  \label{fig:resnet_with_se}
\end{figure}
%\ref{fig:CD}

\begin{figure}[t!]
    \hspace{-1.38em}
    \resizebox{1.035\linewidth}{!}{\usetikzlibrary{positioning,shapes,arrows,arrows.meta,fit,backgrounds,calc}
\definecolor{resnet_block}{RGB}{173, 182, 162}
\definecolor{resnet_conv_block}{RGB}{80,200,120}
\definecolor{resnet_did_block}{RGB}{173, 182, 162}
\definecolor{resnet_id_block}{RGB}{154, 205, 50}
\tikzstyle{process} = [rectangle,minimum width=5cm, minimum height=1cm, text
centered, draw=black, line width = 1pt, fill=white]
 \tikzstyle{waypoint}=[fill,circle,minimum size=3.5pt,inner sep=0pt]
 
\newcommand{\Plus}{\mathord{\begin{tikzpicture}[baseline=0ex, line width=2, scale=0.2]
\draw (1,0) -- (1,2);
\draw (0,1) -- (2,1);
\end{tikzpicture}}}

\begin{tikzpicture}[font=\LARGE,node distance = 1.5cm]
\tikzstyle{arrow} = [-Triangle, line width=1pt]
\node (point) {};

\node (convblock) [draw, rectangle, minimum width=13cm, minimum height=8.8cm,  line width=0pt,draw=none, yshift=0cm,xshift=0cm,fill=resnet_conv_block] at (point)   {};
\node (idblock) [draw, rectangle, minimum width=6.7cm, minimum height=8.801cm,  line width=0pt,draw=none, yshift=0cm,xshift=9cm,fill=resnet_id_block,right of=convblock ]   {};;
\node (didblock) [draw, rectangle, minimum width=6.7cm, minimum height=8.801cm,  line width=0pt,draw=none,yshift=0cm,xshift=5.9cm,fill=resnet_did_block,right of=idblock ]   {};;

\node [above of=convblock,xshift=-4.5cm,yshift=2.4cm, font=\LARGE] (heading) {$\mathbf{Conv \ Block }$};
\node [above of=idblock,xshift=-1.9cm,yshift=2.4cm, font=\LARGE] (heading) {$\mathbf{Id \ Block }$};
\node [above of=didblock,xshift=-2.55cm,yshift=2.4cm, font=\LARGE] (heading) {$\bf{DId}$};
\node [below of=heading,xshift=0.25cm,yshift=0.9cm, font=\LARGE] (heading) {$\bf{Block}$};

\node (cb_conv1_1) [process, above of=convblock,yshift=0.85cm,minimum width=5.6cm,minimum height=0.9cm,xshift=-3.5cm] {$\mathrm{Conv(1 \times 1, }F)$};
\node (cb_conv1_2) [process, right of=cb_conv1_1,above of=convblock,yshift=0.85cm,minimum width=5.6cm,minimum height=0.9cm,xshift=2cm] {$\mathrm{Conv(1 \times 1, }F,\rho=2)$};
\node (cb_conv3_1) [process, below of=cb_conv1_1,yshift=-0.3cm,minimum width=5.6cm,minimum height=0.9cm,] {$\mathrm{Conv(3 \times 3, }F,\rho=2)$};
\node (cb_conv1_3) [process, below of=cb_conv3_1,yshift=-0.3cm,minimum width=5.6cm,minimum height=0.9cm,] {$\mathrm{Conv(1 \times 1, }F)$};

\node (id_conv1_1) [process, above of=idblock,yshift=0.8cm,minimum width=5.6cm,minimum height=0.9cm,xshift=-0.3cm] {$\mathrm{Conv(1 \times 1, }F)$};
\node (id_conv3_1) [process, below of=id_conv1_1,yshift=-0.3cm,minimum width=5.6cm,minimum height=0.9cm,] {$\mathrm{Conv(3 \times 3, }F)$};
\node (id_conv1_3) [process, below of=id_conv3_1,yshift=-0.3cm,minimum width=5.6cm,minimum height=0.9cm,] {$\mathrm{Conv(1 \times 1, }F)$};

\node (did_conv1_1) [process, above of=didblock,yshift=0.8cm,minimum width=5.6cm,minimum height=0.9cm,xshift=-0.3cm] {$\mathrm{Conv(1 \times 1, }F)$};
\node (did_conv3_1) [process, below of=did_conv1_1,yshift=-0.3cm,minimum width=5.6cm,minimum height=0.9cm,] {$\mathrm{DConv(3 \times 3, }F,d)$};
\node (did_conv1_3) [process, below of=did_conv3_1,yshift=-0.3cm,minimum width=5.6cm,minimum height=0.9cm,] {$\mathrm{Conv(1 \times 1, }F)$};

\node (plus_did) [below of=did_conv1_3,draw, fill=white,line width = 1pt, rectangle,xshift=0cm,minimum width=1.2cm,minimum height=0.9cm,yshift=-0.2cm]  {$\Plus$};
\node (out_did) [ below of=plus_did,yshift=-0.3cm] {};

\node (plus_id) [below of=id_conv1_3,draw, fill=white,line width = 1pt, rectangle,xshift=0cm,minimum width=1.2cm,minimum height=0.9cm,yshift=-0.2cm]  {$\Plus$};
\node (out_id) [ below of=plus_id,yshift=-0.3cm] {};

\node(plus)[below of=cb_conv1_3,draw, fill=white,line width = 1pt, rectangle,xshift=0cm,minimum width=1.2cm,minimum height=0.9cm,yshift=-0.2cm] (plus) {$\Plus$};
\node (out) [ below of=plus,yshift=-0.33cm,] {};

\node (way)[waypoint,yshift=1.cm] at ($(cb_conv1_1)!0.5!(cb_conv1_2)$) {};
\node (input) [process, above of=way,yshift=0.3cm,minimum width=1cm,minimum height=0.9cm,xshift=0cm,draw=none,fill=none] {};

\node (way_id)[waypoint,right of=way,xshift=8.7cm]  {};
\node (input_id) [process, right of=input,yshift=0cm,minimum width=1cm,minimum height=0.9cm,xshift=8.7cm,draw=none,fill=none] {};

\node (way_did)[waypoint,right of=way_id,xshift=5.9cm]  {};
\node (input_did) [process, right of=input_id,yshift=0cm,minimum width=1cm,minimum height=0.9cm,xshift=5.9cm,draw=none,fill=none] {};

\draw[arrow,-] (input) to node[midway, xshift=1.8cm,yshift=-0.15cm,font=\LARGE] {$\frac{H}{4} \times \frac{W}{4} \times F$}    (way);

\draw[arrow] (way) -- ++(-3.5,0)  --   (cb_conv1_1.north);
\draw[arrow] (way) -- ++(3.5,0)  --   (cb_conv1_2.north);
\draw[arrow] (cb_conv1_1) to   (cb_conv3_1);
\draw[arrow] (cb_conv3_1) to   (cb_conv1_3);
\draw[arrow] (cb_conv1_3) to   (plus);
\draw[arrow] (cb_conv1_2.south) -- ++(0,0) --++(0,-4.82)  --   (plus.east);
\draw[arrow] (plus) to node[midway, xshift=2cm,yshift=0.15cm,font=\LARGE] {$\frac{H}{8} \times \frac{W}{8} \times F$}   (out);

%id arrows
\draw[arrow,-] (input_id) to node[midway, xshift=1.8cm,yshift=-0.15cm,font=\LARGE] {$\frac{H}{4} \times \frac{W}{4} \times F$}    (way_id);
\draw[arrow] (way_id) to  (id_conv1_1);
\draw[arrow] (id_conv1_1) to  (id_conv3_1);
\draw[arrow] (id_conv3_1) to  (id_conv1_3);
\draw[arrow] (id_conv1_3) to  (plus_id);
\draw[arrow] (plus_id) to node[midway, xshift=2cm,yshift=0.2cm,font=\LARGE] {$\frac{H}{4} \times \frac{W}{4} \times F$}   (out_id);
\draw[arrow] (way_id) -- ++(3.1,0) --++(0,-6.35)  --   (plus_id.east);

%did arrows
\draw[arrow,-] (input_did) to node[midway, xshift=1.8cm,yshift=-0.15cm,font=\LARGE] {$\frac{H}{4} \times \frac{W}{4} \times F$}    (way_did);
\draw[arrow] (way_did) to  (did_conv1_1);
\draw[arrow] (did_conv1_1) to  (did_conv3_1);
\draw[arrow] (did_conv3_1) to  (did_conv1_3);
\draw[arrow] (did_conv1_3) to  (plus_did);
\draw[arrow] (plus_did) to node[midway, xshift=2cm,yshift=0.2cm,font=\LARGE] {$\frac{H}{4} \times \frac{W}{4} \times F$}   (out_did);
\draw[arrow] (way_did) -- ++(3.1,0) --++(0,-6.35)  --   (plus_did.east);

\end{tikzpicture}} 
  \caption{Subblocks of the semantic segmentation image encoder $\bf{E}$, see Figure \ref{fig:resnet_with_se}.  }
  \label{fig:image_encoder_subblocks}
\end{figure}
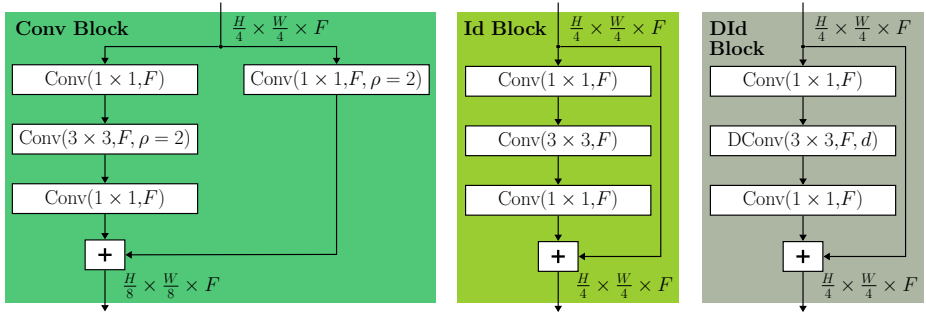

\section{Topology Details}
\label{section:topology_details}
In this section, we show the topological details of the receiver-sided architecture of the current state-of-the-art (SOTA) approach in Figure \ref{fig:decoder} and also our transmitter side in Figure \ref{fig:resnet_with_se}.

% As shown in Figure 1d, inspired by the idea of joint functions in Figure 1b, we 
% propose to perform joint source and task decoding using JD, joint decoding. It 
% allows even lower computational complexity in the cloud, while achieving supe-
% rior rate-distortion (RD) performance. Further, it enables to scale such service 
% to millions of edge devices, keeping the edge device’s computational complexity 
% of our proposed approach the same as the so-far SOTA

As shown in Figure \ref{fig:decoder}, the receiver-sided architecture used by the current SOTA approach \cite{ahuja2023neural} comprises the entire source decoder ($\bf{SD}$) and a separate task decoder ($\bf{D}$).~The $\bf{SD}$ contains in its upper part a grouped transposed convolutional block, with inputs $\bf{\hat{r}}$ and outputs $\bf{\hat{z}}$.~After decoding, the compressed features $\bf{\hat{z}}$ are passed to the task decoder $\bf{D}$, which outputs the semantic segmentation mask $\bf{m}$.~Together, they form the cloud DNN of the method by Ahuja et al. \cite{ahuja2023neural}. Note that this structure deviates from our proposal (Figure 3) due to Ahuja's sequential decoding of compressed features $\bf{\hat{z}}$ and semantic segmentation mask $\bf{m}$, leading to an increased computational complexity and suboptimal performance as compared to our proposal.        

%\ref{fig:CD}

As illustrated in Figure \ref{fig:resnet_with_se}, the edge device (transmitter) comprises a semantic segmentation image encoder $\bf{E}$ and a source encoder $\bf{SE}$. The $\bf{SE}$ produces two bitstreams: the latent bitstream $\bf{b_{\hat{r}}}$ and the hyperprior bitstream $\bf{b_{\hat{h}}}$. These bitstreams are then transmitted to the cloud for processing. Additionally, the image encoder $\bf{E}$ follows the $\texttt{ResNet-50}$ \cite{resnet50} topology, with its subblocks depicted in Figure \ref{fig:image_encoder_subblocks}. The architecture of the identity (Id) block and the dilated identity (DId) blocks is identical, except that the DId block includes one dilated convolution layer. 

\begin{table}[t!]
  \caption{\textbf{Comparison of FLOPs and number of parameters (edge DNNs)}}
  \centering
  \begin{tabular}{@{}lr@{\hskip 0.1in}r@{\hskip 0.1in}r@{\hskip 0.1in}lr@{{\hskip 0.1in}}}
    \toprule
  \multirow{2}{*}{Methods} &\multicolumn{1}{c}{\textbf{COCO}}&\multicolumn{1}{c}{\textbf{Cityscapes}}&\multirow{2}{*}{\textbf{$\#$params (M)}}  \\
    \cmidrule(lr){2-2}\cmidrule(lr){3-3}
      &FLOPs (G)& FLOPs (G) &  \\  
    \midrule
    Song et al.\ \cite{qmapc} (1a)  & 211 &  1687 & 18.77 \\      
    Torfason et al.\ \cite{torfason2018towards} (1b)  & \textbf{28} &  \textbf{210} &\textbf{8.31} \\          
    Ahuja et al.\ \cite{ahuja2023neural} (1c)  & 103 & 801 & 25.14 \\      
    Ours (1d)  & 103  & 801 & 25.14 \\ 
    \bottomrule
  \end{tabular}
\label{table:flops_edge}
\end{table}

%\ref{fig:conventional}
%\ref{fig:conventional2}
%\ref{fig:baseline}
%\ref{fig:compressed}
\begin{table}[t!]
  \caption{\textbf{Comparison of FLOPs and number of parameters (total computational complexity)}}
  \centering
  \begin{tabular}{@{}lr@{\hskip 0.1in}r@{\hskip 0.1in}r@{\hskip 0.1in}r@{}}
    \toprule
  \multirow{2}{*}{Methods} &\multicolumn{2}{c}{\textbf{COCO}}&\multicolumn{2}{c}{\textbf{Cityscapes}} \\
    \cmidrule(lr){2-3}\cmidrule(lr){4-5}
      &FLOPs (G)& $\#$params (M) & FLOPs (G)& $\#$params (M) \\  
    \midrule
    Song et al.\ \cite{qmapc} (1a)  & 1597 & 85.98 & 4521 & 112.07 \\      
    Torfason et al.\ \cite{torfason2018towards} (1b)  & 762 & 51.54 & 1363 & 77.63 \\   
    Ahuja et al.\ \cite{ahuja2023neural} (1c)  & 624 & 41.92 & 2167 & 67.58 \\      
    Ours (1d) & \textbf{113} & \textbf{26.80} & \textbf{840} & \textbf{30.06} \\      
    \bottomrule
  \end{tabular}
\label{table:flops_complete}
\end{table}

%\ref{fig:conventional}
%\ref{fig:conventional2}
%\ref{fig:baseline}
%\ref{fig:compressed}

\section{Computational Complexity}
\label{section:computational_complexity}
In this section, we provide a comparison of the computational complexity incurred by the most relevant approaches on both, the edge device and in total (edge DNNs plus cloud DNNs). 
\begin{figure}[t!]
  \begin{subfigure}{0.5\textwidth}

    \resizebox{\linewidth}{!}{\begin{tikzpicture}
 \tikzstyle{arrow} = [-Triangle, line width=1pt]
\definecolor{color0}{rgb}{0.63921568627451,0.603921568627451,0.552941176470588}
\definecolor{color1}{rgb}{0.913725490196078,0.588235294117647,0.47843137254902}
\definecolor{color2}{rgb}{0.294117647058824,0,0.509803921568627}
\definecolor{darkgoldenrod1811370}{RGB}{181,137,0}
\definecolor{darkgreen}{RGB}{0,100,0}
\definecolor{lightgray}{RGB}{211,211,211}
\definecolor{lightgray204}{RGB}{204,204,204}
\definecolor{rosybrown163154141}{RGB}{163,154,141}
\definecolor{voilet}{RGB}{139,69,19}

\begin{axis}[
width=9.5cm,   % Set the width of the plot
legend cell align={left},
legend cell align={left},
legend cell align={left},
legend style={
  fill opacity=1.0,
  draw opacity=1,
  text opacity=1,
  at={(0.35,0.02)},
  anchor=south west,
  draw=white!80!black
},
legend style={
  fill opacity=1.0,
  draw opacity=1,
  text opacity=1,
  at={(0.35,0.02)},
  anchor=south west,
  draw=white!80!black
},
legend style={
  fill opacity=1.0,
  draw opacity=1,
  text opacity=1,
  at={(0.58,0.02)},
  anchor=south west,
  draw=white!80!black
},
tick align=outside,
tick pos=left,
tick label style={font=\large},
label style={font=\large},
x grid style={lightgray},
xlabel={bits per pixel (bpp) },
xmajorgrids,
xmin=0, xmax=0.41,
xtick style={color=black,font=\Large},
xtick={0,0.1,0.2,0.3,0.4},
xticklabels={0,0.1,0.2,0.3,0.4},
y grid style={white!82.7450980392157!black},
ylabel={\(\displaystyle \textrm{mIoU on $\mathcal{D}_{\mathrm{COCO}}^{\mathrm{val2017}}$}\) (\%)},
 ylabel style={yshift=-6pt},
ymajorgrids,
ymin=54, ymax=68,
ytick style={color=black},
ytick={54,56,58,60,62,64,66,68},
yticklabels={
  \(\displaystyle {54}\),
  \(\displaystyle {56}\),
  \(\displaystyle {58}\),
  \(\displaystyle {60}\),
  \(\displaystyle {62}\),
  \(\displaystyle {64}\),
  \(\displaystyle {66}\),
  \(\displaystyle {68}\)
}
]

\addplot [line width=1.5pt, blue, dashed, forget plot]
table {%
0 67.4
0.41 67.4
};
\label{No_Compression}
\addplot [line width=1.5pt, green!39.2156862745098!black, mark=*, mark size=2, mark options={solid}, forget plot]
table {%
0.059	54 
0.08	59.86 	
0.1018	63.64
0.166	66.30	
0.4045	67.57
};
\label{Ahuja}

\addplot [line width=1.5pt, red, mark=*, mark size=2, mark options={solid}, forget plot]
table {%
0.0412	65.20
0.0489	65.90 
0.0824	66.25
0.1224	66.72
0.2827  67.88
};
\label{ours}

\end{axis}

\node [draw,fill=white,font=\small] at (rel axis cs: 0.587,0.25){\shortstack[l]{
\ref*{No_Compression} No Compression$^{\textcolor{red}{*}}$ \cite{deeplabv3} \\ 
 \ref*{ours} Ours (1d) \\
\ref*{Ahuja} Ahuja et al. \cite{ahuja2023neural} (1c)
}};
\end{tikzpicture}}
    \caption{COCO    }
    \label{fig:coco_results_resnet101}
  \end{subfigure}%
  \hspace{-0.2em}
  \begin{subfigure}{0.5\textwidth}
    \resizebox{\linewidth}{!}{\begin{tikzpicture}

\definecolor{darkgoldenrod1811370}{RGB}{181,137,0}
\definecolor{darkgreen}{RGB}{0,100,0}
\definecolor{lightgray}{RGB}{211,211,211}
\definecolor{lightgray204}{RGB}{204,204,204}
\definecolor{rosybrown163154141}{RGB}{163,154,141}
\definecolor{voilet}{RGB}{139,69,19}
\begin{axis}[
width=9.5cm,   % Set the width of the plot
legend cell align={left},
legend cell align={left},
legend cell align={left},
scaled ticks=false,
tick align=outside,
tick pos=left,
tick label style={font=\large},
label style={font=\large},
x grid style={lightgray},
xlabel={bits per pixel (bpp) },
xmajorgrids,
xmin=0, xmax=0.3,
xtick style={color=black},
x tick label style={/pgf/number format/precision=10},
y grid style={lightgray},
ylabel={\(\displaystyle \textrm{mIoU on $\mathcal{D}_{\mathrm{CS}}^{\mathrm{val}}$}\) (\%)},
ylabel style={yshift=-6pt},
ymajorgrids,
ymin=64, ymax=80,
ytick style={color=black},
ytick={64,66,68,70,72,74,76,78},
yticklabels={
  \(\displaystyle {64}\),
  \(\displaystyle {66}\),
  \(\displaystyle {68}\),  
  \(\displaystyle {70}\),
  \(\displaystyle {72}\),
  \(\displaystyle {74}\),
  \(\displaystyle {76}\),
  \(\displaystyle {78}\)
}
]
\addplot [line width=1.5pt, blue, dashed]
table {%
0 79.41
0.3 79.41
};
\label{No_Compression_cs}
\addplot [line width=1.5pt,darkgreen, mark=*, mark size=1.6, mark options={solid}]
table {%
0.0154 64.53	
0.0195 73.75
0.1031 77.81
0.1385 78.06
0.2589 78.56
};
\label{Ahuja_cityscape}
\addplot [line width=1.5pt,  red, mark=*, mark size=1.6, mark options={solid}]
table {%
0.01   67.67
0.0139 70.3  
0.0206 74.14
0.0861 78.19
0.1055 78.34    
0.1348 78.53
0.2061 78.72
};
\label{ours_cs}

\end{axis}

\node [draw,fill=white,font=\small] at (rel axis cs: 0.587,0.25){\shortstack[l]{
\ref*{No_Compression_cs} No Compression$^{\textcolor{red}{*}}$ \cite{deeplabv3} \\
\ref*{ours_cs} Ours  (1d)  \\ 
\ref*{Ahuja_cityscape} Ahuja et al.\ \cite{ahuja2023neural} (1c) }};

\end{tikzpicture}}
    \caption{Cityscapes    }
    \label{fig:cityscapes_results_resnet101}
  \end{subfigure}
  \caption{ \textbf{Proposed $\bf{JD}$ approach ("Ours") against SOTA approach} with a \texttt{ResNet-101} encoder on the mIoU metric for (a) $\mathcal{D}_{\mathrm{COCO}}^{\mathrm{val2017}}$
and (b) $\mathcal{D}_{\mathrm{CS}}^{\mathrm{val}}$ datasets. The values denoted by $^{\textcolor{red}{*}}$ are taken from respective paper and the identifiers in parentheses (1x) refer to the type of approach in Figures 1c and 1d. On both COCO and Cityscapes datasets, our proposed approach "Ours" achieves better RD trade-off than SOTA baseline at a wide range of bitrates. Note that our proposed approach "Ours" uses $K=1$ (COCO) and $K=3$ (Cityscapes) in the ASPP block.  }
  \label{fig:result_comparison_101}
\end{figure}
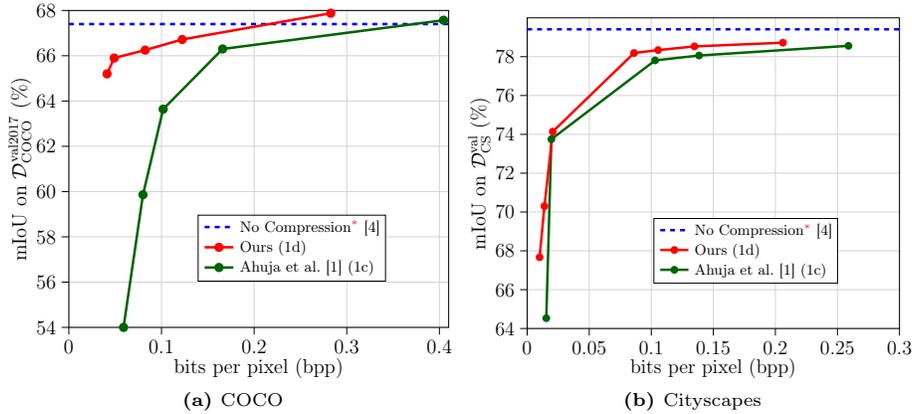

%\ref{fig:baseline} and \ref{fig:compressed}

Table \ref{table:flops_edge} presents the \textit{edge device} computational complexity comparison of our approach against the baseline methods.~Our proposed approach has the same edge device computational complexity as the so-far current state-of-the-art (SOTA) approach by Ahuja et al. \cite{ahuja2023neural}. Compared to Torfason et al. \cite{torfason2018towards}, both methods exhibit a clearly higher computational complexity and number of parameters. However, the decreased edge device computational complexity of Torfason et al. \cite{torfason2018towards} results in inacceptably poor performance, as has been shown in Figures  5a and 5b. Note also that approaches following Figure 1b, such as Torfason et al. \cite{torfason2018towards} transmit an image-coded bitstream which can be decoded at any point during transmission back to a reconstructed image, thereby being susceptible to data privacy fraud.

%\ref{fig:coco_results} and \ref{fig:cityscapes_results}
%pproaches following Figure \ref{fig:conventional2}

Table \ref{table:flops_complete} illustrates a comparison of the \textit{total} computational complexity of our approach against the baseline methods. Our proposed approach utilizes $18\%$/$38\%$ of the FLOPs and $63\%$/$44\%$ of the number of parameters in comparison to the SOTA by Ahuja et al. \cite{ahuja2023neural} on COCO/Cityscapes, respectively. Further, even though our approach exhibits higher edge computational complexity than the baseline by Torfason et al. \cite{torfason2018towards}, it significantly reduces the total computational complexity and utilizes only $14\%$/$61\%$ of the FLOPs and $51\%$/$38\%$ of the number of parameters on COCO/Cityscapes, respectively.
\begin{figure}[t!]
  \begin{subfigure}{0.2455\textwidth}
    \includegraphics[width=\textwidth]{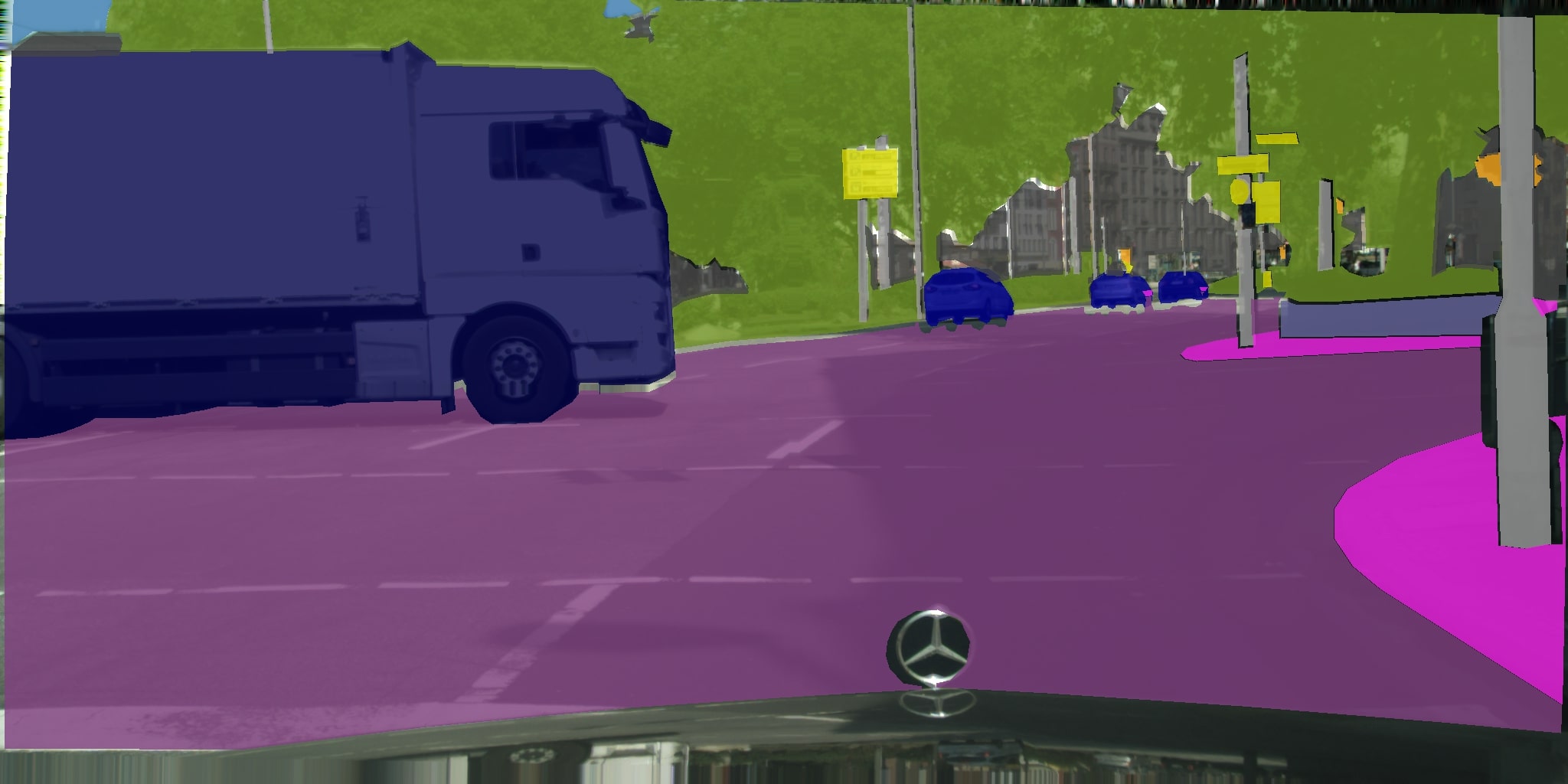} 
    \includegraphics[width=\textwidth]{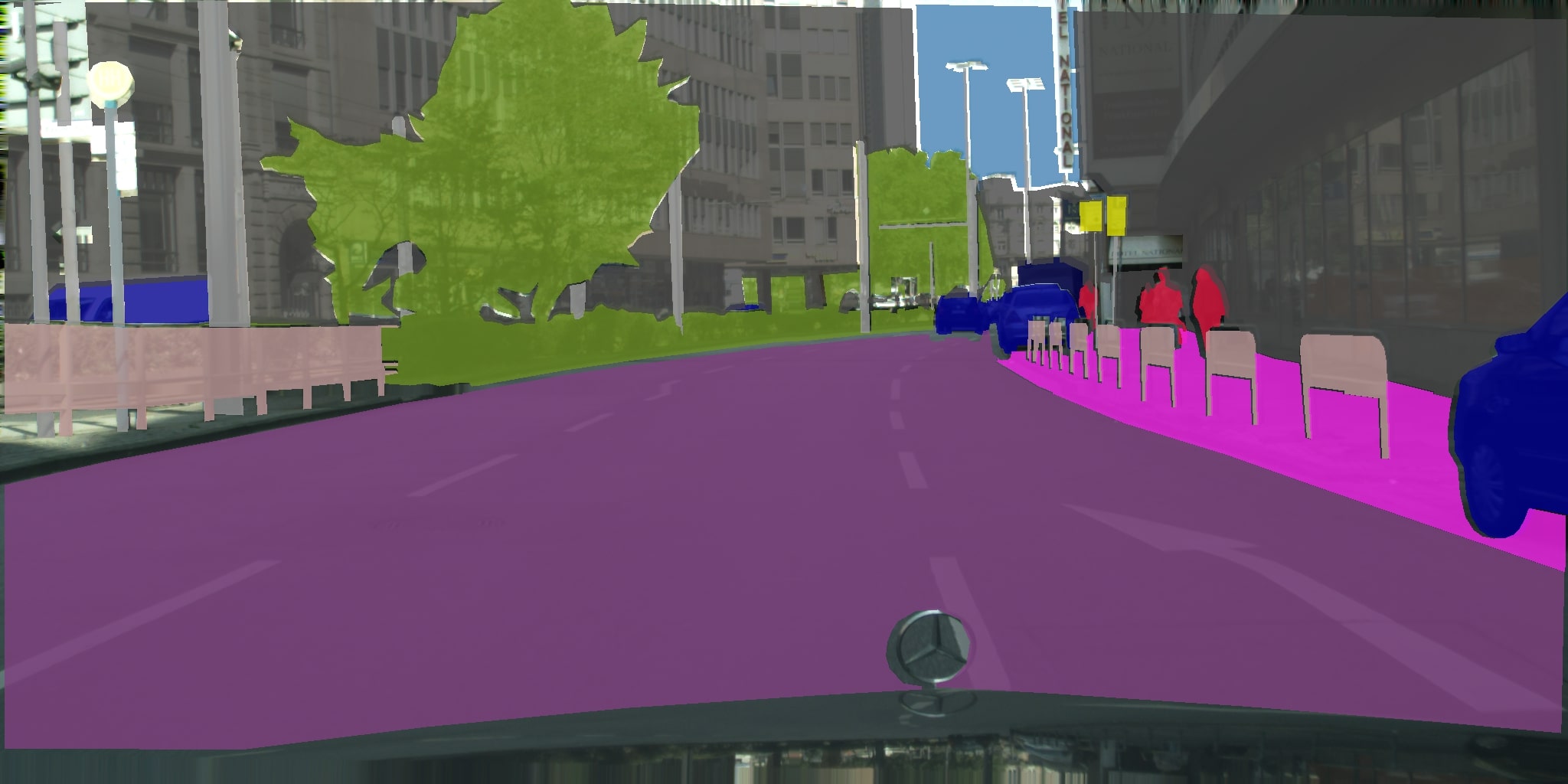} 
    \includegraphics[width=\textwidth]{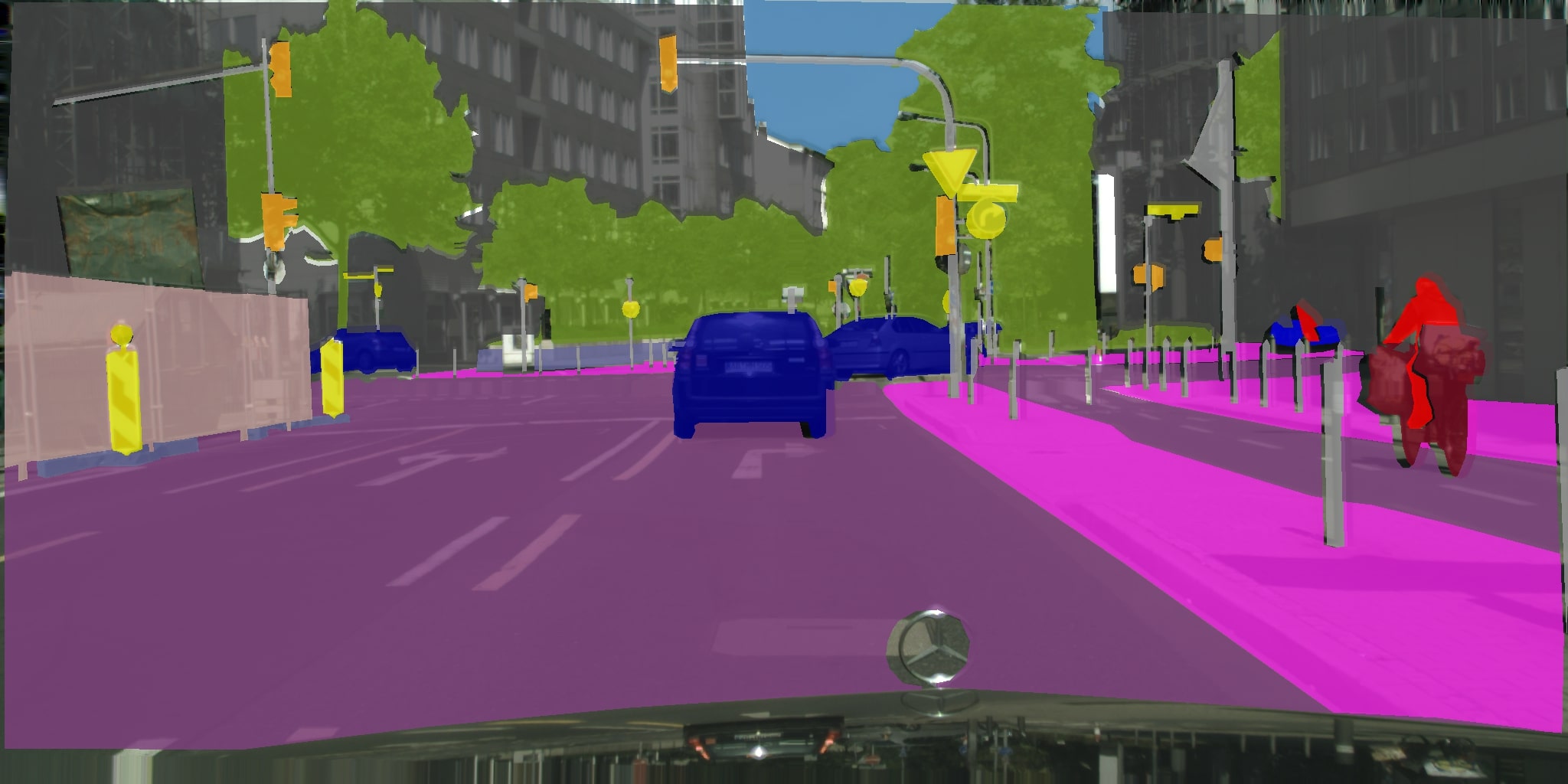} 
    \includegraphics[width=\textwidth]{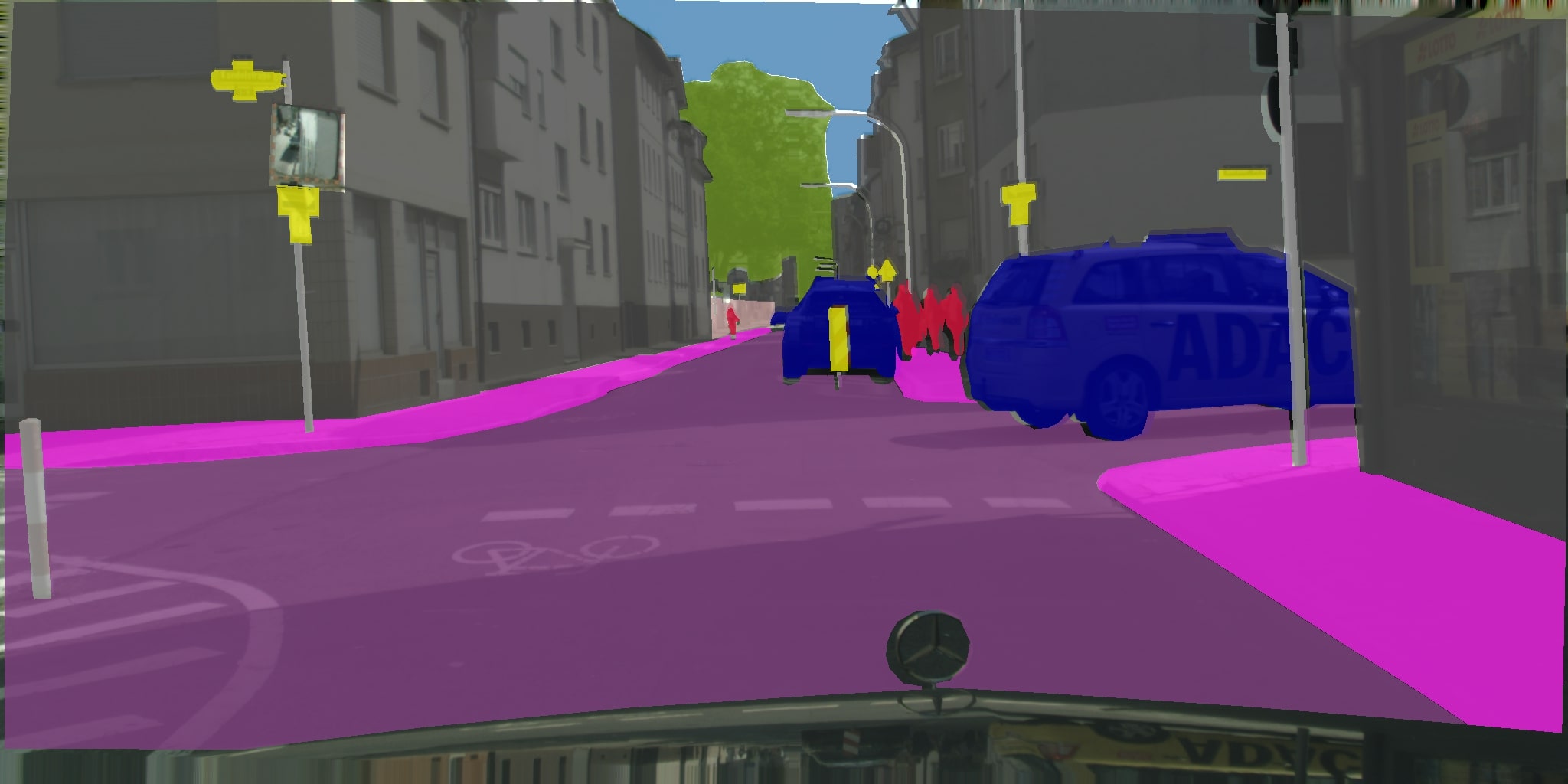} 
    \includegraphics[width=\textwidth]{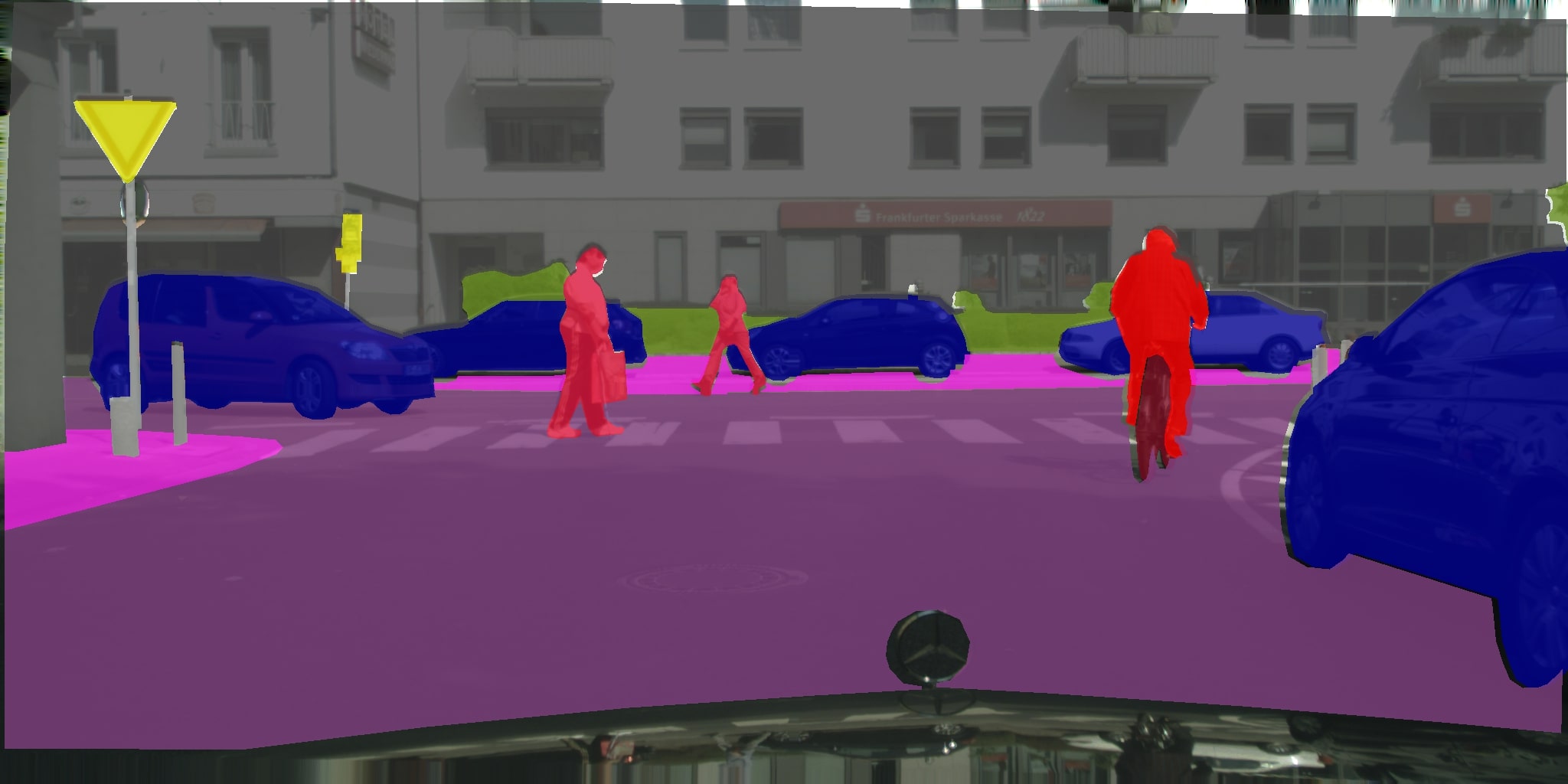}     
    \includegraphics[width=\textwidth]{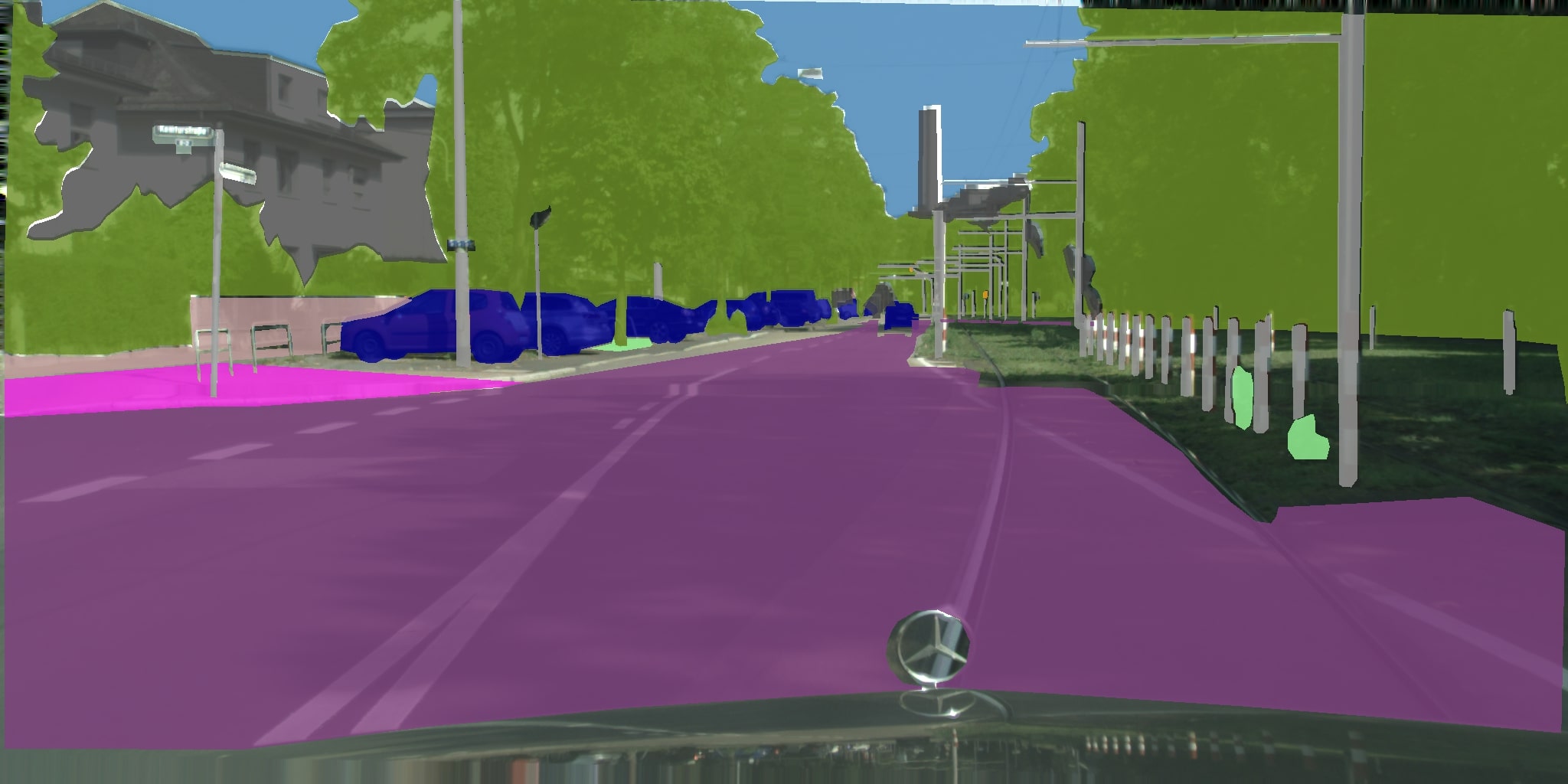} 
    \includegraphics[width=\textwidth]{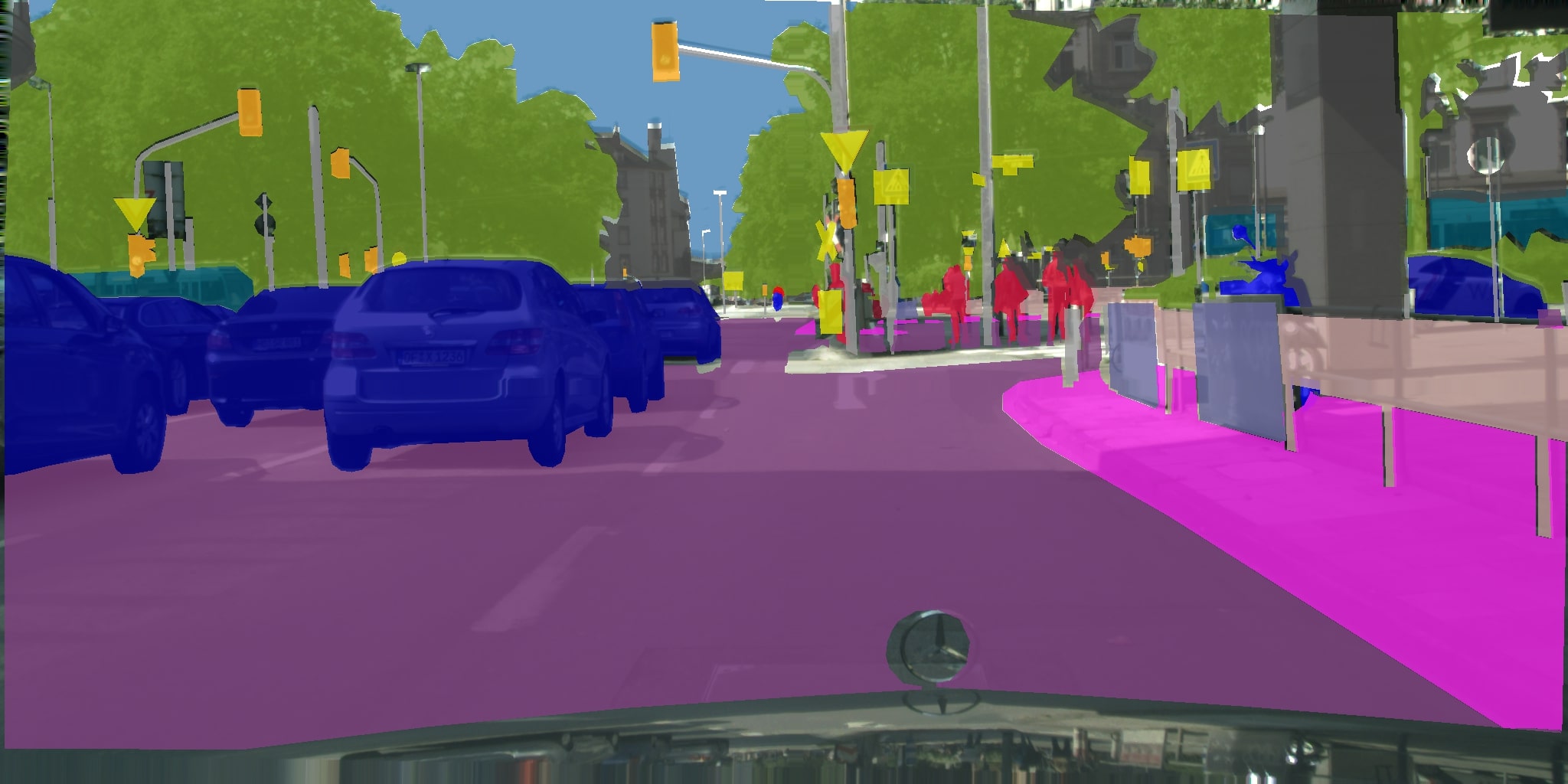} 
    \includegraphics[width=\textwidth]{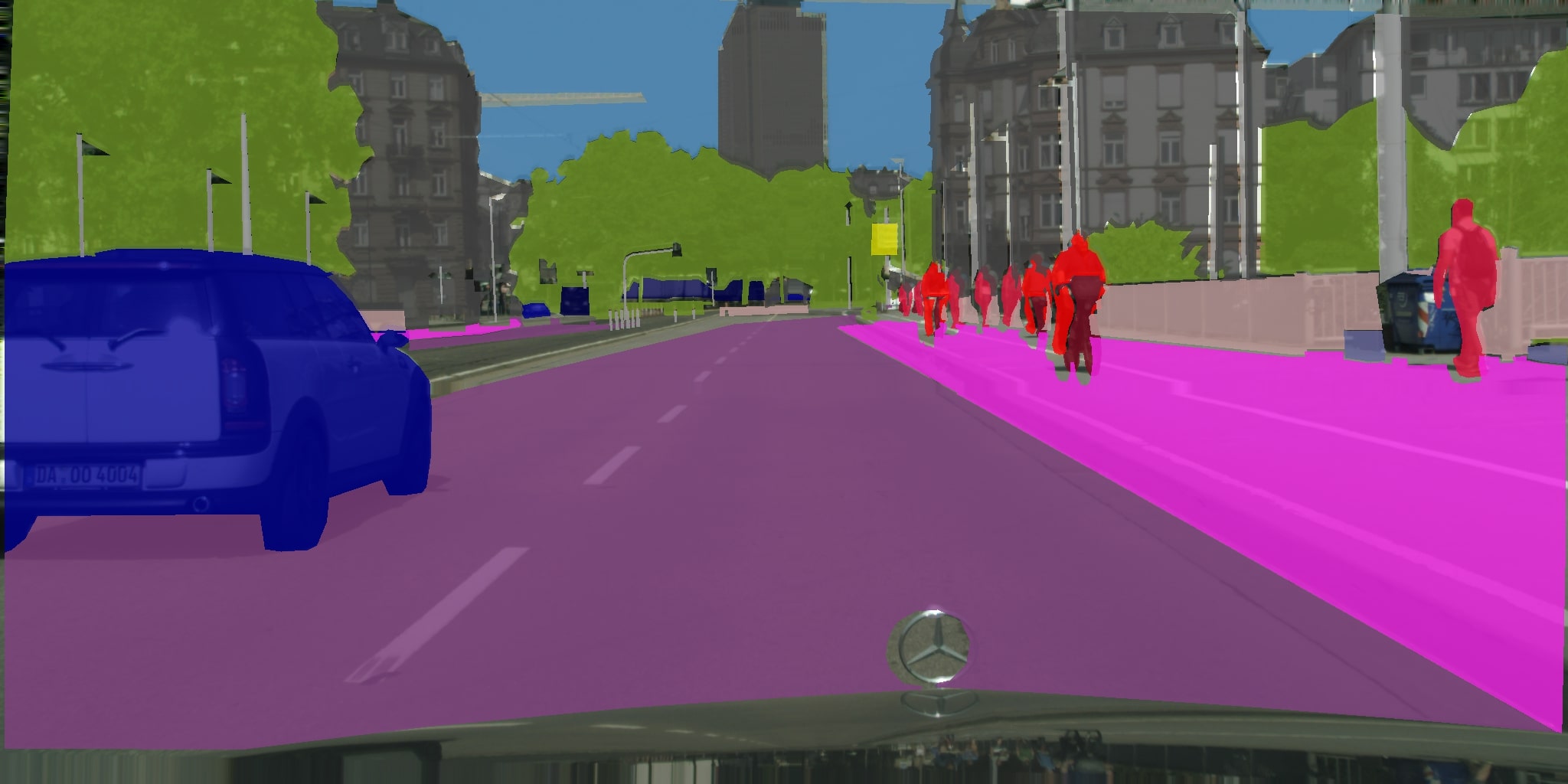} 
    \includegraphics[width=\textwidth]{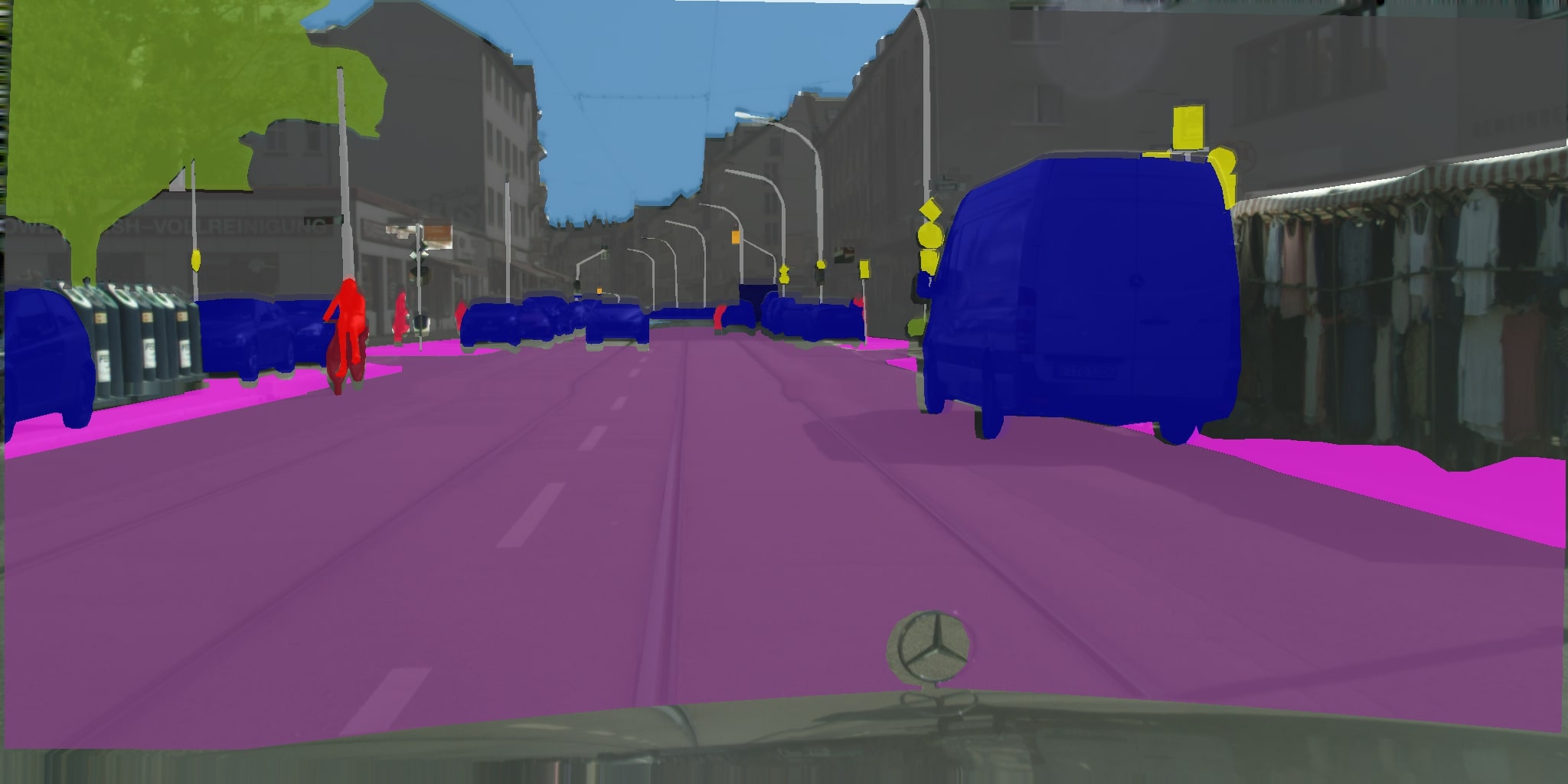}
    \caption{Ground Truth}
  \end{subfigure}
    \hspace{-0.42em}
  \begin{subfigure}{0.2455\textwidth}
    \includegraphics[width=\textwidth]{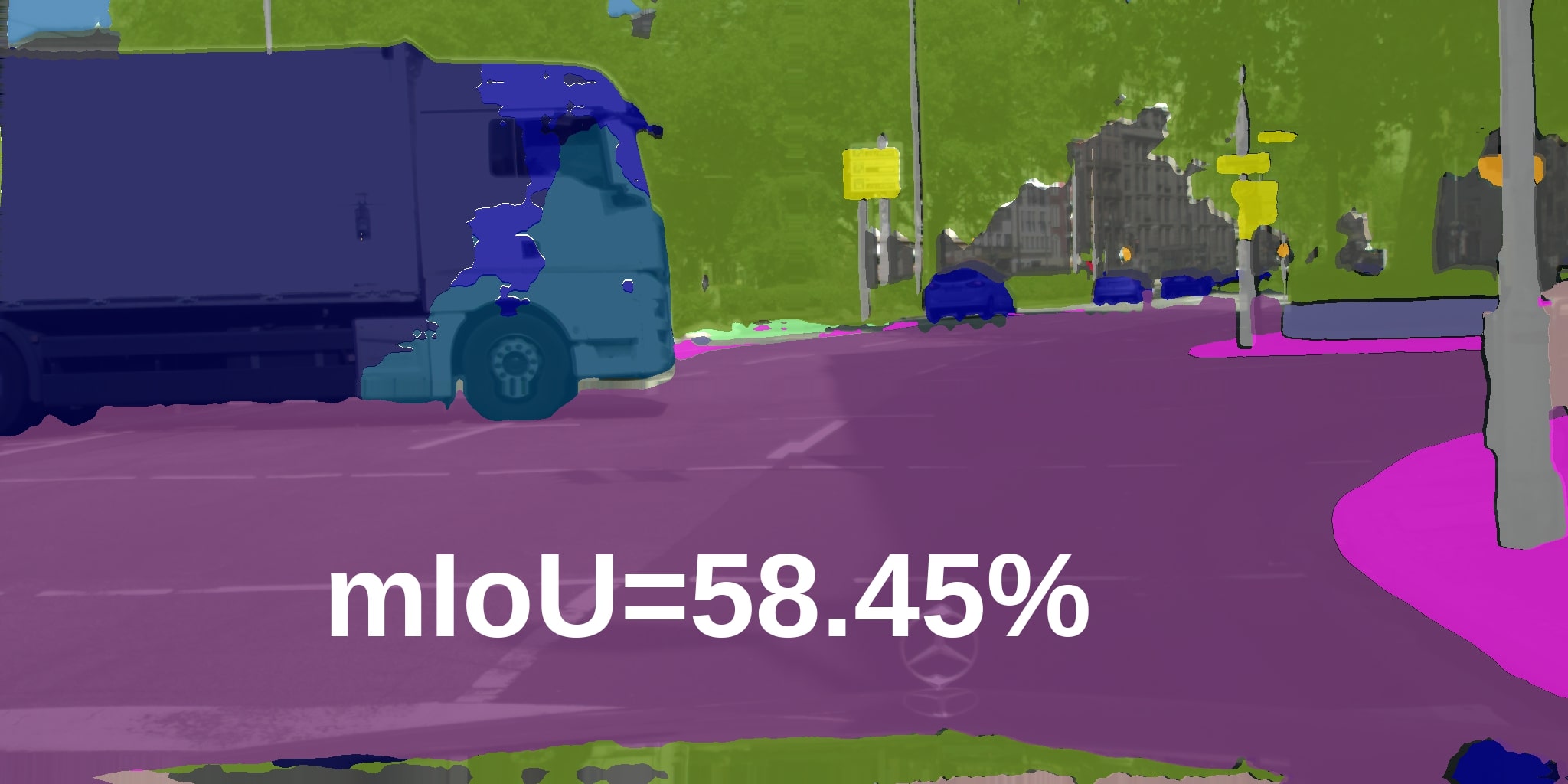} 
    \includegraphics[width=\textwidth]{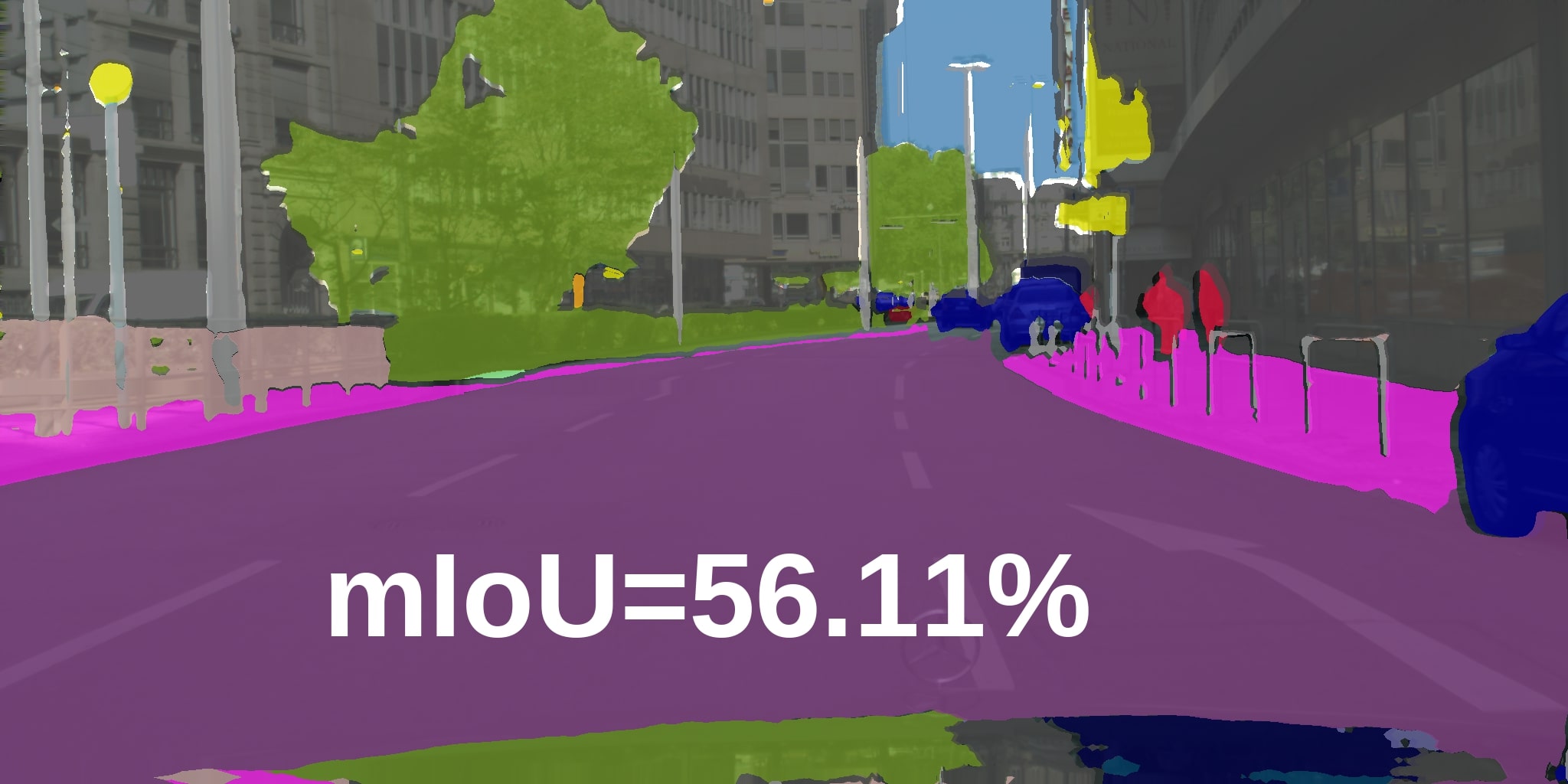} 
    \includegraphics[width=\textwidth]{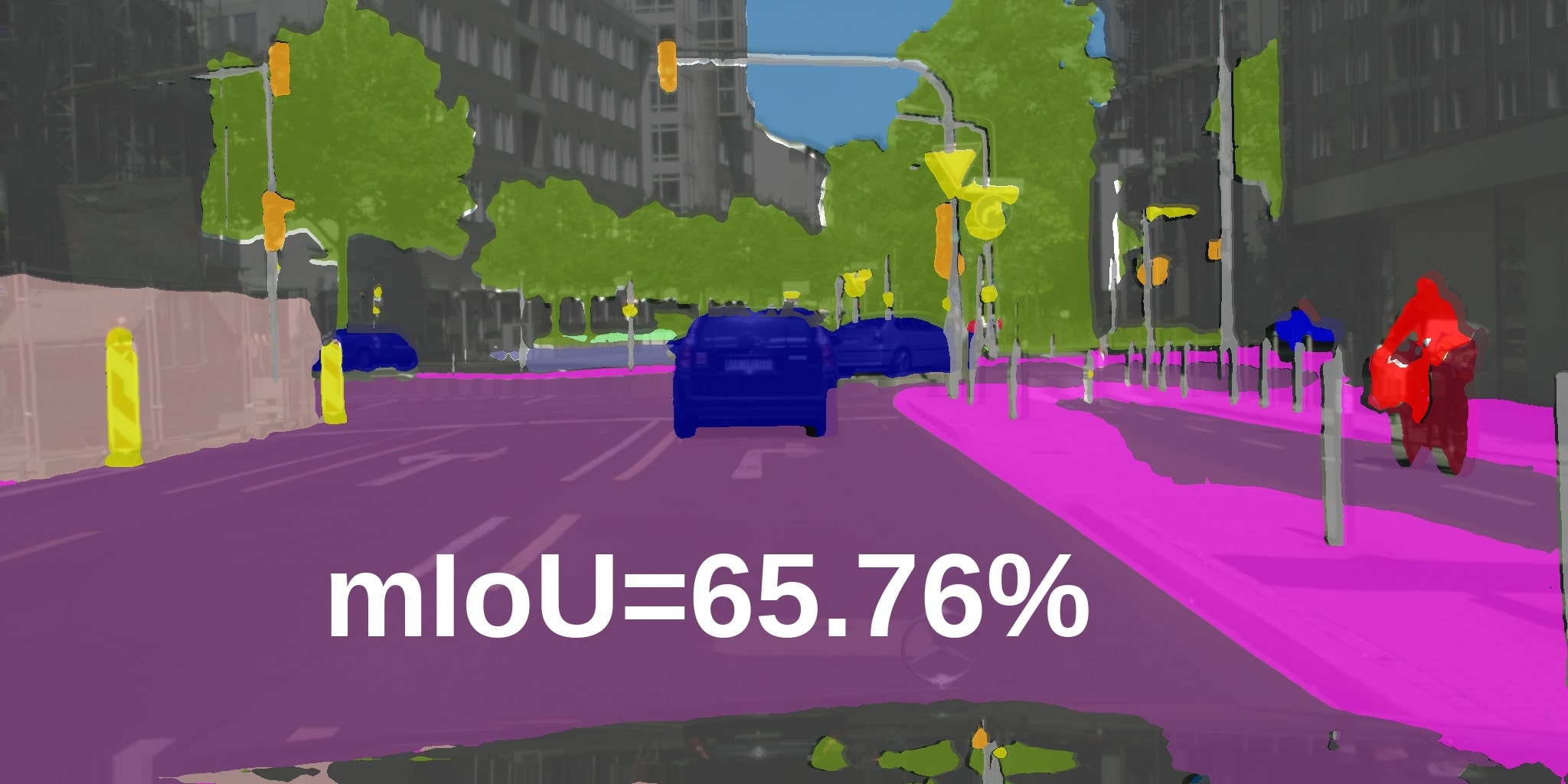}
    \includegraphics[width=\textwidth]{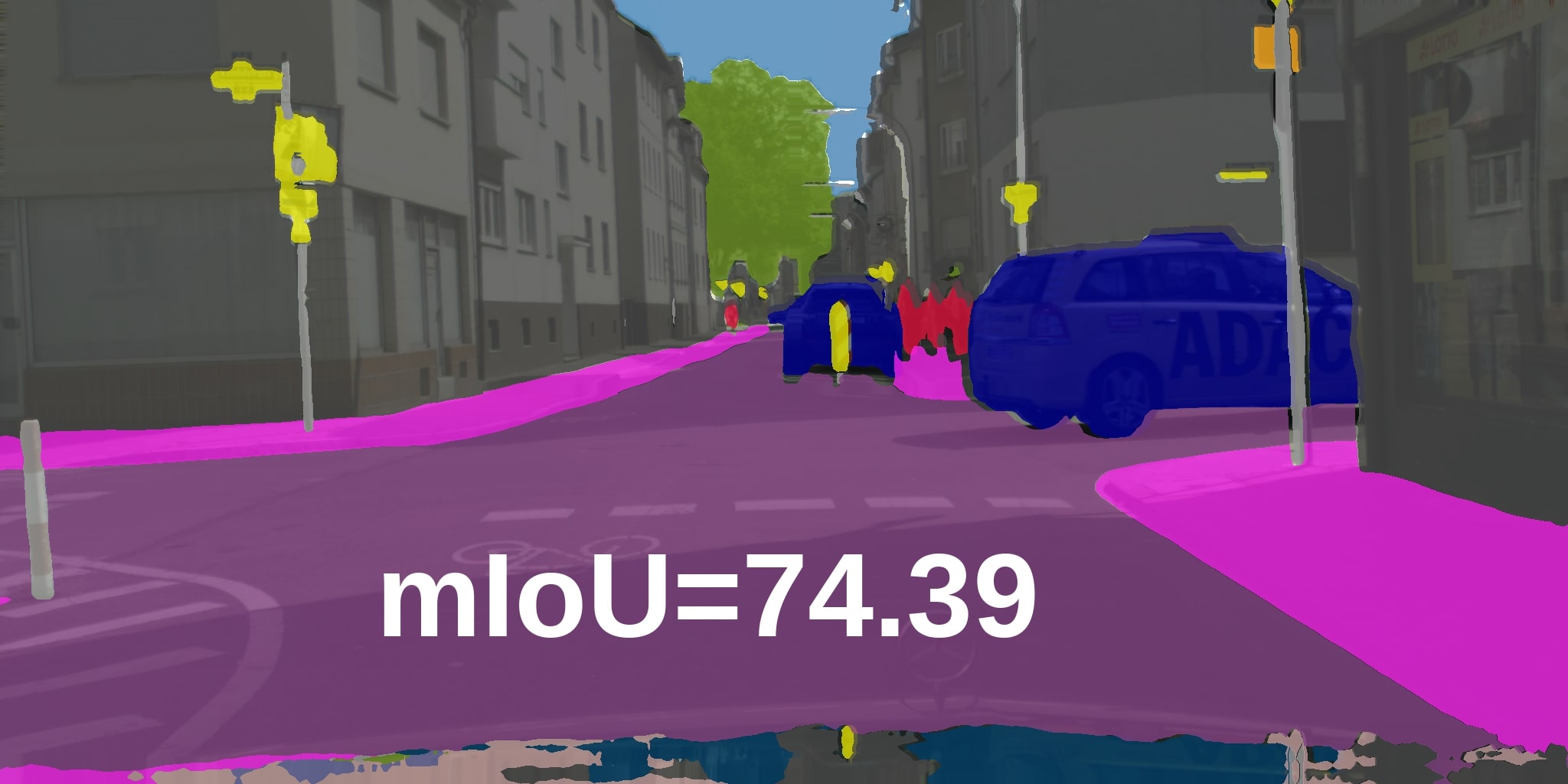} 
    \includegraphics[width=\textwidth]{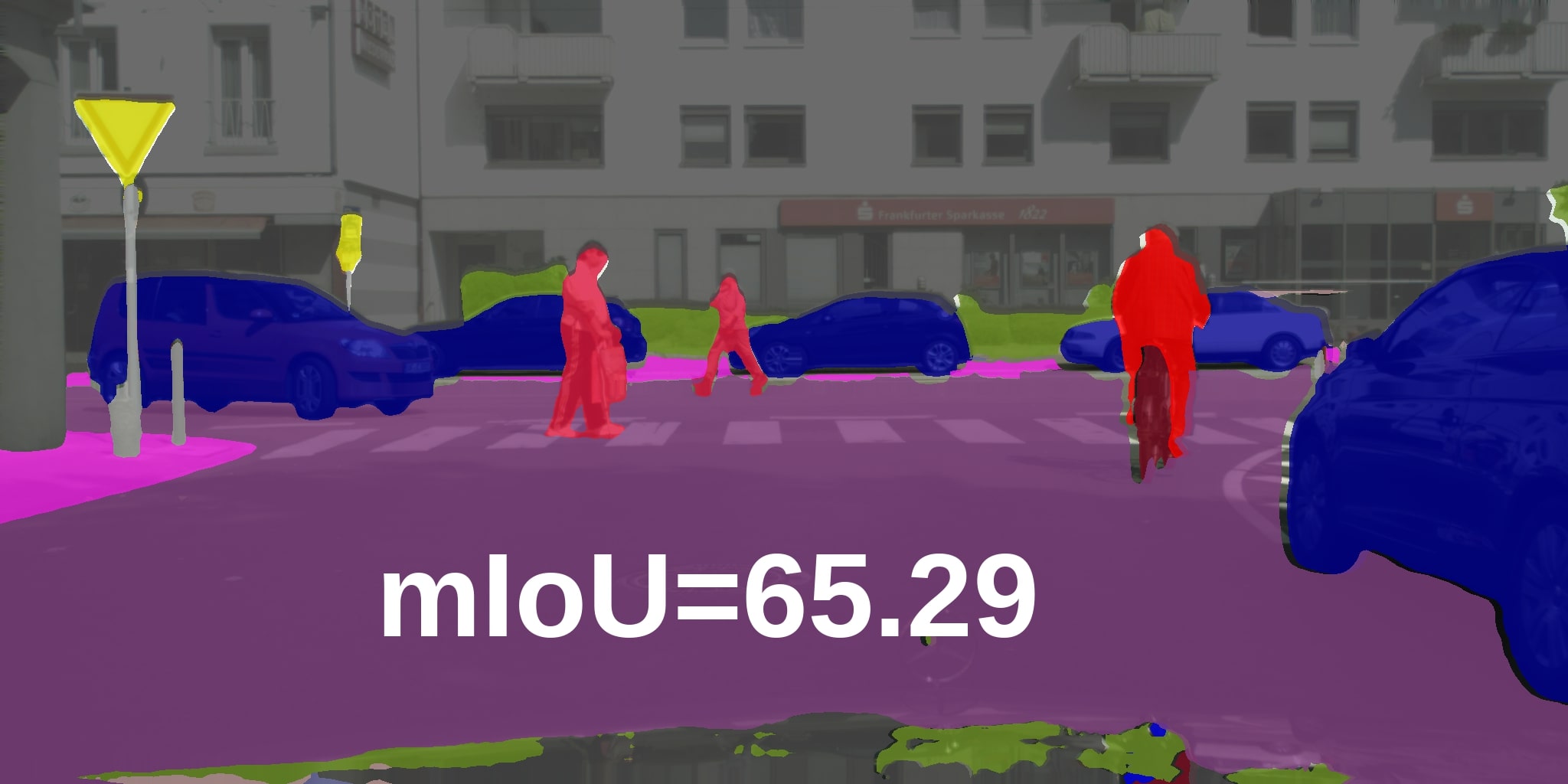}
    \includegraphics[width=\textwidth]{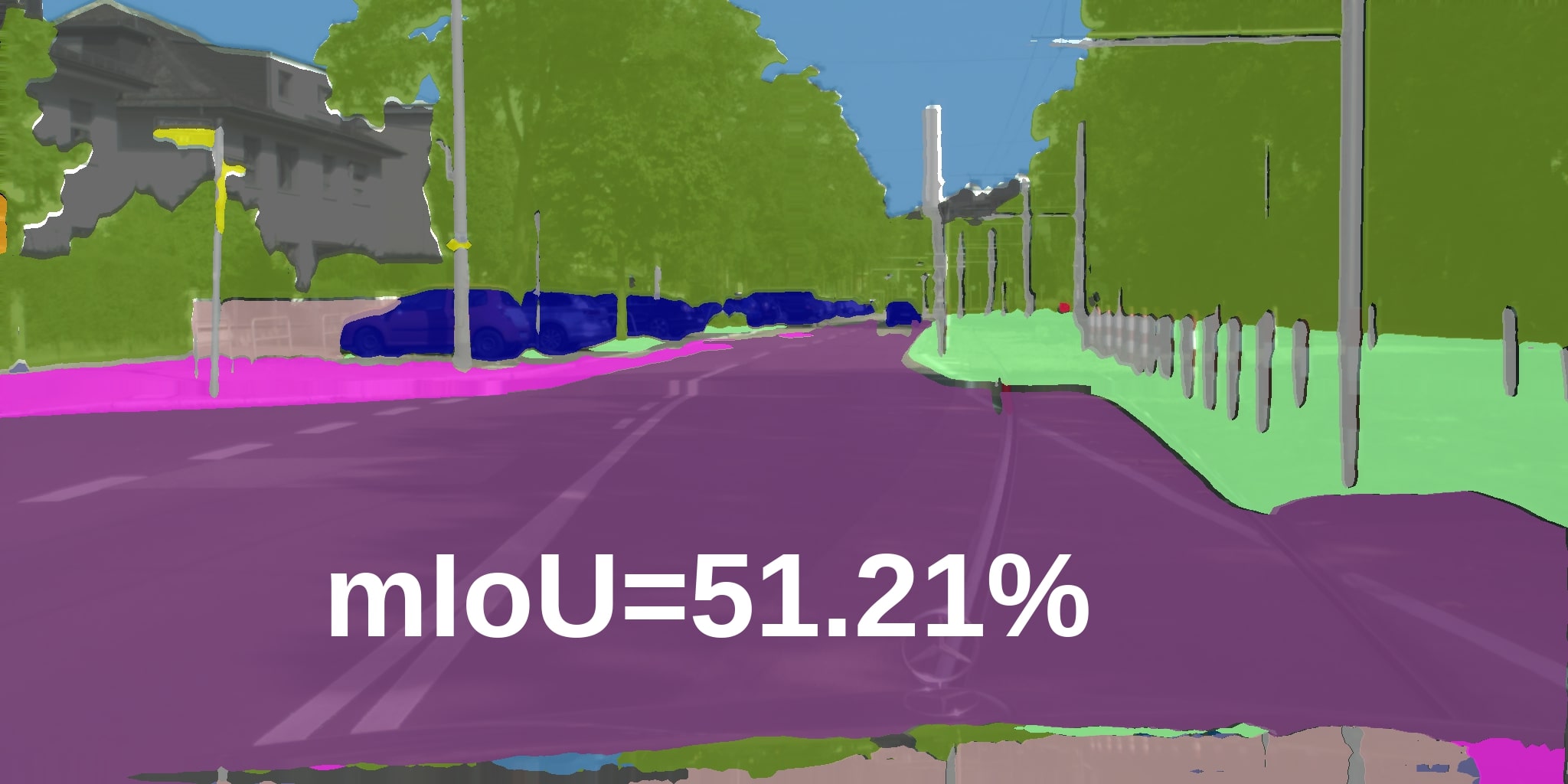} 

    \includegraphics[width=\textwidth]{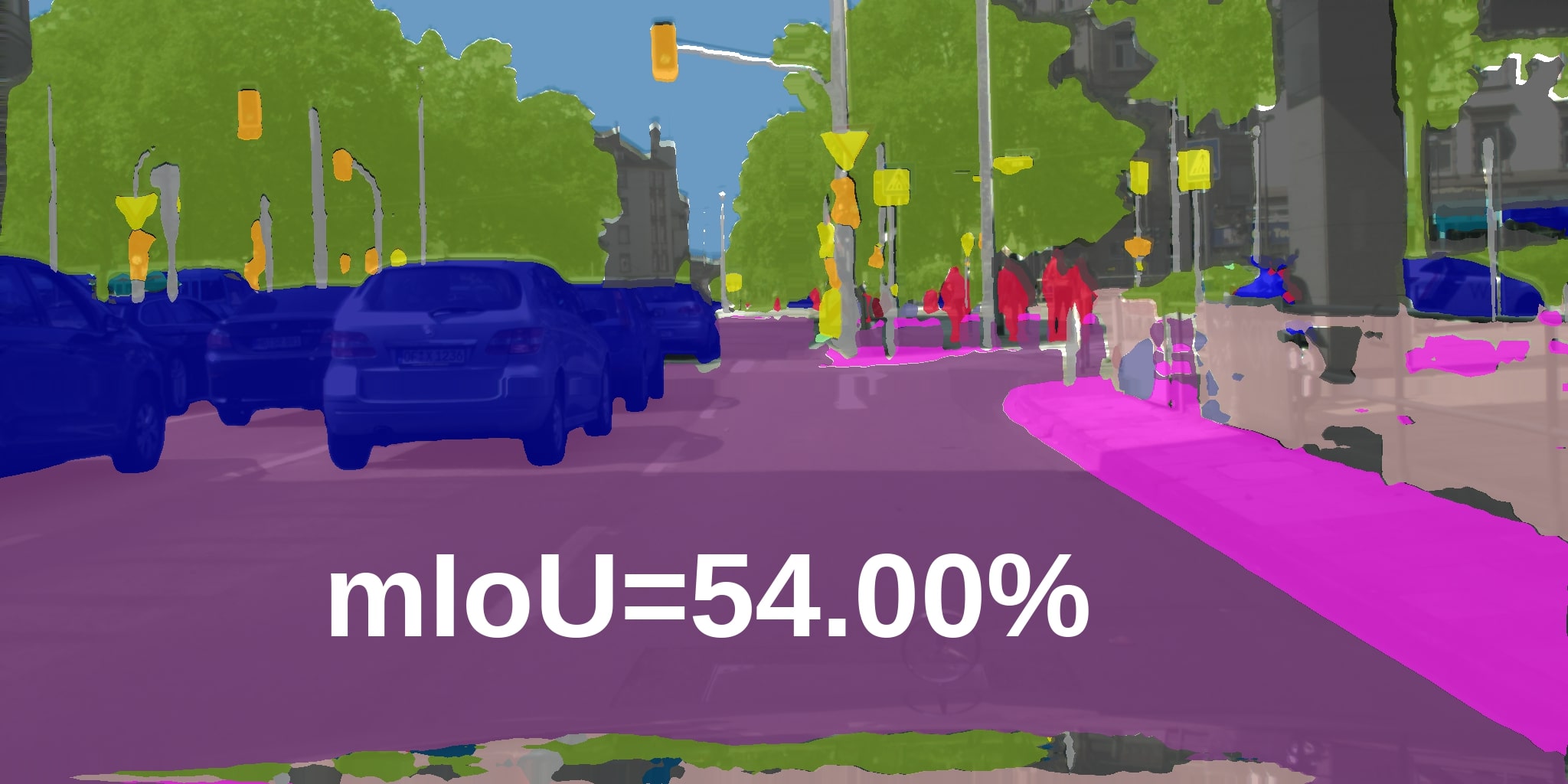} 
    \includegraphics[width=\textwidth]{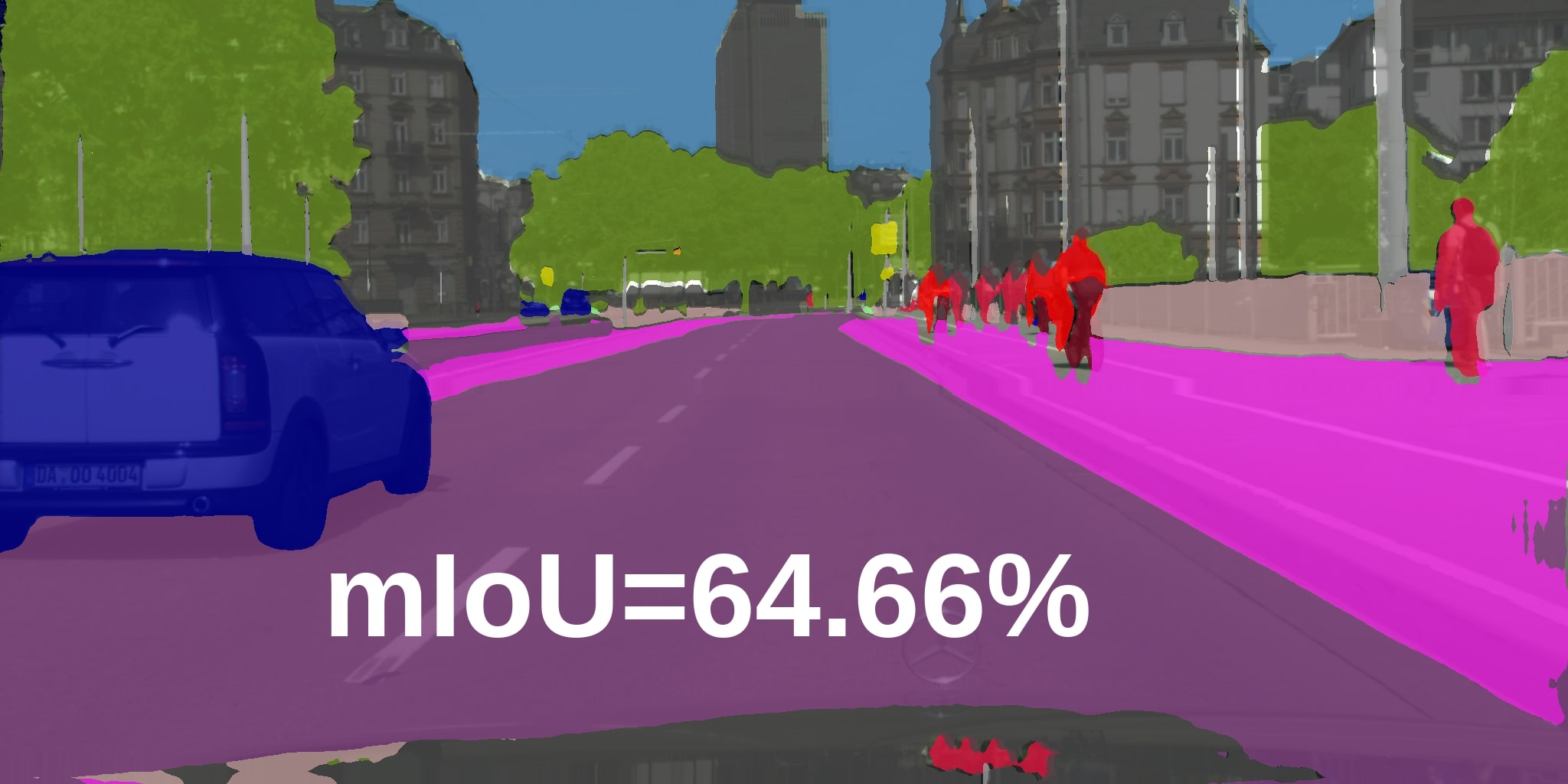} 

    \includegraphics[width=\textwidth]{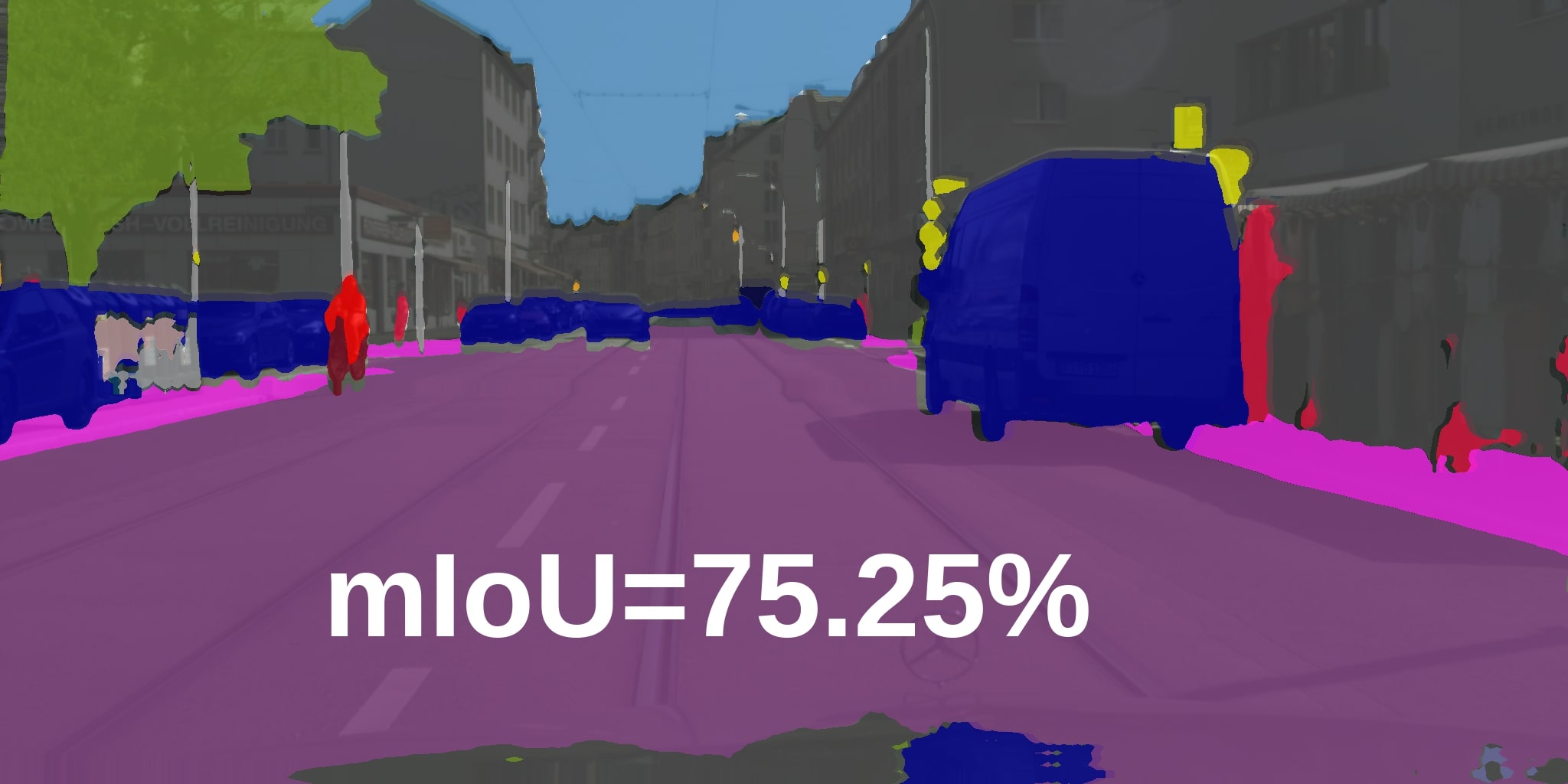}
    \caption{No Compression \cite{deeplabv3}}
  \end{subfigure}
   \hspace{-0.45em}
  \begin{subfigure}{0.2455\textwidth}
    \includegraphics[width=\textwidth]{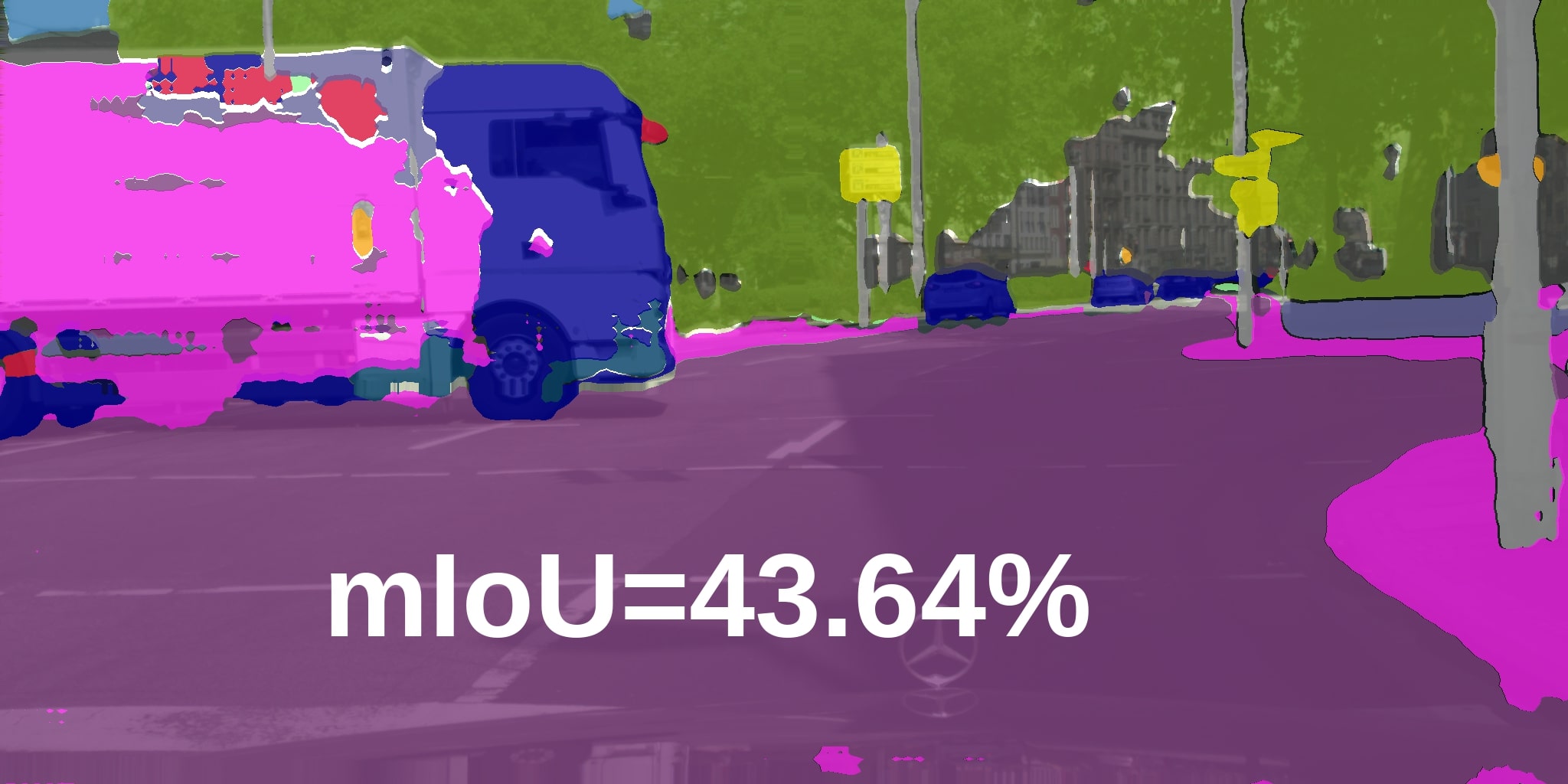} 
    \includegraphics[width=\textwidth]{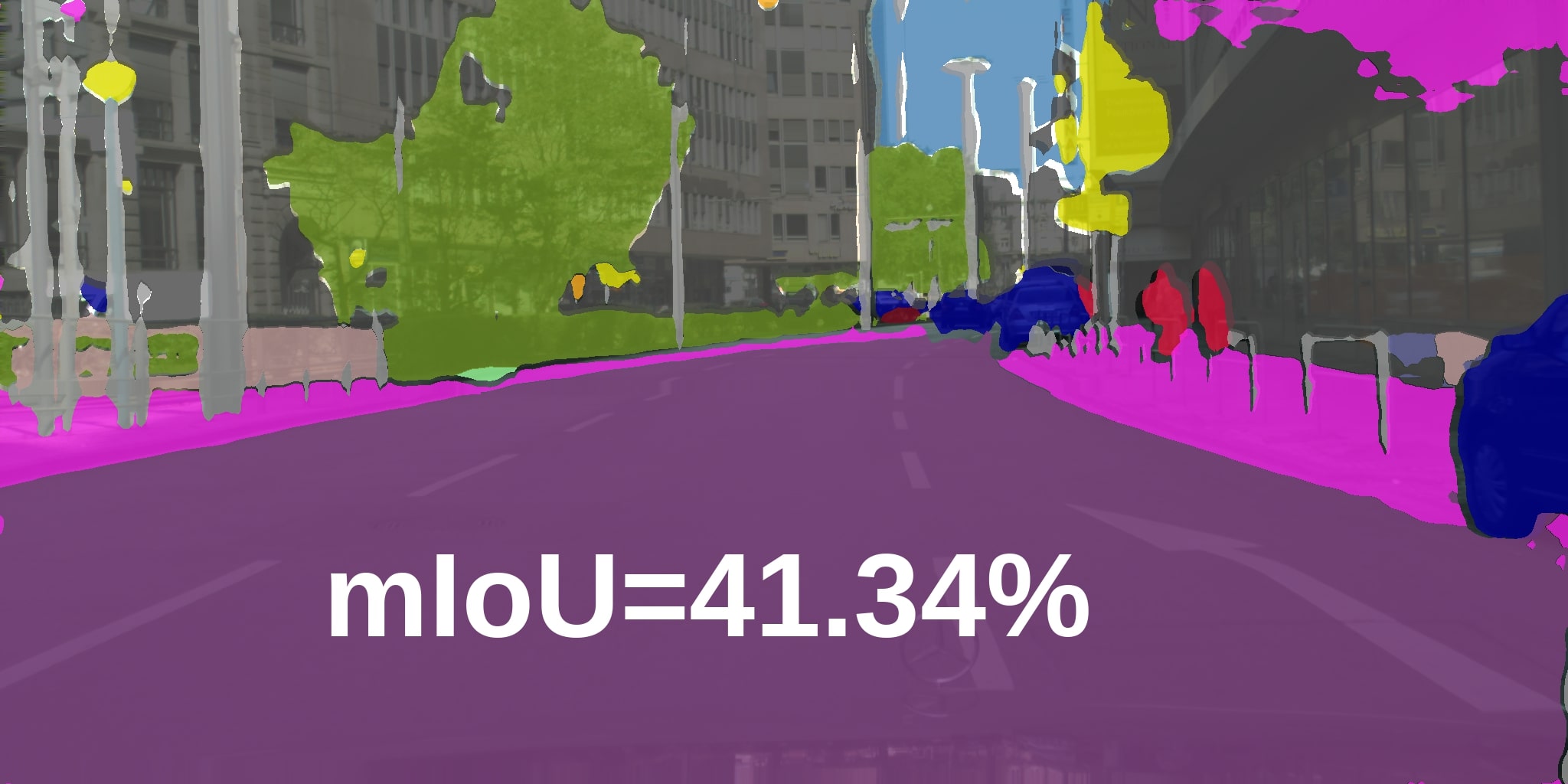} 
    \includegraphics[width=\textwidth]{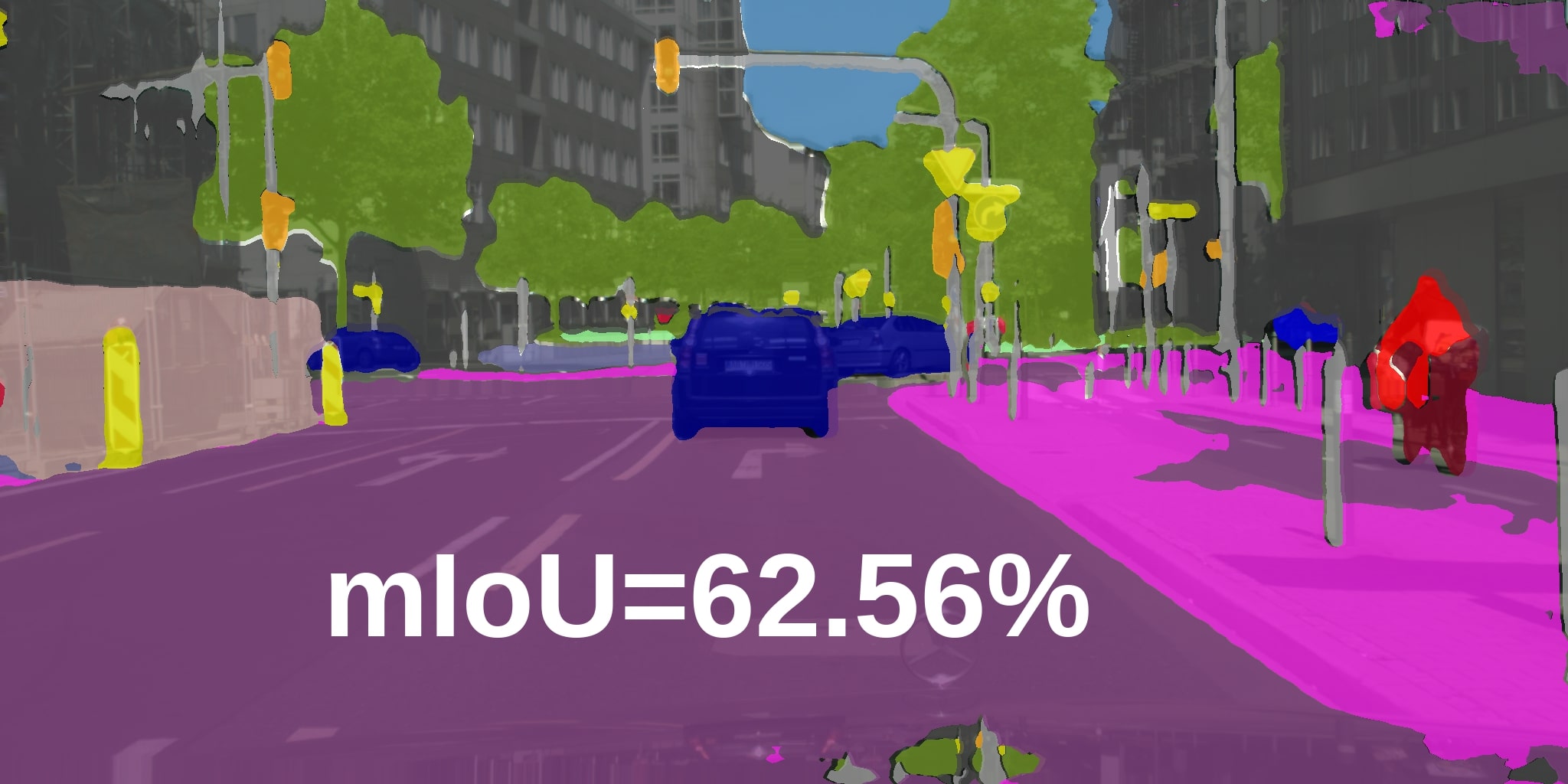} 

    \includegraphics[width=\textwidth]{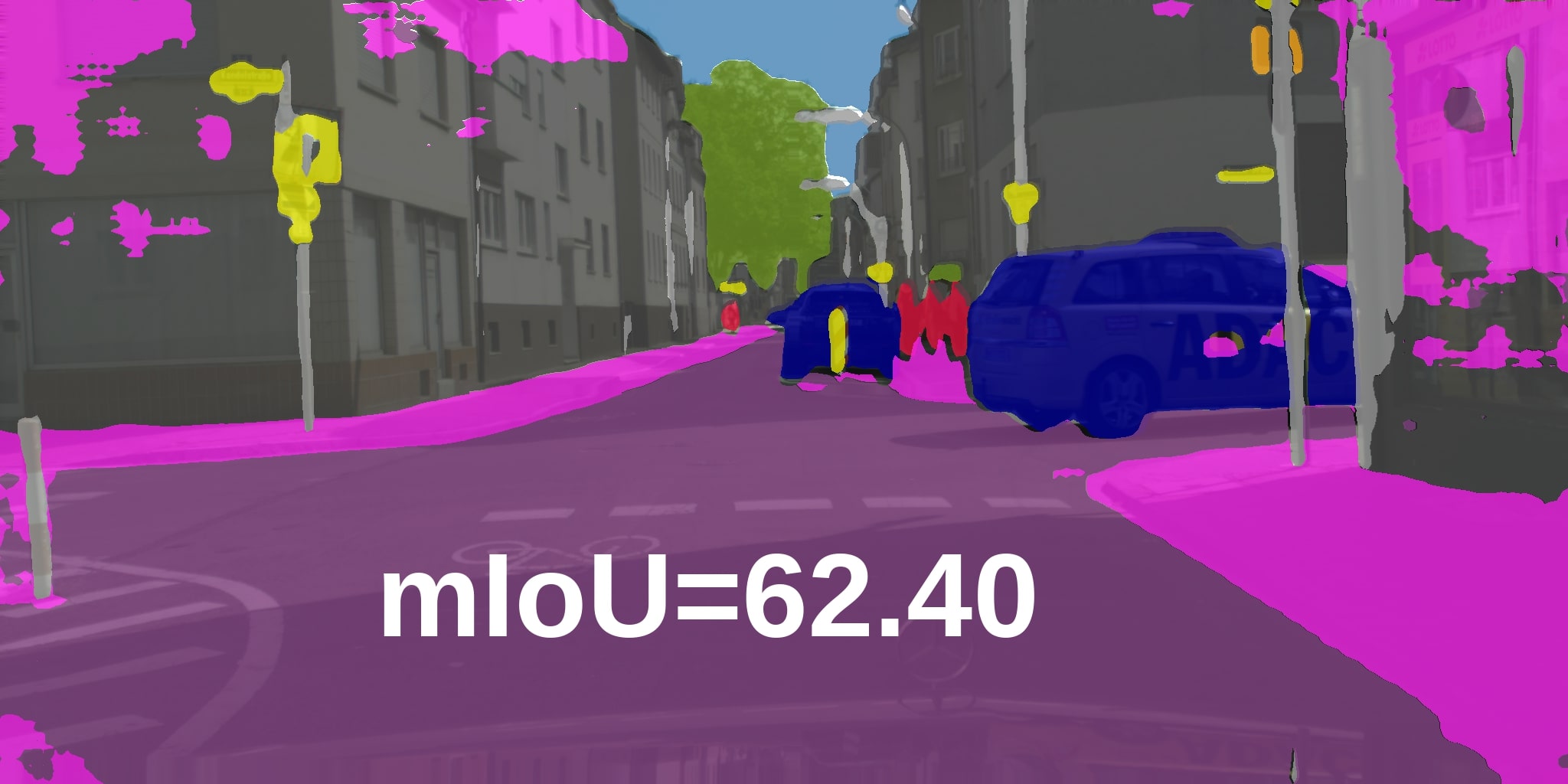} 
    \includegraphics[width=\textwidth]{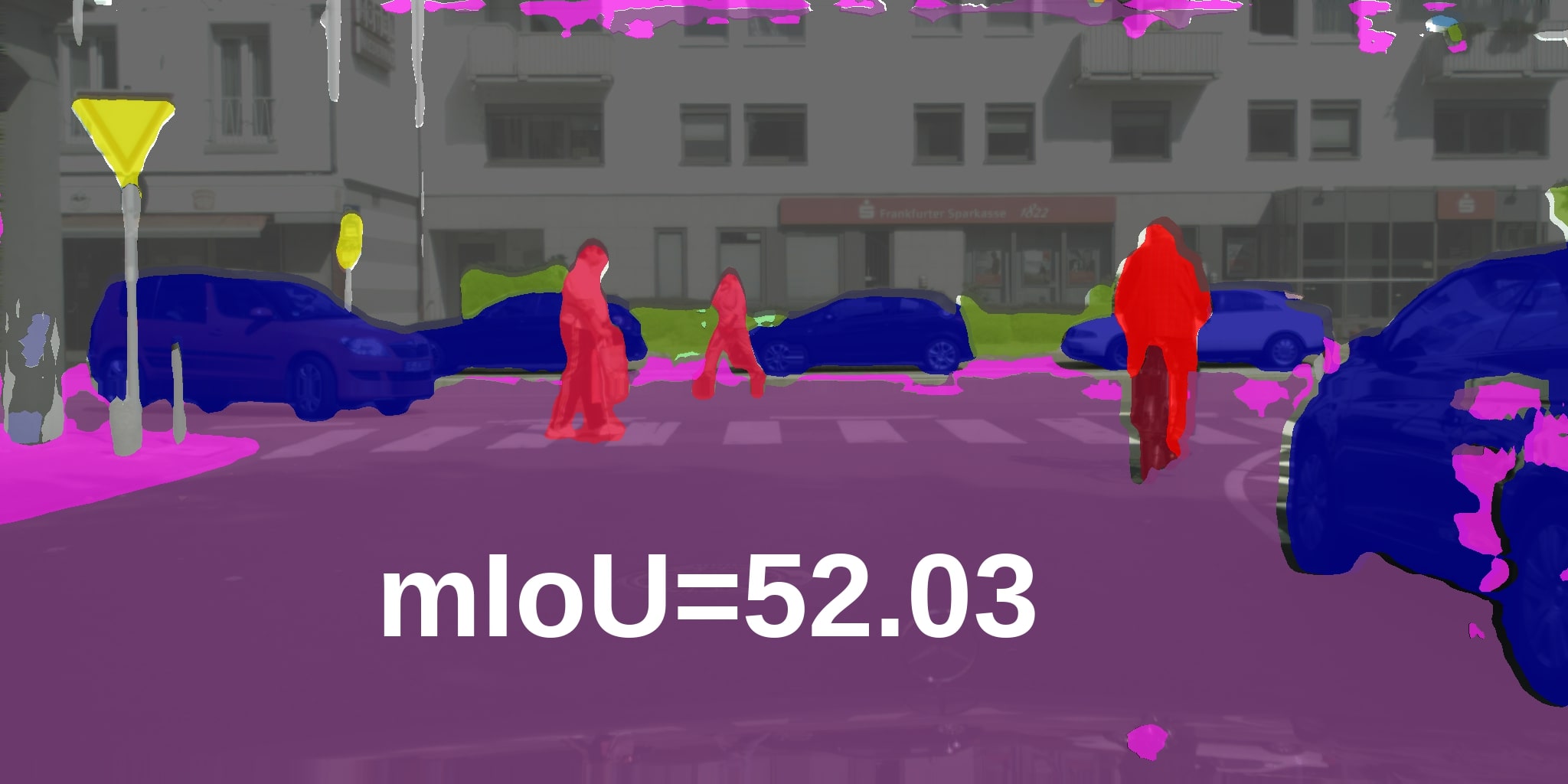}
    
    \includegraphics[width=\textwidth]{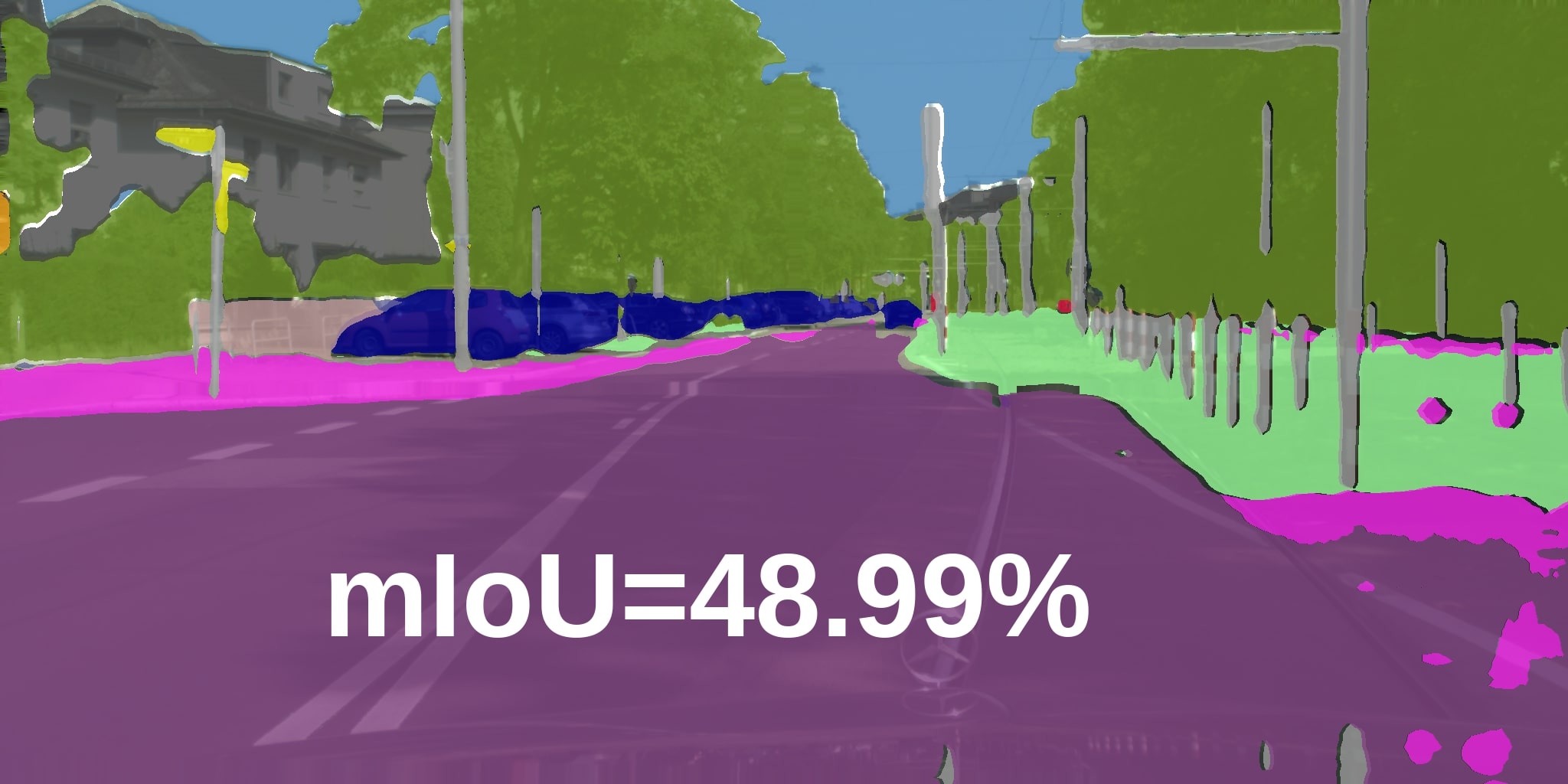} 
    \includegraphics[width=\textwidth]{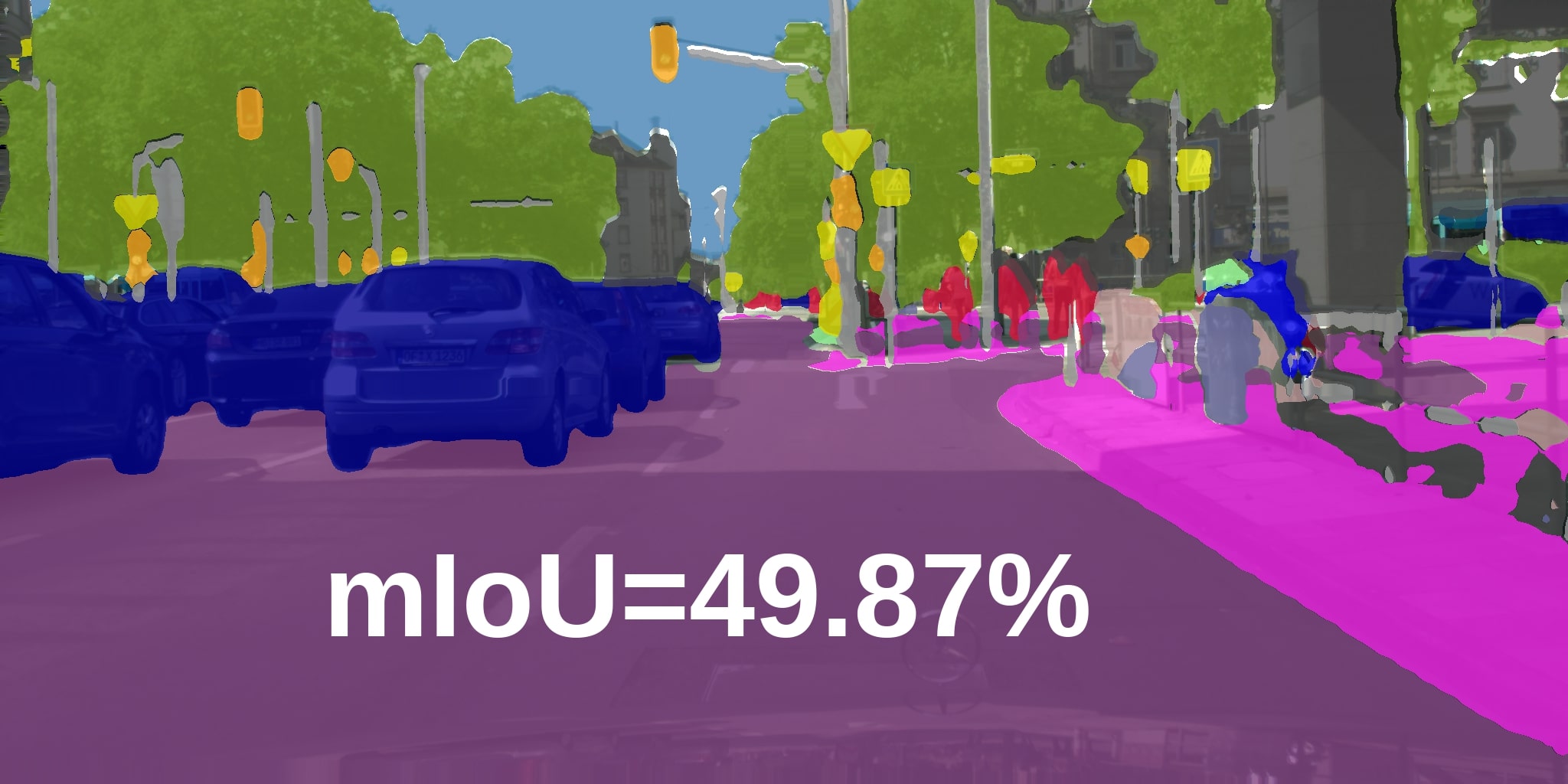} 
    \includegraphics[width=\textwidth]{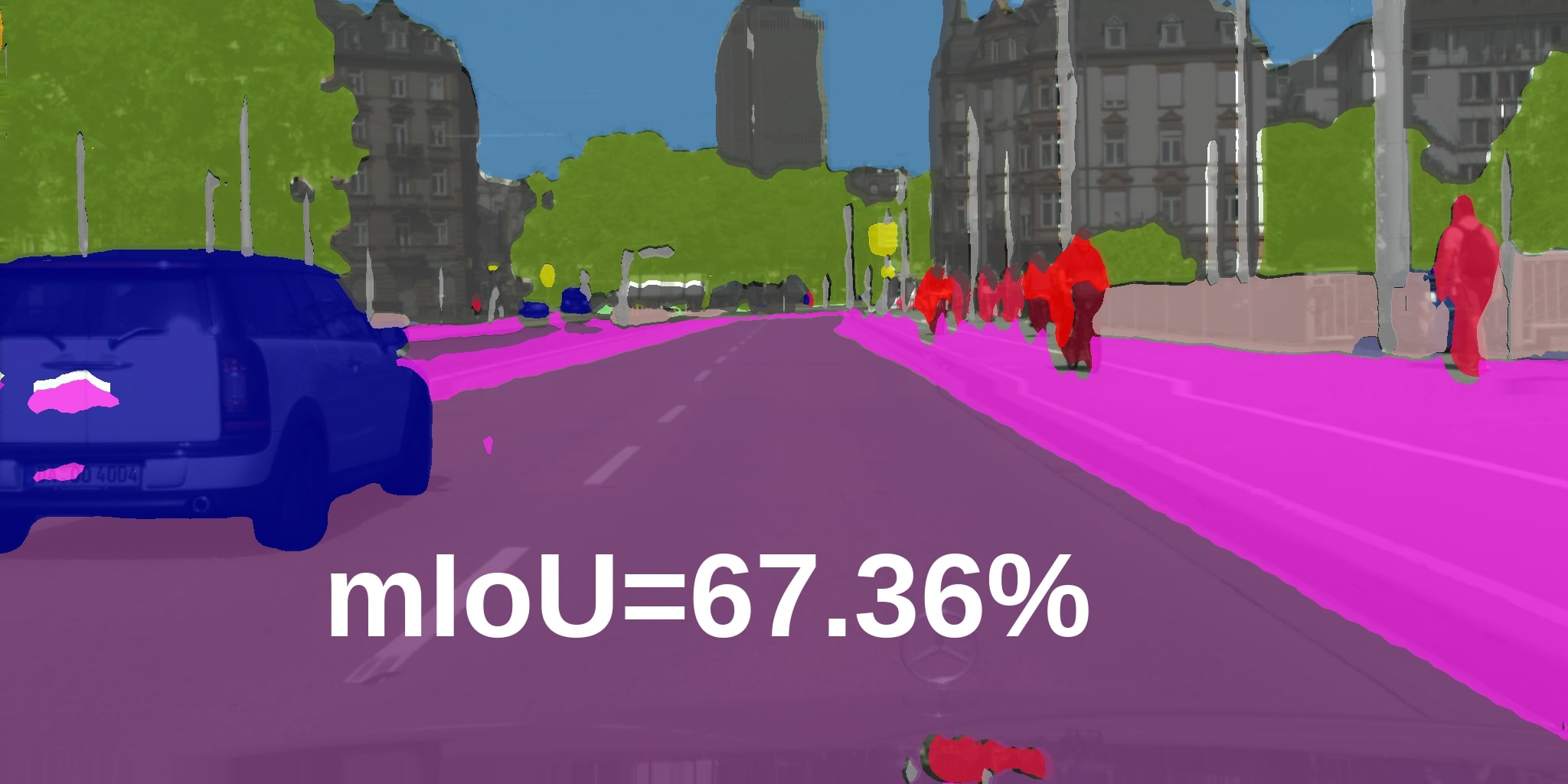} 
    \includegraphics[width=\textwidth]{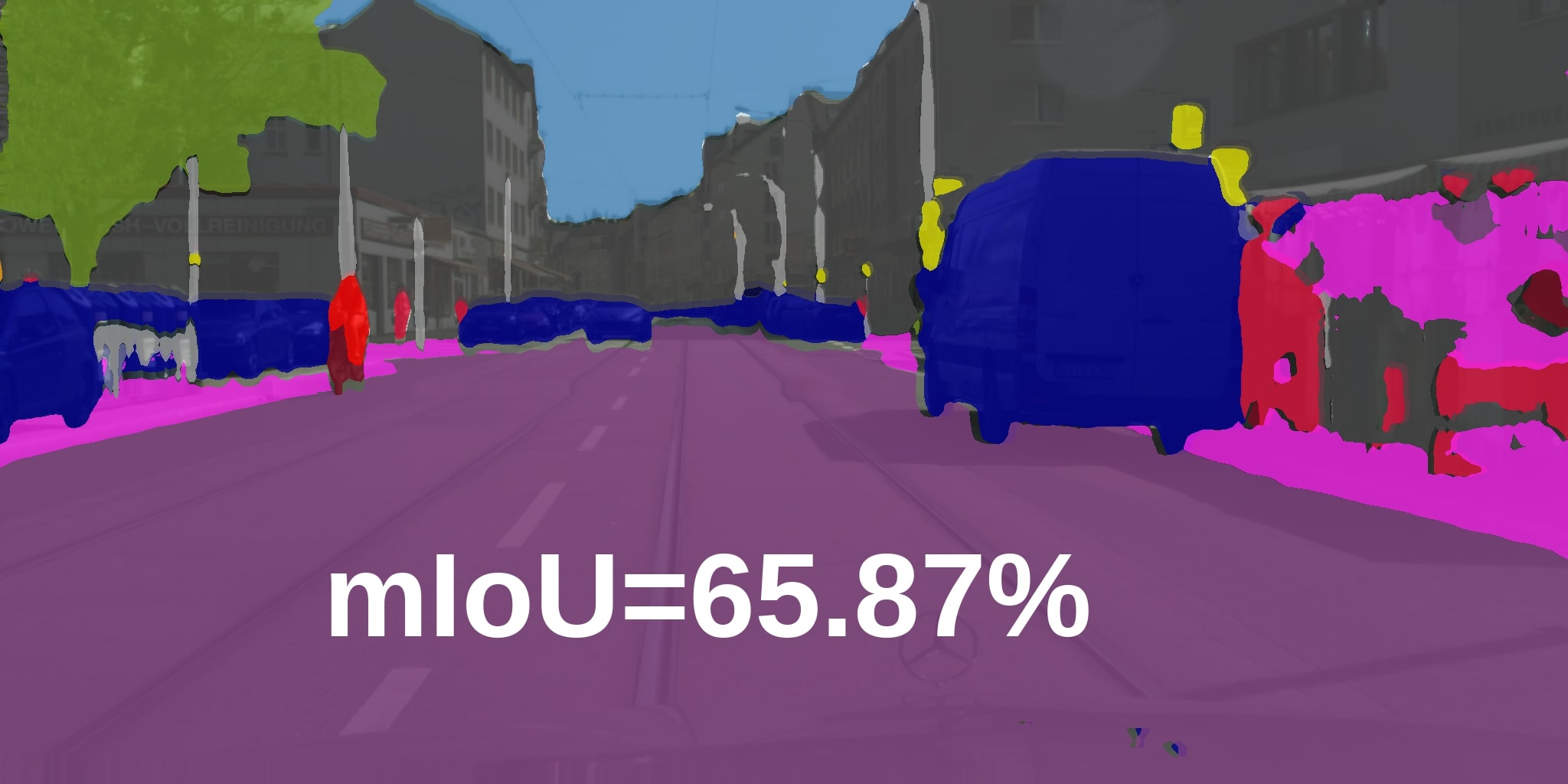} 
    \caption{Ahuja ($0.0195 \ \mathrm{bpp}$)}
    \label{fig:ahuja_qualitative}
  \end{subfigure}%
  \hspace{-0.12em}
  \begin{subfigure}{0.2455\textwidth}
    \includegraphics[width=\textwidth]{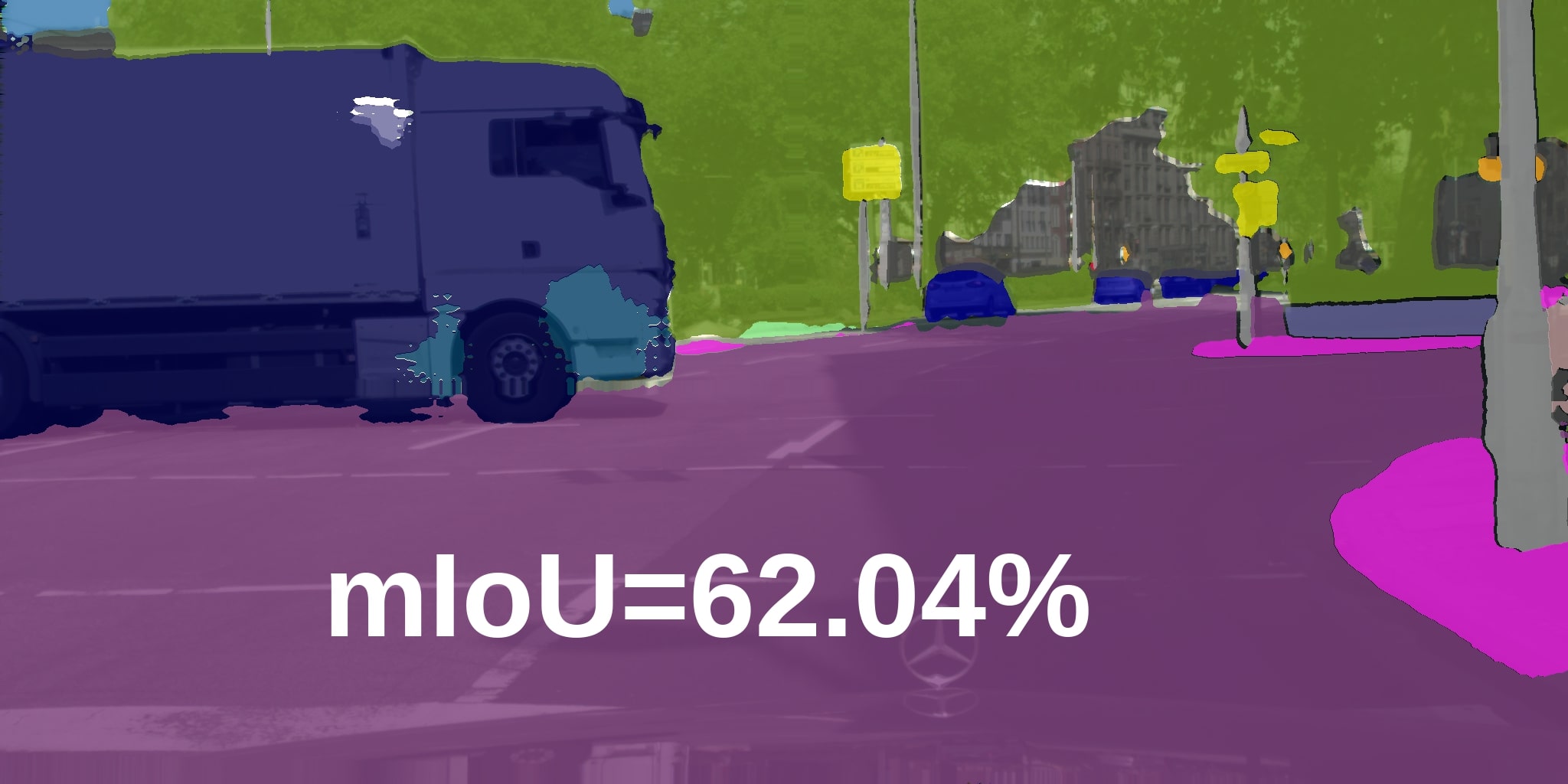} 
    \includegraphics[width=\textwidth]{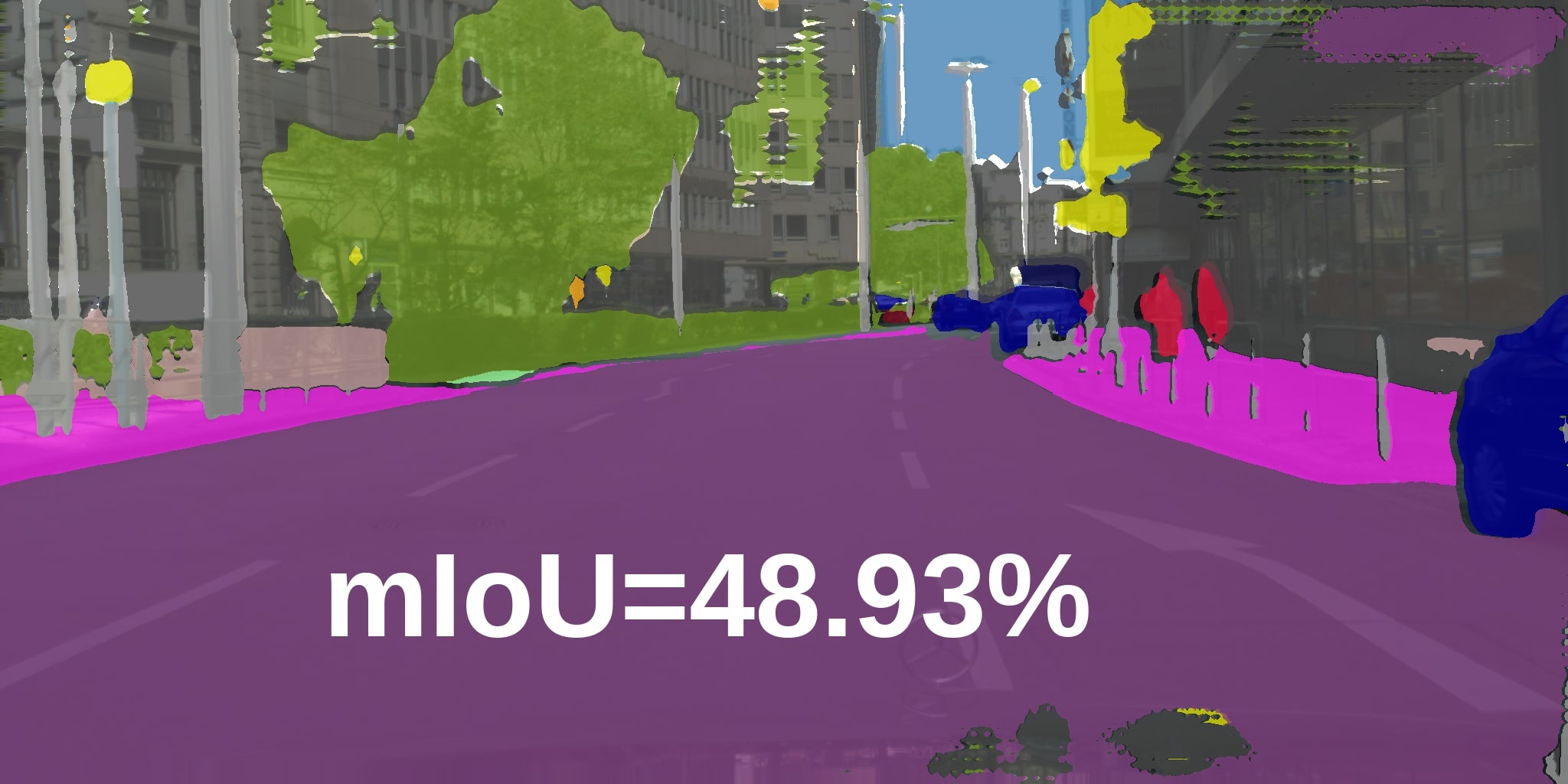} 
    \includegraphics[width=\textwidth]{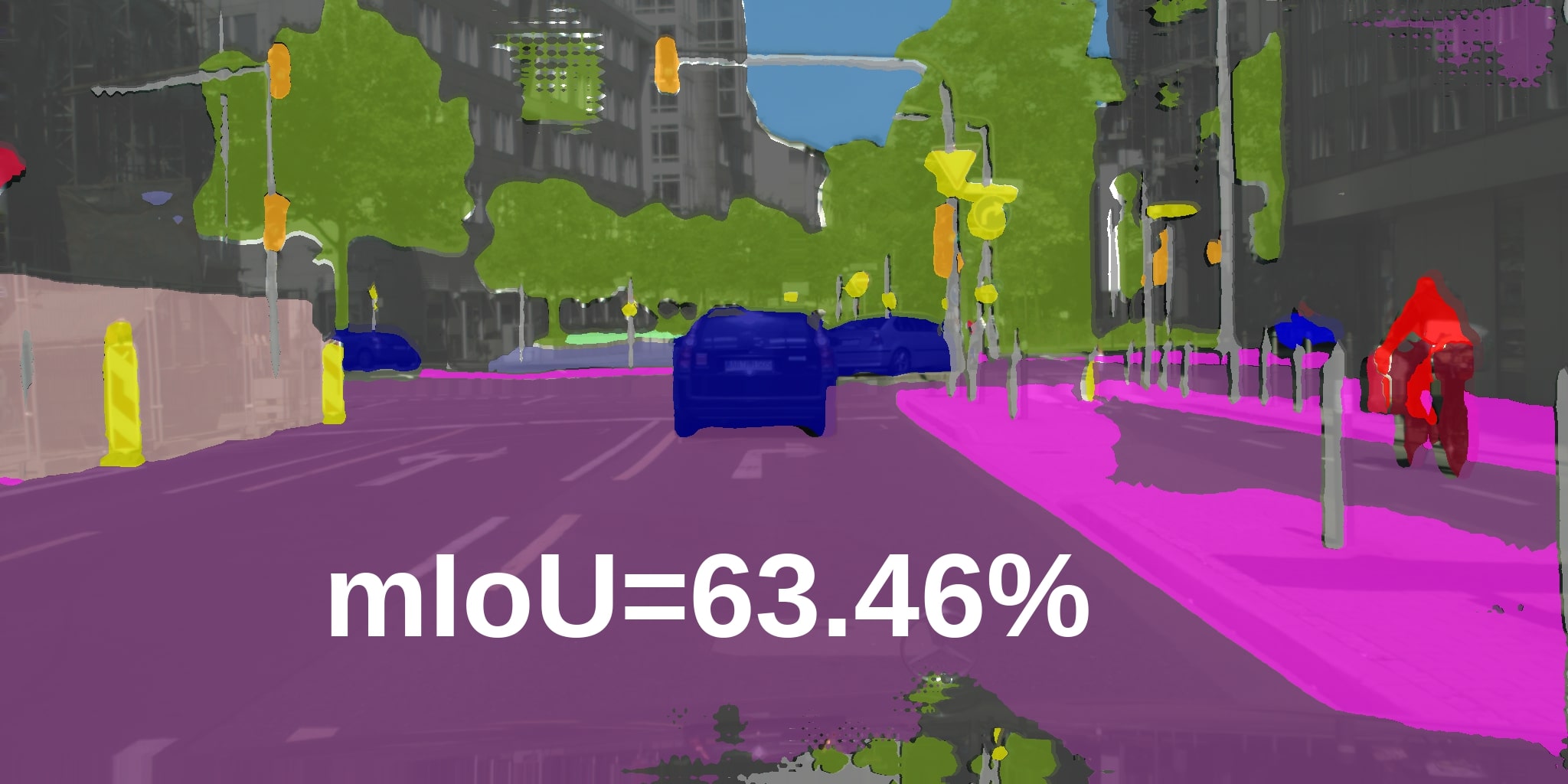} 
    \includegraphics[width=\textwidth]{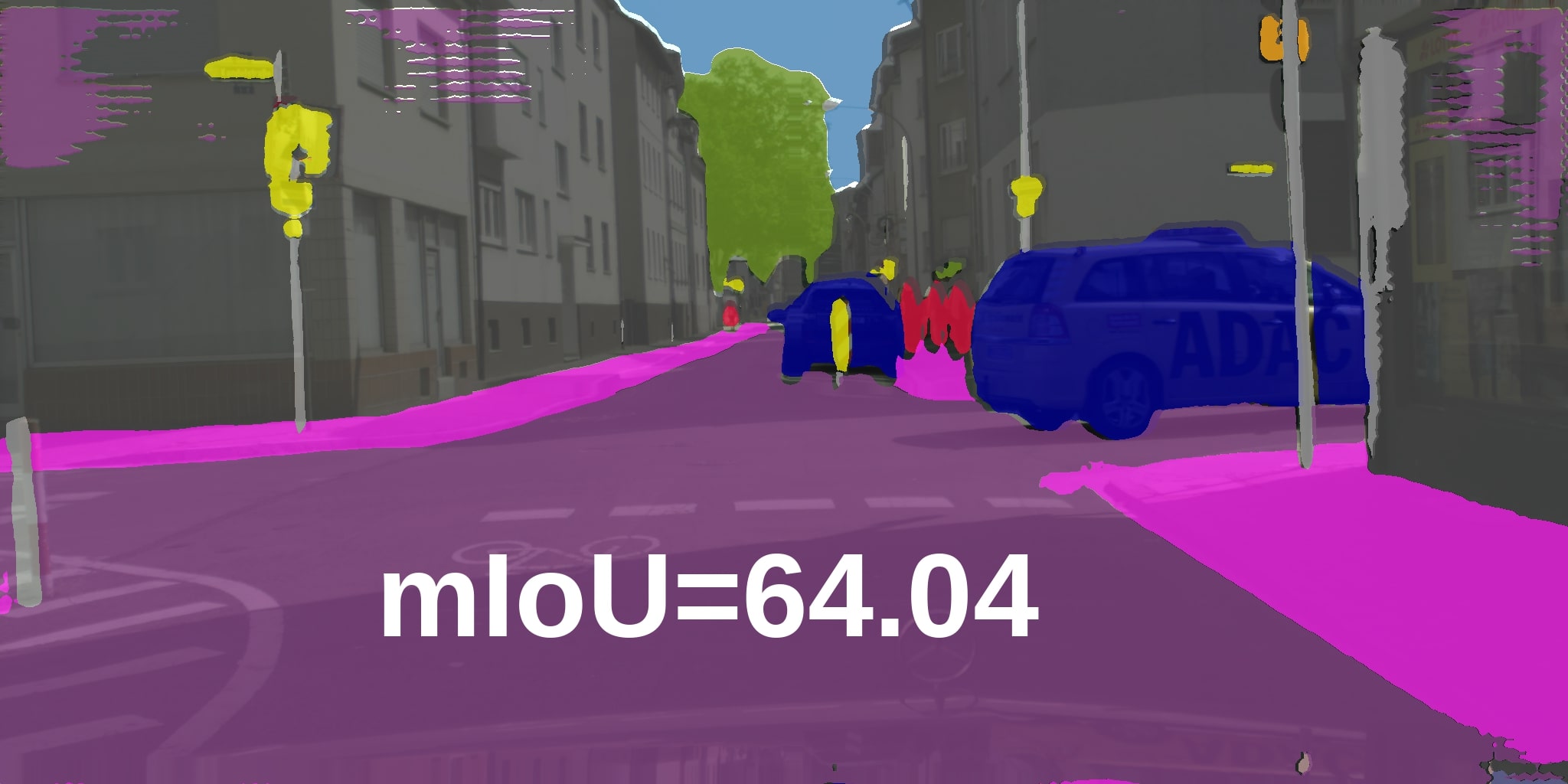} 
    \includegraphics[width=\textwidth]{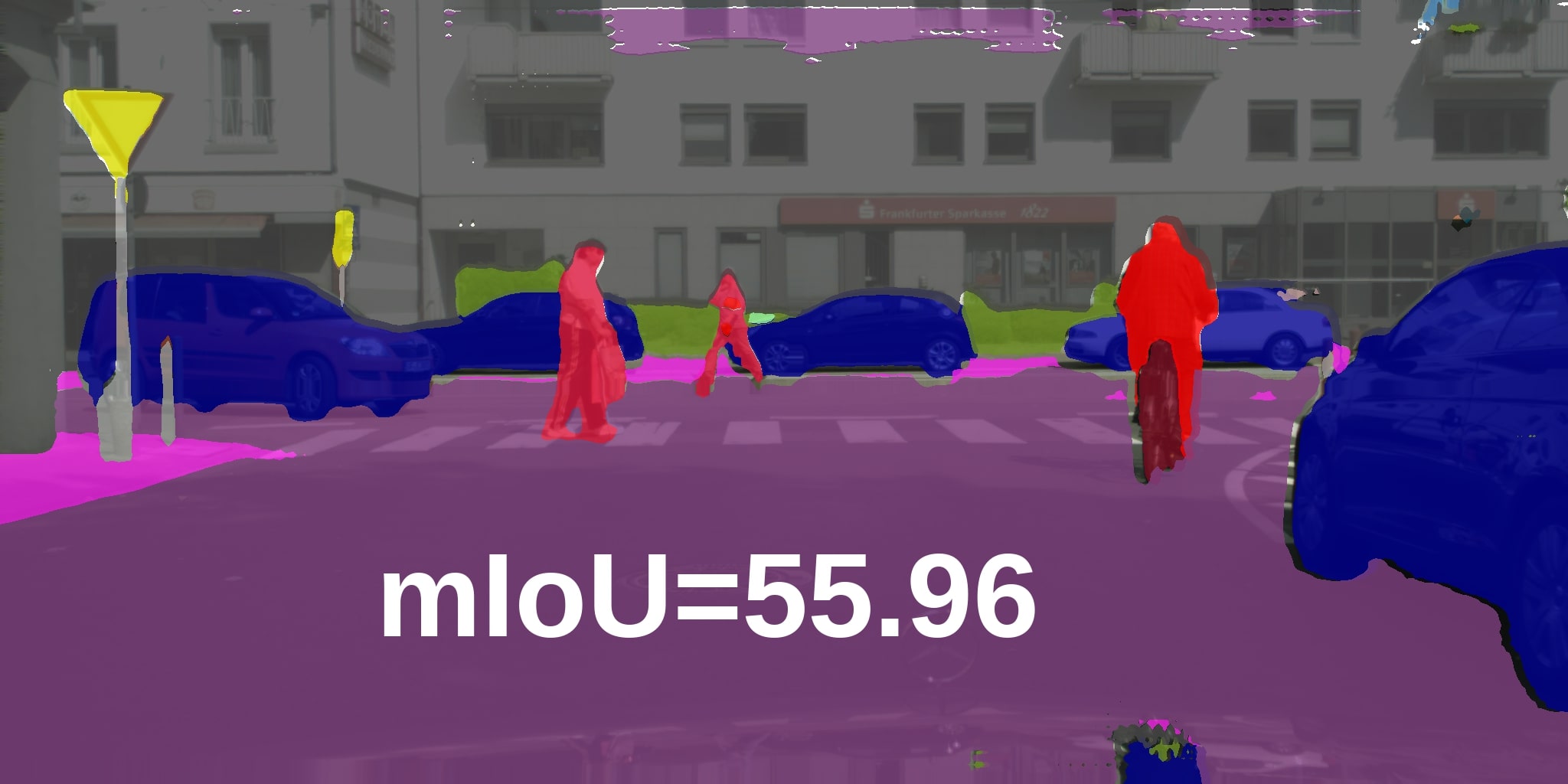}

    \includegraphics[width=\textwidth]{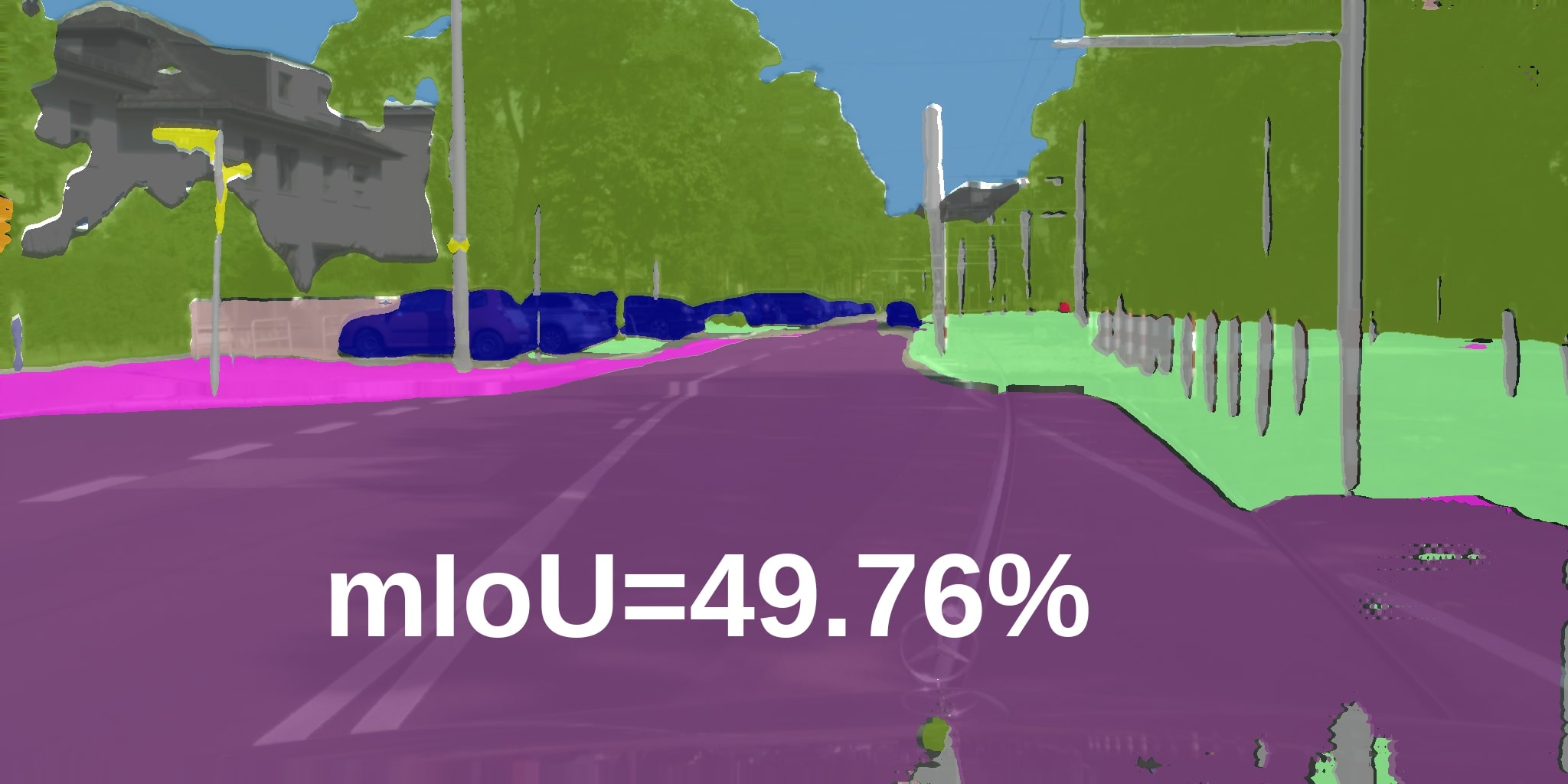} 
     \includegraphics[width=\textwidth]{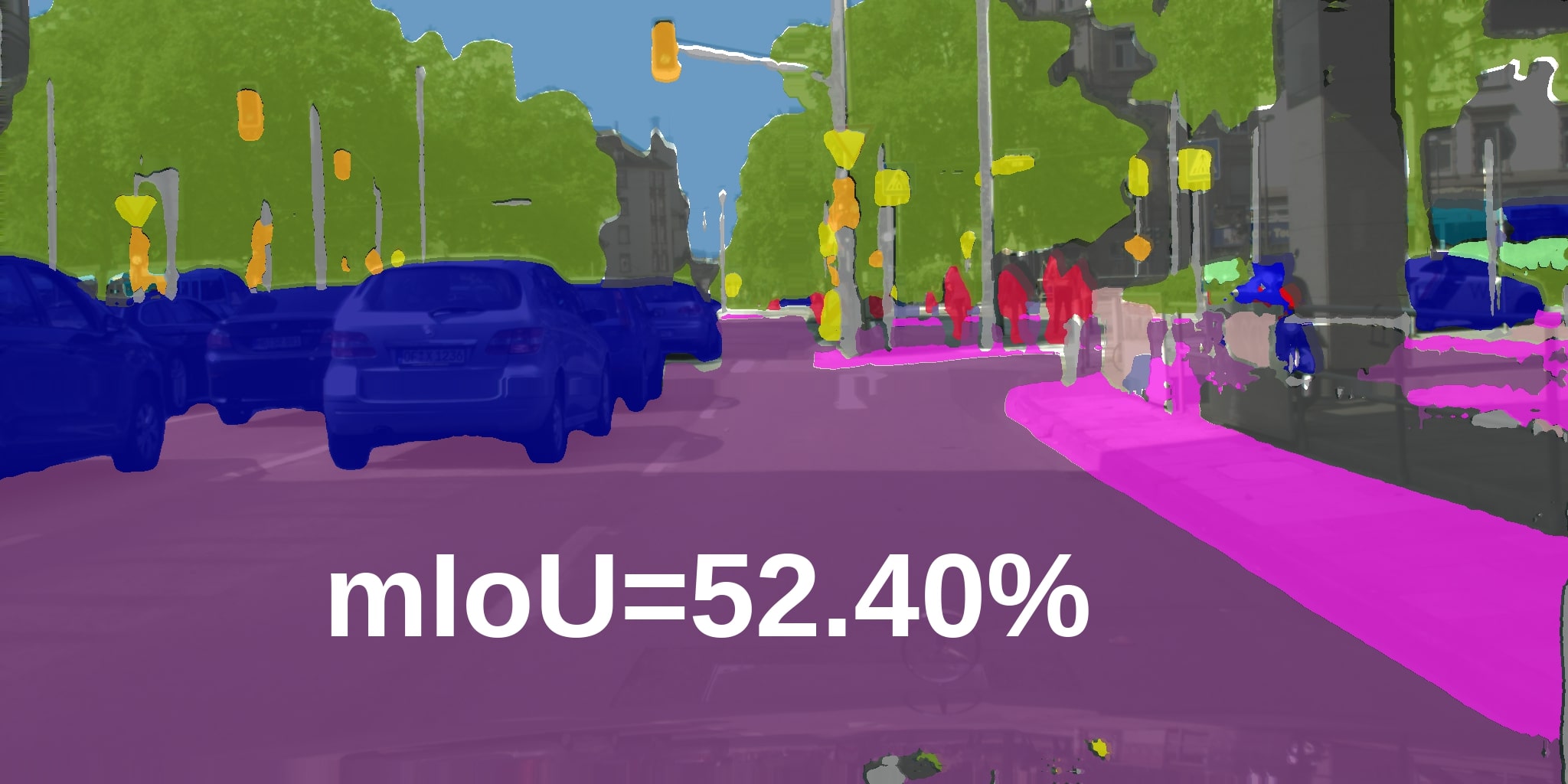} 
    \includegraphics[width=\textwidth]{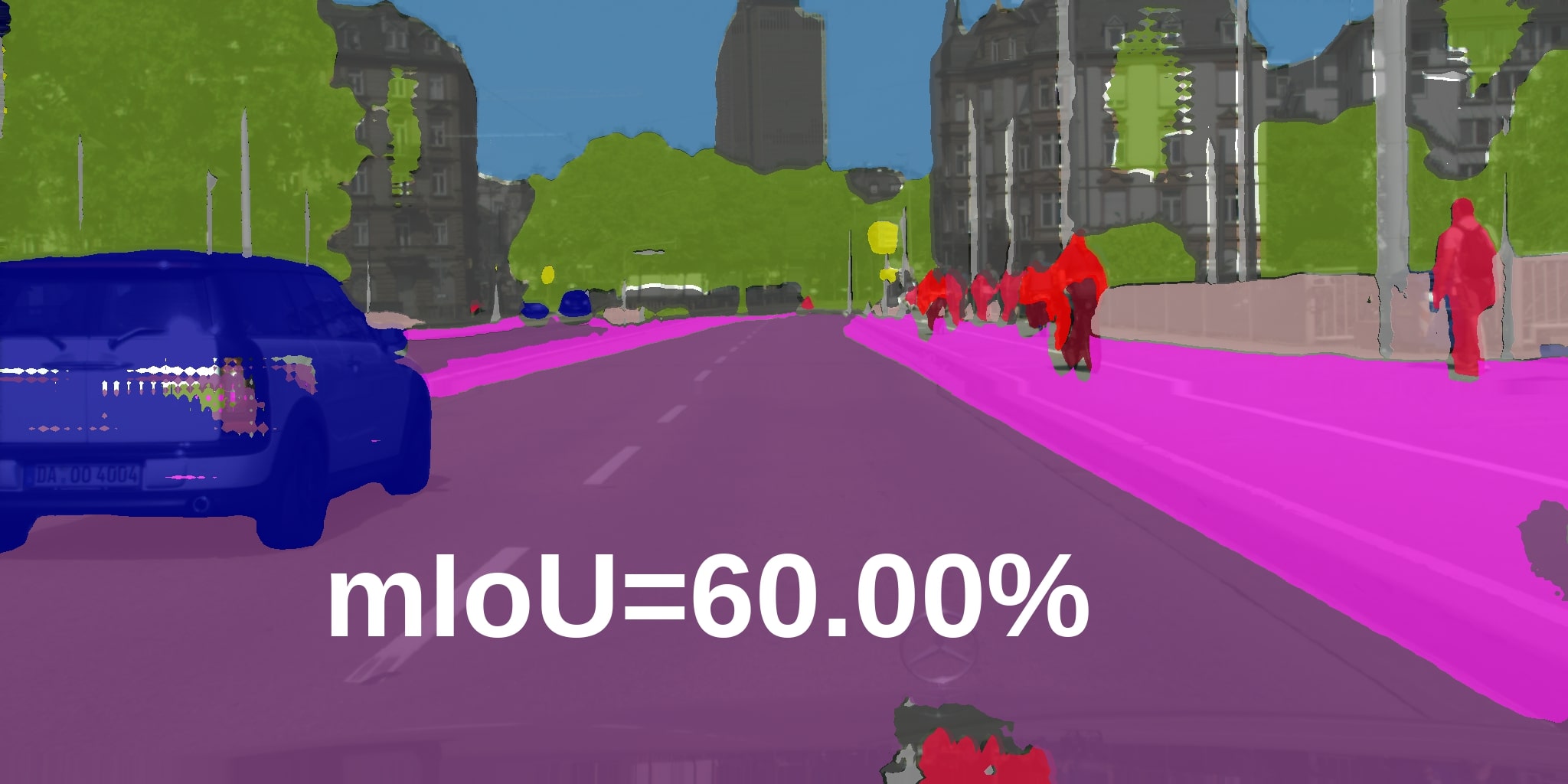} 
    \includegraphics[width=\textwidth]{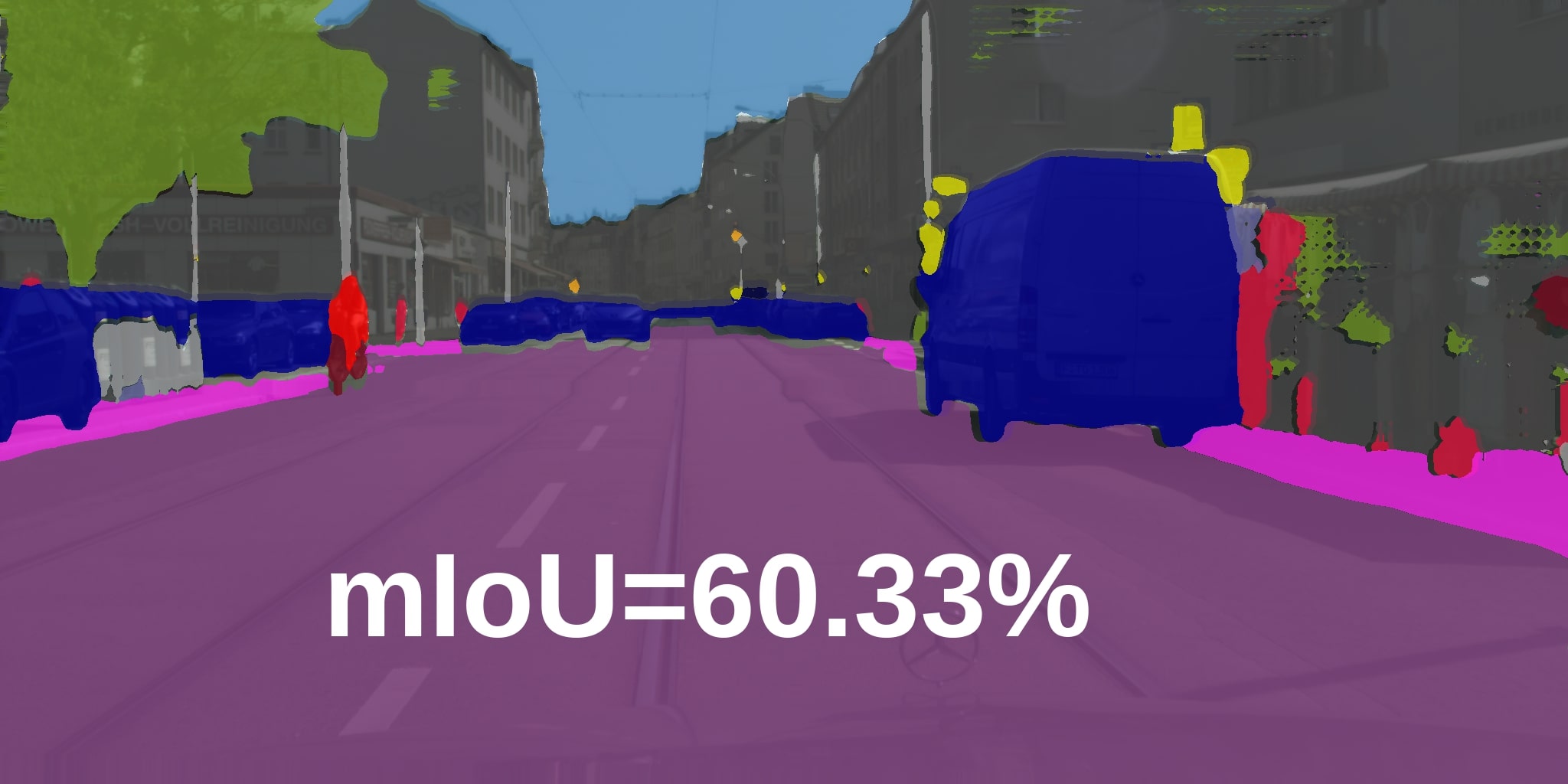} 
    \caption{Ours ($0.0147 \ \mathrm{bpp}$)}
    \label{fig:ours_qualitative}
  \end{subfigure}
  \caption{Qualitative comparison of the \textbf{proposed $\bf{JD}$ approach ("Ours")} against \textbf{Ahuja et al. \cite{ahuja2023neural}} on the Cityscapes dataset \textbf{at low bitrates.}}
  \label{fig:cs_low_qualitative}
  \end{figure}
\begin{figure}[t!]
  \begin{subfigure}{0.2455\textwidth}
    
    \includegraphics[width=\textwidth]{Figures/Supplementary/Cityscapes/GT/test_frankfurt_000000_012009_leftImg8bit.png_0.jpg} 
    \includegraphics[width=\textwidth]{Figures/Supplementary/Cityscapes/GT/test_frankfurt_000000_013942_leftImg8bit.png_0.jpg} 
    \includegraphics[width=\textwidth]{Figures/Supplementary/Cityscapes/GT/test_frankfurt_000000_003025_leftImg8bit.png_0.jpg} 
    \includegraphics[width=\textwidth]{Figures/Supplementary/Cityscapes/GT/test_frankfurt_000000_004617_leftImg8bit.png_0.jpg} 
    \includegraphics[width=\textwidth]{Figures/Supplementary/Cityscapes/GT/test_frankfurt_000000_013240_leftImg8bit.png_0.jpg} 
    \includegraphics[width=\textwidth]{Figures/Supplementary/Cityscapes/GT/test_frankfurt_000000_011810_leftImg8bit.png_0.jpg} 
    \includegraphics[width=\textwidth]{Figures/Supplementary/Cityscapes/GT/test_frankfurt_000000_010351_leftImg8bit.png_0.jpg} 
    \caption{Ground Truth}
  \end{subfigure}
    \hspace{-0.42em}
  \begin{subfigure}{0.2455\textwidth}
    
    \includegraphics[width=\textwidth]{Figures/Supplementary/Cityscapes/No_Compression/test_frankfurt_000000_012009_leftImg8bit.png_0.jpg} 
    \includegraphics[width=\textwidth]{Figures/Supplementary/Cityscapes/No_Compression/test_frankfurt_000000_013942_leftImg8bit.png_0.jpg} 
    \includegraphics[width=\textwidth]{Figures/Supplementary/Cityscapes/No_Compression/test_frankfurt_000000_003025_leftImg8bit.png_0.jpg} 
    \includegraphics[width=\textwidth]{Figures/Supplementary/Cityscapes/No_Compression/test_frankfurt_000000_004617_leftImg8bit.png_0.jpg} 
    \includegraphics[width=\textwidth]{Figures/Supplementary/Cityscapes/No_Compression/test_frankfurt_000000_013240_leftImg8bit.png_0.jpg} 
    \includegraphics[width=\textwidth]{Figures/Supplementary/Cityscapes/No_Compression/test_frankfurt_000000_011810_leftImg8bit.png_0.jpg} 
    \includegraphics[width=\textwidth]{Figures/Supplementary/Cityscapes/No_Compression/test_frankfurt_000000_010351_leftImg8bit.png_0.jpg} 
    \caption{No Compression \cite{deeplabv3}}
  \end{subfigure}
   \hspace{-0.45em}
  \begin{subfigure}{0.2455\textwidth}

    \includegraphics[width=\textwidth]{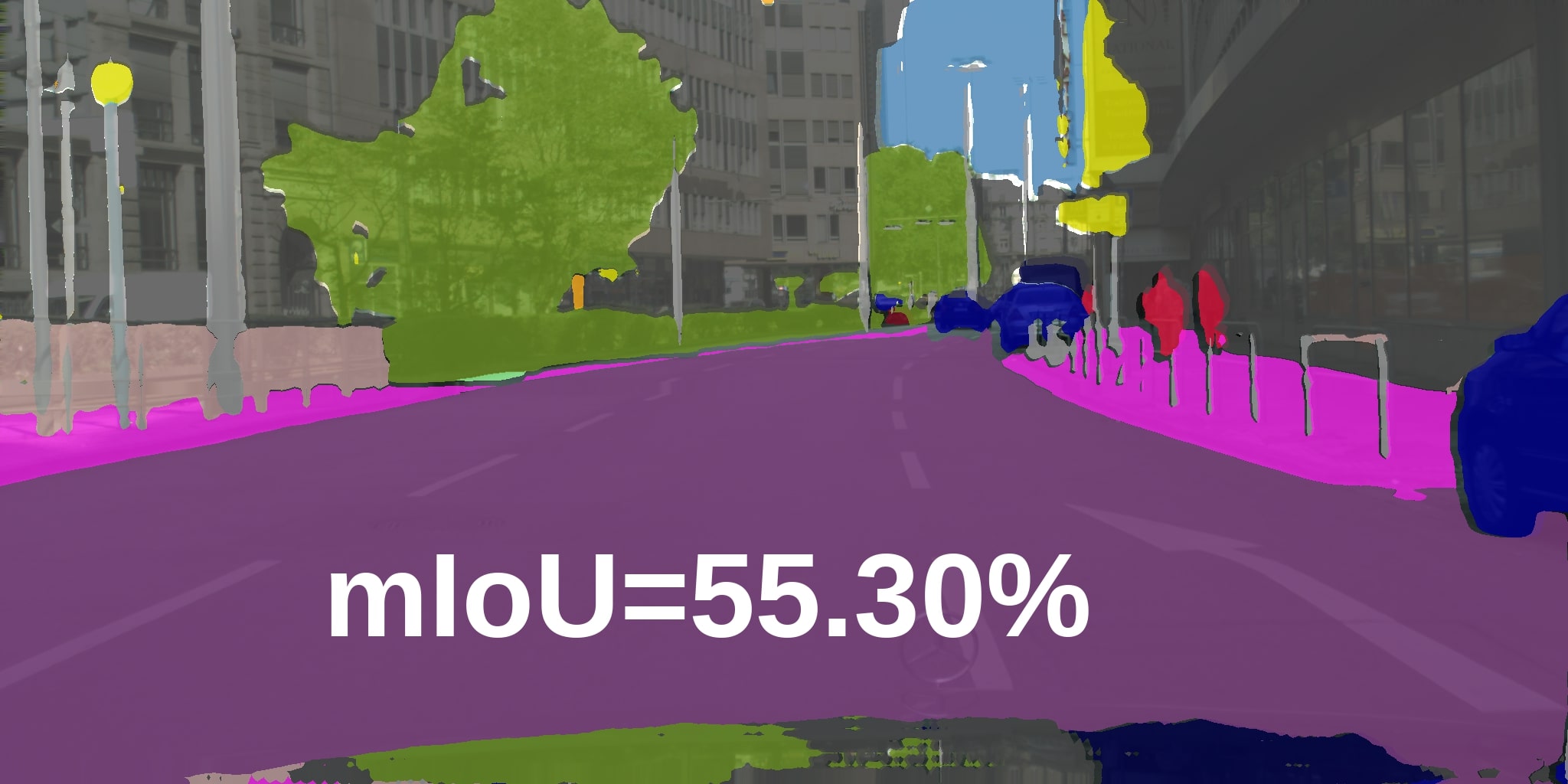} 
    \includegraphics[width=\textwidth]{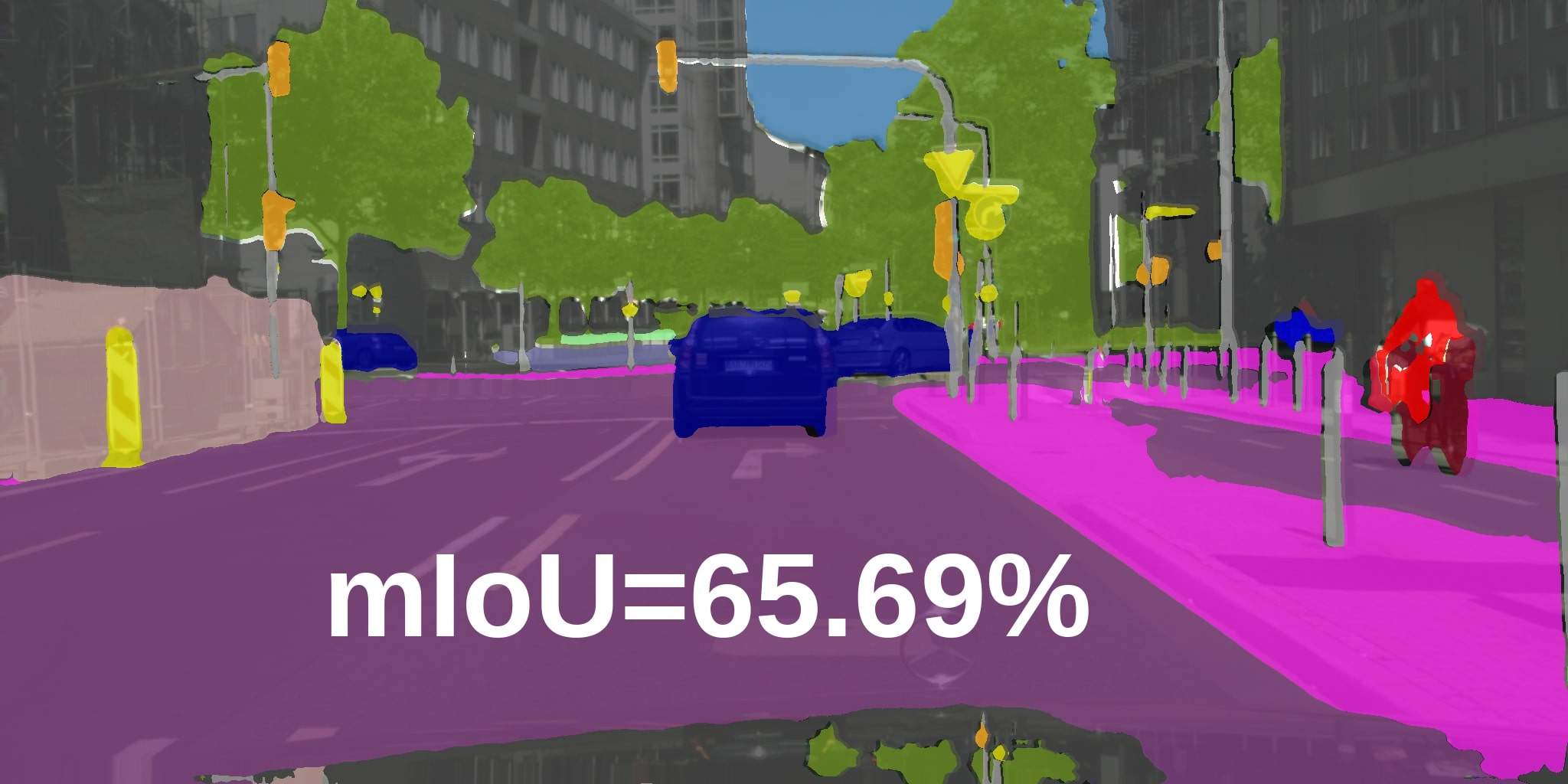} 
    \includegraphics[width=\textwidth]{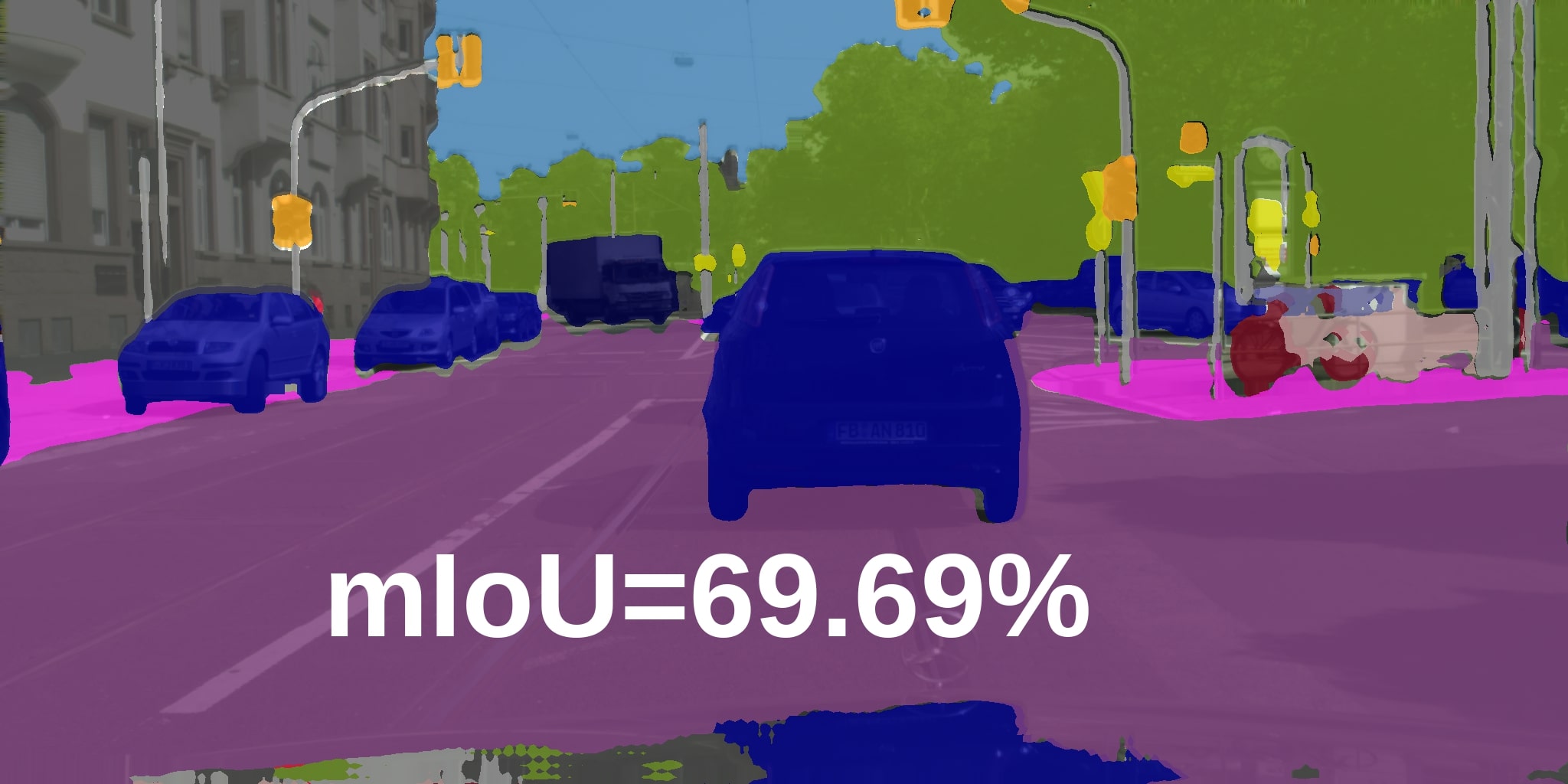} 
    \includegraphics[width=\textwidth]{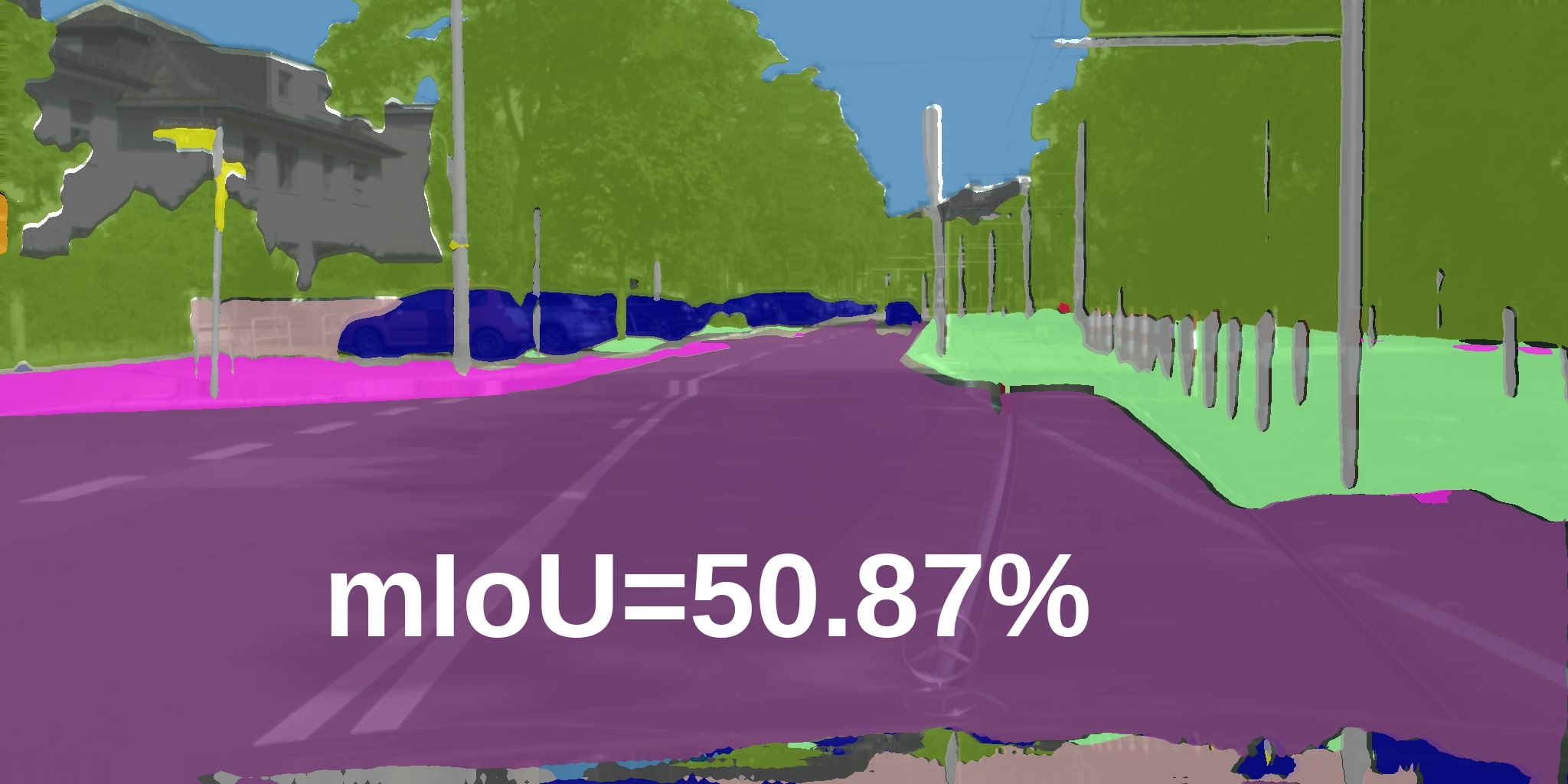} 
    \includegraphics[width=\textwidth]{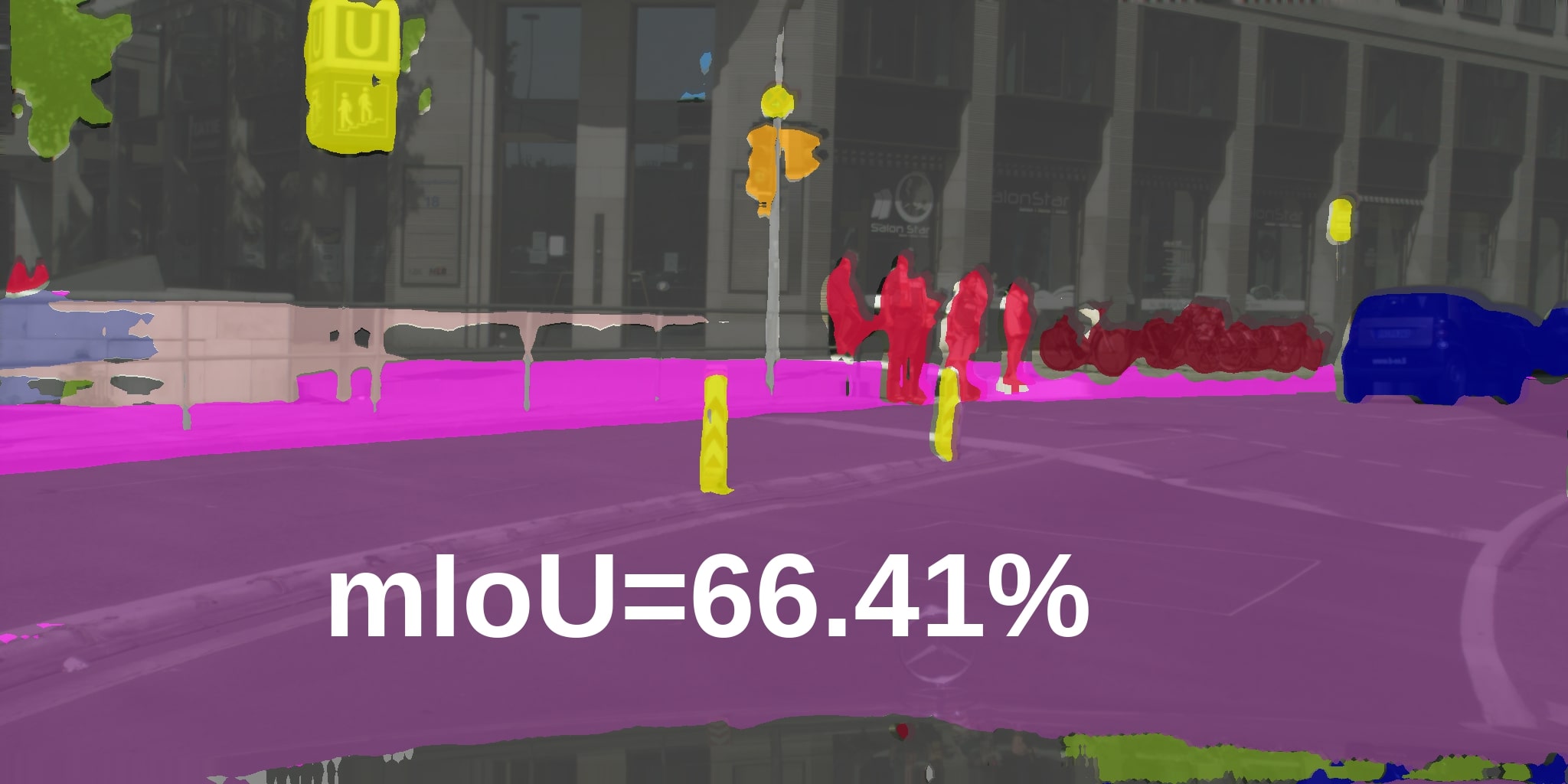} 
    \includegraphics[width=\textwidth]{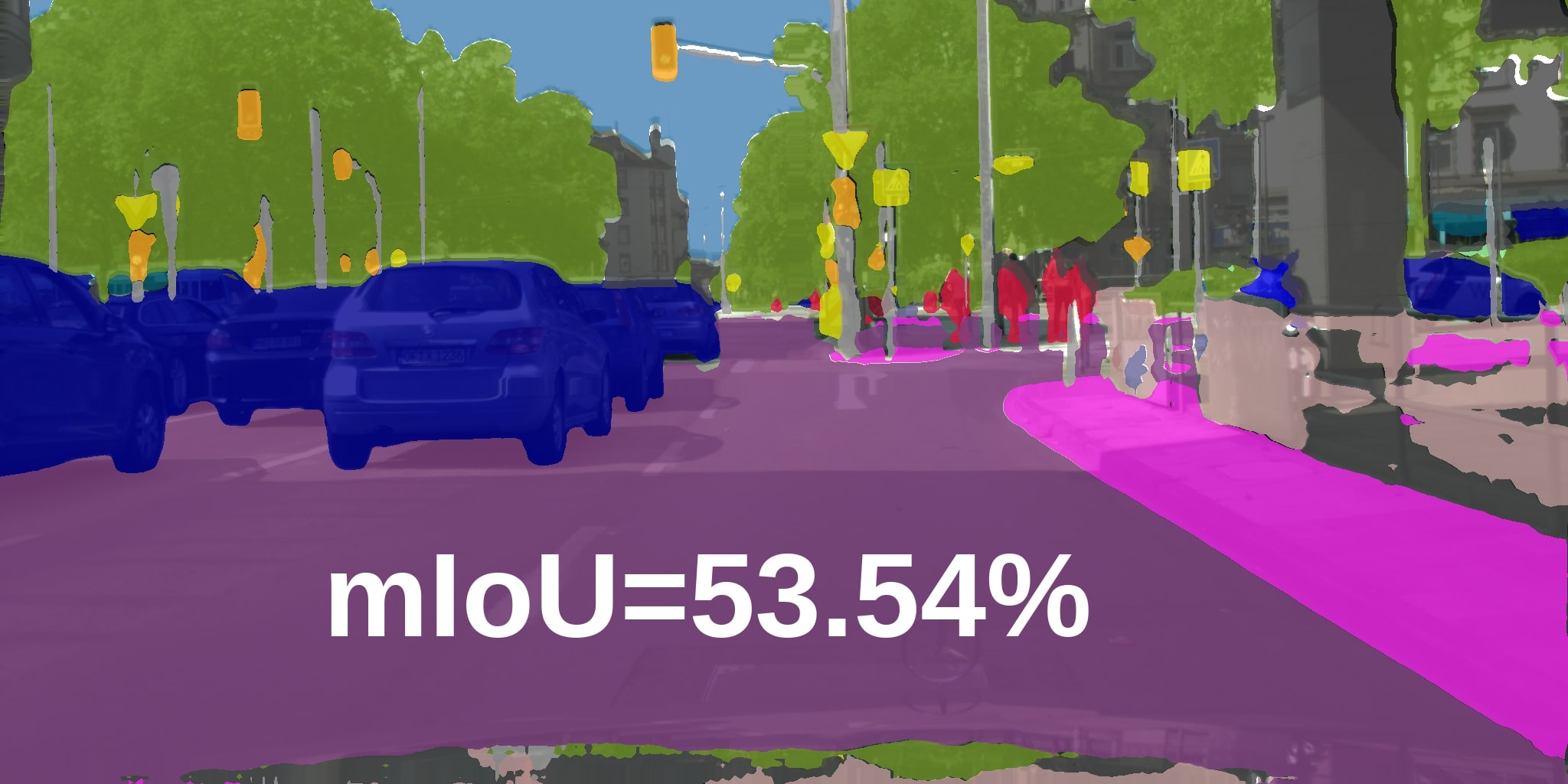} 
    \includegraphics[width=\textwidth]{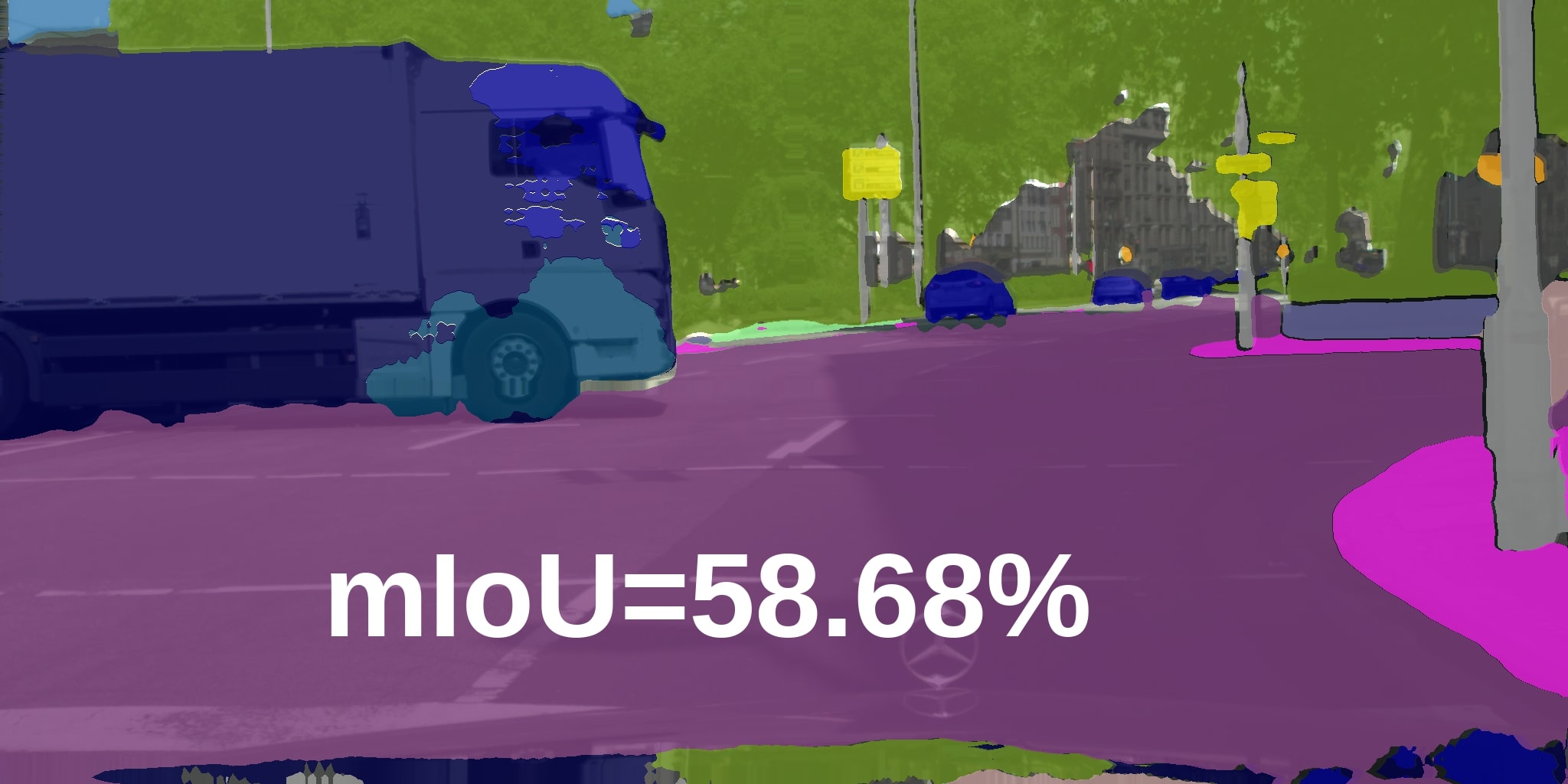} 
    \caption{Ahuja ($0.2856 \ \mathrm{bpp}$)}
    \label{fig:ahuja_qualitative}
  \end{subfigure}%
  \hspace{-0.12em}
  \begin{subfigure}{0.2455\textwidth}
    \includegraphics[width=\textwidth]{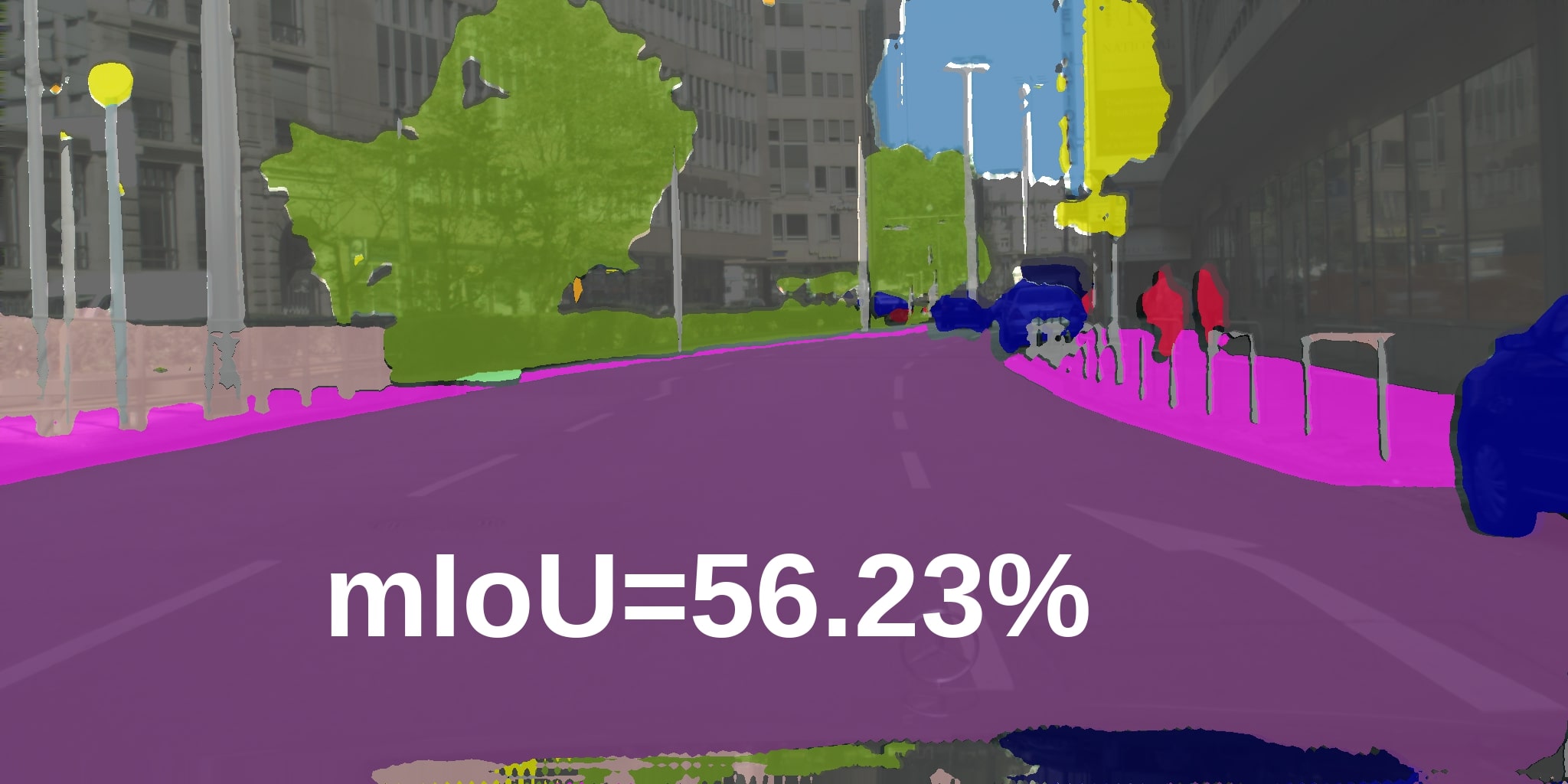} 
    \includegraphics[width=\textwidth]{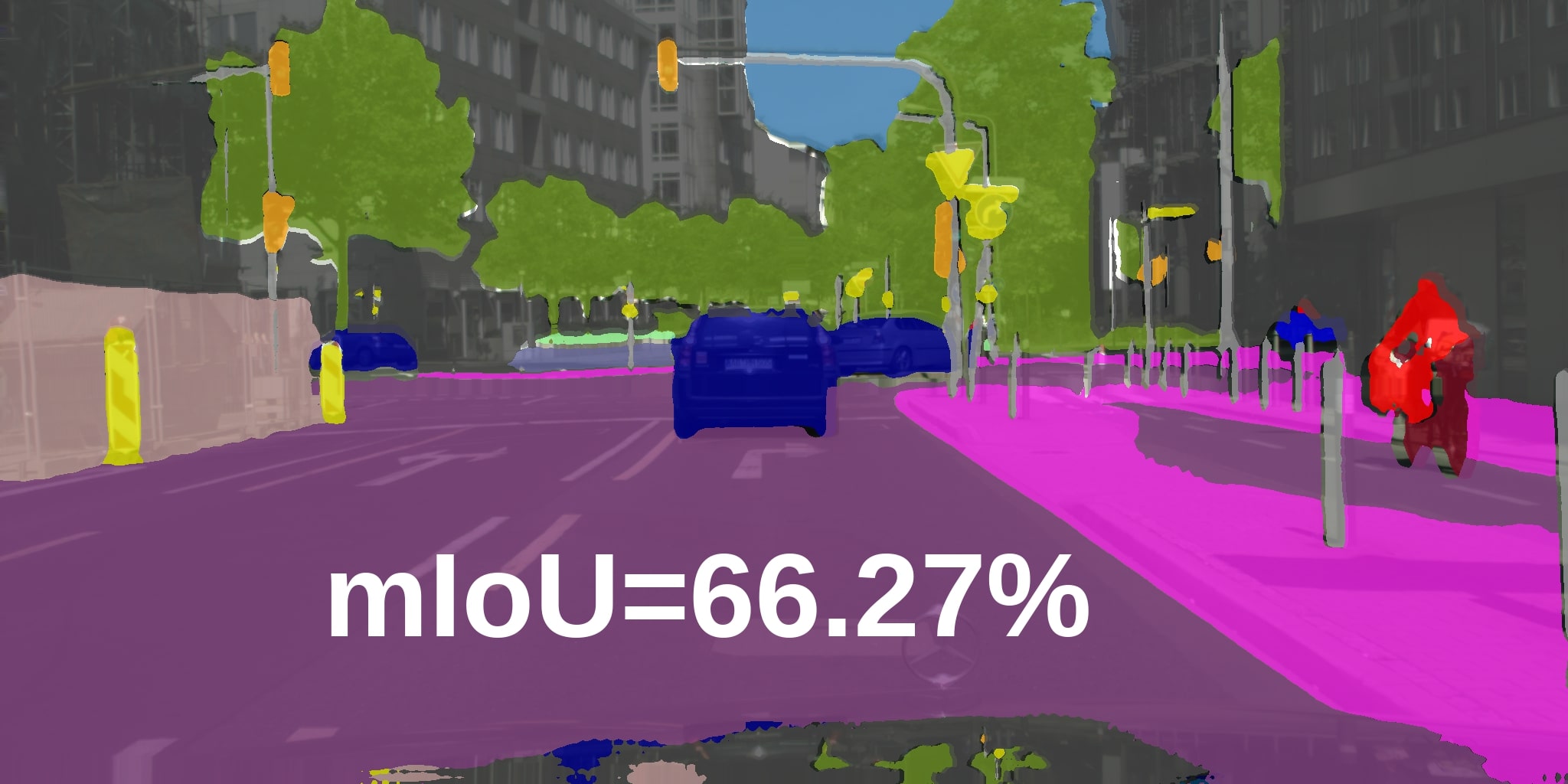} 
    \includegraphics[width=\textwidth]{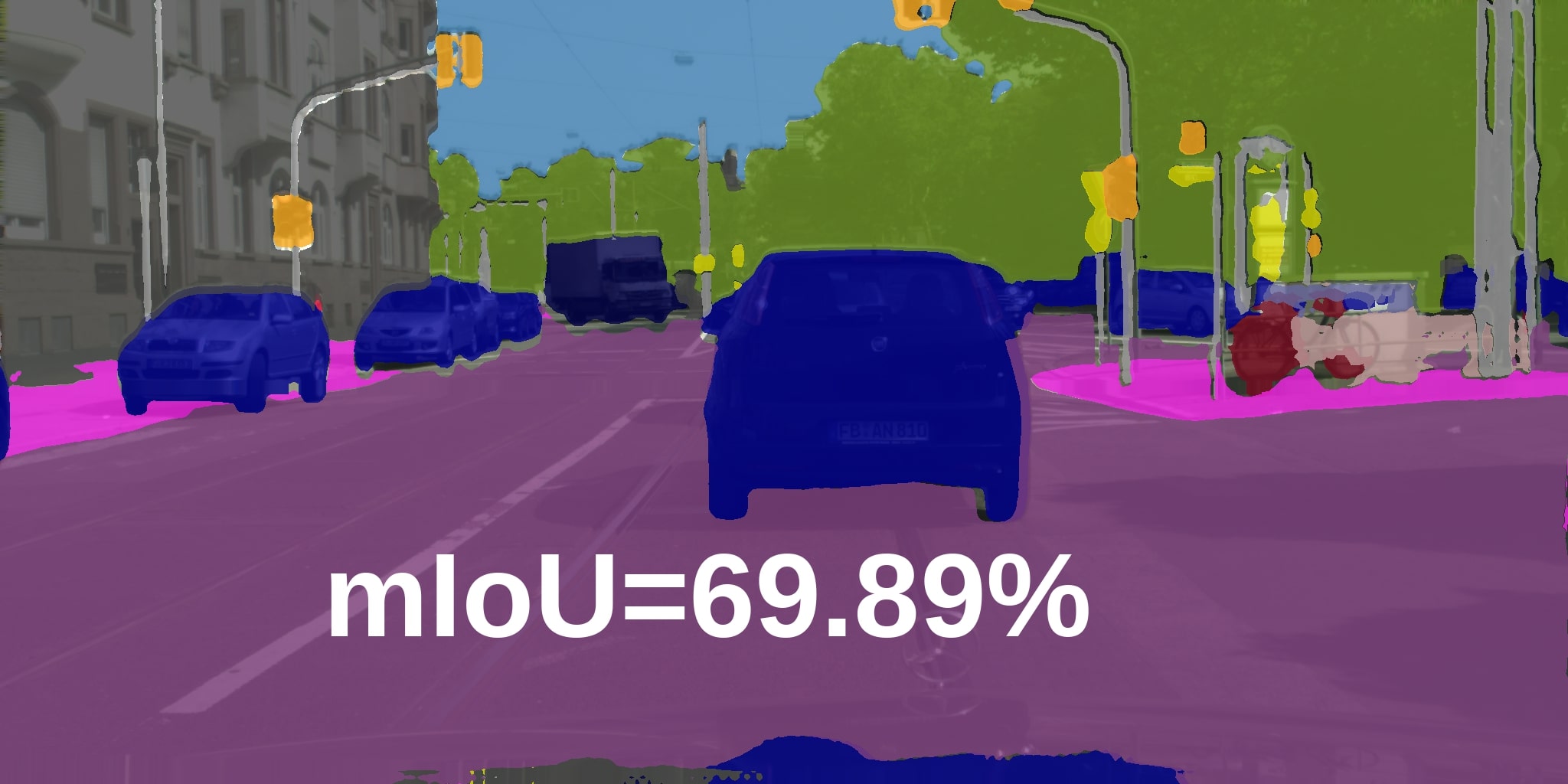} 
    \includegraphics[width=\textwidth]{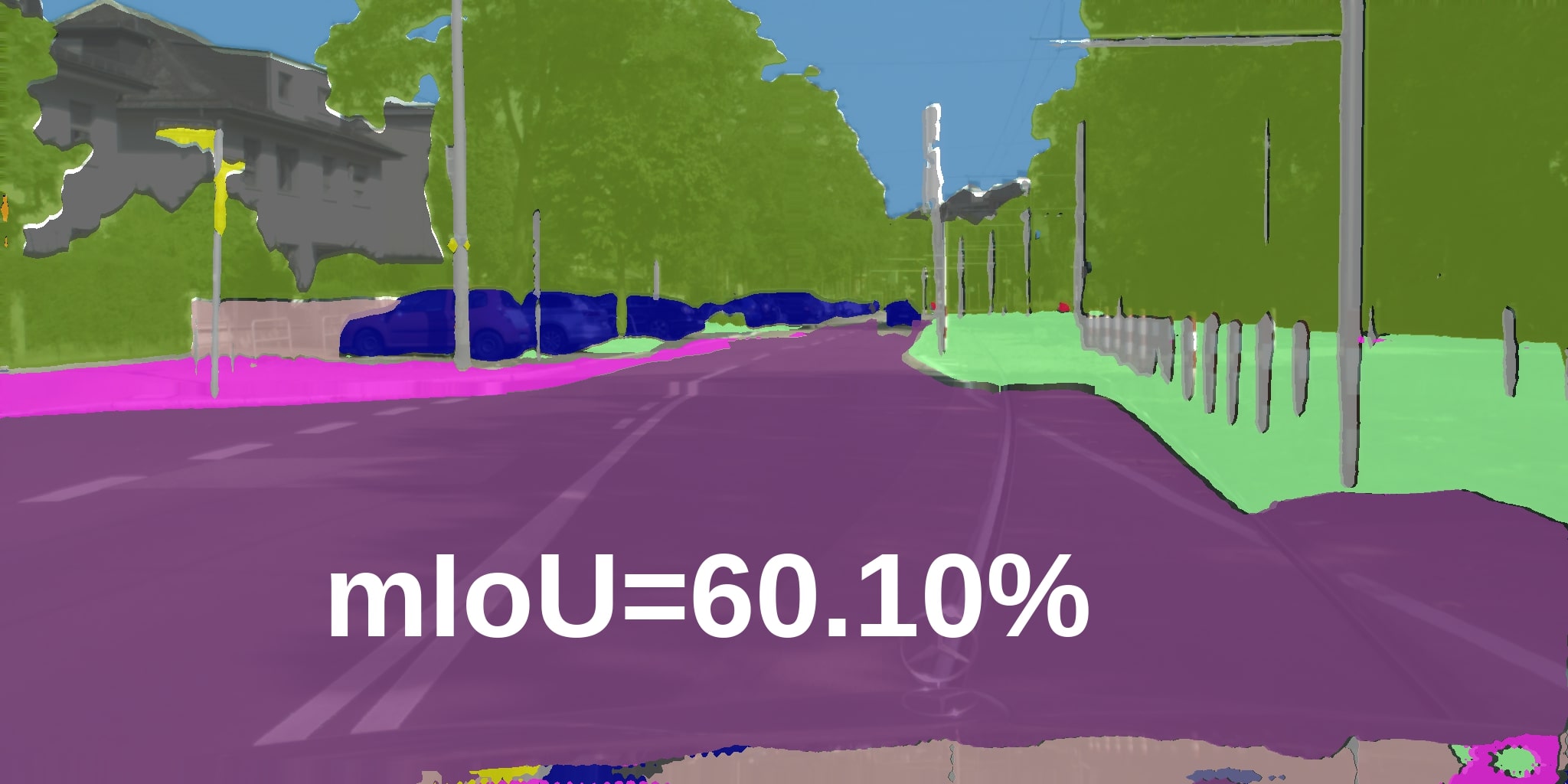} 
    \includegraphics[width=\textwidth]{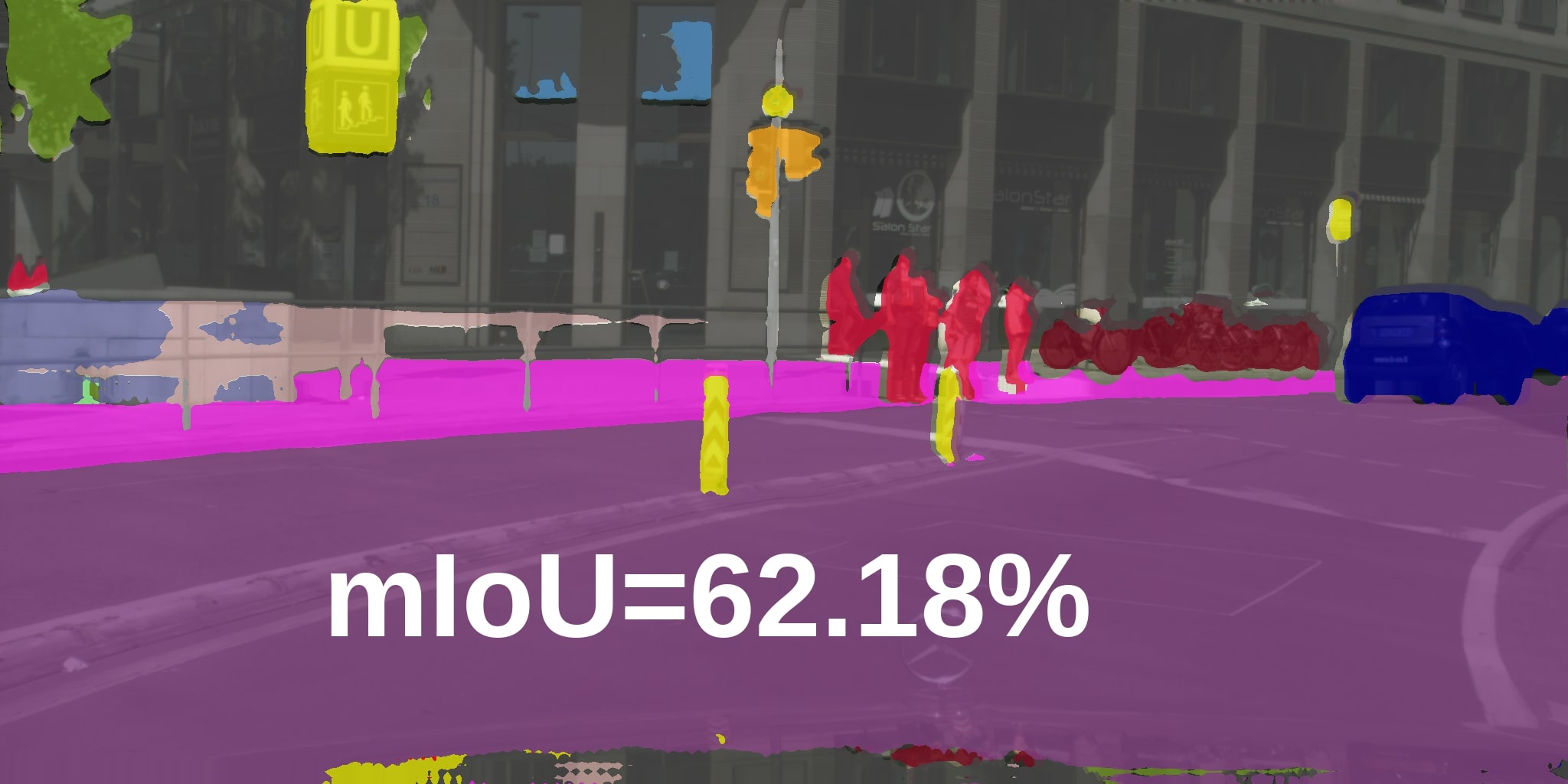} 
    \includegraphics[width=\textwidth]{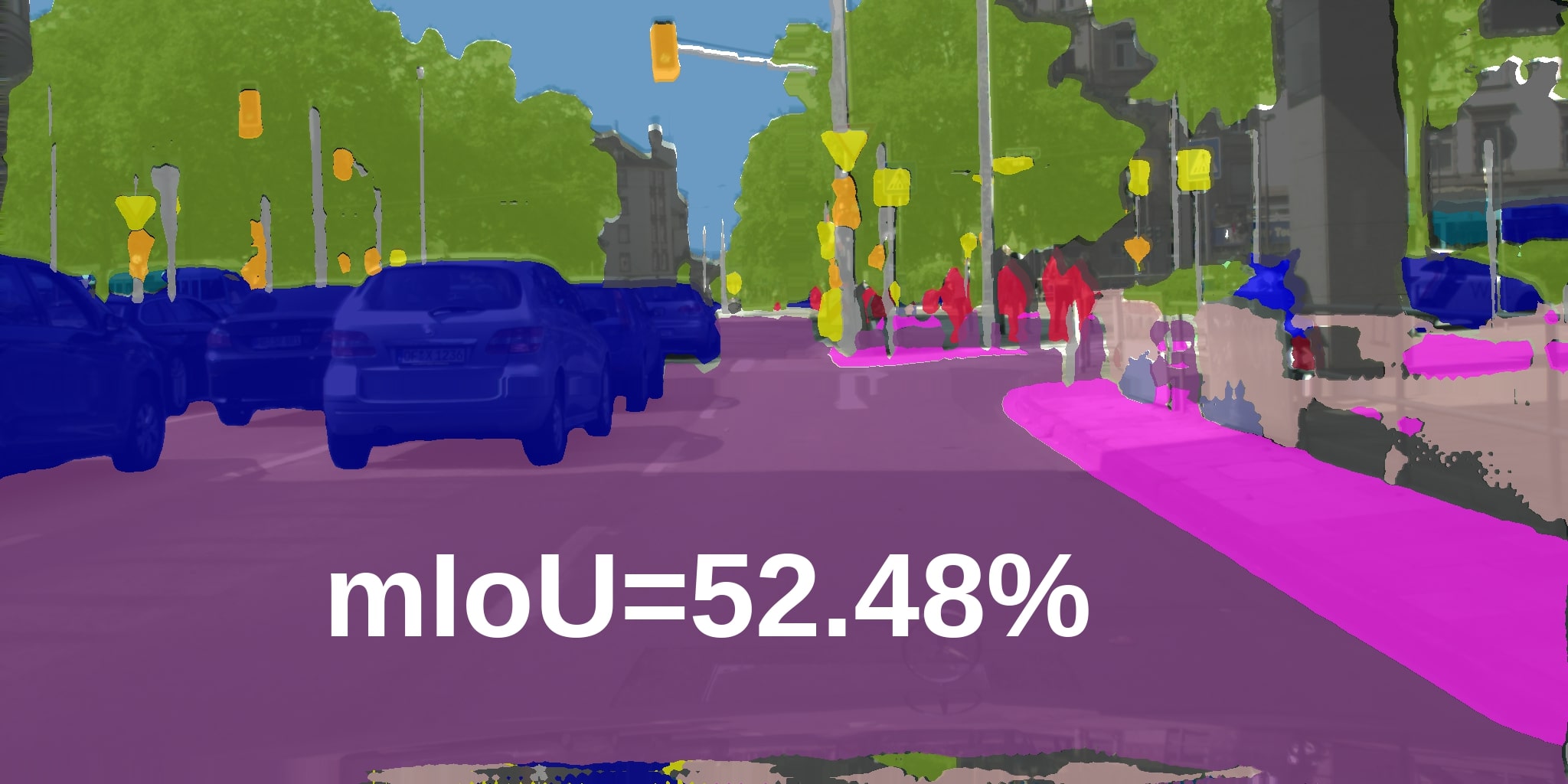} 
    \includegraphics[width=\textwidth]{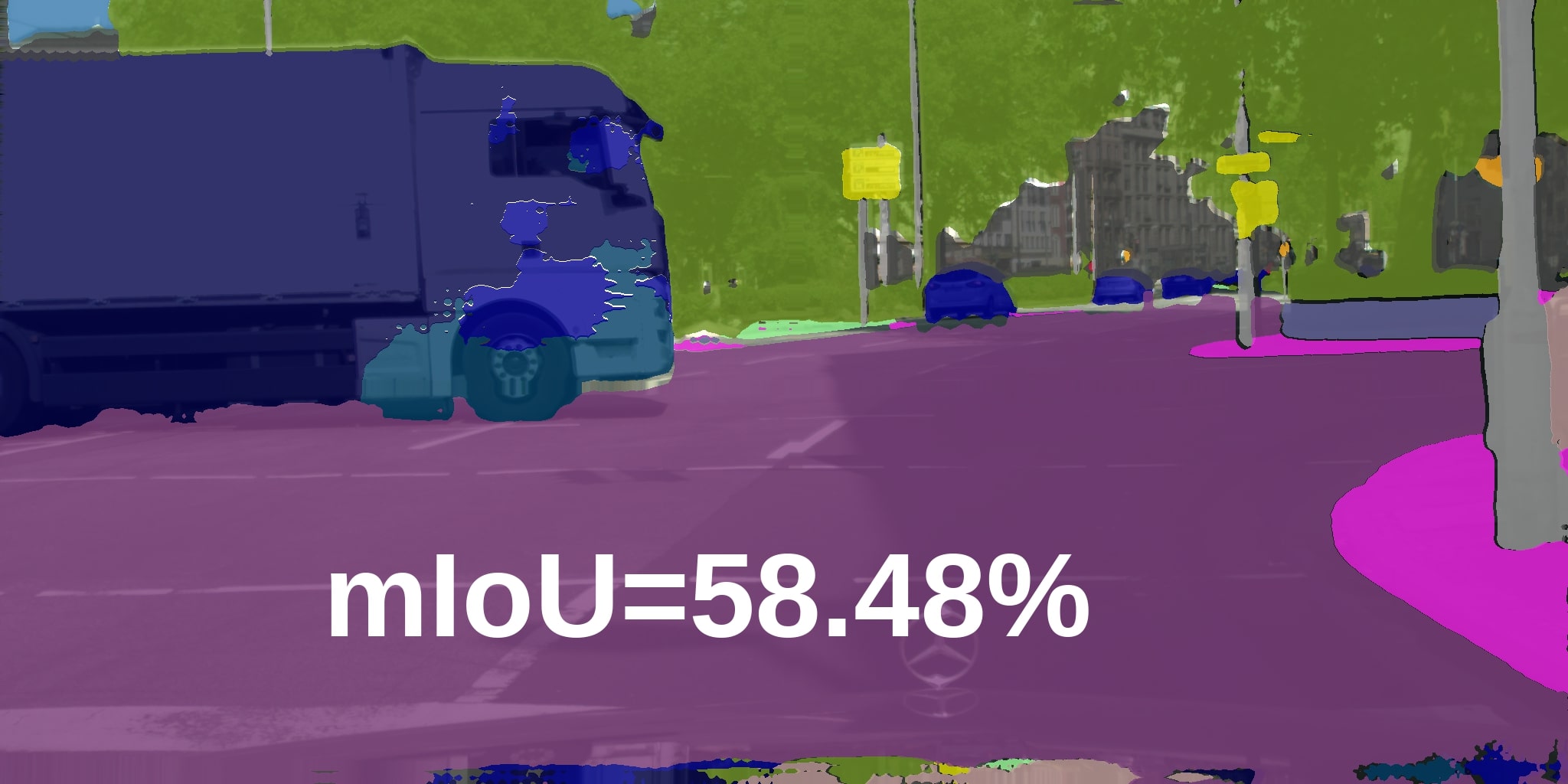} 
    \caption{Ours ($0.2465 \ \mathrm{bpp}$)}
    \label{fig:ours_qualitative}
  \end{subfigure}
  \caption{Qualitative comparison of the \textbf{proposed $\bf{JD}$ approach ("Ours")} against \textbf{Ahuja et al. \cite{ahuja2023neural}} on the Cityscapes dataset \textbf{at high bitrates.}}
\label{fig:cs_high_qualitative}
  \end{figure}
\begin{figure}[t!]
  \begin{subfigure}{0.2455\textwidth}
    \includegraphics[width=\textwidth]{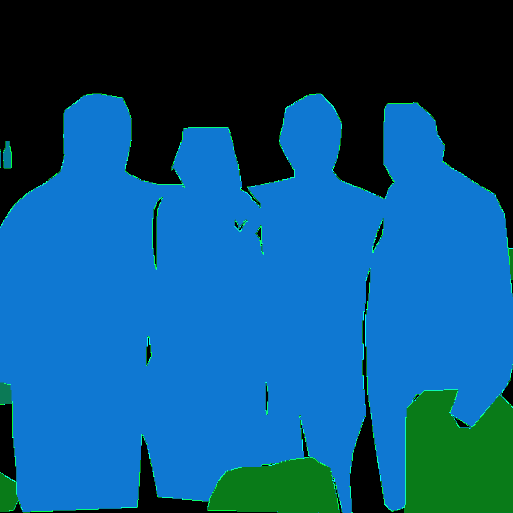} 
    \includegraphics[width=\textwidth]{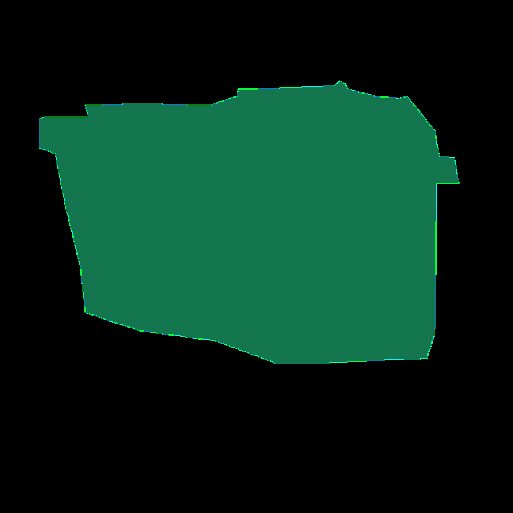} 
    \includegraphics[width=\textwidth]{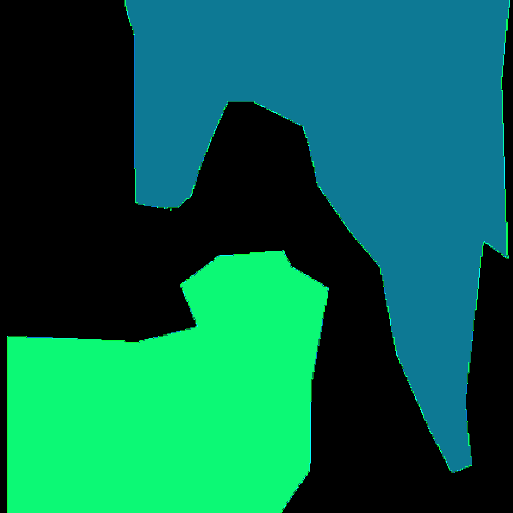}
    \includegraphics[width=\textwidth]{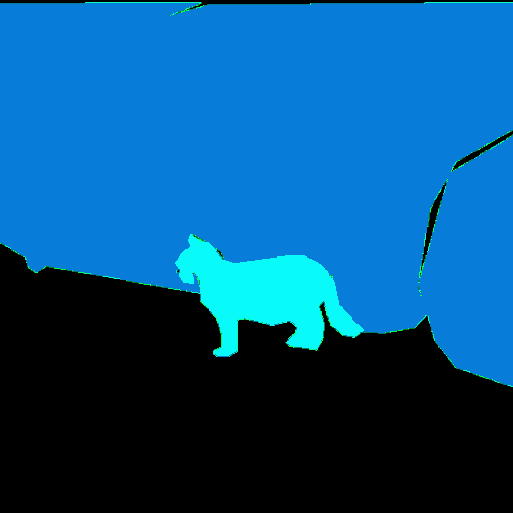}
    \includegraphics[width=\textwidth]{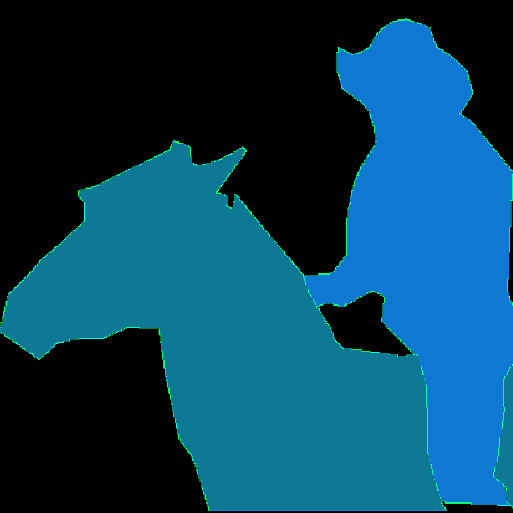}
    \includegraphics[width=\textwidth]{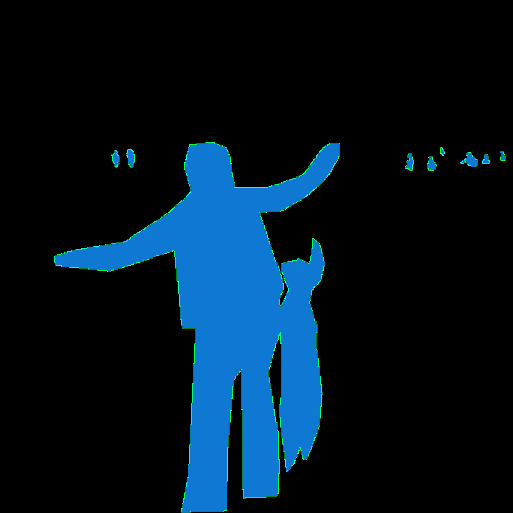} 
    \caption{Ground Truth}
  \end{subfigure}
    \hspace{-0.42em}
  \begin{subfigure}{0.2455\textwidth}
    \includegraphics[width=\textwidth]{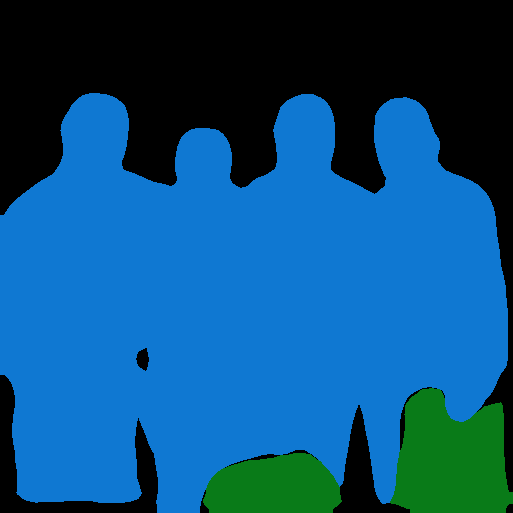} 
    \includegraphics[width=\textwidth]{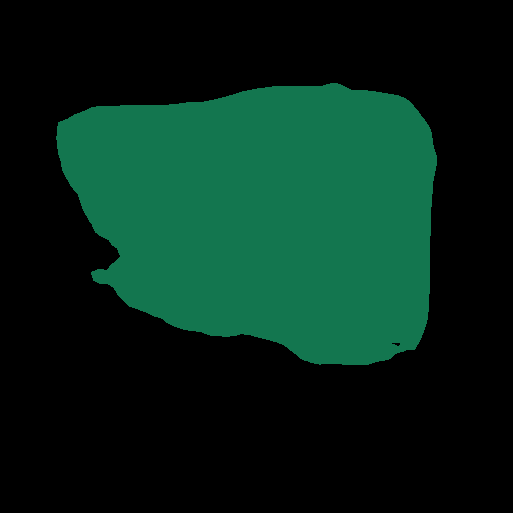} 
    \includegraphics[width=\textwidth]{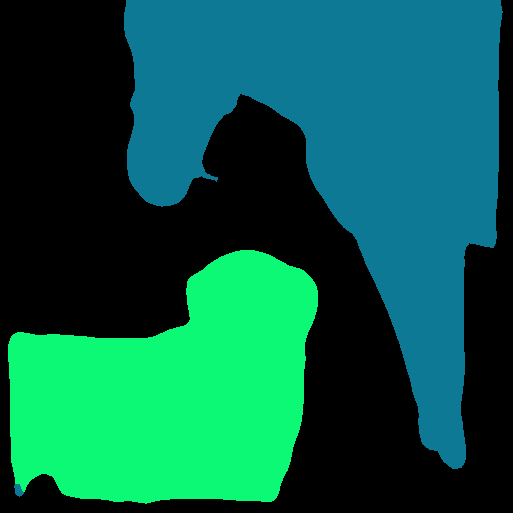}
    \includegraphics[width=\textwidth]{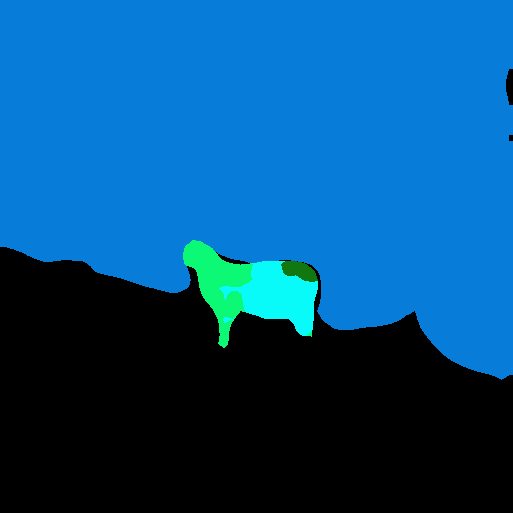}
    \includegraphics[width=\textwidth]{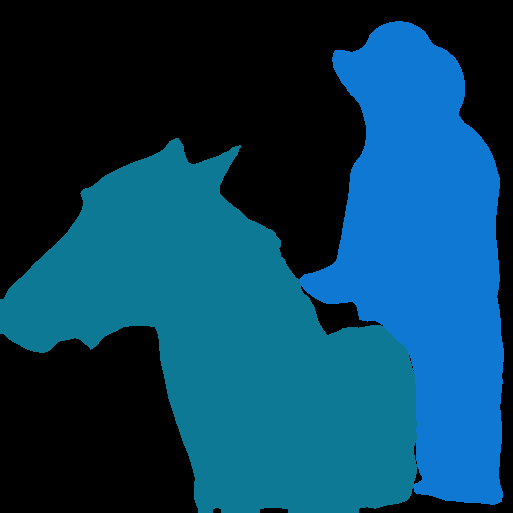}
    \includegraphics[width=\textwidth]{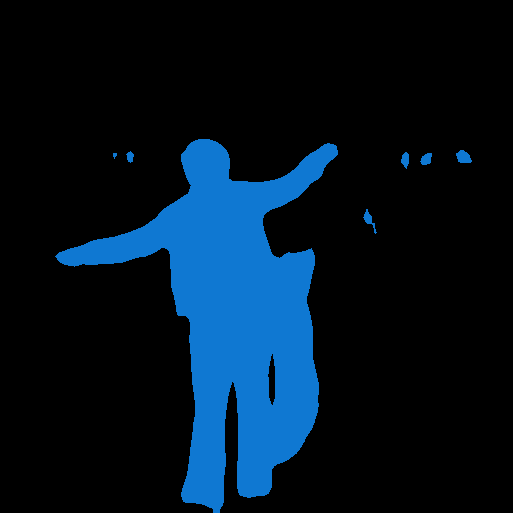} 
    \caption{No Compression \cite{deeplabv3}}
  \end{subfigure}
   \hspace{-0.45em}
  \begin{subfigure}{0.2455\textwidth}
    \includegraphics[width=\textwidth]{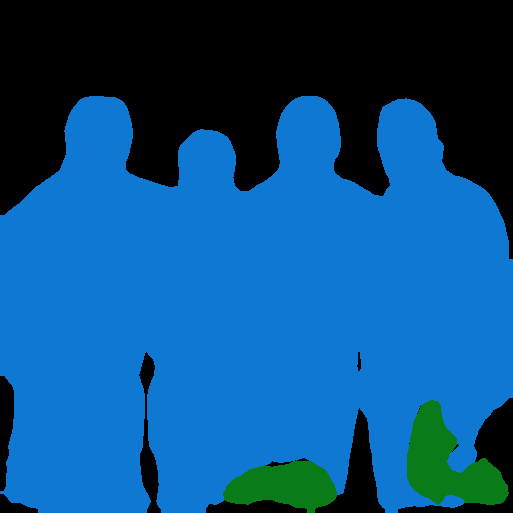} 
    \includegraphics[width=\textwidth]{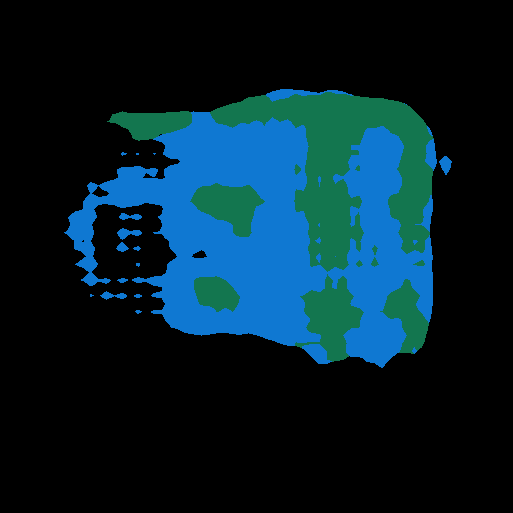} 
    \includegraphics[width=\textwidth]{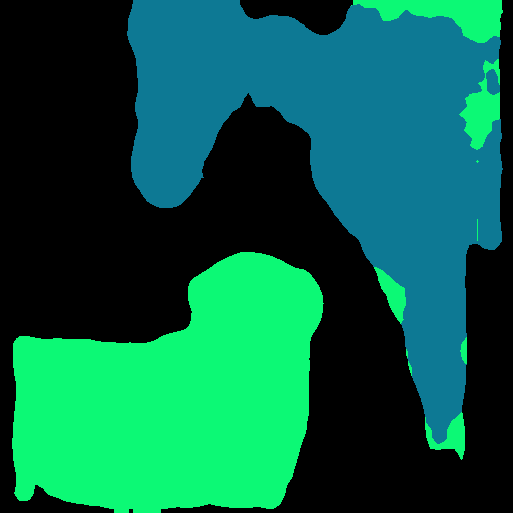}
    \includegraphics[width=\textwidth]{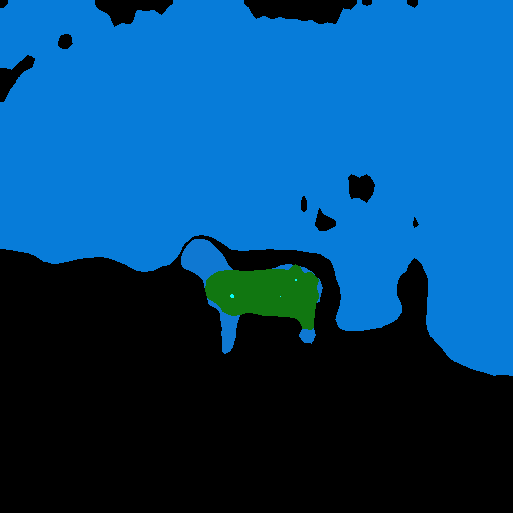}
    \includegraphics[width=\textwidth]{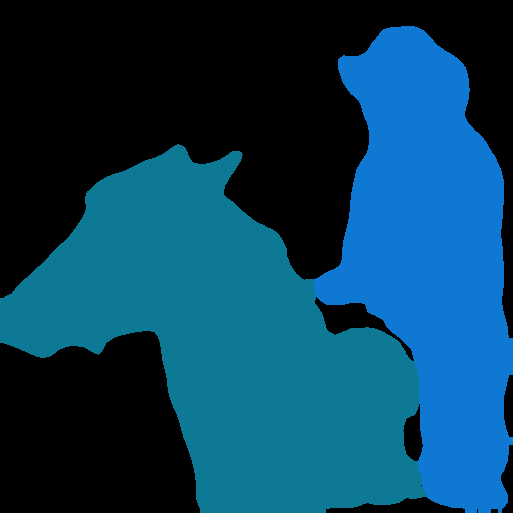}
    \includegraphics[width=\textwidth]{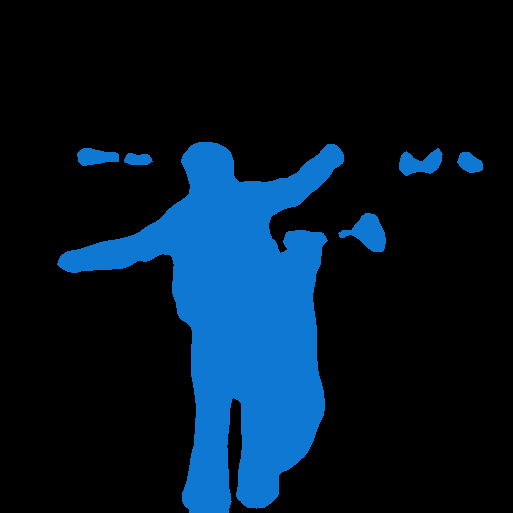} 

    \caption{Ahuja ($0.0724 \ \mathrm{bpp}$)}
    \label{fig:ahuja_qualitative}
  \end{subfigure}%
  \hspace{-0.12em}
  \begin{subfigure}{0.2455\textwidth}
   \includegraphics[width=\textwidth]{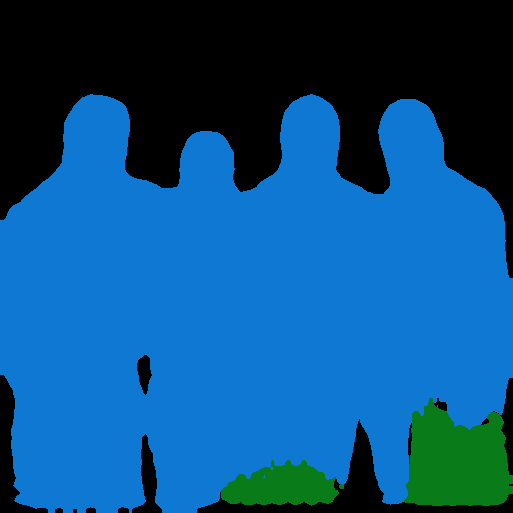} 
    \includegraphics[width=\textwidth]{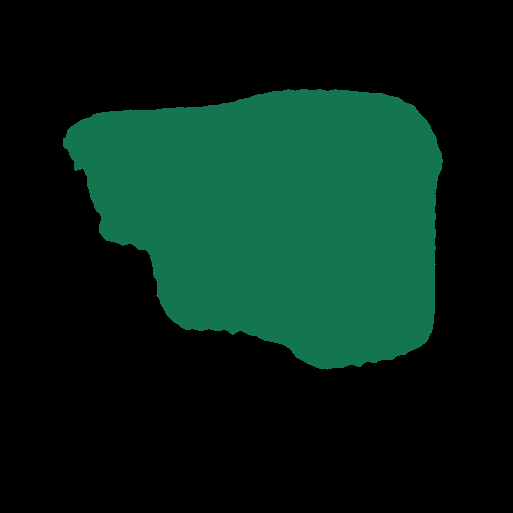} 
    \includegraphics[width=\textwidth]{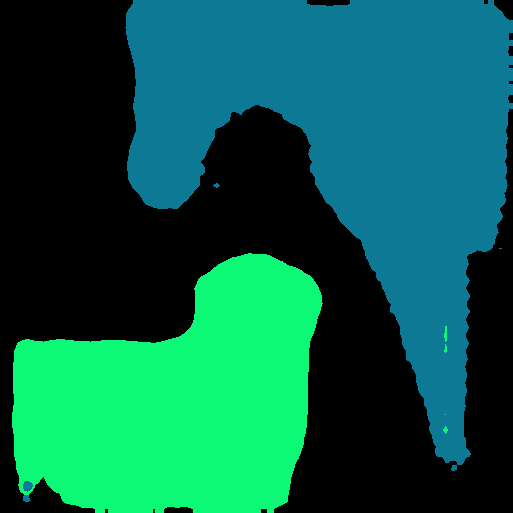}
    \includegraphics[width=\textwidth]{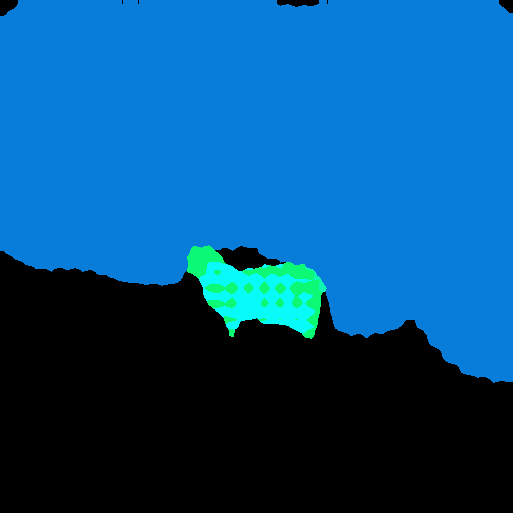}
    \includegraphics[width=\textwidth]{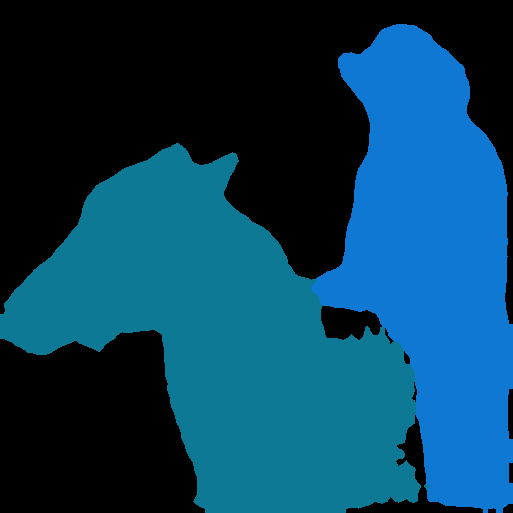}
     \includegraphics[width=\textwidth]{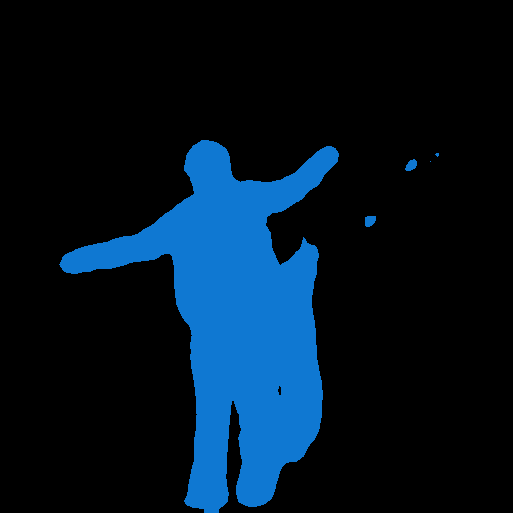} 
    \caption{Ours ($0.0648 \ \mathrm{bpp}$)}
    \label{fig:ours_qualitative}
  \end{subfigure}
  \caption{Qualitative comparison of the \textbf{proposed $\bf{JD}$ approach ("Ours")} against \textbf{Ahuja et al. \cite{ahuja2023neural}} on the COCO dataset \textbf{at low bitrates.}}
  \label{fig:coco_low_qualitative}
  \end{figure}

\begin{figure}[t!]
  \begin{subfigure}{0.2455\textwidth}
    \includegraphics[width=\textwidth]{Figures/Supplementary/COCO/GT/000000049269.png}
    \includegraphics[width=\textwidth]{Figures/Supplementary/COCO/GT/000000172330.png}
    \includegraphics[width=\textwidth]{Figures/Supplementary/COCO/GT/000000032901.png} 
    \includegraphics[width=\textwidth]{Figures/Supplementary/COCO/GT/000000039670.png} 
    \includegraphics[width=\textwidth]{Figures/Supplementary/COCO/GT/000000039914.png} 
    \includegraphics[width=\textwidth]{Figures/Supplementary/COCO/GT/000000183675.png}
    \caption{Ground Truth}
  \end{subfigure}
    \hspace{-0.42em}
  \begin{subfigure}{0.2455\textwidth}
    \includegraphics[width=\textwidth]{Figures/Supplementary/COCO/No_Compression/000000049269.png}
    \includegraphics[width=\textwidth]{Figures/Supplementary/COCO/No_Compression/000000172330.png}
    \includegraphics[width=\textwidth]{Figures/Supplementary/COCO/No_Compression/000000032901.png} 
    \includegraphics[width=\textwidth]{Figures/Supplementary/COCO/No_Compression/000000039670.png} 
    \includegraphics[width=\textwidth]{Figures/Supplementary/COCO/No_Compression/000000039914.png} 
    \includegraphics[width=\textwidth]{Figures/Supplementary/COCO/No_Compression/000000183675.png}
    \caption{No Compression \cite{deeplabv3}}
  \end{subfigure}
   \hspace{-0.45em}
  \begin{subfigure}{0.2455\textwidth}
    \includegraphics[width=\textwidth]{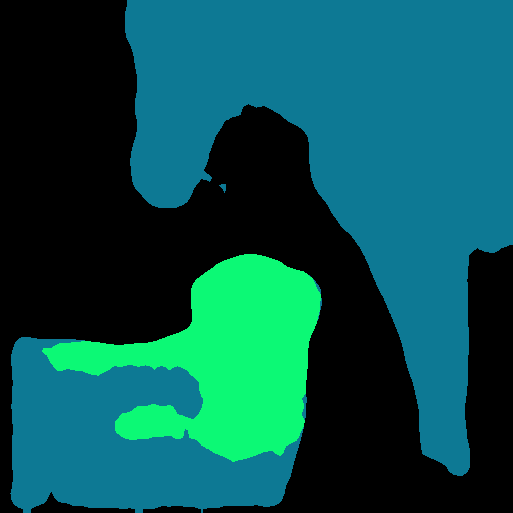}
    \includegraphics[width=\textwidth]{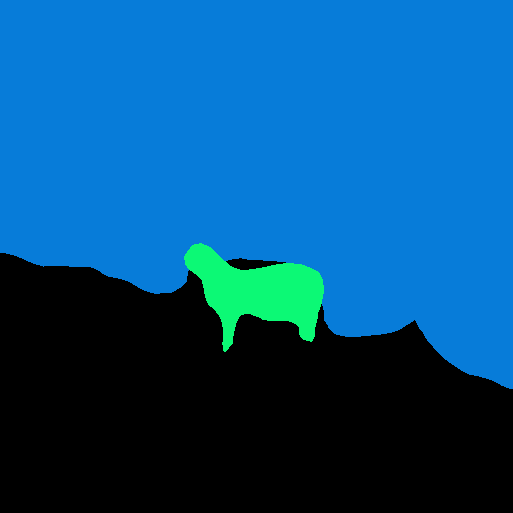}
    \includegraphics[width=\textwidth]{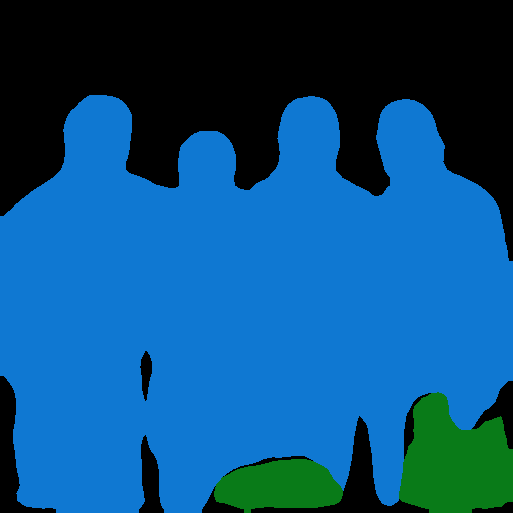} 
    \includegraphics[width=\textwidth]{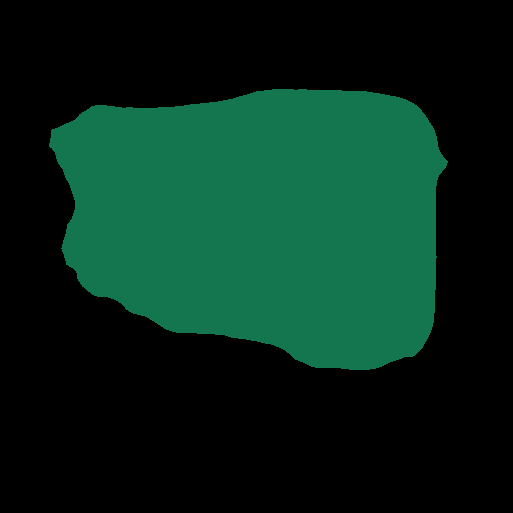} 
    \includegraphics[width=\textwidth]{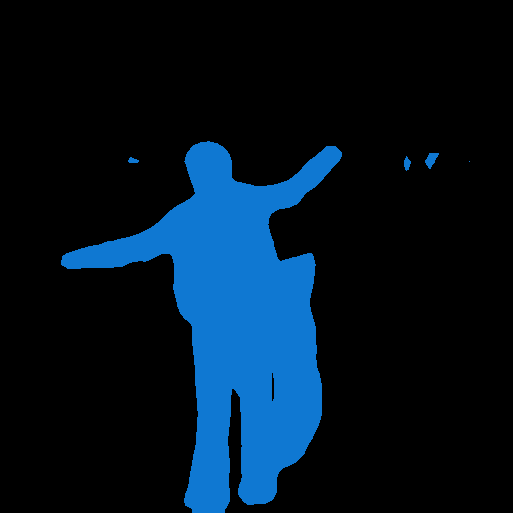} 
    \includegraphics[width=\textwidth]{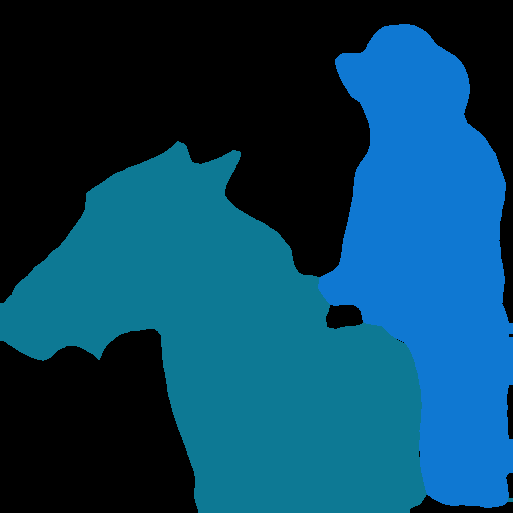}

    \caption{Ahuja ($0.45 \ \mathrm{bpp}$)}
    \label{fig:ahuja_qualitative}
  \end{subfigure}%
  \hspace{-0.12em}
  \begin{subfigure}{0.2455\textwidth}
    \includegraphics[width=\textwidth]{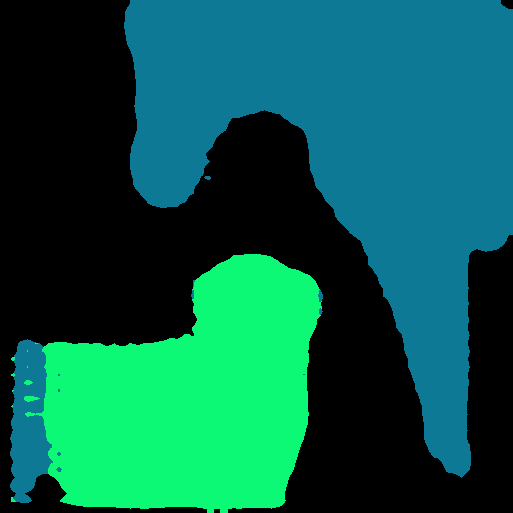}
    \includegraphics[width=\textwidth]{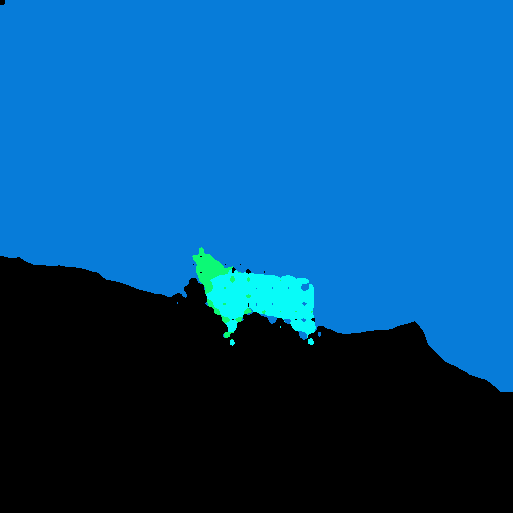}
   \includegraphics[width=\textwidth]{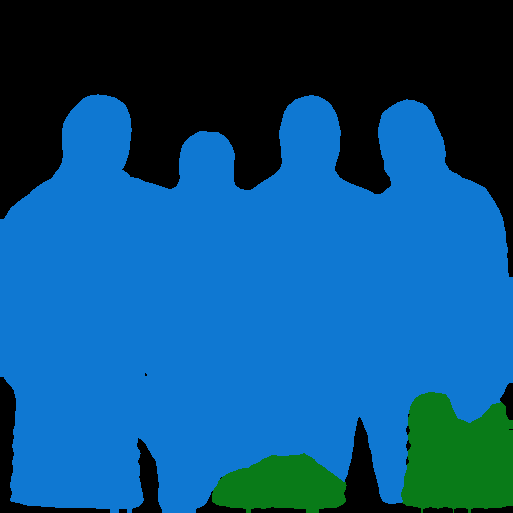} 
    \includegraphics[width=\textwidth]{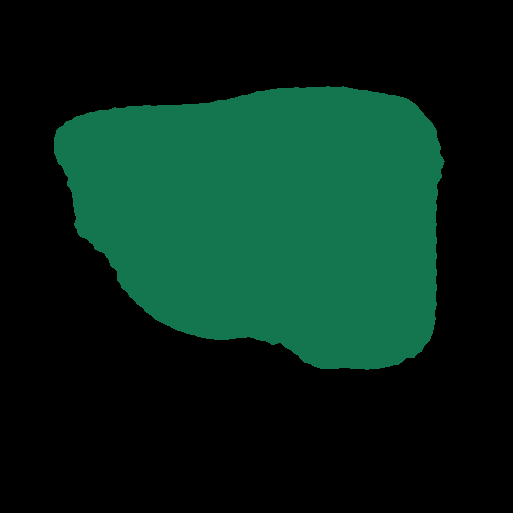} 
    \includegraphics[width=\textwidth]{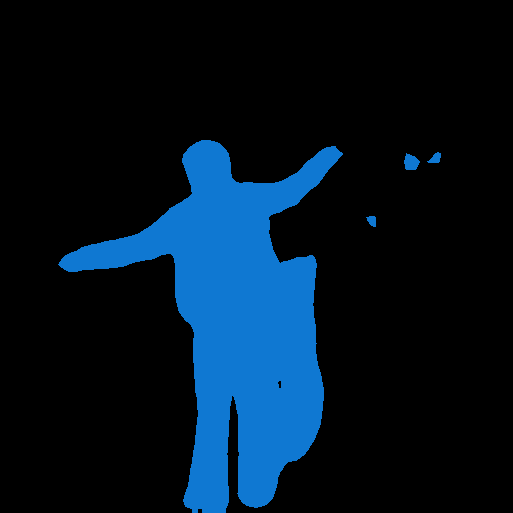} 
    \includegraphics[width=\textwidth]{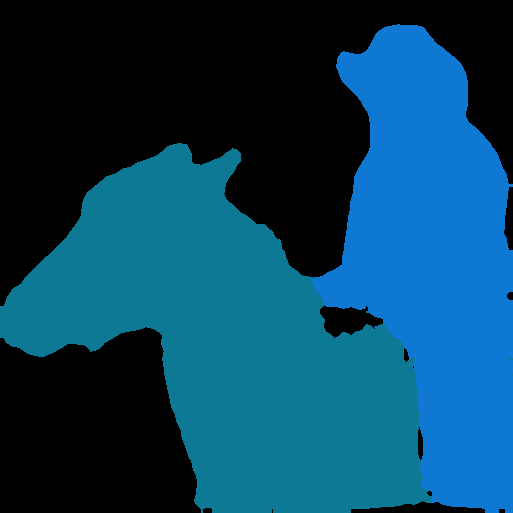}
 
    \caption{Ours ($0.31 \ \mathrm{bpp}$)}
    \label{fig:ours_qualitative}
  \end{subfigure}
  \caption{Qualitative comparison of the \textbf{proposed $\bf{JD}$ approach ("Ours")} against \textbf{Ahuja et al. \cite{ahuja2023neural}} on the COCO dataset \textbf{at high bitrates.}}
  \label{fig:coco_high_qualitative}
  \end{figure}

\section{Additional Results with a \texttt{ResNet-101} Encoder}
\label{section:resnet_101}

In this section, we present additional results with a \texttt{ResNet-101} encoder on both COCO and Cityscapes to better prove the generalizability of our method. Figure \ref{fig:result_comparison_101} shows the rate-distortion (RD) performance of our proposed $\bf{JD}$ against the recent SOTA by Ahuja et al.\ \cite{ahuja2023neural}. Our proposed $\bf{JD}$ outperforms the SOTA approach on both COCO and Cityscapes. Further, it also exceeds the $\texttt{No Compression}$ performance at a bitrate of $0.28 \ \textnormal{bpp}$ on the COCO dataset. Note that we use exactly the same settings (dilation rates and number of repetitions) in $\bf{JD}$ as in the case of the $\texttt{ResNet-50}$ encoder.

\section{Qualitative Results}
\label{section:qualitative}

In this section, we present a qualitative comparison with \texttt{ResNet-50} encoder against current SOTA \cite{ahuja2023neural} by Ahuja et al. \cite{ahuja2023neural} at low and high bitrates on both COCO and Cityscapes. In Figures \ref{fig:cs_low_qualitative} to \ref{fig:coco_high_qualitative}, sample-individual \textit{mIoU} performance is reported, while the \textit{bitrates} of Ahuja et al. \cite{ahuja2023neural} and our proposal are given for the entire respective test set. Note that the bitrates of our method are always chosen to be lower.

Figure \ref{fig:cs_low_qualitative} shows for nine samples (organized in rows) a qualitative comparison of our proposed approach and Ahuja et al. \cite{ahuja2023neural} \textit{at low bitrates on Cityscapes}. Rows 1 to 7 present strong cases for our method. Interestingly, in row 1, we also exceed the \texttt{No Compression} baseline. Finally, rows 8 and 9 show the limitations of our method. In both cases, the mIoU is slightly worse, however, we do not observe missing classes such as \textit{pedestrians}, \textit{signs} and \textit{cars}, which are considered critical for autonomous driving.  

Similarly, Figure \ref{fig:cs_high_qualitative} also illustrates a qualitative comparison, but \textit{at high bitrates on Cityscapes}. Rows 1 to 4 show strong examples for our method, whereas rows 5 to 7 depict limitations with slightly lower mIoU. Importantly, in rows 5 to 7, our approach does not remove important classes. Further, in rows 4 and 7, we also exceed the \texttt{No Compression} baseline.

Figure \ref{fig:coco_low_qualitative} depicts a qualitative comparison of our proposed approach and the method by Ahuja et al. \cite{ahuja2023neural} \textit{at low bitrates on COCO}. We observe that all rows except the last represent the strong cases and our results are far superior than the method by Ahuja et al. \cite{ahuja2023neural}. Even in the last row, our results are comparable but our proposed approach fails to predict the background \texttt{persons} at a very small scale. Note that the approach of Ahuja et al. \cite{ahuja2023neural} also produces only partially correct predictions in the same region.

Figure \ref{fig:coco_high_qualitative} shows a qualitative comparison of our proposed approach and the method by Ahuja et al. \cite{ahuja2023neural} \textit{at high bitrates on COCO}. At higher bitrates, we observe that our proposed approach provides comparable results to Ahuja et al. \cite{ahuja2023neural}. However, in rows 1 and 2, our proposed approach significantly reduces the false predictions of the classes including \texttt{cat} and \texttt{dog}.~Further, it also does not omit the important classes such as \texttt{person} and \texttt{train}.

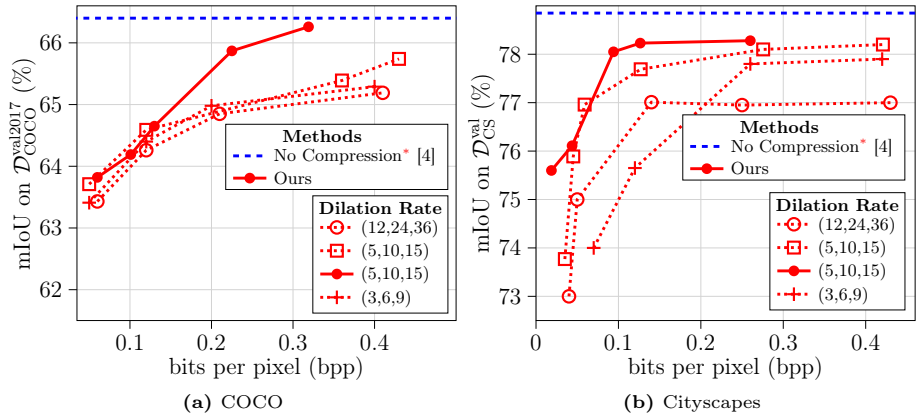
\begin{figure}[t!]
  \begin{subfigure}{0.5\textwidth}
    \centering
    \resizebox{\linewidth}{!}{\begin{tikzpicture}
\definecolor{brightred}{RGB}{238, 75,43}
\definecolor{redbrown}{RGB}{165, 42, 42}
\definecolor{rubyred}{RGB}{224, 17, 95}
\begin{axis}[
legend cell align={left},
legend cell align={left},
legend style={
  fill opacity=1.0,
  draw opacity=1,
  text opacity=1,
  at={(0.5,0.02)},
  anchor=south west,
  draw=white!80!black
},
legend style={
  fill opacity=1.0,
  draw opacity=1,
  text opacity=1,
  at={(0.5,0.02)},
  anchor=south west,
  draw=white!80!black
},
tick align=outside,
tick pos=left,
tick label style={font=\large},
label style={font=\large},
x grid style={white!82.7450980392157!black},
xlabel={bits per pixel (bpp) },
xmajorgrids,
xmin=0.035, xmax=0.5,
xtick style={color=black},
xtick={0,0.1,0.2,0.3,0.4},
xticklabels={0,0.1,0.2,0.3,0.4},
y grid style={white!82.7450980392157!black},
ylabel={\(\displaystyle \textrm{mIoU on $\mathcal{D}_{\mathrm{COCO}}^{\mathrm{val2017}}$}\) (\%)},
ylabel style={yshift=-7pt},
ymajorgrids,
ymin=61.5, ymax=66.6,
ytick style={color=black},
ytick={61,62,63,64,65,66},
yticklabels={
  \(\displaystyle {61}\),
  \(\displaystyle {62}\),
  \(\displaystyle {63}\),
    \(\displaystyle {64}\),
      \(\displaystyle {65}\),
      \(\displaystyle {66}\),
}
]
\addplot [line width=1.5pt, blue, dashed, forget plot]
table {%
0 66.4
1 66.4
};
\label{coco_dilation_no_Compression}
\addplot [line width=1.5pt, color=black,  dotted, draw=none,mark=none]
table {%
0.0095 67.5
};
\label{coco_dilation_fake}
\addplot [line width=1.5pt, red, dotted, mark=+, mark size=3.50, mark options={solid,fill opacity=0,line width=1.1}, forget plot]
table {%
0.05 63.41
0.12 64.4
0.2 64.98
0.4 65.29
};
\label{coco_dilation_k_1};
% \addplot [line width=1.5pt, red, dotted, mark=x, mark size=3.5, mark options={solid,fill opacity=0,line width=1.1}, forget plot]
% table {%
% 0.05 63.11
% 0.08 63.1
% 0.18 63.59
% 0.33 65.08
% 0.65 64.97
% };
% \label{coco_dilation_k_2};
\addplot [line width=1.5pt, red, dotted,mark=square, mark size=3.050, mark options={solid,fill opacity=0,line width=1.1}, forget plot]
table {%
0.05 63.71
0.12 64.59
0.36 65.39
 0.43 65.74
};
\label{coco_dilation_k_3};
\addplot [line width=1.5pt, red, dotted, mark=*, mark size=3.20, mark options={solid,fill opacity=0,line width=1.1}, forget plot]
table {%
0.06 63.43
0.12 64.26
0.21 64.85
0.41 65.19
};
\label{coco_dilation_k_4};
\addplot [line width=1.5pt, red, mark=*, mark size=2.0, mark options={solid}, forget plot]
table {%
%0.0470000505447388 62.6500015258789
0.059999942779541 63.8199996948242
0.10099995136261 64.1900024414062
0.129999995231628 64.6500015258789
0.225000023841858 65.870002746582
0.319000005722046 66.2600021362305
%0.634999990463257 66.4300015258789
};
\label{coco_ablation_ours2}

\end{axis}

%  \node [draw, fill=white, font=\small, align=center] at (rel axis cs: 0.43, 0.165) {
%     \shortstack[l]{\hspace{2.5em} % Adjust the spacing as needed
%       \textbf{Methods} \\
%       \ref*{coco_dilation_no_Compression} No Compression \\
%        \ref*{coco_dilation_k_2} Ours 
%   };
 \node [draw, fill=white, font=\small, align=center] at (rel axis cs: 0.61, 0.52) {
    \shortstack[l]{\hspace{2.5em} % Adjust the spacing as needed
      \textbf{Methods} \\
      \ref*{coco_dilation_no_Compression} No Compression$^{\textcolor{red}{*}}$ \cite{deeplabv3} \\
       \ref*{coco_ablation_ours2} Ours 
    }
  };
\node [draw,fill=white,font=\small] at (rel axis cs: 0.725,0.21){\shortstack[l]{
 \textbf{Dilation Rate}  \\ 
 \ref*{coco_dilation_k_4} (12,24,36) \\   
  \ref*{coco_dilation_k_3} (5,10,15) \\
    \ref*{coco_ablation_ours2} (5,10,15) \\
 \ref*{coco_dilation_k_1} (3,6,9)  
 }};

\end{tikzpicture}}
    \caption{COCO  }
    \label{fig:coco_ablation_dilation}
  \end{subfigure}%
  %\hfill
  %\hspace{0.02em}
  \begin{subfigure}{0.5\textwidth}
    \centering
    \resizebox{\linewidth}{!}{\begin{tikzpicture}
\definecolor{brightred}{RGB}{238, 75,43}
\definecolor{redbrown}{RGB}{165, 42, 42}
\definecolor{rubyred}{RGB}{224, 17, 95}

\begin{axis}[
legend cell align={left},
legend cell align={left},
legend style={
  fill opacity=1.0,
  draw opacity=1,
  text opacity=1,
  at={(0.5,0.02)},
  anchor=south west,
  draw=white!80!black
},
legend style={
  fill opacity=1.0,
  draw opacity=1,
  text opacity=1,
  at={(0.5,0.02)},
  anchor=south west,
  draw=white!80!black
},
tick align=outside,
tick pos=left,
tick label style={font=\large},
label style={font=\large},
x grid style={white!82.7450980392157!black},
xlabel={bits per pixel (bpp) },
xmajorgrids,
xmin=0, xmax=0.46,
xtick style={color=black},
xtick={0,0.1,0.2,0.3,0.4,0.5},
xticklabels={0,0.1,0.2,0.3,0.4,0.5},
y grid style={white!82.7450980392157!black},
ylabel={\(\displaystyle \textrm{mIoU on $\mathcal{D}_{\mathrm{CS}}^{\mathrm{val}}$}\) (\%)},
ylabel style={yshift=-7pt},
ymajorgrids,
ymin=72.5, ymax=79,
ytick style={color=black},
ytick={72,73,74,75,76,77,78},
yticklabels={
   \(\displaystyle {72}\),
  \(\displaystyle {73}\),
  \(\displaystyle {74}\),
  \(\displaystyle {75}\),
  \(\displaystyle {76}\),
  \(\displaystyle {77}\),
  \(\displaystyle {78}\)
}
]
\addplot [line width=1.5pt, blue, dashed, forget plot]
table {%
0 78.85
0.7 78.85
};
\label{dilation_cs_no_Compression}
% \addplot [line width=1.5pt, red, mark=*, mark size=1.5, mark options={solid}, forget plot]
% table {%
% 0.02 72.92
% 0.0436 76.28
% 0.07 77.3
% 0.11 77.67
% 0.28 78.23
% 0.41 78.3
% };
% \label{dilation_ahuja}
\addplot [line width=1.5pt, red, dotted, mark=+, mark size=3.5, mark options={solid,fill opacity=0,line width=1.1}, forget plot]
table {%
0.07 74
0.12 75.65
0.26 77.8
0.42 77.9
};
\label{dilation_cs_k_1}
% \addplot [line width=1.5pt, red, dotted, mark=x, mark size=3.5, mark options={solid,fill opacity=0,line width=1.1}, forget plot]
% table {%
% 0.025 72.66
% 0.05 73.05
% 0.12 77.23
% 0.24 78.08
% 0.42 77.96	
% };
% \label{dilation_cs_k_2}
\addplot  [line width=1.5pt, red, dotted,mark=square, mark size=3.050, mark options={solid,fill opacity=0,line width=1.1}, forget plot]
table {
0.035 73.77
0.0452 75.89
0.059 76.96
0.127 77.69
0.2756 78.1
0.4209 78.2
};
% \addplot [line width=1.5pt, red, mark=*, mark size=2, mark options={solid}, forget plot]
% table {%
% 0.0187 72.595
% 0.04375 76.11
% 0.09405 78.05
% 0.1265 78.23
% 0.26 78.28

% };
\label{dilation_cs_k_3}
\addplot [line width=1.5pt, red, mark=*, mark size=2, mark options={solid}, forget plot]
table {%
0.0187 75.6
0.04375 76.11
0.09405 78.05
0.1265 78.23
0.26 78.28
};
\label{ours_dilation_cs}
\addplot  [line width=1.5pt, red, dotted, mark=*, mark size=3.200, mark options={solid,fill opacity=0,line width=1.1}, forget plot]
table {%
%0.015 70.73
0.04 73
0.05 75
0.14 77.01
0.25 76.95
0.43 77
};
\label{dilation_cs_k_4}
 \addplot [line width=1.5pt, color=black,  dotted, draw=none,mark=none]
 table {%
 0.15 77.5
 };
\label{dilation_cs_fake}
\end{axis}

 \node [draw, fill=white, font=\small, align=center] at (rel axis cs: 0.68, 0.55) {
    \shortstack[l]{\hspace{2.5em} % Adjust the spacing as needed
      \textbf{Methods} \\
      \ref*{dilation_cs_no_Compression} No Compression$^{\textcolor{red}{*}}$ \cite{deeplabv3} \\
       \ref*{ours_dilation_cs} Ours 
    }
  };
\node [draw,fill=white,font=\small] at (rel axis cs: 0.795,0.22){\shortstack[l]{
 \textbf{Dilation Rate}  \\ 
 \ref*{dilation_cs_k_4} (12,24,36) \\   
 \ref*{dilation_cs_k_3} (5,10,15) \\
  \ref*{ours_dilation_cs} (5,10,15) \\
 \ref*{dilation_cs_k_1} (3,6,9)  
 }};

\end{tikzpicture}}
    \caption{Cityscapes }
    \label{fig:cityscapes_ablation_dilation}
  \end{subfigure}
  \caption{\bf{Ablation study on ASPP dilation rates of our proposed $\bf{JD}$ approach} \textnormal{on (a) $\mathcal{D}_{\mathrm{COCO}}^{\mathrm{val2017}}$ and (b) $\mathcal{D}_{\mathrm{CS}}^{\mathrm{val}}$.~Each marker type shows a different dilation rate in the ASPP dilated convolutional subblocks of $\bf{JD}$.~Dilation rates, e.g., ($d=5$, $d=10$, $d=15$), refer to the spacing between the kernel elements in the respective dilated convolutional subblock of ASPP. All the curves---except "Ours"---are without over-parameterization in the ASPP block. This is why there is another ($d=5$, $d=10$, $d=15$) configuration. Note that all methods use $F=256$ (COCO) and $F=512$ (Cityscapes) in the ASPP block, respectively 
  } }
  \label{fig:ablation_dilation}
\end{figure}

\section{Dilation Rate Ablation}
\label{section:dilation}
Figure \ref{fig:ablation_dilation} represents the dilation rate $d$ ablation study in the ASPP dilated convolutional subblocks of $\bf{JD}$.~Each marker type shows a different dilation rate.~Dilation enables the ASPP block to capture information at multiple scales, while achieving lower computational complexity \cite{deeplabv3}.~We ablate over three dilation rates for both datasets. All curves, except "Ours", are without over-parameterization in the ASPP block. As shown in Figures \ref{fig:coco_ablation_dilation} and \ref{fig:cityscapes_ablation_dilation}, we empirically found that the dilation rates ($d=5$, $d=10$, $d=15$) in $\bf{JD}$ produce the best RD performance on both datasets.

%\clearpage  % TODO REVIEW/FINAL: This \clearpage needs to be removed from both review and camera-ready versions.

\bibliographystyle{splncs04}
\bibliography{main}